\documentclass[aps,showpacs,amsmath,amssymb,twocolumn,prapplied,longbibliography,superscriptaddress,notitlepage,floatfix,10pt]{revtex4-2}
\usepackage[dvips]{graphicx}
\usepackage{bm,amsmath,amssymb,amsthm,mathrsfs,amsfonts,dsfont}
\usepackage{epsfig}
\usepackage{braket}
\usepackage{enumerate}
\usepackage{color}
\usepackage{algorithm}
\usepackage{algorithmic}
\usepackage{comment}
\usepackage{appendix}
\usepackage{here}
\usepackage{tabularx}
\usepackage{dcolumn}
\usepackage[caption=false]{subfig}
\usepackage{tikz}
\usetikzlibrary{quantikz}
\usetikzlibrary{positioning}
\usetikzlibrary{arrows}
\usetikzlibrary{arrows.meta}
\usetikzlibrary{shapes.arrows, calc}
\usepackage{adjustbox}
\usepackage{xcolor}
\usepackage[hidelinks]{hyperref}
\usepackage{bookmark}
\tikzset{
    MyLongArrow/.style args={#1 -- #2}{
        insert path={let \p1=($(#1)-(#2)$) in}, 
        single arrow, draw=black, minimum width=8mm, minimum height={veclen(\x1,\y1)}, inner sep=1mm, single arrow head extend=1mm, double arrow head extend=1mm
    }
}

\begin{document}

\title{Resource-efficient Generalized Quantum Subspace Expansion}

\author{Bo Yang}
\email{bo.yang@lip6.fr}
\affiliation{LIP6, Sorbonne Université, 4 Place Jussieu, Paris 75005, France}
\affiliation{Graduate School of Information Science and Technology, The University of Tokyo, 7-3-1 Hongo, Bunkyo-ku, Tokyo 113-8656, Japan}

\author{Nobuyuki Yoshioka}
\affiliation{International Center for Elementary Particle Physics, The University of Tokyo, 7-3-1 Hongo, Bunkyo-ku, Tokyo, 113-0033, Japan}
\affiliation{Theoretical Quantum Physics Laboratory, RIKEN Cluster for Pioneering Research (CPR), Wako-shi, Saitama 351-0198, Japan}
\affiliation{JST, PRESTO, 4-1-8 Honcho, Kawaguchi, Saitama, 332-0012, Japan}

\author{Hiroyuki Harada}
\affiliation{Department of Applied Physics and Physico-Informatics, Keio University, Hiyoshi 3-14-1, Kohoku, Yokohama 223-8522, Japan}

\author{Shigeo Hakkaku}
\affiliation{NTT Computer and Data Science Laboratories, NTT Corporation, 3-9-11 Midori-cho, Musashino-shi, Tokyo 180-8585, Japan}

\author{Yuuki Tokunaga}
\affiliation{NTT Computer and Data Science Laboratories, NTT Corporation, 3-9-11 Midori-cho, Musashino-shi, Tokyo 180-8585, Japan}

\author{Hideaki Hakoshima}
\affiliation{Graduate School of Engineering Science, Osaka University, 1-3 Machikaneyama, Toyonaka, Osaka 560-8531, Japan}
\affiliation{Center for Quantum Information and Quantum Biology, Osaka University, 1-2 Machikaneyama, Toyonaka, Osaka 560-0043, Japan}

\author{Kaoru Yamamoto}
\email{kaoru.yamamoto@ntt.com}
\affiliation{NTT Computer and Data Science Laboratories, NTT Corporation, 3-9-11 Midori-cho, Musashino-shi, Tokyo 180-8585, Japan}

\author{Suguru Endo}
\email{suguru.endou@ntt.com}
\affiliation{NTT Computer and Data Science Laboratories, NTT Corporation, 3-9-11 Midori-cho, Musashino-shi, Tokyo 180-8585, Japan}
\affiliation{JST, PRESTO, 4-1-8 Honcho, Kawaguchi, Saitama, 332-0012, Japan}

\begin{abstract}
Realizing practical quantum computing requires overcoming a number of computation errors and the limitation of device size, which have intensively been tackled by quantum error mitigation (QEM) these days.
As a unified approach of noise-agnostic QEM, generalized quantum subspace expansion (GSE) has lately been proposed to be remarkably robust against stochastic and coherent errors, integrating quantum subspace expansion and virtual state purification.
However, the requirement in GSE to perform entangled measurements between copies of the quantum states remains a significant drawback under the current situation of quantum devices with a restricted number of qubits and their connectivity.
In this work, we propose ``Dual-GSE'', a resource-efficient implementation of GSE to circumvent this overhead by constructing an ansatz of error-mitigated quantum states via dual-state purification without state copies.
Remarkably, the proposed method can further simulate larger quantum systems beyond the size of available quantum hardware, achieved by a suitable ansatz construction inspired by the divide-and-conquer strategy that classically reintroduces the effect of entanglement.
While classically forging the entanglement comes with additional cost, the total sampling overhead can be notably reduced by reusing the same Pauli expectation values among divided-and-conquered subsystems.
We comprehensively analyze the advantages and overhead of Dual-GSE and perform numerical simulations of the eight-qubit transverse-field Ising model under various setups.
Our results demonstrate that Dual-GSE estimates the ground state energy with high accuracy under gate noise with low mitigation overhead and practical sampling cost.
\end{abstract}

\maketitle

\section{Introduction \label{sec:introduction}}

\begin{figure*}[htbp]
    \centering
    \includegraphics[width=1.0\textwidth]{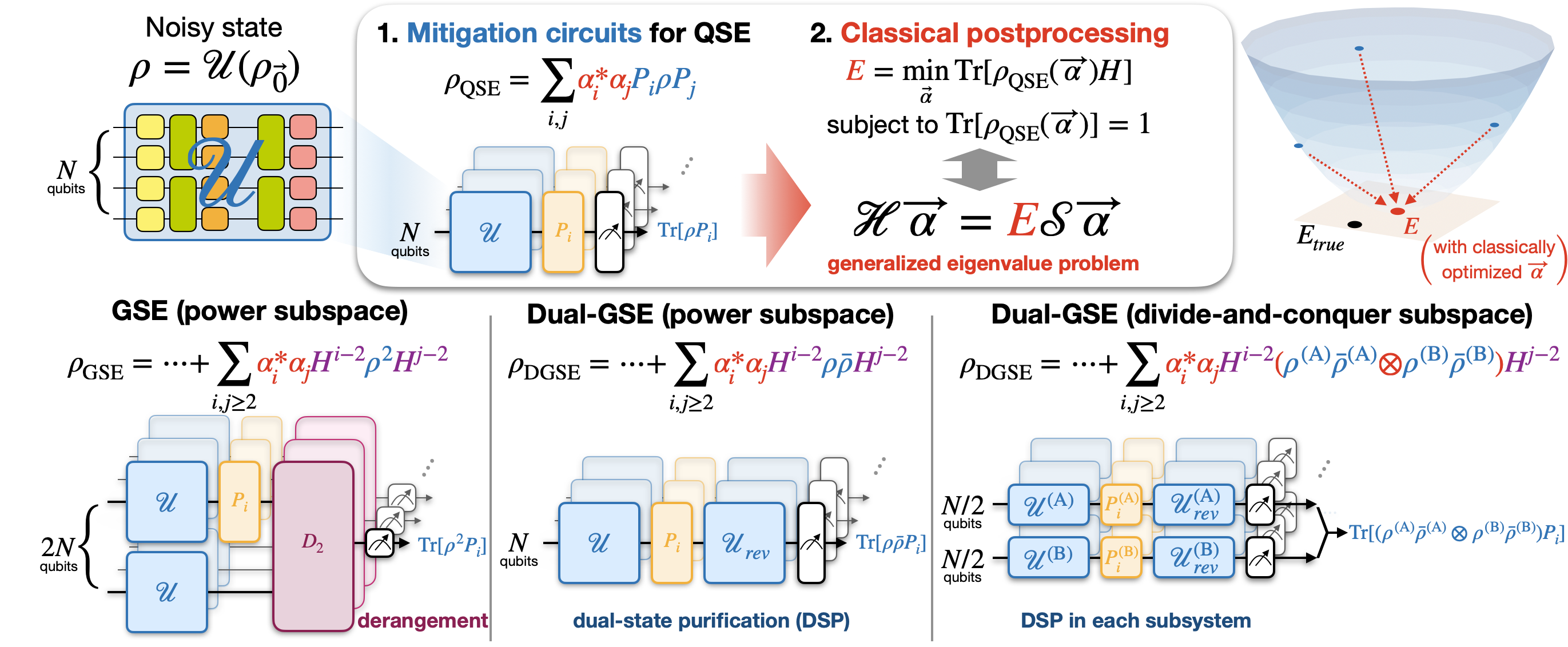}
    \caption{
        Schematic illustration of QSE, GSE, and Dual-GSE. 
        The idea of QSE is to expand the quantum search space classically after obtaining a noisy (optimized) quantum ansatz $\rho$.
        The typical QSE ansatz $\rho_{\mathrm{QSE}}$ is a linear combination of subspaces expanded by Pauli operators.
        The mitigation process is formulated as a constrained optimization problem to find an optimal set of classical parameters $\vec{\alpha}$ for the linear combination of the data retrieved from quantum devices.
        In particular, (step 1) the expectation values of the expanded Pauli subspace are first obtained from quantum devices, and (step 2) then $\vec{\alpha}$ is calculated through a generalized eigenvalue problem $\mathcal{H} \vec{\alpha} = E \mathcal{S} \vec{\alpha}$.
        GSE and Dual-GSE also follow the same workflow as QSE with different quantum ansatz constructions.
        In the power subspace method of the GSE, the mitigation circuits compute powers of noisy ansatz, which requires multiple state copies. 
        Note that we do not show an additional ancilla qubit for conciseness.
        In the power subspace method of our Dual-GSE, the mitigation circuits compute powers of noisy ansatz without state copies using dual-state purification (DSP).
        Remarkably, the divide-and-conquer subspace of Dual-GSE further reduces the required qubits by dividing the whole system into smaller subsystems.
    }
    \label{fig:comparison-qse}
\end{figure*}

Quantum devices are advancing rapidly~\cite{Arute2019-hj, Kim2023, Madsen2022, Zhong2020-bf, Yang2021}, though they remain constrained by size and limited fault tolerance. 
With quantum error correction still out of practical reach, a key challenge is to determine whether such noisy devices offer a computational advantage over classical systems and how to harness their capabilities effectively.
To address this, hardware-efficient error-reduction techniques collectively known as quantum error mitigation (QEM) have been extensively explored~\cite{Endo2021-ku, GiurgicaTiron2020, Strikis2021, Czarnik2021-em, Temme2017-vo, Endo2018-zg, vandenBerg2023, Mari2021, Liu2019-vp, McArdle2019-bj, cai2021resource, Kandala2019-ze, Yang2022, Nation2021, Cai2023}. 
Trading this accuracy for increased sampling costs, QEM methods aim to recover unbiased expectation values by applying additional operations and by postprocessing measurement outcomes~\cite{Takagi2022-fu, Takagi2022-un, Tsubouchi2022, Quek2022}.

The QEM methods can be broadly classified based on whether they are reliant on the noise model.
As represented by probabilistic error cancellation~\cite{Temme2017-vo, vandenBerg2023, Mari2021, Endo2018-zg}, which virtually inverts gate errors based on the noise profile learned in advance~\cite{erhard2019characterizing, Nielsen2021gatesettomography}, the noise characterization in those methods tends to be operationally costly, and time-dependent noise drift would worsen the results.
On the other hand, the other category of ``noise-agnostic'' QEM does not require a noise profile for its QEM procedure.
This includes purification-based methods~\cite{Koczor2021, Huggins2021, Czarnik2021-qe, Huo2022, hakoshima2023localized, ohkura2023leveraging} and quantum subspace expansion (QSE)~\cite{McClean2017-jx, McClean2020, Takeshita2020-ec, Yoshioka2022-vq}.

Namely, the purification-based methods are well-suited for mitigating stochastic errors such as decoherence and depolarizing noise by amplifying the dominant eigenvector of an erroneous state $\rho$ through virtually preparing $\rho^{n}/\operatorname{Tr}\left[\rho^{n}\right], \left(n\geq 2\right)$ using redundant copies of $\rho$.
This approach is initially proposed as exponential suppression by derangement (ESD) and virtual distillation (VD)~\cite{Koczor2021, Huggins2021}.
In contrast, the QSE-based methods effectively suppress coherent errors that may derive from erroneous pulse controls or insufficient optimization of parameters in the variational quantum eigensolver (VQE)~\cite{Peruzzo2014-kp, Farhi2014, Cerezo2021, Tilly2022}.
This is accomplished by constructing a quantum-classical hybrid wavefunction ansatz $|\psi_{\mathrm{hyb}}\rangle=\sum_{i} c_{i} |\psi_{i}\rangle, c_{i} \in \mathbb{C}$ with a set of quantum states $\{|\psi_{i}\rangle\}_{i}$ and parameters $\{c_{i}\}_{i}$ to be classically optimized.
These two approaches are complementary: one is ineffective against the errors the other targets.

To handle this issue, generalized quantum subspace expansion (GSE)~\cite{Yoshioka2022-gq} has recently been proposed as a unified framework of these two approaches.
The remarkably general QSE-based ansatz provided by GSE, covering the purified state through ESD/VD as a special case, allows for mitigating both stochastic and coherent errors.
However, GSE straightforwardly inherits a significant drawback of ESD/VD: the requirements of copies of quantum states and entangled measurements among them.
This would be a crucial overhead against implementing GSE on the current and near-future quantum devices.

To address this device-size overhead, we propose ``Dual-GSE'', a resource-efficient GSE method using fewer or no copies of quantum states, inspired by dual-state purification (DSP)~\cite{Huo2022} and the ``divide-and-conquer'' strategy. 
DSP computes the overlap $\operatorname{Tr}\left[\bar{\rho}\rho\right]$ without redundant state copies by successively uncomputing $\rho$, where $\bar{\rho}$ is the dual state approximating $\rho$.
A unified framework~\cite{cai2021resource} extends this approach to higher-degree overlaps $\operatorname{Tr}\left[(\bar{\rho}\rho)^{n}\right],~\left(n\geq 2\right)$ with fewer copies, by integrating DSP and ESD/VD.
We adapt these methods to the GSE framework and detail two classes of state-copy-free subspace constructions: the power (Krylov) subspace, formed by powers of noisy states, and the fault subspace, formed by noise-amplified states.

To further reduce device-size overhead, we adopt a divide-and-conquer strategy that enables Dual-GSE to compute error-mitigated expectation values for systems beyond the size of available hardware. 
We partition the target system into subsystems and construct a global ansatz as a linear combination of tensored separable subsystems, similar to classical entanglement forging~\cite{eddins2022doubling, Fujii2022, Yuan2021}.
Leveraging the synergy between DSP and divide-and-conquer to minimize qubits, we introduce a divide-and-conquer subspace as a variation of the Krylov-inspired power subspace.
All these methods contribute to enhancing Dual-GSE’s resource efficiency in device-size overhead. 
Figure~\ref{fig:comparison-qse} schematically compares the workflows of QSE, GSE, and Dual-GSE (with or without divide-and-conquer).

We present various numerical simulations of Dual-GSE along with analytical assessments of its overheads.
We estimate the ground state energy of the eight-qubit transverse-field Ising model by VQE ansatz, considering different subspaces, noise levels, and shot counts, showing Dual-GSE works effectively in these setups.
Notably, we observe that Dual-GSE is resource-efficient not only in device-size overhead but also in the sampling cost of the whole process.
This comes from the following two major factors when using DSP and the divide-and-conquer strategy.

First, DSP can reduce the gate overhead of the mitigation process from ESD/VD by avoiding entangled operations between state copies. 
While doubling the circuit depth, DSP consumes much fewer erroneous QEM gates than ESD to compute $\operatorname{Tr}\left[\rho^{2}\right]$ with the same number of noisy ansatz gates.
This advantage is more pronounced under constraints such as linear nearest-neighbor connectivity.
Moreover, the lower QEM overhead of DSP results in a higher post-selection probability, helping moderate variance in normalizing the error-mitigated expectation value.

Second, the divide-and-conquer strategy reduces the rounds of circuit executions by reusing fewer patterns of measurement results in smaller subsystems.
Notably, our numerical simulation suggests that the problem-inspired subsystem division effectively exploits the problem’s structure and achieves high estimation accuracy and precision with a smaller search space, keeping the total sampling cost competitive with full-system simulation.
Besides, smaller subsystems require fewer gates and shallower circuits that further reduce noise-induced sampling costs as well.

The following parts of this paper are organized as follows. 
First, we review the previous works. i.e., purification-based quantum error mitigation methods and the GSE method (Sec.~\ref{sec:preliminaries}). 
Second, we introduce our Dual-GSE framework (Sec.~\ref{sec:dgse}). 
Then, we demonstrate our method via numerical simulations (Sec.~\ref{sec:numerical_simulaion}) and conclude our paper (Sec.~\ref{sec:conclusion}).
In the Appendix sections, we comprehensively analyze in detail the performance and overhead of Dual-GSE under further different setups.

\begin{figure*}[htbp]
    \subfloat[\label{fig:qc_derangement_a}]{
        \begin{adjustbox}{width=0.35\textwidth}\begin{quantikz}
                            & \lstick{$|+\rangle$} & \ctrl{1}                & \ctrl{1} & \meter{$|+\rangle\langle+|$} \\
            \lstick[4]{$n$} & \lstick{$\rho$}      & \gate[4, nwires=3]{D_n} & \gate{O} & \qw       \\
                            & \lstick{$\rho$}      &                         & \qw      & \qw       \\
                            & \vdots               &                         &          &           \\
                            & \lstick{$\rho$}      &                         & \qw      & \qw 
        \end{quantikz}\end{adjustbox}
    }
    \subfloat[\label{fig:qc_derangement_b}]{
        \begin{adjustbox}{width=0.32\textwidth}
            \begin{quantikz}
                \lstick{$|+\rangle$} & \ctrl{3} & \ctrl{2} & \ctrl{1} & \meter{$|+\rangle\langle+|$} \\
                \lstick{$\rho_1$}    & \qw \gategroup[3,steps=2,style={color=red, dashed}, label style={label position=below,anchor=north,yshift=-0.2cm,color=red}]{$D_{3}$} & \swap{1} & \gate{O} & \qw       \\
                \lstick{$\rho_2$}    & \swap{1} & \targX{} & \qw      & \qw       \\
                \lstick{$\rho_3$}    & \targX{} & \qw      & \qw      & \qw 
            \end{quantikz}
        \end{adjustbox}
    }
    \subfloat[\label{fig:qc_derangement_c}]{
        \begin{adjustbox}{width=0.32\textwidth}   
            \tikzset{every picture/.style={line width=1pt}} 
            \begin{tikzpicture}[x=0.75pt,y=0.75pt,yscale=-1,xscale=1,baseline=-120]
                
                \draw  [draw opacity=0] (265,180) .. controls (265,180) and (265,180) .. (265,180) .. controls (229.1,180) and (200,148.66) .. (200,110) .. controls (200,71.34) and (229.1,40) .. (265,40) -- (265,110) -- cycle ; \draw   (265,180) .. controls (265,180) and (265,180) .. (265,180) .. controls (229.1,180) and (200,148.66) .. (200,110) .. controls (200,71.34) and (229.1,40) .. (265,40) ;  
                \draw  [draw opacity=0] (265,160) .. controls (265,160) and (265,160) .. (265,160) .. controls (265,160) and (265,160) .. (265,160) .. controls (253.95,160) and (245,151.05) .. (245,140) .. controls (245,128.95) and (253.95,120) .. (265,120) -- (265,140) -- cycle ; \draw   (265,160) .. controls (265,160) and (265,160) .. (265,160) .. controls (265,160) and (265,160) .. (265,160) .. controls (253.95,160) and (245,151.05) .. (245,140) .. controls (245,128.95) and (253.95,120) .. (265,120) ;  
                \draw  [draw opacity=0] (395,120) .. controls (395,120) and (395,120) .. (395,120) .. controls (395,120) and (395,120) .. (395,120) .. controls (406.05,120) and (415,128.95) .. (415,140) .. controls (415,151.05) and (406.05,160) .. (395,160) -- (395,140) -- cycle ; \draw   (395,120) .. controls (395,120) and (395,120) .. (395,120) .. controls (395,120) and (395,120) .. (395,120) .. controls (406.05,120) and (415,128.95) .. (415,140) .. controls (415,151.05) and (406.05,160) .. (395,160) ;  
                \draw   (265,111) .. controls (265,107.69) and (267.69,105) .. (271,105) -- (289,105) .. controls (292.31,105) and (295,107.69) .. (295,111) -- (295,129) .. controls (295,132.31) and (292.31,135) .. (289,135) -- (271,135) .. controls (267.69,135) and (265,132.31) .. (265,129) -- cycle ;
                \draw   (265,71) .. controls (265,67.69) and (267.69,65) .. (271,65) -- (289,65) .. controls (292.31,65) and (295,67.69) .. (295,71) -- (295,89) .. controls (295,92.31) and (292.31,95) .. (289,95) -- (271,95) .. controls (267.69,95) and (265,92.31) .. (265,89) -- cycle ;
                \draw   (265,31) .. controls (265,27.69) and (267.69,25) .. (271,25) -- (289,25) .. controls (292.31,25) and (295,27.69) .. (295,31) -- (295,49) .. controls (295,52.31) and (292.31,55) .. (289,55) -- (271,55) .. controls (267.69,55) and (265,52.31) .. (265,49) -- cycle ;
                \draw    (295,40) .. controls (336,40) and (325,80) .. (365,80) ;
                \draw    (295,120) .. controls (336,120) and (325,40) .. (365,40) ;
                \draw    (295,80) .. controls (336,80) and (325,120) .. (365,120) ;
                \draw  [color={rgb, 255:red, 255; green, 0; blue, 0 }  ,draw opacity=1 ][dash pattern={on 4.5pt off 4.5pt}] (305,20) -- (355,20) -- (355,130) -- (305,130) -- cycle ;
                \draw   (365,31) .. controls (365,27.69) and (367.69,25) .. (371,25) -- (389,25) .. controls (392.31,25) and (395,27.69) .. (395,31) -- (395,49) .. controls (395,52.31) and (392.31,55) .. (389,55) -- (371,55) .. controls (367.69,55) and (365,52.31) .. (365,49) -- cycle ;
                \draw  [draw opacity=0] (265,170) .. controls (240.15,170) and (220,149.85) .. (220,125) .. controls (220,100.15) and (240.15,80) .. (265,80) -- (265,125) -- cycle ; \draw   (265,170) .. controls (240.15,170) and (220,149.85) .. (220,125) .. controls (220,100.15) and (240.15,80) .. (265,80) ;  
                \draw    (265,170) -- (395,170) ;
                \draw  [draw opacity=0] (395,80) .. controls (419.85,80) and (440,100.15) .. (440,125) .. controls (440,149.85) and (419.85,170) .. (395,170) -- (395,125) -- cycle ; \draw   (395,80) .. controls (419.85,80) and (440,100.15) .. (440,125) .. controls (440,149.85) and (419.85,170) .. (395,170) ;  
                \draw    (265,160) -- (395,160) ;
                \draw    (265,180) -- (395,180) ;
                \draw  [draw opacity=0] (395,40) .. controls (395,40) and (395,40) .. (395,40) .. controls (430.9,40) and (460,71.34) .. (460,110) .. controls (460,148.66) and (430.9,180) .. (395,180) -- (395,110) -- cycle ; \draw   (395,40) .. controls (395,40) and (395,40) .. (395,40) .. controls (430.9,40) and (460,71.34) .. (460,110) .. controls (460,148.66) and (430.9,180) .. (395,180) ;  
                \draw    (365,80) -- (395,80) ;
                \draw    (365,120) -- (395,120) ;
                \draw  [draw opacity=0] (240,300) .. controls (240,300) and (240,300) .. (240,300) .. controls (240,300) and (240,300) .. (240,300) .. controls (228.95,300) and (220,291.05) .. (220,280) .. controls (220,268.95) and (228.95,260) .. (240,260) -- (240,280) -- cycle ; \draw   (240,300) .. controls (240,300) and (240,300) .. (240,300) .. controls (240,300) and (240,300) .. (240,300) .. controls (228.95,300) and (220,291.05) .. (220,280) .. controls (220,268.95) and (228.95,260) .. (240,260) ;  
                \draw  [draw opacity=0] (420,260) .. controls (420,260) and (420,260) .. (420,260) .. controls (420,260) and (420,260) .. (420,260) .. controls (431.05,260) and (440,268.95) .. (440,280) .. controls (440,291.05) and (431.05,300) .. (420,300) -- (420,280) -- cycle ; \draw   (420,260) .. controls (420,260) and (420,260) .. (420,260) .. controls (420,260) and (420,260) .. (420,260) .. controls (431.05,260) and (440,268.95) .. (440,280) .. controls (440,291.05) and (431.05,300) .. (420,300) ;  
                \draw   (255,251) .. controls (255,247.69) and (257.69,245) .. (261,245) -- (279,245) .. controls (282.31,245) and (285,247.69) .. (285,251) -- (285,269) .. controls (285,272.31) and (282.31,275) .. (279,275) -- (261,275) .. controls (257.69,275) and (255,272.31) .. (255,269) -- cycle ;
                \draw    (240,300) -- (420,300) ;
                \draw   (295,251) .. controls (295,247.69) and (297.69,245) .. (301,245) -- (319,245) .. controls (322.31,245) and (325,247.69) .. (325,251) -- (325,269) .. controls (325,272.31) and (322.31,275) .. (319,275) -- (301,275) .. controls (297.69,275) and (295,272.31) .. (295,269) -- cycle ;
                \draw   (375,251) .. controls (375,247.69) and (377.69,245) .. (381,245) -- (399,245) .. controls (402.31,245) and (405,247.69) .. (405,251) -- (405,269) .. controls (405,272.31) and (402.31,275) .. (399,275) -- (381,275) .. controls (377.69,275) and (375,272.31) .. (375,269) -- cycle ;
                \draw   (335,251) .. controls (335,247.69) and (337.69,245) .. (341,245) -- (359,245) .. controls (362.31,245) and (365,247.69) .. (365,251) -- (365,269) .. controls (365,272.31) and (362.31,275) .. (359,275) -- (341,275) .. controls (337.69,275) and (335,272.31) .. (335,269) -- cycle ;
                \draw   (295,200) -- (330,190) -- (365,200) -- (347.5,200) -- (347.5,220) -- (365,220) -- (330,230) -- (295,220) -- (312.5,220) -- (312.5,200) -- cycle ;
                \draw    (240,260) -- (255,260) ;
                \draw    (285.04,260) -- (295,260) ;
                \draw    (324.96,260) -- (334.91,260) ;
                \draw    (365.04,260) -- (375,260) ;
                \draw    (405,260) -- (420,260) ;
                
                \draw (273,33) node [anchor=north west][inner sep=0.75pt]  [font=\large]  {$\rho _{1}$};
                \draw (273,73) node [anchor=north west][inner sep=0.75pt]  [font=\large]  {$\rho _{2}$};
                \draw (273,113) node [anchor=north west][inner sep=0.75pt]  [font=\large]  {$\rho _{3}$};
                \draw (373,33) node [anchor=north west][inner sep=0.75pt]  [font=\large]  {$O$};
                \draw (263,253) node [anchor=north west][inner sep=0.75pt]  [font=\large]  {$\rho _{1}$};
                \draw (303,253) node [anchor=north west][inner sep=0.75pt]  [font=\large]  {$\rho _{2}$};
                \draw (343,253) node [anchor=north west][inner sep=0.75pt]  [font=\large]  {$\rho _{3}$};
                \draw (383,253) node [anchor=north west][inner sep=0.75pt]  [font=\large]  {$O$};
                
            \end{tikzpicture}
        \end{adjustbox}
    }
    \hfill
    \caption{
        (a) Quantum circuit for the purification-based QEM in Ref.~\cite{Koczor2021} where the derangement operator $D_n$ over $n$ state copies is controlled by an ancillary qubit.
        This circuit computes the value $\operatorname{Tr}\left[\rho^nO \right]$ for an observable $O$.
        (b) The purification circuit over three state copies for computing $\operatorname{Tr}\left[\rho_1\rho_2\rho_3 O \right]$.
        The dashed red box implements the derangement operation $D_3$ with SWAP gates.
        (c) The tensor network representation in Fig.~\ref{fig:qc_derangement}(b).
        The derangement operation $D_3$ corresponds to the dashed red box, visualizing the cyclic permutation among $\rho_1$, $\rho_2$, and $\rho_3$.
        By solving the knots in the upper wire, this can be flattened into the lower wire with one circle, representing the expectation value $\operatorname{Tr}\left[\rho_1\rho_2\rho_3 O \right]$.
    }
    \label{fig:qc_derangement}
\end{figure*}

\section{Preliminaries \label{sec:preliminaries}}
In this section, we review purification-based QEM methods, i.e., ESD and VD using multiple copies of noisy quantum states, and the DSP and resource-efficient purification methods significantly reducing the overhead of state copies at the cost of the circuit depth. 
Then, we review the GSE method, which allows for a general ansatz construction.

\subsection{Exponential suppression by derangement and virtual distillation \label{sec: virtual distillation}}

ESD~\cite{Koczor2021} and VD~\cite{Huggins2021} use multiple copies of quantum states to exponentially suppress the bias of estimators from the correct expectation values.
While ESD and VD are implemented differently, these methods are quite similar in that they compute the following error-mitigated expectation value
\begin{equation}\label{eq:expval_esdvd}
    \langle O \rangle_{\mathrm{PB}} = \frac{\operatorname{Tr}\left[\rho^n O \right]}{\operatorname{Tr}\left[\rho^n\right]},
\end{equation}
regarding an observable $O$ with $n$ state copies.

Here, we review the ESD method. 
The numerator $\operatorname{Tr}\left[\rho^n O \right]$ can be computed with the quantum circuit with the controlled-derangement operation shown in Fig.~\ref{fig:qc_derangement}.
The derangement unitary operation $D_n$ permutes all the copies of the quantum state in a way that they do not return to their original location. 
A simple example is the cyclic shift, i.e., $D_n |\psi^{(1)}\rangle\otimes |\psi^{(2)}\rangle ...\otimes |\psi^{(n)}\rangle= |\psi^{(n)}\rangle \otimes |\psi^{(1)}\rangle...\otimes |\psi^{(n-1)}\rangle$.
The quantum circuit of $D_3$ and its tensor network representation are shown in Fig.~\ref{fig:qc_derangement}(b) and Fig.~\ref{fig:qc_derangement}(c), respectively.
The denominator can be computed similarly by removing the controlled-$O$ operation from the quantum circuit.

The spectral decomposition of the density matrix $\displaystyle \rho = \sum_{k} p_{k} |\psi_{k}\rangle \langle \psi_{k}|$ helps us understand how ESD/VD suppresses the stochastic errors.
Taking the power of $\rho$ to the degree $n$, we have
\begin{equation}\label{eq:rho_esdvd}
\begin{split}
    \rho^n &= \sum_{k} p_{k}^n |\psi_{k}\rangle\langle \psi_{k}| \\ 
    &= p_{0}^n \left(\left|\psi_{0}\right\rangle\left\langle\psi_{0}\right| + \sum_{k \geq 1}\left(\frac{p_{k}}{p_{0}}\right)^n |\psi_{k}\rangle\langle \psi_{k}| \right),
\end{split}
\end{equation}
where $\displaystyle p_{0} = \max_{k} p_{k}$, and $\left|\psi_{0}\right\rangle$ is its corresponding eigenvector, which we call the dominant eigenvector.
We can clearly see that the contribution of $|\psi_{k}\rangle $ for $k\geq 1$ is exponentially suppressed with $n$. 
Note that the dominant eigenvector approximates the noiseless state well with only a small coherent error in the presence of stochastic errors~\cite{Koczor2021dominant}. 
Substituting $\rho^n$ in Eq.~\eqref{eq:expval_esdvd} for Eq.~\eqref{eq:rho_esdvd} results in
\begin{equation}
\begin{split}
\langle O \rangle_{\mathrm{PB}}
&= \frac{\displaystyle \left\langle\psi_{0}\right| O \left|\psi_{0}\right\rangle+\sum_{k \geq 1} \left(\frac{p_{k}}{p_{0}}\right)^n \langle \psi_{k}| O |\psi_{k}\rangle}{\displaystyle 1 + \sum_{k \geq 1} \left(\frac{p_{k}}{p_{0}}\right)^n} \\
&= \left\langle\psi_{0}\right| O \left|\psi_{0}\right\rangle \\
&+ \sum_{k \geq 1} \left(\frac{p_{k}}{p_{0}}\right)^n \left(\langle \psi_{k}| O |\psi_{k}\rangle -\left\langle\psi_{0}\right| O \left|\psi_{0}\right\rangle\right) + \mathcal{O}(x^2), 
\end{split}
\end{equation}
where $x=\sum_{k \geq 1} \left(\frac{p_{k}}{p_{0}}\right)^n$ and $\mathcal{O}(x^2)$ represents the terms of $x$ with degree $2$ and higher. 
Therefore, the effect of stochastic noise is exponentially suppressed to the number of copies $n$.

\subsection{Dual-state purification \label{sec:dual_state_purification}}

DSP~\cite{Huo2022} is another purification-based QEM method that evaluates the expectation value of a physical observable using powers of the density matrix.
In ESD/VD, we have seen that it implements this power of density matrix in charge of state copies, which is a considerably high cost for the near-term quantum devices.
In contrast, DSP avoids this overhead by introducing a ``dual'' quantum state that can be realized by connecting the inverse operation of the original noisy density matrix in series within the same circuit.

In DSP, we can compute the error-mitigated expectation values for an observable $O$ expressed by
\begin{equation}\label{eq:expval_dsp}
    \langle O \rangle_{\mathrm{DSP}}
    = \frac{\operatorname{Tr}\left[(\bar{\rho}\rho+\rho\bar{\rho})O/2\right]}{\operatorname{Tr}\left[(\bar{\rho}\rho+\rho\bar{\rho})/2\right]}
    = \operatorname{Tr}\left[\frac{\bar{\rho}\rho+\rho\bar{\rho}}{2\times\operatorname{Tr}\left[\bar{\rho}\rho\right]}O \right].
\end{equation}
Here, $\bar{\rho}$ is the dual state of the original state $\rho$, and $\langle O \rangle_{\mathrm{DSP}}$ approximates the expectation value of the observable $O$ on a virtually purified state $\rho^{2}/\operatorname{Tr}\left[\rho^{2}\right]$.
Since any observable $O$ can be expressed by a linear combination of Pauli operators as $O=\sum_{k}{c}_{k}P_{k}$, we can assume $O$ is a Pauli operator without losing generality.

To formally define the dual state $\bar{\rho}$, we first introduce $|\vec{0}\rangle$ or $\rho_{\vec{0}} = |\vec{0}\rangle\langle\vec{0}|$ as the initial state, and $\mathcal{U}$ as the computation process to make $\rho=\mathcal{U}(\rho_{\Vec{0}})$.
Note that $\mathcal{U}$ is generally a noisy quantum process that may not be described by a unitary operator.
Then, suppose that $\mathcal{U}_{\mathrm{rev}}(\cdot)=\sum_{k} E_{k} (\cdot) E_{k}^{\dagger}$ denotes the operation that uncomputes $\mathcal{U}$. Denoting the dual channel as $\mathcal{U}_{\mathrm{dual}}(\cdot) = \sum_{k} E_{k}^{\dagger} (\cdot) E_{k}$, we introduce the dual state as
\begin{equation}
    \bar{\rho} = \mathcal{U}_{\mathrm{dual}}(\rho_{\vec{0}})
\end{equation}
with the initial state $\rho_{\vec{0}}=|\vec{0}\rangle\langle\vec{0}|$.
In the absence of noise, we have $\mathcal{U}_{\mathrm{rev}} (\cdot) = U^{\dagger} (\cdot) U$ for $\mathcal{U}=U(\cdot) U^{\dagger}$ with $U$ being a unitary operator for computation giving $\bar{\rho}= U |\vec{0}\rangle\langle\vec{0}| U^{\dagger}$, which is equivalent to the ideal target state.

When the noise process is represented by a gate-independent Markovian process, we can describe the noise channels as below.
Let $\mathcal{G}_{l}$ and $\mathcal{F}_{l}$ denote the $l$-th gate process and its accompanying noise among $L$ gate operations in total.
Without loss of generality, we can assume $\mathcal{G}_{l}$ can be decomposed into (a set of) single-qubit and two-qubit unitary operations.
This describes the computation process $\mathcal{U}$ and its dual process as
\begin{equation} \label{eq:mathcal_U}
    \mathcal{U}=\mathcal{F}_{L} \circ \mathcal{G}_{L} \circ \mathcal{F}_{L-1} \circ \mathcal{G}_{L-1} \circ\cdots\circ \mathcal{F}_{1} \circ \mathcal{G}_{1},
\end{equation}
\begin{equation}\label{eq:mathcal_U_dual}
    \mathcal{U}_{\mathrm{dual}} = \mathcal{G}_{L} \circ \bar{\mathcal{F}}_{L} \circ \mathcal{G}_{L-1} \circ \bar{\mathcal{F}}_{L-1} \circ\cdots\circ \mathcal{G}_{1} \circ \bar{\mathcal{F}}_{1},
\end{equation}
where $\bar{\mathcal{F}}_{l}(\cdot)= \sum_{n} F_{ln}^{\dagger} (\cdot) F_{ln}$ for $\mathcal{F}_{l}(\cdot)= \sum_{n} F_{ln}(\cdot) F_{ln}^{\dagger}$ in Kraus representation.
Note that any noise can be converted into stochastic Pauli noise via Pauli twirling techniques~\cite{Dankert2009, Li2017, Cai2019}, and $\bar{\mathcal{F}}_l=\mathcal{F}_{l}$ holds under this noise model.
We numerically demonstrate the state $\rho$ and its dual state $\bar{\rho}$ get closer to each other as the depth of the quantum circuit increases under stochastic Pauli errors.
Moreover, as a special case, we have $\bar{\rho}=\rho$ under the global depolarizing channel and a particular local depolarizing noise commuting with single-qubit and two-qubit operations we are considering in ansatz circuits.
The detailed discussion can be found in Appendix \ref{sec:appendix_diff_rho_rho_dual}.
We may further convert stochastic Pauli errors into depolarizing errors using error simplification methods~\cite{mills2023simplifying} when the circuit has a certain symmetry.

The DSP method can be performed either with an intermediate direct measurement in the quantum circuit or an indirect measurement using an ancillary qubit. 
We next describe the detailed implementations of each option.

\subsubsection{Dual-state purification with direct measurement}

The circuit implementation of DSP with a direct intermediate measurement is shown in Fig.~\ref{fig:qc_dsp}(a), which consists of the computation process $\mathcal{U}$, a mid-circuit measurement gadget for the target observable $O$, and the uncomputation process $\mathcal{U}_{\mathrm{rev}}$.
Then, the joint probability of obtaining $b\in\{0,1\}$ from the intermediate measurement and bitstring $\vec{0}$ from measuring the register qubits on a computational basis is described as
\begin{equation}
    \mathrm{p}_{b,\vec{0}}=\operatorname{Tr}\left[\mathcal{U}_{\mathrm{rev}} \left(\frac{I + (-1)^{b} O}{2}\mathcal{U}(\rho_{\vec{0}})\frac{I + (-1)^{b} O}{2} \right)\operatorname{P}_{\vec{0}}\right], 
\end{equation}
where $\operatorname{P}_{\vec{0}} = |\vec{0}\rangle\langle\vec{0}|$ is a projector corresponding to the post-selection regarding $\vec{0}$ on the main quantum registers.
Recalling $\mathcal{U}_{\mathrm{rev}}(\cdot)=\sum_{k} E_{k} (\cdot) E_{k}^{\dagger}$ and $\bar{\rho}= \mathcal{U}_{\mathrm{dual}}(\rho_{\vec{0}}) =\sum_{k} E_{k}^{\dagger} \rho_{\vec{0}} E_{k}$ and assuming that the observable $O$ is a Pauli observable, we obtain
\begin{equation}
\begin{split}
    \mathrm{p}_{0,\vec{0}} - \mathrm{p}_{1,\vec{0}} &= \operatorname{Tr}\left[\frac{\bar{\rho}\rho+\rho\bar{\rho}}{2} O     \right],
\end{split}
\end{equation}
which gives the quantity $\langle O \rangle_{\mathrm{DSP}}$ in Eq.~\eqref{eq:expval_dsp} by
\begin{equation}
    \langle O \rangle_{\mathrm{DSP}} = \frac { \mathrm{p}_{0,\vec{0}} - \mathrm{p}_{1,\vec{0}} } { \mathrm{p}_{\vec{0}} },
\end{equation}
where $\mathrm{p}_{\vec{0}}$ is defined as
\begin{equation}\label{eq:p_vec0}
\mathrm{p}_{\vec{0}}=\operatorname{Tr}\left[\mathcal{U}_{\mathrm{rev}} (\mathcal{U}(\rho_{\vec{0}}))\operatorname{P}_{\vec{0}}\right]=\operatorname{Tr}\left[\bar{\rho}\rho\right],
\end{equation}
which can be implemented by removing the mid-circuit measurement, $\mathcal{U}_{O}$, and $\mathcal{U}_{O}^{\dagger}$ from the circuit in Fig.~\ref{fig:qc_dsp}(a).

\begin{figure}[htbp]
    \subfloat[\label{fig:qc_dsp_a}]{
        \begin{adjustbox}{width=0.48\textwidth}\begin{quantikz}
            \lstick{$|0\rangle$} & \gate[5, nwires=3]{\mathcal{U}} & \gate[5, nwires=3]{\mathcal{U}_{O}^{\dagger}} & \meter{Z} & \gate[5, nwires=3]{\mathcal{U}_{O}} & \gate[5, nwires=3]{\mathcal{U}_{\mathrm{rev}}} & \push{\ \langle 0|} \\
            \lstick{$|0\rangle$} &                                 &                                             & \qw       &                                   &                                                & \push{\ \langle 0|} \\
            \vdots               &                                 &                                             &           &                                   &                                                & \vdots \\
            \lstick{$|0\rangle$} &                                 &                                             & \qw       &                                   &                                                & \push{\ \langle 0|} \\
            \lstick{$|0\rangle$} &                                 &                                             & \qw       &                                   &                                                & \push{\ \langle 0|} \\
        \end{quantikz}\end{adjustbox}
    }
    \hfill
    \subfloat[\label{fig:qc_dsp_b}]{
        \begin{adjustbox}{width=0.48\textwidth}\begin{quantikz}
            \lstick{$|+\rangle$} & \qw                             & \qw                                         & \ctrl{1} & \qw                               & \qw                                            & \meter{X} \\
            \lstick{$|0\rangle$} & \gate[5, nwires=3]{\mathcal{U}} & \gate[5, nwires=3]{\mathcal{U}_{O}^{\dagger}} & \gate{Z} & \gate[5, nwires=3]{\mathcal{U}_{O}} & \gate[5, nwires=3]{\mathcal{U}_{\mathrm{rev}}} & \push{\ \langle 0|} \\
            \lstick{$|0\rangle$} &                                 &                                             & \qw      &                                   &                                                & \push{\ \langle 0|} \\
            \vdots               &                                 &                                             &          &                                   &                                                & \vdots \\
            \lstick{$|0\rangle$} &                                 &                                             & \qw      &                                   &                                                & \push{\ \langle 0|} \\
            \lstick{$|0\rangle$} &                                 &                                             & \qw      &                                   &                                                & \push{\ \langle 0|} \\
        \end{quantikz}\end{adjustbox}
    }
    \caption{
        Quantum circuits for dual-state purification (DSP) with (a) direct intermediate measurement and (b) indirect measurement.
        Since we can assume $O$ is a Pauli operator here without losing generality, the circuits are designed based on the fact that any Pauli operator $O$ can be conjugated into $Z_1$ via a Clifford operator $U_{O}$ as $O = U_{O} Z_1 U_{O}^{\dagger}$. 
        The gates $\mathcal{U}_{O}$ and $\mathcal{U}_{O}^{\dagger}$ in the figures denote the noisy operations of $U_{O}$ and $U_{O}^{\dagger}$ respectively.
    }
    \label{fig:qc_dsp}
\end{figure}

\subsubsection{Dual-state purification with an ancillary qubit}

The DSP procedure can also be realized with the Hadamard test, with which we indirectly measure the observable $O$ using an ancillary qubit. 
As shown in Fig.~\ref{fig:qc_dsp}(b), the computation process $\mathcal{U}$ and the uncomputation process $\mathcal{U}_{\mathrm{rev}}$ are applied in the same way as an intermediate measurement circuit, whereas the operation regarding the observable $O$ on the register qubits is controlled by the ancillary qubit.

Denoting the probability of measuring the ancillary and the register qubits in $b~(b=0,1)$ and $\vec{0}$ by $\tilde{\mathrm{p}}_{b,\vec{0}}$, we can show $\tilde{\mathrm{p}}_{b,\vec{0}} = \mathrm{p}_{b,\vec{0}}~(b=0,1)$ and hence
\begin{equation}\label{eq:trace_dsp_ancillary}
    \tilde{\mathrm{p}}_{0 ,\vec{0}}-\tilde{\mathrm{p}}_{1 ,\vec{0}}=\left\langle X \otimes \operatorname{P}_{\vec{0}}\right\rangle_{O} = \operatorname{Tr}\left[\frac{\bar{\rho}\rho+\rho\bar{\rho}}{2} O \right], 
\end{equation}
which is obtained by measuring observable $X\otimes\operatorname{P}_{\vec{0}}$ without the mid-circuit measurement at the cost of one additional ancillary qubit, as shown in Fig.~\ref{fig:qc_dsp}(b). 
Therefore, the following quantity
\begin{equation}\label{eq:expval_dsp_ancillary}
    \langle O \rangle_{\mathrm{DSP}} = \frac{\tilde{\mathrm{p}}_{0,\vec{0}}-\tilde{\mathrm{p}}_{1,\vec{0}}}{\mathrm{p}_{\vec{0}}}
\end{equation}
gives Eq.~\eqref{eq:expval_dsp}, where $\tilde{\mathrm{p}}_{\vec{0}}$ is also defined as
\begin{equation}\label{eq:tilde_p_vec0}
    \tilde{\mathrm{p}}_{\vec{0}}=\operatorname{Tr}\left[\mathcal{U}_{\mathrm{rev}} (\mathcal{U}(\rho_{\vec{0}}))\operatorname{P}_{\vec{0}}\right]=\operatorname{Tr}\left[\bar{\rho}\rho\right],
\end{equation}
obtained from the as by removing the operations controlled-Z, $\mathcal{U}_{O}$, and $\mathcal{U}_{O}^{\dagger}$ from the circuit in Fig.~\ref{fig:qc_dsp}(b).

\subsection{Resource-efficient purification}

By combining the conventional purification method with DSP, Ref.~\cite{cai2021resource} proposed a resource-efficient purification framework. 
An example of the quantum circuit for resource-efficient purification using two copies of quantum states is shown in Fig.~\ref{fig:qc_dsp_resource_efficient}(a), where we have
\begin{equation}
\begin{split}
\tilde{\mathrm{p}}_{0, \vec{0}} - \tilde{\mathrm{p}}_{1, \vec{0}} &= \langle X \otimes \operatorname{P}_{\vec{0}} \otimes \operatorname{P}_{\vec{0}} \rangle_{O} \\
&= \operatorname{Tr}\left[\frac{\bar{\rho} \rho \bar{\rho} \rho + \rho \bar{\rho} \rho \bar{\rho}}{2} O \right],
\end{split}
\end{equation}
which approximates $\operatorname{Tr}\left[\rho^4O \right]$.
Here $\tilde{\mathrm{p}}_{b, \vec{0}}$ represents the joint probability of obtaining $b~(b=0,1)$ from the ancillary qubit and post-selecting $\vec{0}$ from all copies.
By eliminating the controlled-$O$ operation, we can compute $\operatorname{Tr}\left[ (\bar{\rho} \rho \bar{\rho} \rho + \rho \bar{\rho} \rho \bar{\rho})/2 \right]$ and thus obtain the approximation of $\operatorname{Tr}\left[\rho^4O \right]/\operatorname{Tr}\left[\rho^4\right]$, corresponding to the purification with four copies in ESD/VD.

More generally, when we have $n$ copies and use a controlled-derangement operation among the copies, as shown in Fig.~\ref{fig:qc_dsp_resource_efficient}(b), we can compute the following term
\begin{equation}
\begin{split}
\mathrm{p}_{0, \vec{0}}- \mathrm{p}_{1, \vec{0}}&=\langle X \otimes \operatorname{P}_{\vec{0}}^{\otimes n}\rangle_{O} \\
&= \operatorname{Tr}\left[\frac{(\bar{\rho} \rho)^n+ (\rho \bar{\rho})^n}{2} O \right],
\end{split}
\end{equation}
corresponding to the approximation of $\operatorname{Tr}\left[\rho^{2n}O \right] /\operatorname{Tr}\left[\rho^{2n}\right]$. 

\begin{figure}[htbp]
    \subfloat[\label{fig:qc_dsp_resource_efficient_a}]{
        \begin{adjustbox}{width=0.48\textwidth}\begin{quantikz}
            \lstick{$|+\rangle$}      & \qw                & \ctrl{1} & \ctrl{2} & \qw                               & \meter{X}               \\
            \lstick{$|\vec{0}\rangle$} & \gate{\mathcal{U}} & \gate{O} & \swap{1} & \gate{\mathcal{U}_{\mathrm{rev}}} & \push{\ \langle\vec{0}|} \\
            \lstick{$|\vec{0}\rangle$} & \gate{\mathcal{U}} & \qw      & \targX{} & \gate{\mathcal{U}_{\mathrm{rev}}} & \push{\ \langle\vec{0}|}
        \end{quantikz}\end{adjustbox}
    }
    \hfill
    \subfloat[\label{fig:qc_dsp_resource_efficient_b}]{
        \begin{adjustbox}{width=0.48\textwidth}\begin{quantikz}
            \lstick{$|+\rangle$}      & \qw                & \ctrl{1} & \ctrl{1}                & \qw                               & \meter{X}               \\
            \lstick{$|\vec{0}\rangle$} & \gate{\mathcal{U}} & \gate{O} & \gate[3, nwires=2]{D_n} & \gate{\mathcal{U}_{\mathrm{rev}}} & \push{\ \langle\vec{0}|} \\
            \vdots                    & \vdots             &          &                         & \vdots                            & \vdots                  \\
            \lstick{$|\vec{0}\rangle$} & \gate{\mathcal{U}} & \qw      &                         & \gate{\mathcal{U}_{\mathrm{rev}}} & \push{\ \langle\vec{0}|}
        \end{quantikz}\end{adjustbox}
    }
    \caption{
        Quantum circuits for resource-efficient purification using Hadamard test on (a) $2$ state copies and (b) $n$ state copies.
    }
    \label{fig:qc_dsp_resource_efficient}
\end{figure}

\subsection{Generalized quantum subspace expansion\label{sec:generalized_subspace_expansion}}

Generalized quantum subspace expansion (GSE) introduced in Ref.~\cite{Yoshioka2022-gq} is an error-agnostic QEM method that integrates QSE~\cite{McClean2020} and ESD~\cite{Koczor2021} or VD~\cite{Huggins2021}.
Given a Hamiltonian $H$, GSE mitigates the noisy approximation of its eigenstate by constructing the following QEM ansatz
\begin{equation}\label{eq:rho_gse_general}
    \rho_{\mathrm{GSE}}=\frac{\mathcal{P}^{\dagger} \mathcal{A} \mathcal{P}}{\operatorname{Tr}\left[\mathcal{P}^{\dagger} \mathcal{A} \mathcal{P}\right]},
\end{equation}
using a positive semidefinite Hermitian operator $\mathcal{A}$ and the linear combination of subspaces $\displaystyle\mathcal{P} = \sum_{i} \alpha_{i} \sigma_{i}$, $\alpha_{i} \in \mathbb{C}$ where $\sigma_{i}$ can be generally a non-Hermitian operator.
According to the process of the QSE, $\sigma_{i}$ can be seen as a basis of the subspace that should be prepared and measured on quantum devices, and $\alpha_{i}$ should be determined through classical postprocessing to fulfill the desired property with respect to $H$.
In GSE, $\sigma_{i}$ takes quite a general form
\begin{equation}\label{eq:subspace_general}
    \sigma_{i}=\sum_{k} h_{k}^{(i)} \prod_{l=1}^{L_{ik}} V_{kl}^{(i)} \rho_{kl}^{(i)} W_{kl}^{(i)\dagger},
\end{equation}
where $h_{k}^{(i)} \in \mathbb C$, $\rho_{kl}^{(i)}$ is a quantum state in the $k$-th term, $V_{kl}^{(i)}$ and $W_{kl}^{(i)\dagger}$ are operators that preserve the efficient measurement process, and $L_{ik}$ denotes the number of multiplied quantum states.
To compute the set of classical coefficients $\{\alpha_{i}\}_{i}$ that keeps $\operatorname{Tr}\left[\rho_{\mathrm{GSE}}\right] = 1$, GSE solves the following generalized eigenvalue problem 
\begin{equation}\label{eq:generalized_eigenvalue_problem}
\begin{split}
    \mathcal{H} \vec{\alpha} = E \mathcal{S} \vec{\alpha},
    ~\text{ with }~
    \begin{cases}
        \displaystyle \mathcal{H}_{ij} = \operatorname{Tr}\left[\sigma_{i}^{\dagger} \mathcal{A} \sigma_{j} H\right], \\[5pt]
        \displaystyle \mathcal{S}_{ij} = \operatorname{Tr}\left[\sigma_{i}^{\dagger} \mathcal{A} \sigma_{j}\right],
    \end{cases}
\end{split}
\end{equation}
where the matrix elements $\mathcal{S}_{ij}$ and $\mathcal{H}_{ij}$ are supposed to be retrieved efficiently from quantum devices.
With the optimized coefficients ${\alpha}_{i}$, the expectation value of $\rho_{\mathrm{GSE}}$ for an arbitrary observable $O$ can be computed by replacing $H$ with $O$ as
\begin{equation}\label{eq:expval_gse}
    \langle O \rangle_{\mathrm{GSE}} = \sum_{i, j} \alpha_{i}^{*} \alpha_{j} \operatorname{Tr}\left[\sigma_{i}^{\dagger} \mathcal{A} \sigma_{j} O \right].
\end{equation}

The advantage of GSE lies mainly in its extensive and flexible subspace construction of $\{\sigma_{i}\}_{i}$.
This subspace includes the power of the density matrix, whose expectation value can be retrieved by the ESD/VD.
In particular, suppose the subspaces are set to $\mathcal{A}=I$ and $\sigma_{i}=\rho^{i},~i\in \{1,\ldots,M\}$ where $\rho^{i}$ is the power of $\rho$ with degree $i$, the GSE ansatz becomes
\begin{equation}\label{eq:rho_gse_ps}
    \rho_{\mathrm{ps}}=\sum_{i, j=1}^M \alpha_{i}^{*} \alpha_{j} \rho^{i+j-2},
\end{equation}
which is called ``power subspace'' ansatz. 
This construction can be extended to the set of subspaces including the factor of a Hamiltonian, such as $\{I, \rho, \rho H\}$, which is denoted by ``GSE+'' in Ref.~\cite{Yoshioka2022-gq}.
By adding the Hamiltonian factor, the power subspace method can cover a richer search space, performing better with the same power degree of the density matrix.

GSE also takes extrapolation-based QEM methods into account.
Let the noisy state $\rho\left(\epsilon\right)$ be a noisy quantum state with an error rate $\epsilon$.
Assuming $\mathcal{A}=I$ and $\sigma_{i}=\rho\left(\lambda_{i} \epsilon\right),~i\in \{0, 1,\ldots,M\}$ where the error rate is $\lambda_{i} \geq 1$ times amplified from $\rho\left(\epsilon\right)$, the GSE ansatz becomes
\begin{equation}\label{eq:rho_gse_fs}
    \rho_{\mathrm{fs}}=\sum_{i, j = 1}^{M} \alpha_{i}^{*} \alpha_{j} \rho\left(\lambda_{i} \epsilon\right) \rho\left(\lambda_{j} \epsilon\right),
\end{equation}
which is called ``fault subspace'' ansatz~\cite{Yoshioka2022-gq}.

Whereas QSE is less effective for mitigating incoherent errors, and ESD/VD may cause the problem of coherent mismatch, which stems from coherent errors~\cite{Koczor2021dominant}, GSE can mitigate both the coherent errors and stochastic errors by adopting the linear combination of noisy density matrix with different power degrees and different error rates in its expanded subspaces.
However, in the power subspace of GSE, multiple state copies are still required, which makes the whole procedure less practical.

\section{The Framework of Dual-GSE \label{sec:dgse}}

To alleviate the requirement of state copies in GSE, we use DSP to compute expectation values of the density matrix multiplication.
In this section, we first provide the general QEM ansatz for our resource-efficient GSE, namely, Dual-GSE, to see how the resource-efficient circuit is integrated as a subroutine of GSE.
We then see two practical subspace constructions introduced in Ref.~\cite{Yoshioka2022-gq}, i.e. the power subspace and the fault subspace.
In addition, we also introduce a novel subspace construction for Dual-GSE, which allows for simulating the physical models whose size is beyond that of given quantum devices in a divide-and-conquer way.

\subsection{The general ansatz for Dual-GSE \label{sec:dgse_The_General_Ansatz_for_Dual-GSE}}

To simulate a broader class of ansatz corresponding to Eq.~\eqref{eq:subspace_general}, we consider generalizing the quantum circuits for DSP and resource-efficient purification in the previous section. 
Assuming $\rho_{\mathrm{GSE}}$ is properly normalized by the suitable choice of $\{\alpha_{i}\}_{i}$, the GSE ansatz of Eq.~\eqref{eq:rho_gse_general} is expanded with Eq.~\eqref{eq:subspace_general} as
\begin{equation}\label{eq:rho_gse_general_expanded}
\begin{split}
    \rho_{\mathrm{GSE}}
    &=\sum_{i,i'}\alpha_{i}^{*}\alpha_{i'} \sum_{k,k'} h_{k}^{(i)*}h_{k'}^{(i')} \\
    &\quad\quad \times \left(\tau_{kL_{ik}}^{(i)\dagger}\cdots\tau_{k2}^{(i)\dagger}\tau_{k1}^{(i)\dagger}\right) \mathcal{A} \left(\tau_{k'1}^{(i')}\tau_{k'2}^{(i')}\cdots\tau_{k'L_{i'k'}}^{(i')}\right), 
\end{split}
\end{equation}
where we define $\tau_{kl}^{(i)}=V_{kl}^{(i)}\rho_{kl}^{(i)}W_{kl}^{(i)\dagger}$ and $\tau_{kl}^{(i)\dagger}=W_{kl}^{(i)}\rho_{kl}^{(i)}V_{kl}^{(i)\dagger}$ using operators $V_{kl}^{(i)}$ and $W_{kl}^{(i)}$.

We here discuss only the case where $\mathcal{A} = I$ and the power degrees $L_{ik}$ in Eq.~\eqref{eq:rho_gse_general_expanded} are all odd.
Other general cases are provided in Appendix~\ref{sec:appendix_quantum_circuits_for_the_generalized_ansatz}.
Defining $\bar{\tau}_{kl}^{(i)}=V_{kl}^{(i)}\bar{\rho}_{kl}^{(i)}W_{kl}^{(i)\dagger}$ and $\bar{\tau}_{kl}^{(i)\dagger}=W_{kl}^{(i)}\bar{\rho}_{kl}^{(i)}V_{kl}^{(i)\dagger}$, the Dual-GSE ansatz is described as
\begin{equation}\label{eq:rho_dgse_general_odd}
\begin{split}
    \rho_{\mathrm{DGSE}}
    &=\sum_{i,i'}\alpha_{i}^{*}\alpha_{i'} \sum_{k,k'} h_{k}^{(i)*}h_{k'}^{(i')} \\
    &\quad\quad \times \left(\tau_{kL_{ik}}^{(i)\dagger} \cdots \tau_{k3}^{(i)\dagger}\bar{\tau}_{k2}^{(i)\dagger}\tau_{k1}^{(i)\dagger}\right) \\
    &\quad\quad \times \left(\bar{\tau}_{k'1}^{(i')}\tau_{k'2}^{(i')}\bar{\tau}_{k'3}^{(i')} \cdots \bar{\tau}_{k'L_{i'k'}}^{(i')}\right).
\end{split}
\end{equation}
This approximates the GSE ansatz $\rho_{\mathrm{GSE}}$ as we see that $\bar{\rho} \approx \rho$ as circuit depth grows under the noise level where QEM methods are supposed to be applied effectively (see Appendix~\ref{sec:appendix_diff_rho_rho_dual}).

To compute the coefficients in a generalized linear eigenvalue problem (\ref{eq:generalized_eigenvalue_problem}), we need to evaluate the following quantity:
\begin{equation}\label{eq:trace_dgse_general_odd}
    \operatorname{Tr}\left[ \left(\tau_{kL_{ik}}^{(i)\dagger} \cdots \tau_{k3}^{(i)\dagger}\bar{\tau}_{k2}^{(i)\dagger}\tau_{k1}^{(i)\dagger}\right) \left(\bar{\tau}_{k'1}^{(i')}\tau_{k'2}^{(i')}\bar{\tau}_{k'3}^{(i')} \cdots \bar{\tau}_{k'L_{i'k'}}^{(i')}\right) O \right].
\end{equation}
Here, we assume $O$ is a Pauli operator that constitutes the linear decomposition of the Hamiltonian or an identity operator. 
Eq.~\eqref{eq:trace_dgse_general_odd} can be obtained by generalized resource-efficient quantum circuits with $\displaystyle \left\lceil\frac{L_{ik}+L_{i'k'}}{2}\right\rceil$ state copies (see Appendix~\ref{sec:appendix_quantum_circuits_for_the_generalized_ansatz}).
For example, when $L_{ik}=1$, the quantum circuit to compute $\operatorname{Tr}\left[ \tau_{k1}^{(i)\dagger}\bar{\tau}_{k'1}^{(i')}O \right]$ is shown in Fig.~\ref{fig:qc_dgse_general_one_copy}.
The quantum circuits for $\mathcal{A} \neq I$ can also be found in Appendix \ref{sec:appendix_quantum_circuits_for_the_generalized_ansatz}.

\begin{figure}[htbp]
    \centering
    \begin{adjustbox}{width=\columnwidth}\begin{quantikz}
            \lstick{$|+\rangle$}       & \qw                                & \ctrl{1} & \octrl{1} & \ctrl{1} & \ctrl{1}         & \octrl{1}        & \qw                                 & \meter{X,Y}    \\
            \lstick{$|\vec{0}\rangle$} & \gate{\mathcal{U}^{\mathrm{(in)}}} & \gate{W} & \gate{V}  & \gate{O} & \gate{W^{\dagger}} & \gate{V^{\dagger}} & \gate{\mathcal{U}^{\mathrm{(out)}}} & \push{\ \langle\vec{0}|}
    \end{quantikz}\end{adjustbox}
    \caption{
        The quantum circuit for the expectation value $\operatorname{Tr}\left[\tau^{\dagger}\bar{\tau}O \right]=\operatorname{Tr}\left[(W\rho V^{\dagger})(V\bar{\rho}W^{\dagger})O\right]$, where $\rho = \mathcal{U}^{\mathrm{(in)}}(\rho_{\vec{0}})$, $\bar{\rho} = \mathcal{U}_{\mathrm{dual}}^{\mathrm{(out)}}(\rho_{\vec{0}})$, and $\mathcal{U}^{(\mathrm{out})}_{\mathrm{dual}}$ is the dual process of $\mathcal{U}_{\mathrm{rev}}^{(\mathrm{out})}$.
        We can obtain this quantity as $\langle X\otimes \operatorname{P}_{\vec{0}}\rangle_{O} + i \langle Y\otimes \operatorname{P}_{\vec{0}}\rangle_{O}$ by postprocessing the measurement results on $X$ and $Y$ bases on the ancillary qubit while post-selecting the main quantum register by $\operatorname{P}_{\vec{0}}$.
        Note that the superscripts and subscripts attached to the symbols such as $\tau_{k1}^{(i)\dagger}$ and $\tau_{k'1}^{(i')}$ are omitted into $\tau^{\dagger}$ and $\tau$ to keep notations simple in this figure.
    }
    \label{fig:qc_dgse_general_one_copy}
\end{figure}

While we have introduced the general ansatz construction described above, we will explain how to design resource-efficient implementation of particularly useful ansatz, i.e., the power subspace and the fault subspace methods, and the novel divide-and-conquer subspace method allowing for the simulation of quantum systems larger than the scale of available quantum hardware.
We also remark that these subspace constructions are noise-resilient because they do not require controlled-$V$ and -$W$ gates shown in Fig.~\ref{fig:qc_dgse_general_one_copy}, but only use the original and slightly modified resource-efficient DSP circuits in Fig.~\ref{fig:qc_dsp_resource_efficient}.

\subsection{Power subspace \label{sec:dgse_power_subspace}}

Let us first discuss the power subspace ansatz that contains density matrices with degrees up to two.
Setting $\mathcal{A}=I$ and $\{\sigma_{i}\}_{i}=\{I, \rho\}$ in Eq.~\eqref{eq:rho_gse_general} and Eq.~\eqref{eq:subspace_general}, the ansatz becomes
\begin{equation}
\begin{split}
    \rho_{\mathrm{ps}} 
    &= (\alpha_{1}^{*} I + \alpha_{2}^{*} \rho) (\alpha_{1} I + \alpha_{2} \rho) \\
    &=|\alpha_{1}|^{2} I + (\alpha_{1}^{*}\alpha_{2} + \alpha_{2}^{*}\alpha_{1}) \rho + |\alpha_{2}|^{2} \rho^{2} \\
    &=f_{1} I + f_{2} \rho + f_{3} \rho^{2},
\end{split}
\end{equation}
by replacing $|\alpha_{1}|^2$, 
$\alpha_{1}^{*}\alpha_{2} + \alpha_{2}^{*}\alpha_{1}$,
and $|\alpha_{2}|^2$ 
with $f_{1}$, $f_{2}$, and $f_{3}$.
To evaluate the coefficients for solving the generalized linear equation (\ref{eq:generalized_eigenvalue_problem}), we need to approximate, for example,  $\operatorname{Tr}\left[\rho^{2} O \right]$ as
\begin{equation}
    \operatorname{Tr}\left[\frac{\bar{\rho}\rho+\rho\bar{\rho}}{2} O \right] = \langle X\otimes \operatorname{P}_{\vec{0}}\rangle_{O},
\end{equation}
using the DSP circuit in Fig.~\ref{fig:qc_dsp}, circumventing the use of copies of the quantum state $\rho$.

For the ansatz with higher power degrees up to three, we can assign $\mathcal{A}=\rho$ and $\{\sigma_{i}\}_{i}=\{I, \rho\}$, which gives
\begin{equation}
\begin{split}
    \rho_{\mathrm{ps}} 
    &= (\alpha_{1}^{*} I + \alpha_{2}^{*} \rho) \rho (\alpha_{1} I + \alpha_{2} \rho) \\
    &=|\alpha_{1}|^{2} \rho + (\alpha_{1}^{*}\alpha_{2} + \alpha_{2}^{*}\alpha_{1}) \rho^{2} + |\alpha_{2}|^{2} \rho^{3}.
\end{split}
\end{equation}
Now, $\operatorname{Tr}\left[\rho^3 O \right]$ can be evaluated by performing post-selection on the partial system of the DSP circuit such as Fig.~\ref{fig:removemeasurement} that computes $\operatorname{Tr}\left[\rho \bar{\rho} \rho O \right]$.
Similarly, because the approximation of $\operatorname{Tr}\left[\rho^{2n} O \right]$ and $\operatorname{Tr}\left[\rho^{2n-1} O \right]$ can be obtained from resource-efficient purification circuits with $n$ copies of quantum states, we can straightforwardly generalize this argument to the power subspace ansatz with higher power degrees.

\begin{figure}[htbp]
    \begin{adjustbox}{width=0.48\textwidth}\begin{quantikz}
        \lstick{$|+\rangle$}      & \qw                & \ctrl{1} & \ctrl{2} & \qw                               & \meter{X,Y}               \\
        \lstick{$|\vec{0}\rangle$} & \gate{\mathcal{U}} & \gate{O} & \swap{1} & \gate{\mathcal{U}_{\mathrm{rev}}} & \push{\ \langle\vec{0}|} \\
        \lstick{$|\vec{0}\rangle$} & \gate{\mathcal{U}} & \qw      & \targX{} & \qw                               & \qw
    \end{quantikz}\end{adjustbox}
    \caption{
        Quantum circuit with two state copies for approximating $\operatorname{Tr}\left[\rho^3 O \right]$ by eliminating the dual-state preparation by $\mathcal{U}_{\mathrm{rev}}$ and $\operatorname{P}_{\vec{0}}$ on one quantum register.
        The quantity $\operatorname{Tr}\left[\rho \bar{\rho} \rho O \right]$ is obtained by measuring $\langle X \otimes \operatorname{P}_{\vec{0}} \otimes I\rangle - i \langle Y \otimes \operatorname{P}_{\vec{0}} \otimes I\rangle$.
    }
    \label{fig:removemeasurement}
\end{figure}

We can also consider the case where an unphysical state is included in the subspace, e.g., $\{\sigma_{i}\}_{i}=\{I, \rho, \rho B \} $ for some operator $B$.
By linearly decomposing $B=\sum_\beta {b}_{\beta} P_\beta$ with ${b}_{\beta} \in \mathbb{C}$ and Pauli operator ${P}_{\beta}$, we need to evaluate, for example: 
\begin{equation}
    \mathcal{H}_{22} = \operatorname{Tr}\left[B^{\dagger} \rho \mathcal{A} \rho B H\right] = \sum_{h, \beta', \beta} {c}_{h} {b}_{\beta'}^{*} {b}_{\beta} \operatorname{Tr}\left[\rho \mathcal{A} \rho P_{\beta}P_{h} P_{\beta'} \right], 
\end{equation}
which can be obtained by computing $\operatorname{Tr}\left[\rho \mathcal{A} \rho P_{\beta} P_{h} P_{\beta'}\right]$ with a DSP circuit for $\mathcal{A}=I$ and with a resource-efficient purification circuit for $\mathcal{A}=\rho$.
Here, we have decomposed $H=\sum_{h} c_{h} P_{h}$, $c_{h} \in \mathbb{C}$.

As a specific example, we consider a power subspace ansatz $\rho_{\mathrm{ps}}$ that has the power degree of density matrix maximally up to two by setting $\mathcal{A}=I$ and $\{\sigma_{i}\}_{i}=\{I\}\cup\{\rho H^{k-2}\}_{k=2,\ldots,M}$, inspired by the ``GSE+'' option in Ref.~\cite{Yoshioka2022-gq}.
This ansatz construction expands the subspace by gradually adding new Pauli observables showing up in higher powers of the Hamiltonian.
The ansatz $\rho_{\mathrm{ps}}$ can be written as
\begin{equation}
\begin{split}
    \rho_{\mathrm{ps}} 
    &= \left(\alpha_{1}^{*}I + \sum_{k=2}^{M} \alpha_{k}^{*} H^{k-2} \rho \right) \left(\alpha_{1}I + \sum_{k=2}^{M} \alpha_{k}\rho H^{k-2} \right) \\
    &= \alpha_{1}^{*}\alpha_{1}I + \sum_{i=2}^{M}\left( \alpha_{1}^{*}\alpha_{i}\rho H^{i-2} + \alpha_{i}^{*}\alpha_{1}H^{i-2}\rho \right) \\
    &\quad + \sum_{i,j=2}^{M}\alpha_{i}^{*}\alpha_{j} H^{i-2}\rho\bar{\rho}H^{j-2},
\end{split}
\end{equation}
where we approximate $\rho^{2}$ with $\rho\bar{\rho}$.
We can replace $\rho\bar{\rho}$ with $\bar{\rho}\rho$ or symmetrized state $(\bar{\rho}\rho + \rho\bar{\rho}) / 2$, which also approximate $\rho^{2}$.

Supposing we use $(\bar{\rho}\rho + \rho\bar{\rho}) / 2$ in the power subspace ansatz $\rho_{\mathrm{ps}}$, the matrix elements $\mathcal{S}_{ij}$ and $\mathcal{H}_{ij}$ for the generalized eigenvalue linear problem Eq.~\eqref{eq:generalized_eigenvalue_problem} can be obtained as
\begin{equation}\label{eq:S_ij}
\begin{split}
    \mathcal{S}_{ij}
    = \displaystyle\operatorname{Tr}\left[\frac {\bar{\rho}\rho + \rho\bar{\rho}} {2} H^{i+j-4} \right]
    = \sum_{k}{c}_{k}\langle X\otimes \operatorname{P}_{\vec{0}}\rangle_{P_{k}},
\end{split}
\end{equation}
\begin{equation}\label{eq:H_ij}
\begin{split}
    \mathcal{H}_{ij}
    = \displaystyle\operatorname{Tr}\left[\frac {\bar{\rho}\rho + \rho\bar{\rho}} {2} H^{i+j-3} \right]
    = \sum_{k}{c}_{k}^{\prime}\langle X\otimes \operatorname{P}_{\vec{0}}\rangle_{P_{k}},
\end{split}
\end{equation}
for $i,j\geq2$, where we assume the powers of Hamiltonian are decomposed as $H^{i+j-4} = \sum_{k} {c}_{k}P_{k}$ and $H^{i+j-3} = \sum_{k}{c}_{k}^{\prime}P_{k}$.
The matrix elements with $i=1$ or $j=1$ are described as
\begin{equation}\label{eq:S_{i}=0_or_{j}=0}
    \mathcal{S}_{ij} = \displaystyle\operatorname{Tr}\left[\frac{\rho + \bar{\rho}}{2} H^{i+j-3}\right] = \sum_{k}{c}_{k}^{\prime}\operatorname{Tr}\left[\frac{\rho + \bar{\rho}}{2} P_{k}\right],
\end{equation}
\begin{equation}\label{eq:H_{i}=0_or_{j}=0}
    \mathcal{H}_{ij} = \displaystyle\operatorname{Tr}\left[\frac{\rho + \bar{\rho}}{2} H^{i+j-2}\right] = \sum_{k}{c}_{k}^{\prime\prime}\operatorname{Tr}\left[\frac{\rho + \bar{\rho}}{2} P_{k}\right],
\end{equation}
with $H^{i+j-3} = \sum_{k}{c}_{k}^{\prime}P_{k}$ and $H^{i+j-2} = \sum_{k}{c}_{k}^{\prime\prime}P_{k}$.
Since $\rho$ approximates  $(\rho + \bar{\rho})/2$, Eq.~\eqref{eq:S_{i}=0_or_{j}=0} and Eq.~\eqref{eq:H_{i}=0_or_{j}=0} can be computed by measuring the corresponding Pauli operators over the original quantum circuit.
Finally, the top-left matrix elements of $\mathcal{S}$ and $\mathcal{H}$, the case of $i = 1$ and $j = 1$, can be further simplified as $\mathcal{S}_{11} = \operatorname{Tr}\left[I\right]$ and $\mathcal{H}_{11} = \operatorname{Tr}\left[H\right]$ respectively, which is trivially available without quantum computers.

While one might point out that the mitigation overhead in terms of the necessary number of measurement queries $Q$ to run circuits on quantum computers scales inefficiently as the number of subspaces increases, this overhead can be suppressed by reusing the same combinations of Pauli observables since they may appear cumulatively many times when computing each matrix element $\mathcal{S}_{ij}$ and $\mathcal{H}_{ij}$.
This is likely to hold for the Hamiltonian, for which the Pauli observables in $H^{m}=\sum_{k}{c}_{mk}P_{k}, m\leq M$ are mostly included in $H^{2M}=\sum_{k}{\tilde{c}}_{k}P_{k}$.
In this case, while the measurement overhead will scale roughly $O(|H^{2M}|)$ with $M$ subspaces where $|H|$ represents the number of terms when expanding $H$ into the sum of Pauli observables, $Q$ is roughly $M^2$ times reduced from the accumulative measurement of the same Pauli observables.
We numerically analyze the mitigation overhead of the measurement query by the power subspace in Section~\ref{sec:Leveraging_mitigation_overheads}.

Another point one might be concerned about is the sampling overhead relevant to the post-selection overhead when using DSP circuits to compute the trace of the square of density matrices in the power subspace.
We see that the sampling cost scales only quadratically to the post-selection probability of DSP circuits, which would not be an additional overhead compared to using ESD circuits as a subroutine.
We discuss the relationship between the sampling overhead and the post-selection probability in detail in Appendix~\ref{sec:appendix_Leveraging_sampling_cost_regarding_the_post-selection_probability_of_DSP_circuits} and demonstrate in Appendix~\ref{sec:appendix_Comparison_of_GSE_with_Different_Noisy_Subroutines} the advantage of using DSP over ESD under the noisy execution of both circuits to compute $\operatorname{Tr}\left[\rho^{2}\right]$.

\subsection{Fault subspace \label{sec:dgse_fault_subspace}}

\begin{figure}[htbp]
    \subfloat[\label{fig:dualfault_a}]{
        \begin{adjustbox}{width=0.48\textwidth}\begin{quantikz}
            \lstick{$|0\rangle$} & \gate[5, nwires=3]{\mathcal{U}^{(\mathrm{in})}} & \gate[5, nwires=3]{\mathcal{U}_{O}^{\dagger}} & \meter{Z} & \gate[5, nwires=3]{\mathcal{U}_{O}} & \gate[5, nwires=3]{\mathcal{U}^{(\mathrm{out})}} & \push{\ \langle 0|} \\
            \lstick{$|0\rangle$} &                                                 &                                             & \qw       &                                   &                                                  & \push{\ \langle 0|} \\
            \vdots               &                                                 &                                             &           &                                   &                                                  & \vdots \\
            \lstick{$|0\rangle$} &                                                 &                                             & \qw       &                                   &                                                  & \push{\ \langle 0|} \\
            \lstick{$|0\rangle$} &                                                 &                                             & \qw       &                                   &                                                  & \push{\ \langle 0|} \\
        \end{quantikz}\end{adjustbox}
    }
    \hfill
    \subfloat[\label{fig:dualfault_b}]{
        \begin{adjustbox}{width=0.48\textwidth}\begin{quantikz}
            \lstick{$|+\rangle$} & \qw                                             & \qw                                         & \ctrl{1} & \qw                               & \qw                                             & \meter{X,Y} \\
            \lstick{$|0\rangle$} & \gate[5, nwires=3]{\mathcal{U}^{(\mathrm{in})}} & \gate[5, nwires=3]{\mathcal{U}_{O}^{\dagger}} & \gate{Z} & \gate[5, nwires=3]{\mathcal{U}_{O}} & \gate[5, nwires=3]{\mathcal{U}^{(\mathrm{out})}} & \push{\ \langle 0|} \\
            \lstick{$|0\rangle$} &                                                 &                                             & \qw      &                                   &                                                  & \push{\ \langle 0|} \\
            \vdots               &                                                 &                                             &          &                                   &                                                  & \vdots \\
            \lstick{$|0\rangle$} &                                                 &                                             & \qw      &                                   &                                                  & \push{\ \langle 0|} \\
            \lstick{$|0\rangle$} &                                                 &                                             & \qw      &                                   &                                                  & \push{\ \langle 0|} \\
        \end{quantikz}\end{adjustbox}
    }
    \caption{
        The DSP circuits to obtain the matrix elements of $\mathcal{H}$ and $\mathcal{S}$ in the fault subspace method,
        (a) with direct mid-circuit measurement, and 
        (b) with indirect measurement.
        In these circuits, we can also assume the operator $O$ is a Pauli operator that can be conjugated into $Z_1$ via a Clifford operator $U_{O}$ as $O = U_{O} Z_1 U_{O}^{\dagger}$.
        The gates $\mathcal{U}_{O}$ and $\mathcal{U}_{O}^{\dagger}$ in the figures denote the noisy operations of $U_{O}$ and $U_{O}^{\dagger}$ respectively.
    }
    \label{fig:dualfault}
\end{figure}

We can also perform the fault subspace type GSE with only one copy of the quantum state using DSP quantum circuits.
Let $\rho(\epsilon)=\mathcal{U}(\epsilon)(\rho_{\vec{0}})$ and $\bar{\rho}(\epsilon )=\mathcal{U}_{\mathrm{dual}}(\rho_{\vec{0}})$ denote the noisy quantum state with error rate $\epsilon$ and its dual state, two types of the fault subspace ansatz are provided,
\begin{equation}\label{eq:rho_dgse_fs}
\begin{split}
    \rho_{\mathrm{fs1}} &= \sum_{i,j} \alpha_{i}^{*} \alpha_{j} \frac{\bar{\rho}(\lambda_{i} \epsilon) \rho(\lambda_{j} \epsilon)+\rho(\lambda_{j}\epsilon)\bar{\rho}(\lambda_{i} \epsilon)}{2}, \\ 
    \rho_{\mathrm{fs2}} &=  \sum_{i,j} \alpha_{i}^{*} \alpha_{j} \bar{\rho}(\lambda_{i} \epsilon) \rho(\lambda_{j} \epsilon). 
\end{split}
\end{equation}
Here we call $\rho_{\mathrm{fs1}}$ the symmetrized ansatz and $\rho_{\mathrm{fs2}}$ the simplified ansatz.
Note that $\rho_{\mathrm{fs1}}$ and $\rho_{\mathrm{fs2}}$ are equivalent when $\bar{\rho}(\lambda_{i} \epsilon)$ and $\rho(\lambda_{j} \epsilon)$ commute with each other.

To compute the expectation value of physical quantities using the fault subspace ansatz, we use the quantum circuits in Fig.~\ref{fig:dualfault}.
Setting $\mathcal{U}^{(\mathrm{in})}= \mathcal{U}(\lambda_{\mathrm{in}} \epsilon)$ and $\mathcal{U}^{(\mathrm{out})}= \mathcal{U}_{\mathrm{rev}}(\lambda_{\mathrm{out}} \epsilon)$, these circuits give
\begin{equation}\label{eq:expval_dgse_fs_symmetrized}
\begin{split}
    \left\langle X\otimes \operatorname{P}_{\vec{0}}\right\rangle_{O} = \operatorname{Tr}\left[\frac{\bar{\rho}(\lambda_{\mathrm{out}} \epsilon) \rho(\lambda_{\mathrm{in}} \epsilon) + \rho(\lambda_{\mathrm{in}} \epsilon)\bar{\rho}(\lambda_{\mathrm{out}} \epsilon)}{2} O \right], \\
    \left\langle Y\otimes \operatorname{P}_{\vec{0}}\right\rangle_{O} = \operatorname{Tr}\left[\frac{\bar{\rho}(\lambda_{\mathrm{out}} \epsilon) \rho(\lambda_{\mathrm{in}} \epsilon) - \rho(\lambda_{\mathrm{in}} \epsilon)\bar{\rho}(\lambda_{\mathrm{out}} \epsilon)}{-2i} O \right],
\end{split}
\end{equation}
for the symmetrized ansatz $\rho_{\mathrm{fs1}}$.
This also induces the following asymmetry form for the simplified ansatz $\rho_{\mathrm{fs2}}$.
\begin{equation}\label{eq:expval_dgse_fs_asymmetrized}
\begin{split}
    \langle X \otimes \operatorname{P}_{\vec{0}}\rangle_{O} + i\langle Y \otimes \operatorname{P}_{\vec{0}}\rangle_{O} = \operatorname{Tr}\left[\rho(\lambda_{\mathrm{in}}\epsilon) \bar{\rho}(\lambda_{\mathrm{out}}\epsilon) O \right], \\
    \langle X \otimes \operatorname{P}_{\vec{0}}\rangle_{O} - i\langle Y \otimes \operatorname{P}_{\vec{0}}\rangle_{O} = \operatorname{Tr}\left[\bar{\rho}(\lambda_{\mathrm{out}}\epsilon) \rho(\lambda_{\mathrm{in}}\epsilon) O \right].
\end{split}
\end{equation}
Note that the fault subspace ansatz also allows unphysical subspace expansions discussed in the power subspace ansatz.

The typical fault subspace might be $\{\rho(\lambda_{k}\epsilon)\}_{k=1,2,\ldots,M}$ with $\lambda_{k} \geq 1$, as introduced in Ref.~\cite{Yoshioka2022-gq}, which covers the noise extrapolation methods as well.
Then the matrix elements $\mathcal{S}_{ij}$ and $\mathcal{H}_{ij}$ in Eq.~\eqref{eq:generalized_eigenvalue_problem} are expressed as
\begin{equation}\label{eq:S_ij_fs}
\begin{split}
    \mathcal{S}_{i j} 
    =\operatorname{Tr}\left[\frac{\bar{\rho}_{j} \rho_{i}+\rho_{i} \bar{\rho}_{j}}{2}\right]
    =\operatorname{Tr}\left[\bar{\rho}(\lambda_{j}\epsilon) \rho(\lambda_{i}\epsilon) \right],
\end{split}
\end{equation}
\begin{equation}\label{eq:H_ij_fs}
\begin{split}
    \mathcal{H}_{i j} 
    &=\operatorname{Tr}\left[\frac{\bar{\rho}_{j} \rho_{i}+\rho_{i} \bar{\rho}_{j}}{2}H\right] \\
    &=\sum_{h} c_{h}\operatorname{Tr}\left[\frac{\bar{\rho}(\lambda_{j}\epsilon) \rho(\lambda_{i}\epsilon)+\rho(\lambda_{i}\epsilon) \bar{\rho}(\lambda_{j}\epsilon)}{2}P_{h}\right],
\end{split}
\end{equation}
assuming $H = \sum_{h}c_{h}P_{h}$.
The noise amplification can be implemented in the same way as used in the noise extrapolation methods either at the hardware~\cite{Temme2017-vo, Kandala2019-ze} or software~\cite{GiurgicaTiron2020} level.

The mitigation overhead of this subspace construction scales $O(M^2|H|)$ to the number of subspaces $M$, for examining the overlap of input and output quantum states with every combination of error scales $\lambda_{\mathrm{in}}, \lambda_{\mathrm{out}}\in\{\lambda_{1}, \lambda_{2}, \ldots, \lambda_{M}\}$.

\subsection{Divide-and-conquer subspace \label{sec:dgse_divide-and-conquer_subspace}}

We can further design subspaces for Dual-GSE to simulate larger quantum systems while recovering the purified expectation values as well.
Our subspace design is motivated by the ``divide-and-conquer'' strategy~\cite{eddins2022doubling, Mizuta2021-pw, Fujii2022, sun2022perturbative}, where the computation task on a large quantum device is first divided into fractions of small problems that can be run on smaller quantum devices, and then the fractions are classically combined through postprocessing measurement outcomes, facilitating a virtual simulation of the entanglement among the divided subsystems.
In a similar spirit, we construct a divide-and-conquer subspace ansatz to induce entanglement classically among the subsystems that follow the subvolume law.
In the following part, we introduce an example of the divide-and-conquer strategy in our Dual-GSE.

To begin with, let us consider a quantum state in the composite system of subsystems $\mathrm{A}$ and $\mathrm{B}$, 
\begin{equation}
    |\psi^{(\mathrm{AB})}\rangle=\sum_{i} \alpha_{i} C_{i} \left(|\psi^{(\mathrm{A})}\rangle \otimes |\psi^{(\mathrm{B})}\rangle\right),
\label{Eq: forge}
\end{equation}
where $\alpha_{i} \in \mathbb{C}$ are classical coefficients, $|\psi^{(\mathrm{A})}\rangle$ and $|\psi^{(\mathrm{B})}\rangle$ are quantum states in the subsystem $\mathrm{A}$ and $\mathrm{B}$, and $C_{i}$ is a general operator that can be linearly expanded
\begin{equation}
    C_{i}=\sum_{\gamma} g_{i\gamma} \left(P_{i\gamma}^{(\mathrm{A})} \otimes P_{i\gamma}^{(\mathrm{B})}\right),
\end{equation}
for $g_{i\gamma} \in \mathbb{C}$ with $C_{1}=I$.
Here, $|\psi^{(\mathrm{AB})}\rangle$ is generally an entangled state and is assumed to be normalized as a state vector by adjusting $\{\alpha_{i}\}_{i}$ properly.
For a Hamiltonian $H=\sum_{h} c_{h} \left(P_{h}^{(\mathrm{A})} \otimes P_{h}^{(\mathrm{B})}\right)$, we can compute the expectation value as:
\begin{equation}
\begin{split}
    \langle H\rangle
    &=\sum_{i, i', \gamma, \gamma', h} c_{h} \alpha_{i'}^{*} \alpha_{i} g_{i'\gamma'}^{*} g_{i\gamma} \\
    &\quad\quad\quad\quad \times \langle \psi^{(\mathrm{A})}| P_{i'\gamma'}^{(\mathrm{A})} P_{h}^{(\mathrm{A})} P_{i\gamma}^{(\mathrm{A})} |\psi^{(\mathrm{A})}\rangle \\
    &\quad\quad\quad\quad \times \langle \psi^{(\mathrm{B})}| P_{i'\gamma'}^{(\mathrm{B})} P_{h}^{(\mathrm{B})} P_{i\gamma}^{(\mathrm{B})} |\psi^{(\mathrm{B})}\rangle.
\end{split}
\end{equation}
This indicates that the expectation value of the entangled state can be computed via the Pauli measurements on each subsystem independently.

The ansatz state corresponding to Eq.~\eqref{Eq: forge} in density matrix representation is
\begin{equation}\label{eq:rho_dgse_dc_degree_1}
    \rho^{(\mathrm{AB})} = \sum_{i,j}\alpha_{i} \alpha_{j}^{*} C_{i} \left(\rho^{(\mathrm{A})} \otimes \rho^{(\mathrm{B})}\right) C_{j}^{\dagger},
\end{equation}
where $\rho^{(\mathrm{A})} = |\psi^{(\mathrm{A})}\rangle\langle\psi^{(\mathrm{A})}|$ and $\rho^{(\mathrm{B})} = |\psi^{(\mathrm{B})}\rangle\langle\psi^{(\mathrm{B})}|$.

We first consider the ansatz with a squared density matrix in each subsystem following the original GSE procedure:
\begin{equation}\label{eq:rho_dgse_dc_degree_2}
    \sum_{i,j} \alpha_{i}^{*} \alpha_{j} C_{i}^{\dagger} \left(\left(\rho^{(\mathrm{A})}\right)^2 \otimes \left(\rho^{(\mathrm{B})}\right)^2\right) C_{j}.
\end{equation}
This corresponds to setting the subspace to $\mathcal{A}=I$ and $\sigma_{i}=\left(\rho^{(\mathrm{A})} \otimes \rho^{(\mathrm{B})}\right) C_{i}$ in Eq.~\eqref{eq:rho_gse_general}.
To solve the generalized eigenvalue problem, we have to measure the matrix elements of
\begin{equation}\label{eq:S_ij_gse_dc}
\begin{split}
    \mathcal{S}_{ij}^{\prime}
    &= \operatorname{Tr}\left[ C_{i}^{\dagger} \left(\left(\rho^{(\mathrm{A})}\right)^2 \otimes \left(\rho^{(\mathrm{B})}\right)^2\right) C_{j} \right] \\
    &=\sum_{\gamma, \gamma'}  g_{i\gamma'}^{*} g_{j\gamma} \operatorname{Tr}\left[ \left(\rho^{(\mathrm{A})}\right)^2 P_{j\gamma}^{(\mathrm{A})} P_{i\gamma'}^{(\mathrm{A})} \right] \\
    &\quad\quad\quad\quad\quad\quad \times\operatorname{Tr}\left[ \left(\rho^{(\mathrm{B})}\right)^2 P_{j\gamma}^{(\mathrm{B})}  P_{i\gamma'}^{(\mathrm{B})} \right],
\end{split}
\end{equation}
\begin{equation}\label{eq:H_ij_gse_dc}
\begin{split}
    \mathcal{H}_{ij}^{\prime}
    &= \operatorname{Tr}\left[ C_{i}^{\dagger} \left(\left(\rho^{(\mathrm{A})}\right)^2 \otimes \left(\rho^{(\mathrm{B})}\right)^2\right) C_{j} H\right] \\
    &=\sum_{h,\gamma, \gamma'} c_{h} g_{i\gamma'}^{*} g_{j\gamma} \operatorname{Tr}\left[ \left(\rho^{(\mathrm{A})}\right)^2 P_{j\gamma}^{(\mathrm{A})} P_{h}^{(\mathrm{A})} P_{i\gamma'}^{(\mathrm{A})} \right] \\
    &\quad\quad\quad\quad\quad\quad\quad\quad \times \operatorname{Tr}\left[ \left(\rho^{(\mathrm{B})}\right)^2 P_{j\gamma}^{(\mathrm{B})} P_{h}^{(\mathrm{B})} P_{i\gamma'}^{(\mathrm{B})}\right].
    \end{split}
\end{equation}
Here, every trace term can be obtained as an expectation value by the ESD/VD subroutine in GSE.

This illustrates the divide-and-conquer strategy for GSE to classically reintroduce the entanglement between $\mathrm{A}$ and $\mathrm{B}$ to the expectation value by including $H$ or nonlocal $C_{i}$ to subspaces. 
Therefore, we can virtually enlarge the simulated system size beyond that of actual quantum devices.
However, the requirement of state copies in ESD/VD counteracts the advantage of this divide-and-conquer strategy.

To avoid this drawback in GSE, we again employ the DSP method in Dual-GSE, accordingly changing Eq.~\eqref{eq:S_ij_gse_dc} and Eq.~\eqref{eq:H_ij_gse_dc} to
\begin{equation}\label{eq:dgse_dc_S_ij}
\begin{split}
    \mathcal{S}_{ij}^{\prime\prime}
    &= \operatorname{Tr}\left[ C_{i}^{\dagger} \left(\frac{\bar{\rho}^{(\mathrm{A})} \rho^{(\mathrm{A})} + \rho^{(\mathrm{A})} \bar{\rho}^{(\mathrm{A})}}{2} \right.\right. \\
    &\quad\quad\quad\quad\quad \left.\left.\otimes \frac{\bar{\rho}^{(\mathrm{B})} \rho^{(\mathrm{B})} + \rho^{(\mathrm{B})} \bar{\rho}^{(\mathrm{B})}}{2}\right) C_{j} \right] \\
    &=\sum_{\gamma, \gamma'}  g_{i\gamma'}^{*} g_{j\gamma} \operatorname{Tr}\left[ \frac{\bar{\rho}^{(\mathrm{A})} \rho^{(\mathrm{A})} + \rho^{(\mathrm{A})} \bar{\rho}^{(\mathrm{A})}}{2} P_{j\gamma}^{(\mathrm{A})} P_{i\gamma'}^{(\mathrm{A})} \right] \\
    &\quad\quad\quad\quad\quad\quad \times \operatorname{Tr}\left[ \frac{\bar{\rho}^{(\mathrm{B})} \rho^{(\mathrm{B})} + \rho^{(\mathrm{B})} \bar{\rho}^{(\mathrm{B})}}{2} P_{j\gamma}^{(\mathrm{B})} P_{i\gamma'}^{(\mathrm{B})} \right],
\end{split}
\end{equation}
\begin{equation}\label{eq:dgse_dc_H_ij}
\begin{split}
    \mathcal{H}_{ij}^{\prime\prime}
    &= \operatorname{Tr}\left[ C_{i}^{\dagger} \left(\frac{\bar{\rho}^{(\mathrm{A})} \rho^{(\mathrm{A})} + \rho^{(\mathrm{A})} \bar{\rho}^{(\mathrm{A})}}{2} \right.\right. \\
    &\quad\quad\quad\quad\quad \left.\left.\otimes \frac{\bar{\rho}^{(\mathrm{B})} \rho^{(\mathrm{B})} + \rho^{(\mathrm{B})} \bar{\rho}^{(\mathrm{B})}}{2}\right) C_{j} H \right]  \\
    &=\sum_{h,\gamma, \gamma'} c_{h} g_{i\gamma'}^{*} g_{j\gamma} \\
    &\quad\quad\quad\quad \times \operatorname{Tr}\left[ \frac{\bar{\rho}^{(\mathrm{A})} \rho^{(\mathrm{A})} + \rho^{(\mathrm{A})} \bar{\rho}^{(\mathrm{A})}}{2} P_{j\gamma}^{(\mathrm{A})} P_{h}^{(\mathrm{A})} P_{i\gamma'}^{(\mathrm{A})}\right] \\
    &\quad\quad\quad\quad \times \operatorname{Tr}\left[ \frac{\bar{\rho}^{(\mathrm{B})} \rho^{(\mathrm{B})} + \rho^{(\mathrm{B})} \bar{\rho}^{(\mathrm{B})}}{2} P_{j\gamma}^{(\mathrm{B})} P_{h}^{(\mathrm{B})} P_{i\gamma'}^{(\mathrm{B})} \right],
\end{split}
\end{equation}
which corresponds to the simulation of the quantum state
\begin{equation}\label{Eq: twice}
\begin{split}
    &\sum_{i,j} \alpha_{i}^{*} \alpha_{j} C_{i}^{\dagger} \left(\frac{\bar{\rho}^{(\mathrm{A})} \rho^{(\mathrm{A})}+ \rho^{(\mathrm{A})} \bar{\rho}^{(\mathrm{A})}}{2} \right. \\
    &\quad\quad\quad\quad\quad\quad\quad\left.\otimes \frac{\bar{\rho}^{(\mathrm{B})} \rho^{(\mathrm{B})} + \rho^{(\mathrm{B})} \bar{\rho}^{(\mathrm{B})}}{2}\right) C_{j}.
\end{split}
\end{equation}
This allows for simulating a system that is twice the size of quantum devices without additional state copy overhead, as well as mitigating the effect of computational errors and device noise.

We can further design an ansatz that performs the power subspace in a divide-and-conquer way.
Choosing the set of subspaces as $\{\sigma_{i} \}_{i}=\{I^{\mathrm{(AB)}}\} \cup \left\{\left(\rho^{(\mathrm{A})} \otimes \rho^{(\mathrm{B})}\right) C_{k} \right\}_{k\geq 2}$, we define the following ansatz
\begin{equation}
\begin{split}
    \rho_{\mathrm{dc}}^{(\mathrm{AB})} &= \left(\alpha_{1}^{*} I^{\mathrm{(AB)}} + \sum_{k\geq2} \alpha_{k}^{*} C_{k}^{\dagger} \left(\rho^{(\mathrm{A})} \otimes \rho^{(\mathrm{B})}\right) \right)\\
    &\quad \times \left(\alpha_{1} I^{\mathrm{(AB)}} + \sum_{k\geq2} \alpha_{k} \left(\rho^{(\mathrm{A})} \otimes \rho^{(\mathrm{B})}\right) C_{k}\right),
\end{split}
\end{equation}
which we name ``divide-and-conquer subspace''.
Since the maximum power degree of $\rho^{(\mathrm{A})}$ and $\rho^{(\mathrm{B})}$ is two, the matrix elements of $\mathcal{H}$ and $\mathcal{S}$ can be approximated without state copies thanks to the DSP circuits in Fig.~\ref{fig:qc_dsp}.

The above example using two subsystems can be trivially generalized into multiple subsystems.
By dividing the system $\mathrm{S}$ covering the whole Hilbert space into a set of subsystems $\{\mathrm{S}_{i}\}_{i}$, the divide-and-conquer subspace can be represented as
\begin{equation}
    \{\sigma_{i}\}_{i} = \{I^{\mathrm{(S)}}\}\cup\left\{\left(\bigotimes_{i}\tau^{(\mathrm{S}_{i})}\right)C_{k}\right\}_{k\geq 2},
\end{equation}
where $\tau^{(\mathrm{S}_{i})} = V^{(\mathrm{S}_{i})}\rho^{(\mathrm{S}_{i})}W^{(\mathrm{S}_{i})\dagger}$.
Note that different subsystems $\mathrm{S}_{i}, \mathrm{S}_{j}$ can have different dimensions, and the effective operation in $C_{k}$ does not necessarily cover the whole system $\mathrm{S}$.

As a specific example, the subspace construction with $C_{k}=H^{k-2}$, $M$ subspaces, and $N_{\mathrm{dc}}$ subsystems $\{\mathrm{S}_{1}, \ldots, \mathrm{S}_{N_{\mathrm{dc}}}\}$ offers
\begin{equation}
    \{\sigma_{i}\}_{i} = \{I^{(\mathrm{S})}\}\cup\left\{\left(\bigotimes_{l=1}^{N_{\mathrm{dc}}}\rho^{(\mathrm{S}_{l})}\right)H^{k-2}\right\}_{k=2,\ldots,M}.
\end{equation}
Defining $\operatorname{P}_{\vec{0}}^{(\mathrm{S}_{l})}=|\vec{0}^{(\mathrm{S}_{l})}\rangle\langle\vec{0}^{(\mathrm{S}_{l})}|$ as a projector onto the all-zero bitstring in each subsystem, the matrix elements $\mathcal{S}_{ij}$ and $\mathcal{H}_{ij}$ for $i, j \geq 2$ in Eq.~\eqref{eq:generalized_eigenvalue_problem} become
\begin{equation}\label{eq:S_ij_dc}
\begin{split}
    \mathcal{S}_{i j}
    &=\operatorname{Tr}\left[\left(\bigotimes_{l=1}^{N_{\mathrm{dc}}}\frac{\bar{\rho}_{j}^{(\mathrm{S}_{l})} \rho_{i}^{(\mathrm{S}_{l})}+\rho_{i}^{(\mathrm{S}_{l})} \bar{\rho}_{j}^{(\mathrm{S}_{l})}}{2}\right)H^{i+j-4}\right] \\
    &=\sum_{k} c_{k}\prod_{l=1}^{N_{\mathrm{dc}}}\langle X \otimes \operatorname{P}_{\vec{0}}^{(\mathrm{S}_{l})}\rangle_{P_{k}^{(\mathrm{S}_{l})}},
\end{split}
\end{equation}
\begin{equation}\label{eq:H_ij_dc}
\begin{split}
    \mathcal{H}_{i j}
    &=\operatorname{Tr}\left[\left(\bigotimes_{l=1}^{N_{\mathrm{dc}}}\frac{\bar{\rho}_{j}^{(\mathrm{S}_{l})} \rho_{i}^{(\mathrm{S}_{l})}+\rho_{i}^{(\mathrm{S}_{l})} \bar{\rho}_{j}^{(\mathrm{S}_{l})}}{2}\right)H^{i+j-3}\right] \\
    &=\sum_{k}{c}_{k}^{\prime}\prod_{l=1}^{N_{\mathrm{dc}}}\langle X \otimes \operatorname{P}_{\vec{0}}^{(\mathrm{S}_{l})}\rangle_{P_{k}^{(\mathrm{S}_{l})}},
\end{split}
\end{equation}
where we assume the powers of Hamiltonian are decomposed as $H^{i+j-4} = \sum_{k} c_{k}\left(\bigotimes_{l=1}^{N_{\mathrm{dc}}} P_{k}^{(\mathrm{S}_{l})}\right)$ and $H^{i+j-3} = \sum_{k}{c}_{k}^{\prime}\left(\bigotimes_{l=1}^{N_{\mathrm{dc}}} P_{k}^{(\mathrm{S}_{l})}\right)$.
Other matrix elements also follow the same argument we provided for the power subspace.

The advantage of the divide-and-conquer subspace can also be found in significantly smaller query overhead compared to the power subspace.
Let us define $Q_{\mathrm{ps}}$ and $Q_{\mathrm{dc}}$ as the number of measurement queries required in the power subspace and the divide-and-conquer subspace.
For a given Hamiltonian $H$ described as $H=\sum_{h}c_{h}P_{h}$ with Pauli operators $P_{h}\in\{I,X,Y,Z\}^{\otimes N}$ over an $N$-site system, $Q_{\mathrm{ps}}$ scales in $O(|H^{2M}|)$ to the number of subspaces $M$ as discussed before.
When expanding the subspaces, $|H^{2M}|$ covers $4^N$ Pauli observables in the worst case.
On the other hand, $Q_{\mathrm{dc}}$ scales at worst in $O\left(4^{N/N_{\mathrm{dc}}}\right)$. 
Thus, by carefully designing the subspaces, the divide-and-conquer strategy has the potential to significantly ease the blow-up of measurement queries thanks to the smaller size of divided subsystems.

Note that reducing the number of measured Pauli operators consequently contributes to suppressing the estimation variance of each measurement query, and thus further lowering the sampling cost.
In Section~\ref{sec:Leveraging_mitigation_overheads}, we numerically show that $Q_{\mathrm{dc}}$ with two divided subsystems increases much less than $Q_{\mathrm{ps}}$ by up to two orders of magnitude for executing the same number of subspaces.
Therefore, in addition to executing the same number of subspaces with fewer measurement queries by the reusing strategy as discussed for the power subspace, the divide-and-conquer strategy further contributes to reducing the measurement queries.

The divide-and-conquer overhead can be evaluated with the optimized classical parameters $\vec{\alpha}^{\prime}$ associated with the normalized overlap matrix $\tilde{\mathcal{S}}$ where its diagonal elements are unity.
The effect of the divide-and-conquer strategy and the impact of gate noise and shot noise can be discussed separately.
Without the gate noise, the sampling cost scales in $O\left(\left\|\vec{\alpha}^{\prime}\right\|_{2}^{4}\right)$ when dividing the whole system to two symmetric subsystems.
The noise effect worsens this scale by a $\left(1-p\right)^{-8}$ when each subsystem suffers from the global depolarizing error rate $p$.
We derive these evaluations in Appendix~\ref{sec:appendix_Leveraging_Sampling_Cost_Regarding_Divide-and-conquer_Overhead} and see the reusing strategy of measurement results will reduce the divide-and-conquer overhead.

It should also be mentioned that the efficacy of this divide-and-conquer strategy would be scoped to the case where the entanglement among subsystems follows the area law or subvolume law.
In other words, the divide-and-conquer strategy would effectively reintroduce the entanglement into expectation values when the number of nontrivial Schmidt coefficients among subsystems does not scale exponentially.
This is also a common restriction among other divide-and-conquer methods~\cite{eddins2022doubling, Mizuta2021-pw, Fujii2022, sun2022perturbative,Yuan2021}.
Nevertheless, since many physical models of our interest have a Hamiltonian fitting this local condition, Dual-GSE still has a huge potential to enhance the accuracy of physical simulations.

\section{Numerical Simulation \label{sec:numerical_simulaion}}

\begin{figure}
    \centering
    \subfloat[\label{fig:path_graph_8}]{
        \begin{tikzpicture}
            \node[shape=circle,draw=black] (0) at (0,0) {};
            \node[shape=circle,draw=black] (1) at (1,0) {};
            \node[shape=circle,draw=black] (2) at (2,0) {};
            \node[shape=circle,draw=black] (3) at (3,0) {};
            \node[shape=circle,draw=black] (4) at (4,0) {};
            \node[shape=circle,draw=black] (5) at (5,0) {};
            \node[shape=circle,draw=black] (6) at (6,0) {};
            \node[shape=circle,draw=black] (7) at (7,0) {};
            \path [-] (0) edge (1);
            \path [-] (1) edge (2);
            \path [-] (2) edge (3);
            \path [-] (3) edge (4);
            \path [-] (4) edge (5);
            \path [-] (5) edge (6);
            \path [-] (6) edge (7);
        \end{tikzpicture}
    }
    \hfill
    \subfloat[\label{fig:path_graph_4_4}]{
        \begin{tikzpicture}
            \node[shape=circle,draw=black] (0) at (0,0) {};
            \node[shape=circle,draw=black] (1) at (1,0) {};
            \node[shape=circle,draw=black] (2) at (2,0) {};
            \node[shape=circle,draw=black] (3) at (3,0) {};
            \node[shape=circle,draw=black] (4) at (4,0) {};
            \node[shape=circle,draw=black] (5) at (5,0) {};
            \node[shape=circle,draw=black] (6) at (6,0) {};
            \node[shape=circle,draw=black] (7) at (7,0) {};
            \path [-] (0) edge (1);
            \path [-] (1) edge (2);
            \path [-] (2) edge (3);
            \path [dotted] (3) edge (4);
            \path [-] (4) edge (5);
            \path [-] (5) edge (6);
            \path [-] (6) edge (7);
        \end{tikzpicture}
    }
    \caption{
        The graph structure of the Ising model used in the numerical simulation. 
        (a) The eight-qubit system is used to simulate the power subspace and fault subspace. 
        (b) Two four-qubit systems are used for the divide-and-conquer subspace simulation, where the entanglement between these two systems is recovered classically.
        }
    \label{fig:path_graphs}
\end{figure}

\begin{figure}[htbp]
    \begin{adjustbox}{width=0.48\textwidth}
    \begin{quantikz}
        \qw & \gate{R_X\left(\theta_{0}^{(l)}\right)} \gategroup[6,steps=4,style={dashed, rounded corners, inner sep=6pt}, label style={label position=below, anchor=north, yshift=-0.2cm}]{\Large $\times L$ layers} 
                                             & \gate{R_Z\left(\theta_{N}^{(l)}\right)}    & \ctrl{1}   & \qw        & \qw    & \gate{R_X\left(\theta_{0}^{(L+1)}\right)}   & \gate{R_Z\left(\theta_{N}^{(L+1)}\right)}    & \qw \\
        \qw & \gate{R_X\left(\theta_{1}^{(l)}\right)}   & \gate{R_Z\left(\theta_{N+1}^{(l)}\right)}  & \control{} & \ctrl{1}   & \qw    & \gate{R_X\left(\theta_{1}^{(L+1)}\right)}   & \gate{R_Z\left(\theta_{N+1}^{(L+1)}\right)}  & \qw \\
        \qw & \gate{R_X\left(\theta_{2}^{(l)}\right)}   & \gate{R_Z\left(\theta_{N+2}^{(l)}\right)}  & \ctrl{1}   & \control{} & \qw    & \gate{R_X\left(\theta_{2}^{(L+1)}\right)}   & \gate{R_Z\left(\theta_{N+2}^{(L+1)}\right)}  & \qw \\
            & \vdots                         & \vdots                          & \vdots     &            &        & \vdots                           & \vdots                            &     \\
        \qw & \gate{R_X\left(\theta_{N-2}^{(l)}\right)} & \gate{R_Z\left(\theta_{2N-2}^{(l)}\right)} & \ctrl{-1}  & \ctrl{1}   & \qw    & \gate{R_X\left(\theta_{N-2}^{(L+1)}\right)} & \gate{R_Z\left(\theta_{2N-2}^{(L+1)}\right)} & \qw \\
        \qw & \gate{R_X\left(\theta_{N-1}^{(l)}\right)} & \gate{R_Z\left(\theta_{2N-1}^{(l)}\right)} & \qw        & \control{} & \qw    & \gate{R_X\left(\theta_{N-1}^{(L+1)}\right)} & \gate{R_Z\left(\theta_{2N-1}^{(L+1)}\right)} & \qw
    \end{quantikz}
    \end{adjustbox}
    \caption{
        We estimate the ground state energy of $H$ with the parameterized quantum circuit with $L=8$ layers. Each layer consists of parameterized $R_X$ and $R_Z$ gates, followed by controlled-$Z$ gates.
    }
    \label{fig:qc_vqe}
\end{figure}

\begin{figure*}[htbp]
    \centering
    \includegraphics[width=1.0\textwidth]{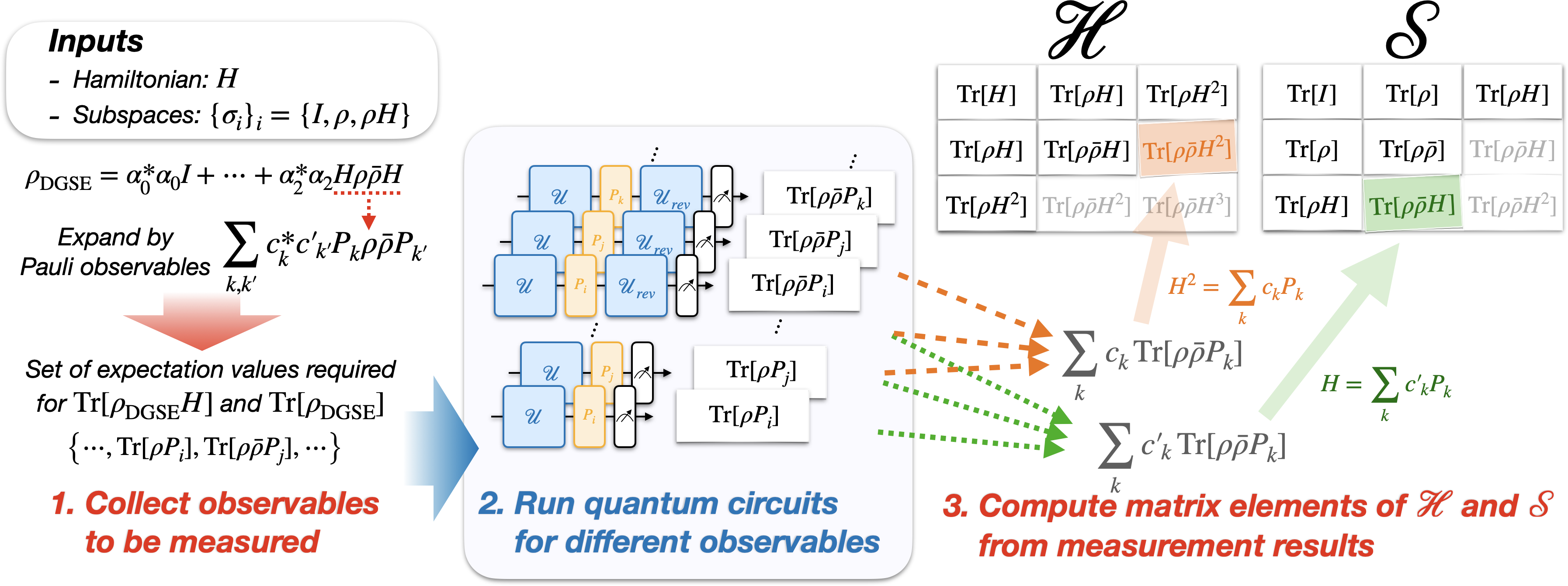}
    \caption{
        The schematic workflow of the classical pre- and postprocessing in Dual-GSE. 
        To save the number of queries of using the quantum device, the expectation value of each combination of a Pauli observable and quantum ansatz is computed in advance. 
        Note that the procedures except for step 2 are performed on classical computers.
    }
    \label{fig:procedure_dgse}
\end{figure*}

We numerically demonstrate our proposed method with an eight-qubit transverse-field Ising model whose Hamiltonian is given by
\begin{equation}\label{eq:hamiltonian}
    H= - \sum_{(i,j)\in E}Z_{i}Z_{j} - \sum_{i\in V}X_{i}
\end{equation}
for a path graph structure $G=(V, E)$ shown in Fig.~\ref{fig:path_graphs}(a).
Here, we define $N$ as the number of sites in the Ising model.
In our case, $N=8$.
Mapping each site to each qubit on a quantum computer, our task is to improve the estimated ground state energy of the Hamiltonian $H$ by first optimizing the variational quantum circuit as shown in Fig.~\ref{fig:qc_vqe} and then performing Dual-GSE.

\begin{table}[htbp]
    \centering
    \begin{tabular}{ ll }
        \hline
         & subspaces \\
        \hline
        \hline
        power              & $\{I\}\cup\{\rho H^{k-2}\}_{k=2,\ldots,M}$ \\
        fault              & $\{\rho(\lambda\epsilon)\}_{\lambda=1,\ldots,M}$ \\
        divide-and-conquer & $\{I^{(\mathrm{AB})}\}\cup\{(\rho^{(\mathrm{A})}\otimes\rho^{(\mathrm{B})})H^{k-2}\}_{k=2,\ldots,M}$ \\
        \hline
    \end{tabular}
    \caption{
        The subspaces used in the numerical simulation.
    }
    \label{tab:subspaces}
\end{table}

\begin{figure*}[htbp]
    \raggedright
    \subfloat[power subspace\label{fig:local-stochastic-pauli_power-subspace_subspace-to-diff}]{
        \includegraphics[width=0.32\textwidth]{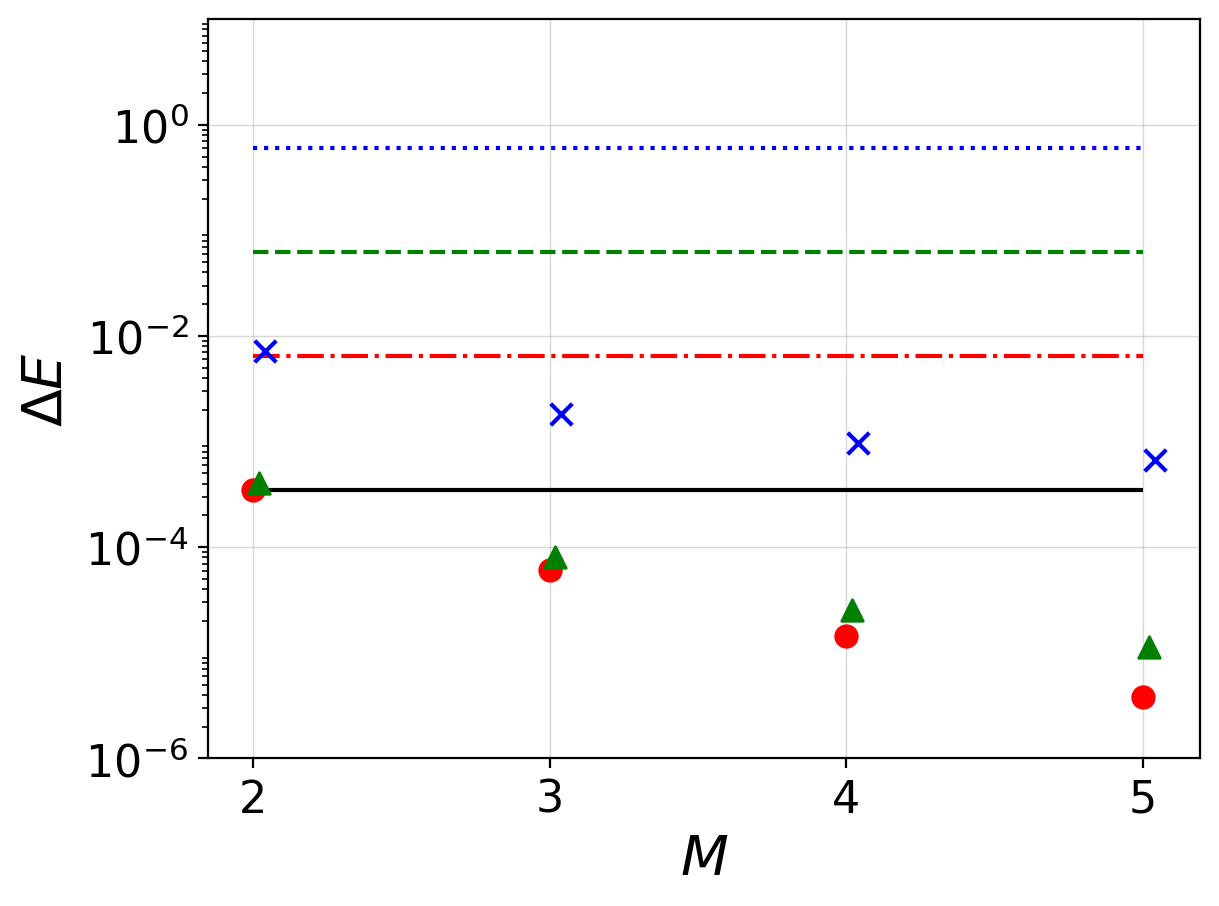}
    }
    \subfloat[fault subspace\label{fig:local-stochastic-pauli_fault-subspace_subspace-to-diff}]{
        \includegraphics[width=0.32\textwidth]{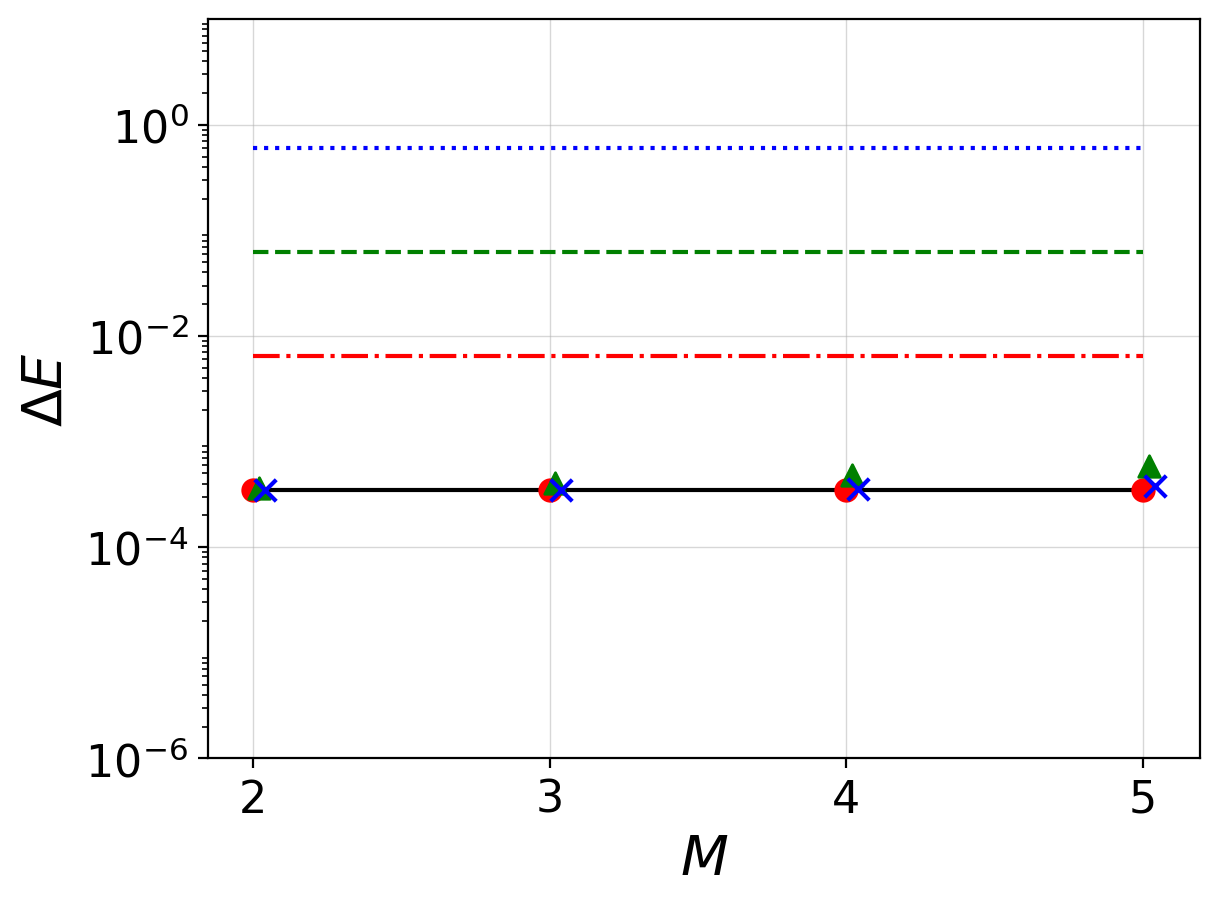}
    }
    \subfloat[divide-and-conquer subspace\label{fig:local-stochastic-pauli_cfe-subspace_subspace-to-diff}]{
        \includegraphics[width=0.32\textwidth]{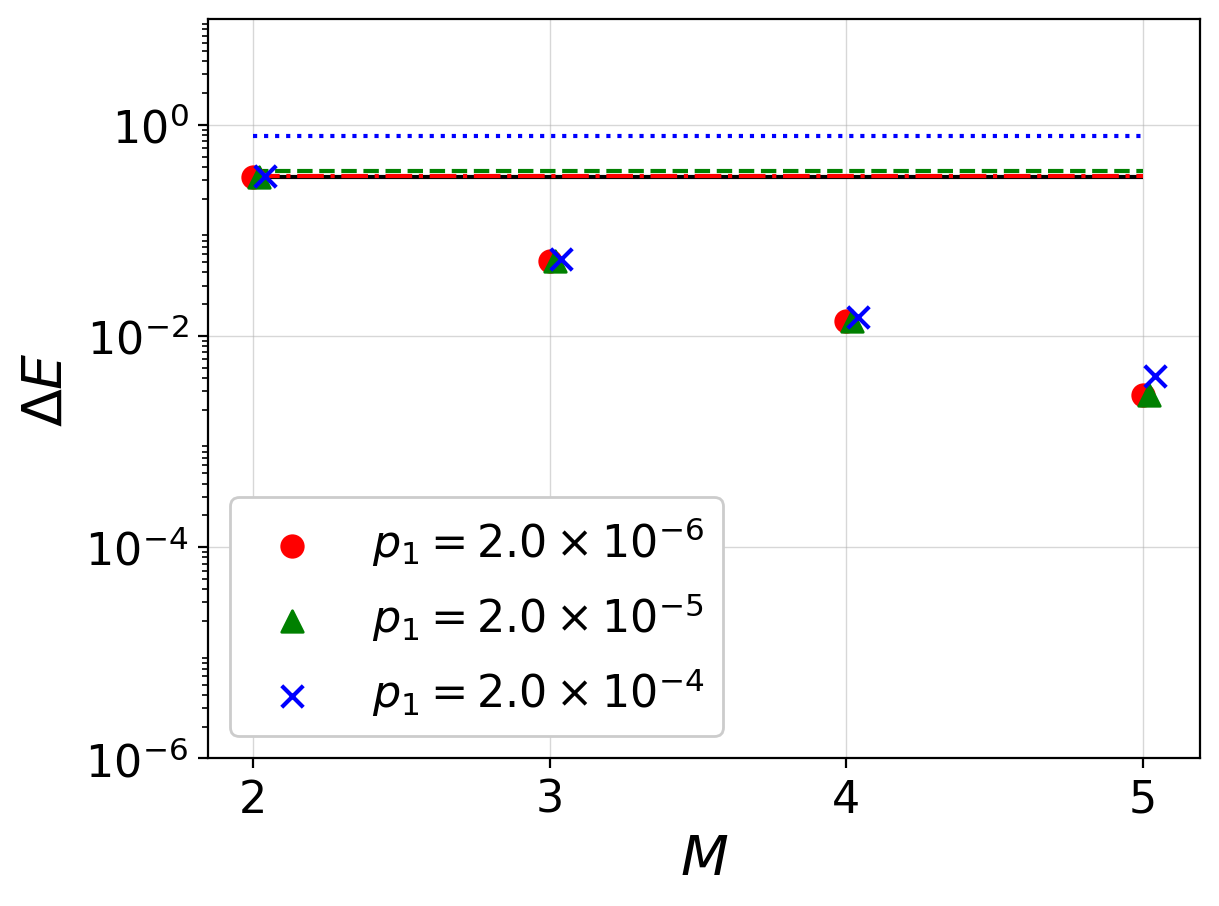}
    }
    \caption{
        The estimation bias $\Delta E$ from the theoretical ground state energy, computed by Dual-GSE with (a) the power subspace, (b) the fault subspace, and (c) the divide-and-conquer subspace.
        In each figure, the results under different noise levels are shown with different colors and symbols.
        The horizontal black line denotes the inherent estimation bias under the noise-free optimization of variational quantum circuits using the ansatz (a, b) over the whole eight-qubit system and (c) over the tensor product of two four-qubit subsystems.
        In every figure, the cross blue markers and the dotted blue line represent the estimation bias with and without Dual-GSE under the local stochastic noise with magnitude $2.0\times 10^{-4}$, the triangle green markers and the dashed green line with magnitude $2.0\times 10^{-5}$, and the circle red markers and dash-dot red line with magnitude $2.0\times 10^{-6}$.
    }
    \label{fig:local-stochastic-pauli_subspace-to-diff}
\end{figure*}

For the first step, the noise-free optimization with the variational quantum circuit updates the parameters up to 500 iterations with the BFGS option in SciPy.
Fixing this set of optimized parameters, we examine how well Dual-GSE can improve the estimated ground state energy under the gate and finite shot noise.
We assume the gate noise is from local stochastic Pauli channels that occur right after each single-qubit or two-qubit gate operation.
The stochastic Pauli error in single-qubit gate operation is set to $\displaystyle\left\{1-p_1, 0.2 p_1, 0.2 p_1, 0.6 p_1\right\}$ for each Pauli error in $\{I, X, Y, Z\}$, i.e., the channel represented as
\begin{equation} \label{eq:lsp_1}
    \mathcal{E}_{\mathrm{lsp},p_{1}}^{(1)}\left(\rho\right) = (1-p_{1})\rho + p_{1}\left(p_{x} X\rho X + p_{y} Y\rho Y + p_{z} Z\rho Z\right)
\end{equation}
with $p_{x} = 0.2, p_{y} = 0.2, p_{z} = 0.6$.
In two-qubit gate operations, the error channel is set as the tensor product of two local single-qubit error channels for each qubit:
\begin{equation} \label{eq:lsp_2}
    \mathcal{E}_{\mathrm{lsp},p_{2}}^{(2)} = \mathcal{E}_{\mathrm{lsp},p_{2}}^{(1)} \otimes \mathcal{E}_{\mathrm{lsp},p_{2}}^{(1)}.
\end{equation}

In our numerical simulations, we set $p_2 = 10 p_1$ with $p_1$ taking $\{2.0 \times 10^{-6}, 2.0\times 10^{-5}, 2.0\times 10^{-4}\}$, which is equivalent to adding $0.0704$, $0.704$, and $7.04$ errors to one quantum circuit with eight qubits and eight layers respectively.
Most significantly, we set the noise level for two-qubit operations ten times larger than that for single-qubit operations throughout our numerical simulation.

To examine the shot noise effect, we sample the expectation values of Pauli observables by generating Gaussian noise in accordance with the variance of the Pauli observable.
The local stochastic Pauli noise in the circuit is fixed to $p_1 = 2.0 \times 10^{-6}$, and the total shot counts $N_{\mathrm{s}}$ are distributed equally to each measurement query.
On the basis of the noisy expectation values, we compose $\mathcal{S}$ and $\mathcal{H}$ matrices for the generalized eigenvalue problem $\mathcal{H}\vec\alpha = E\mathcal{S}\vec\alpha$, and sample $N_{\mathrm{samples}}=1000$ samples of the estimated energy $E$ as a solution of $\mathcal{H}\vec\alpha = E\mathcal{S}\vec\alpha$.
The detailed procedure for adding shot noise is described in Appendix \ref{sec:appendix_effect_of_shot_noise_on_observable_estimation}.

To reuse the measurement results with the same Pauli observable on the same ansatz, we take the following steps for classical pre- and postprocessing, illustrated in Fig.~\ref{fig:procedure_dgse}.
\begin{enumerate}
    \item Collect in advance the expectation value of all possible combinations of Pauli observables and quantum states required by the given Hamiltonian and subspace construction.
    \item Run quantum circuits for those combinations of observables and quantum states.
    \item Compute matrix elements $\mathcal{S}_{ij}$ and $\mathcal{H}_{ij}$ for the generalized eigenvalue problem $\mathcal{H}\vec\alpha = E\mathcal{S}\vec\alpha$, by summing up the precomputed expectation values of corresponding Pauli observables.
    \item Regularize the matrices $\mathcal{S}$ and $\mathcal{H}$ using Algorithm 1 in Ref.~\cite{Epperly2022} to make the solution of $\mathcal{H}\vec\alpha = E\mathcal{S}\vec\alpha$ more stable.
    Here, we set the threshold defined in Ref.~\cite{Epperly2022} to $10.0/\sqrt{N_s}$.
    \item Solve $\mathcal{H}\vec\alpha = E\mathcal{S}\vec\alpha$ and choose the minimal eigenvalue in the restricted valid range around the theoretical ground state $E_{0}<0$, which can be upper- and lower-bounded through efficient classical algorithms for the ground state approximation~\cite{FroeseFischer1987, verstraete2004renormalization, White1992, Bansal2009, Carleo2017, Choo2020, Yoshioka2021}.
    We assume the classical approximation bounds in advance in the range of the valid estimation into $1.1E_{\mathrm{true}}<E<0.9E_{\mathrm{true}}$ around the noise-free theoretical ground state energy $E_{\mathrm{true}} < 0$.
\end{enumerate}

Under these settings, we examine three practical subspaces discussed in the previous section: the power subspace, the fault subspace, and the divide-and-conquer subspace.
The ansatz for each subspace construction is listed in Table~\ref{tab:subspaces}.
For the fault subspace, we set $\epsilon$ in Eq.~\eqref{eq:rho_dgse_fs} to the noise level of single-gate stochastic error $p_1$.
For the divide-and-conquer subspace, we use two four-qubit subsystems $\mathrm{A}$ and $\mathrm{B}$ divided as Fig.~\ref{fig:path_graphs}(b),
where we first perform the four-qubit noise-free optimization on parameterized circuits with a Hamiltonian in the same form as Eq.~\eqref{eq:hamiltonian} but only covering a four-qubit subsystem.

\subsection{Simulation results without shot noise \label{sec:Simulation_results_without_shot_noise}}

\begin{figure}[htbp]
    \centering
    \includegraphics[width=1.0\columnwidth]{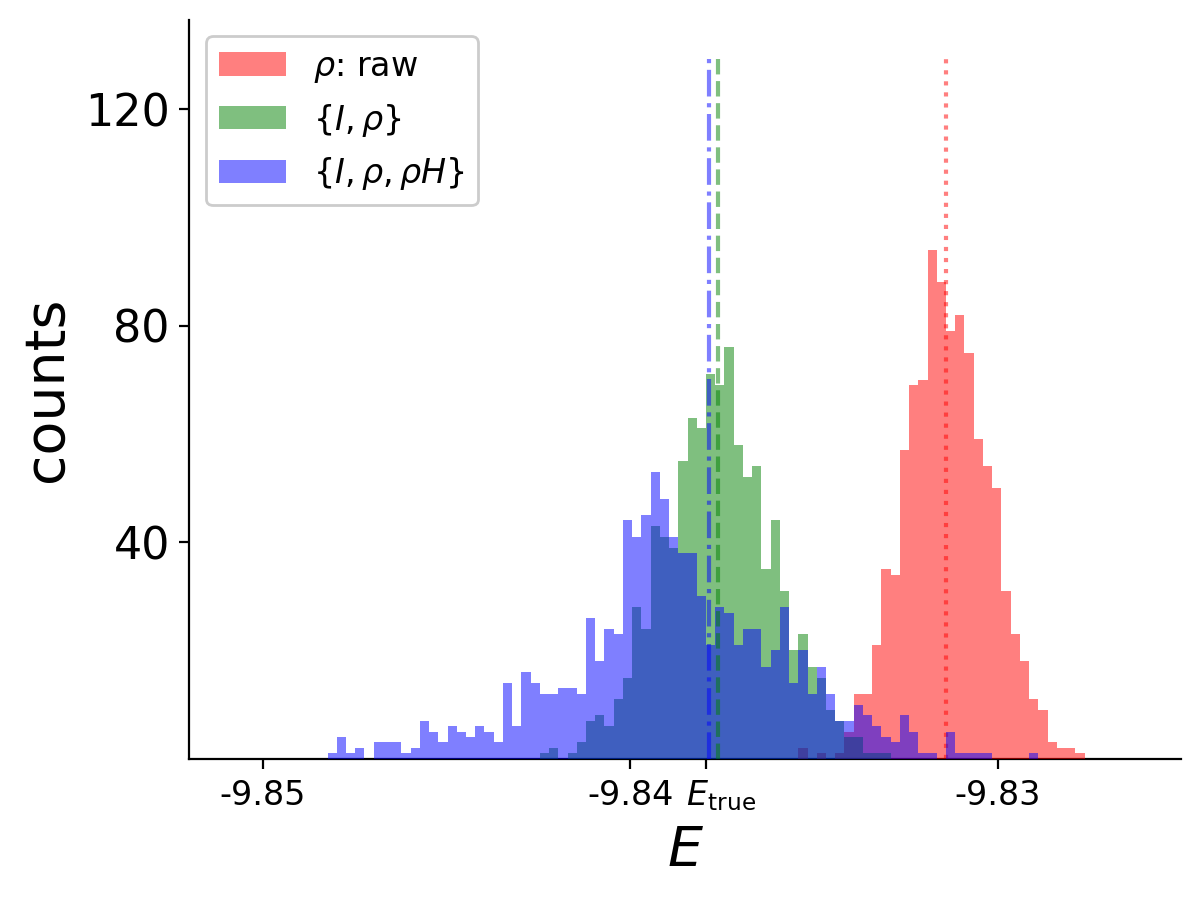}
    \caption{
        The histogram of estimated ground state energy $E$ with and without Dual-GSE under the local stochastic Pauli noise and finite shot noise, which are set to $p_1=2.0 \times 10^{-6}$ and $10^{8}$ shots. 
        The blue distribution shows the estimation without Dual-GSE. 
        The green and red distributions show the estimations with Dual-GSE using the power subspace method with two and three subspaces. 
        The label $E_{\mathrm{true}}$ on the x-axis shows the analytically computed noise-free energy, and each vertical dotted colored line shows the mean value of its distribution.
    }
    \label{fig:local-stochastic-pauli_power-subspace_histograms}
\end{figure}

\begin{figure*}[htbp]
    \raggedright
    \subfloat[power subspace\label{fig:local-stochastic-pauli_power-subspace_shots-to-diff}]{
        \includegraphics[width=0.32\textwidth]{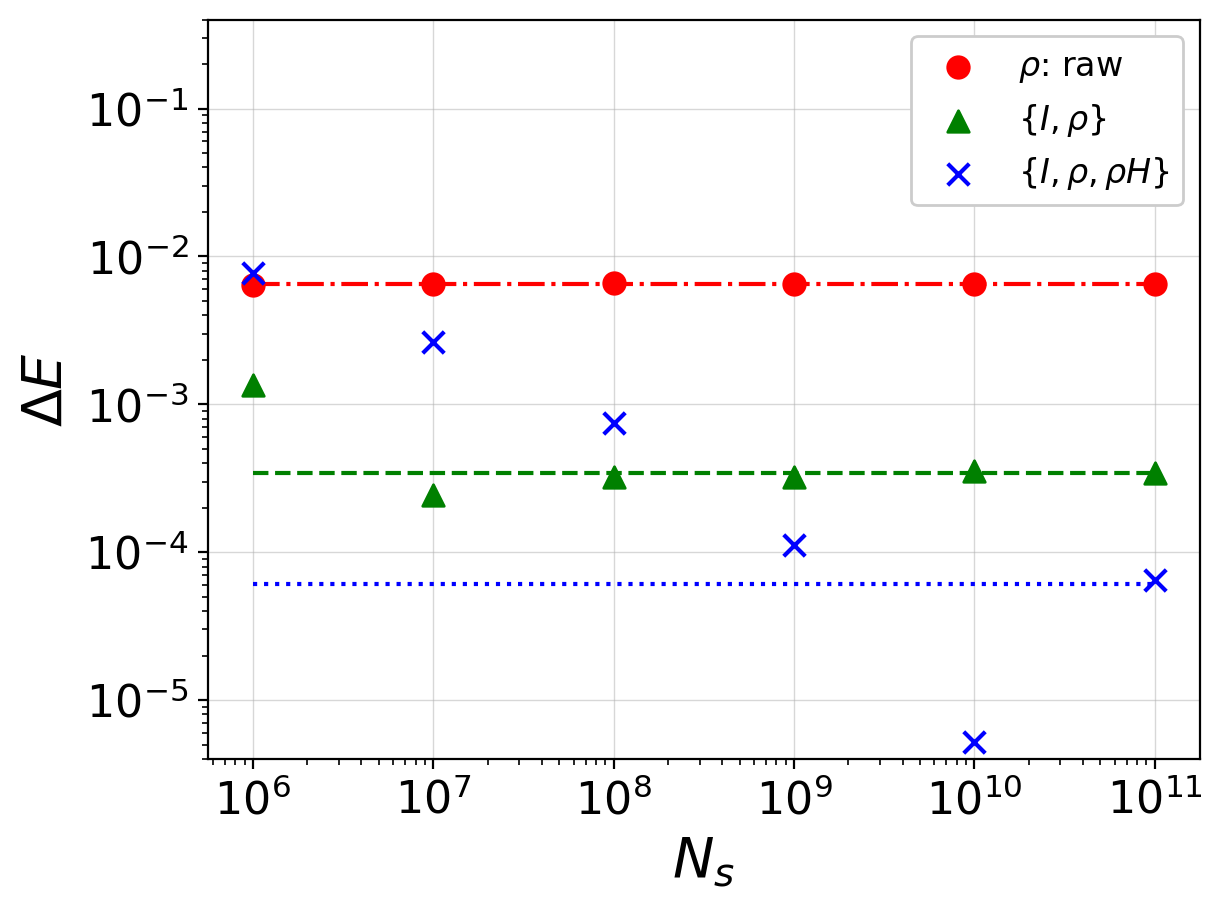}
    }
    \hfill
    \subfloat[fault subspace\label{fig:local-stochastic-pauli_fault-subspace_shots-to-diff}]{
        \includegraphics[width=0.32\textwidth]{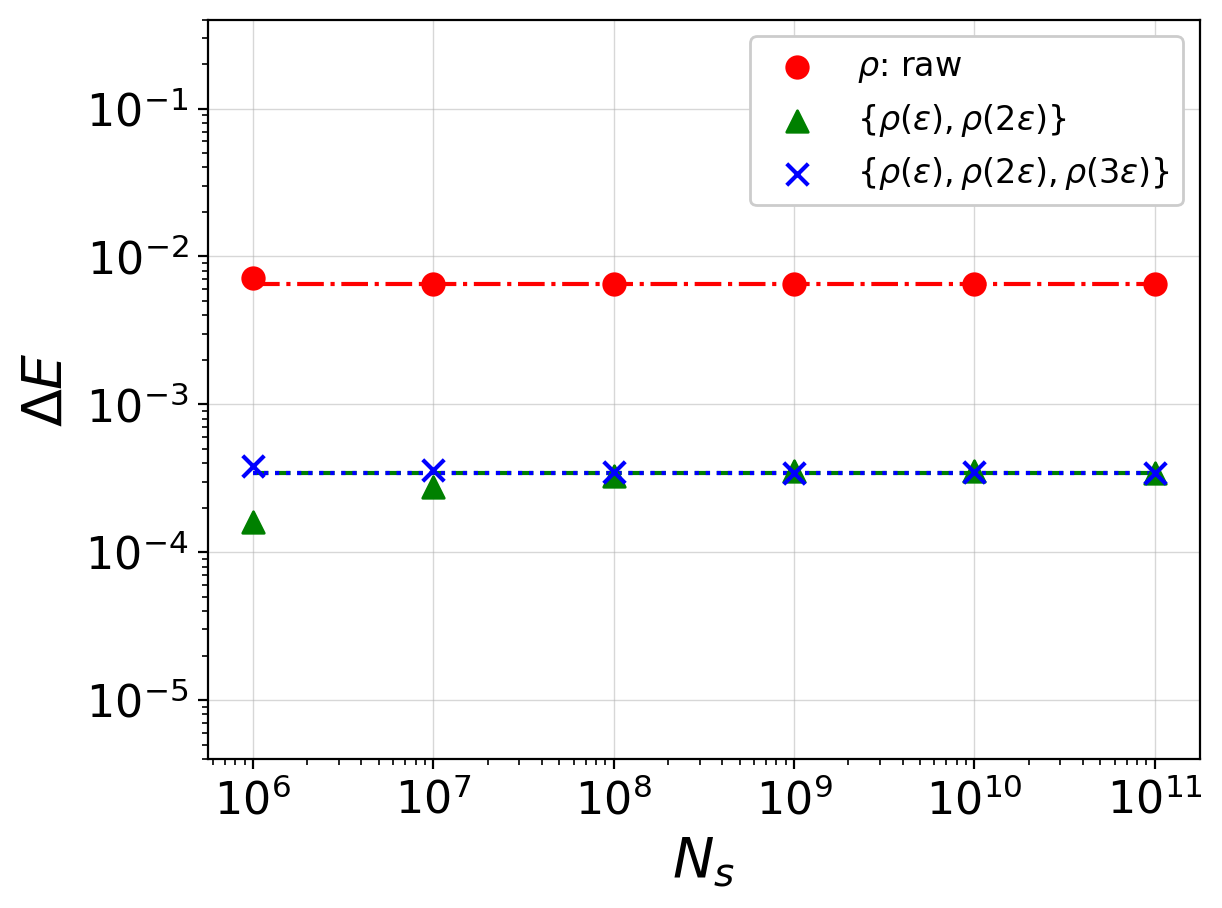}
    }
    \hfill
    \subfloat[divide-and-conquer subspace\label{fig:local-stochastic-pauli_cfe-subspace_shots-to-diff}]{
        \includegraphics[width=0.32\textwidth]{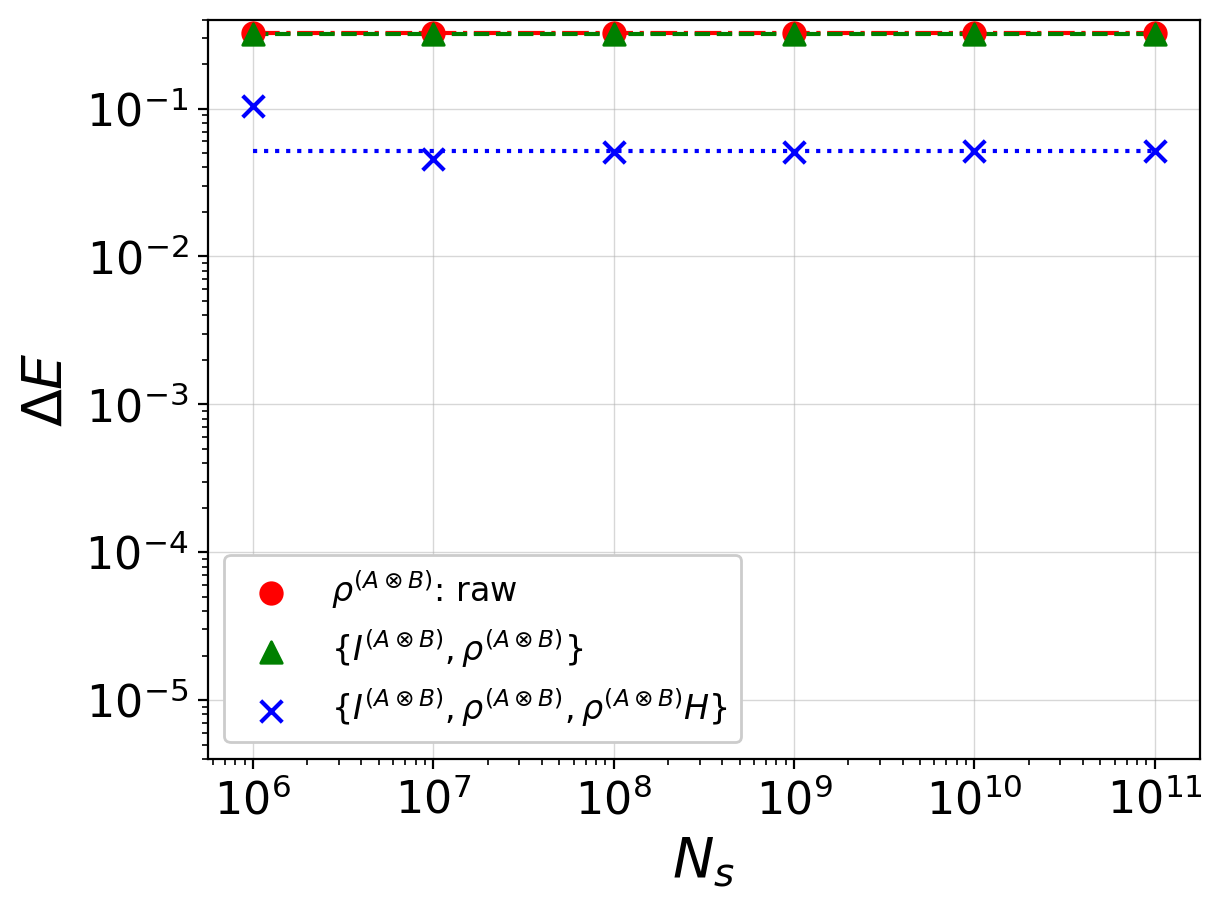}
    }
    \caption{
        The estimation bias $\Delta E$ caused by the finite shot noise in addition to the gate noise.
        The number of shot counts $N_{\mathrm{s}}$ is changed in magnitude from $10^{6}$ to $10^{11}$.
        The local stochastic Pauli noise applied to the gates in the circuit is fixed to $p_1 = 2.0 \times 10^{-6}$.
        The horizontal dotted lines represent the estimated value by Dual-GSE under the gate noise but without finite shot noise, whose colors correspond to different numbers of subspaces.
        Note that in (b), the horizontal dotted green line overlaps with the blue one.
        To save the space in (c), we use $\rho^{(\mathrm{A}\otimes\mathrm{B})}$ as a label to represent the separable state $\rho^{(\mathrm{A})}\otimes\rho^{(\mathrm{B})}$.
    }
    \label{fig:local-stochastic-pauli_shots-to-diff}
\end{figure*}

We first discuss the simulation results without shot noise.
Here, the estimation bias $\Delta E$ is examined between the theoretical ground state energy $E_{0}$ and the estimated energy $E$ with different numbers of subspace.
The result for each subspace construction is shown in Fig.~\ref{fig:local-stochastic-pauli_subspace-to-diff}.

From the results in Fig.~\ref{fig:local-stochastic-pauli_subspace-to-diff}(a) for the power subspace, the error is exponentially suppressed in accordance with the number of subspaces $M$.
Thanks to the linear combination of quantum states with different power degrees $I$, $\rho$, and $\rho^{2}$, the mitigated ground state energies are likely to be made more precise than the original estimated energy by variational optimization without error mitigation under the noise-free simulation.
Therefore, the power subspace seems effective in mitigating the algorithmic error inherent to the variational optimization.

The results by the fault subspace are shown in Fig.~\ref{fig:local-stochastic-pauli_subspace-to-diff}(b).
Regardless of the number of subspaces and the noise level we impose, the fault subspace achieves almost the same estimation precision as the noise-free simulation. 
However, the estimation precision does not clearly surpass the noise-free simulation because it does not contain subspace components that essentially expand search space, such as $\rho H^k$.
This implies the fault subspace is more suitable with a few subspaces to suppress stochastic errors.

Finally, the estimation bias by the divide-and-conquer subspace is shown in Fig.~\ref{fig:local-stochastic-pauli_subspace-to-diff}(c).
With the choice of two subspaces $\{I^{(\mathrm{AB})}, \rho^{(\mathrm{A})}\otimes \rho^{(\mathrm{B})}\}$, the estimation bias remains almost in the same precision as the noise-free simulation with the separable ansatz because the entanglement between subsystem $\mathrm{A}$ and $\mathrm{B}$ is not taken into account.
When it comes to the subspace $\{I^{(\mathrm{AB})} \} \cup \{(\rho^{(\mathrm{A})}\otimes\rho^{(\mathrm{B})}) H^{k-2}\}_{k=2}^{M}$ with the number of subspaces $M\geq 3$, the estimation precision surpasses the precision of separable ansatz since the nonlocal effect of the Hamiltonian between $\mathrm{A}$ and $\mathrm{B}$ is included. 
Besides, we also witness a clear exponential error suppression with $M$ in this range. 
Overall, these results imply that the divide-and-conquer subspace can effectively simulate the entangled physical model with separated small subsystems.

Further numerical experiments on another graph structure of the Ising Hamiltonian, another noise model, and other noise levels for the ansatz state can be found in Appendix.~\ref{sec:appendix_Numerical_Simulation_on_Different_Model_and_Noise_Settings}.
These results also support the practicality and versatility of our proposed method.

\begin{figure*}[htbp]
    \subfloat[power subspace\label{fig:local-stochastic-pauli_power-subspace_shots-to-stddev}]{
        \includegraphics[width=0.32\textwidth]{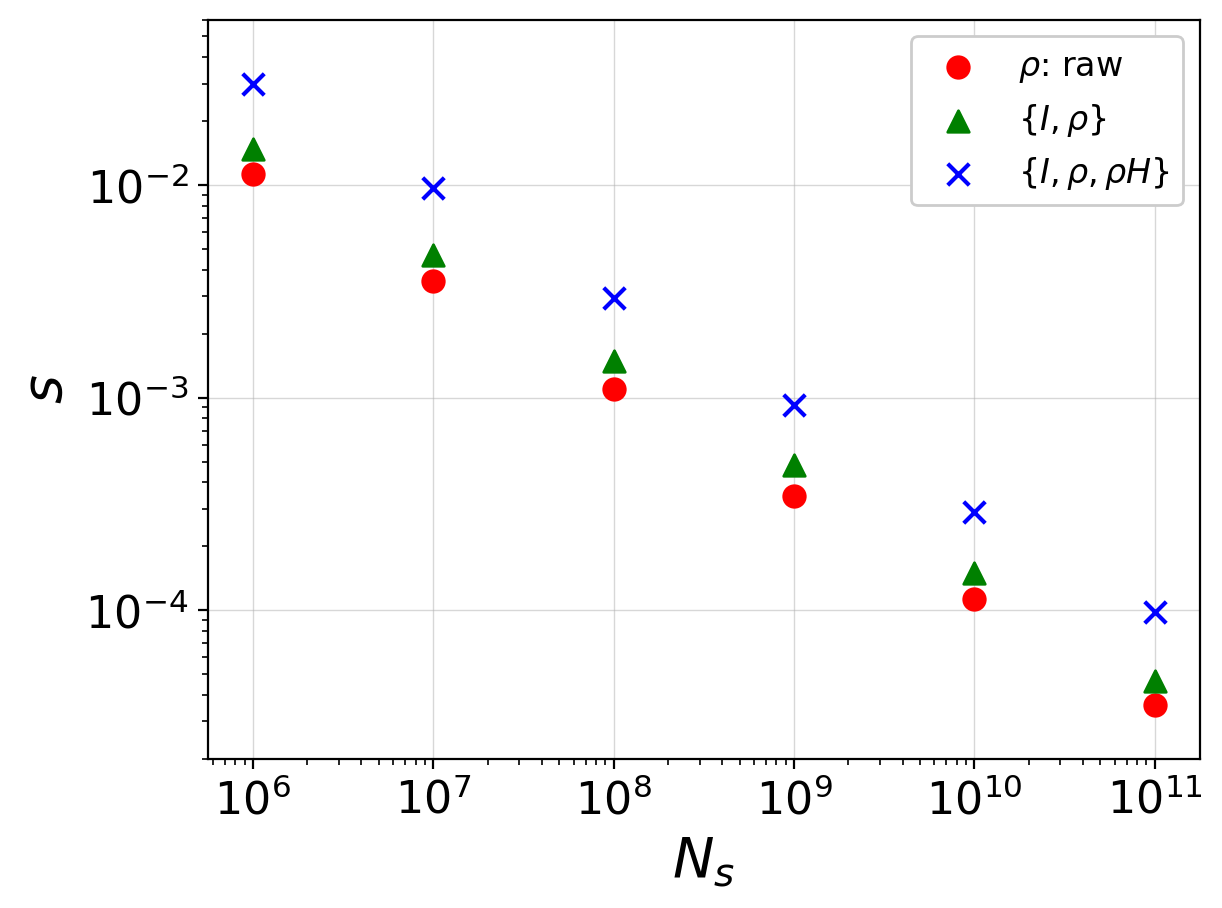}
    }
    \subfloat[fault subspace\label{fig:local-stochastic-pauli_fault-subspace_shots-to-stddev}]{
        \includegraphics[width=0.32\textwidth]{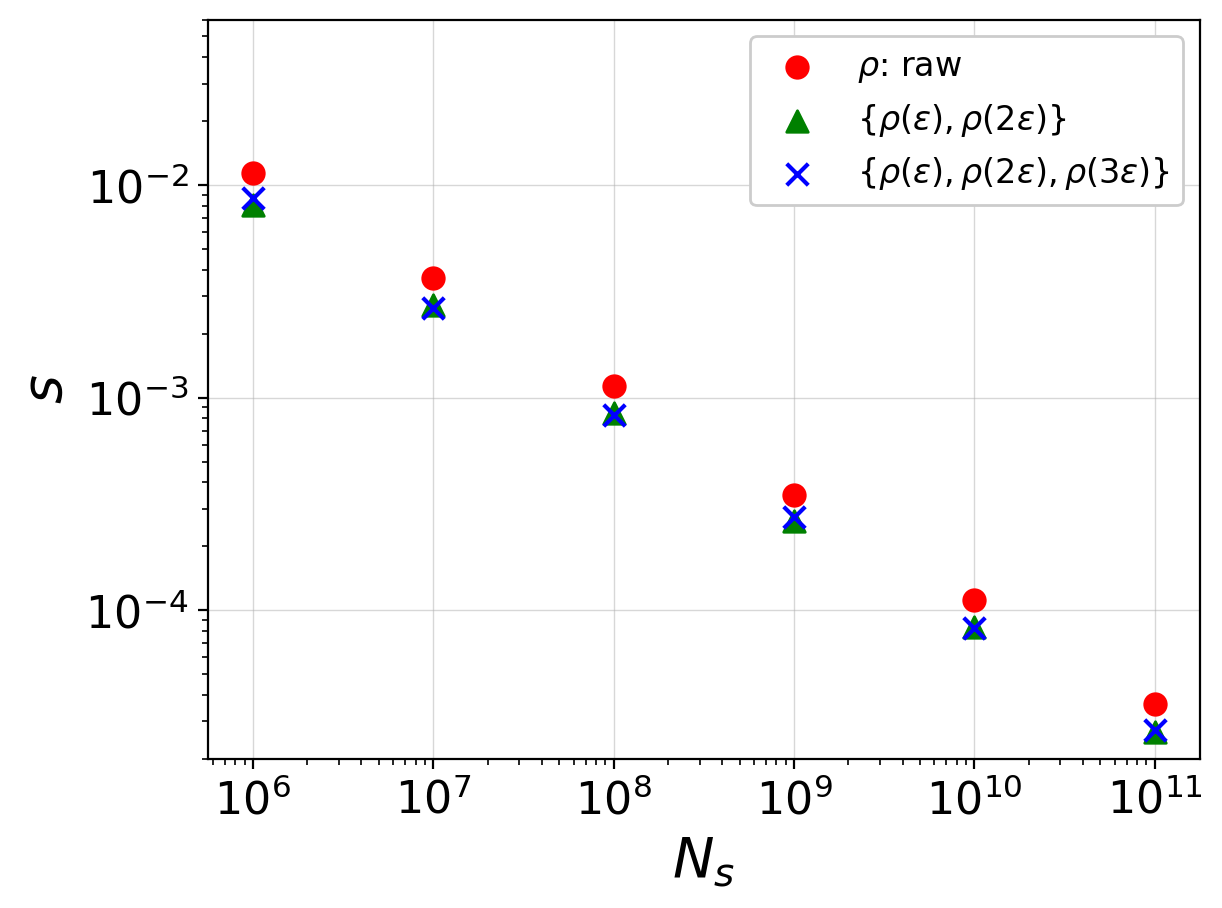}
    }
    \subfloat[divide-and-conquer subspace\label{fig:local-stochastic-pauli_cfe-subspace_shots-to-stddev}]{
        \includegraphics[width=0.32\textwidth]{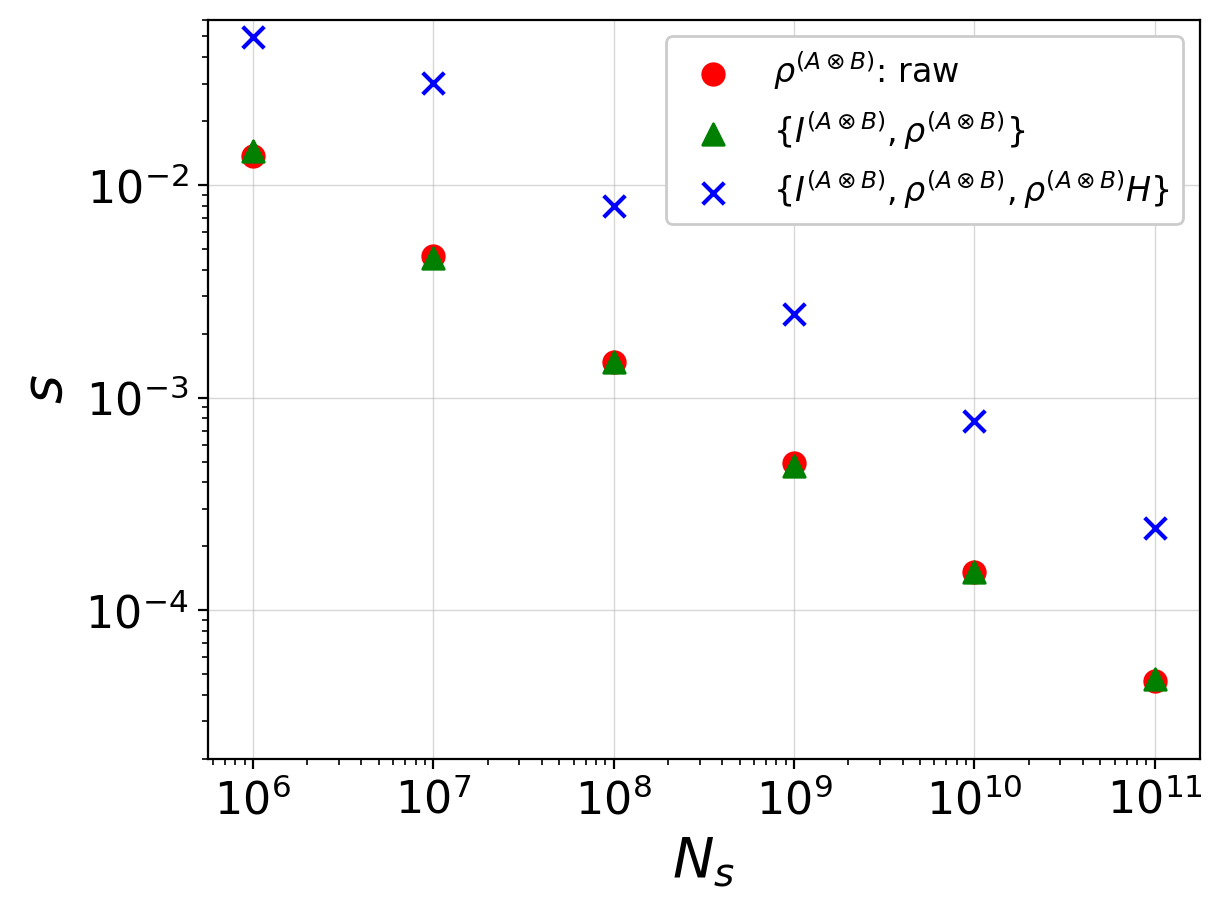}
    }
    \caption{
        The standard deviation $s$ caused by the finite shot noise. 
        The number of shot counts $N_{\mathrm{s}}$ is changed in magnitude from $10^{6}$ to $10^{11}$.
        The local stochastic Pauli noise applied to the gates in the circuit is fixed to $p_{1} = 2.0 \times 10^{-6}$. 
    }
    \label{fig:local-stochastic-pauli_shots-to-stddev}
\end{figure*}

\subsection{Simulation results under finite shot noise}

Here, we investigate the effect of finite shot noise in Dual-GSE. 
Figure~\ref{fig:local-stochastic-pauli_power-subspace_histograms} visually demonstrates the trade-off between the accuracy and estimation variance in accordance with the number of subspaces used in the power subspace.
We can also show that appending subspaces will enhance the estimation precision in exchange for a higher sampling cost.

We also examine the performance of Dual-GSE with different shot counts for each type of subspace construction.
The corresponding results are shown in Fig.~\ref{fig:local-stochastic-pauli_shots-to-diff}, where we fix the gate noise level to $p_{1} = 2.0 \times 10^{-6}$.
The additional estimation bias caused by finite shot counts converges to the estimation bias without shot noise as the shot counts increase.
The convergence speed is affected by the type and the number of subspaces.
Since the power subspace tends to use more measurement queries for different Pauli observables, the estimation bias by the power subspace converges more slowly than that by the fault subspace.

Figure~\ref{fig:local-stochastic-pauli_shots-to-stddev} shows the standard deviation in the estimation by each subspace under the finite shot noise.
In the power subspace and divide-and-conquer subspace, the standard deviation increases subject to the number of subspaces $M$ because $N_{\mathrm{s}}$ shots are distributed to more measurement queries as $M$ increases.
In the fault subspace, there seems to be no significant difference in the standard deviation between performing Dual-GSE or not.
In accordance with the central limit theorem, the standard deviation in Fig.~\ref{fig:local-stochastic-pauli_shots-to-stddev} converges with the speed of $1/\sqrt{N_{\mathrm{s}}}$ in every type of subspace.

\begin{figure*}[htbp]
    \raggedright
    \subfloat[accumulative\label{fig:subspace-to-pauli_accumulative}]{
        \includegraphics[width=0.48\textwidth]{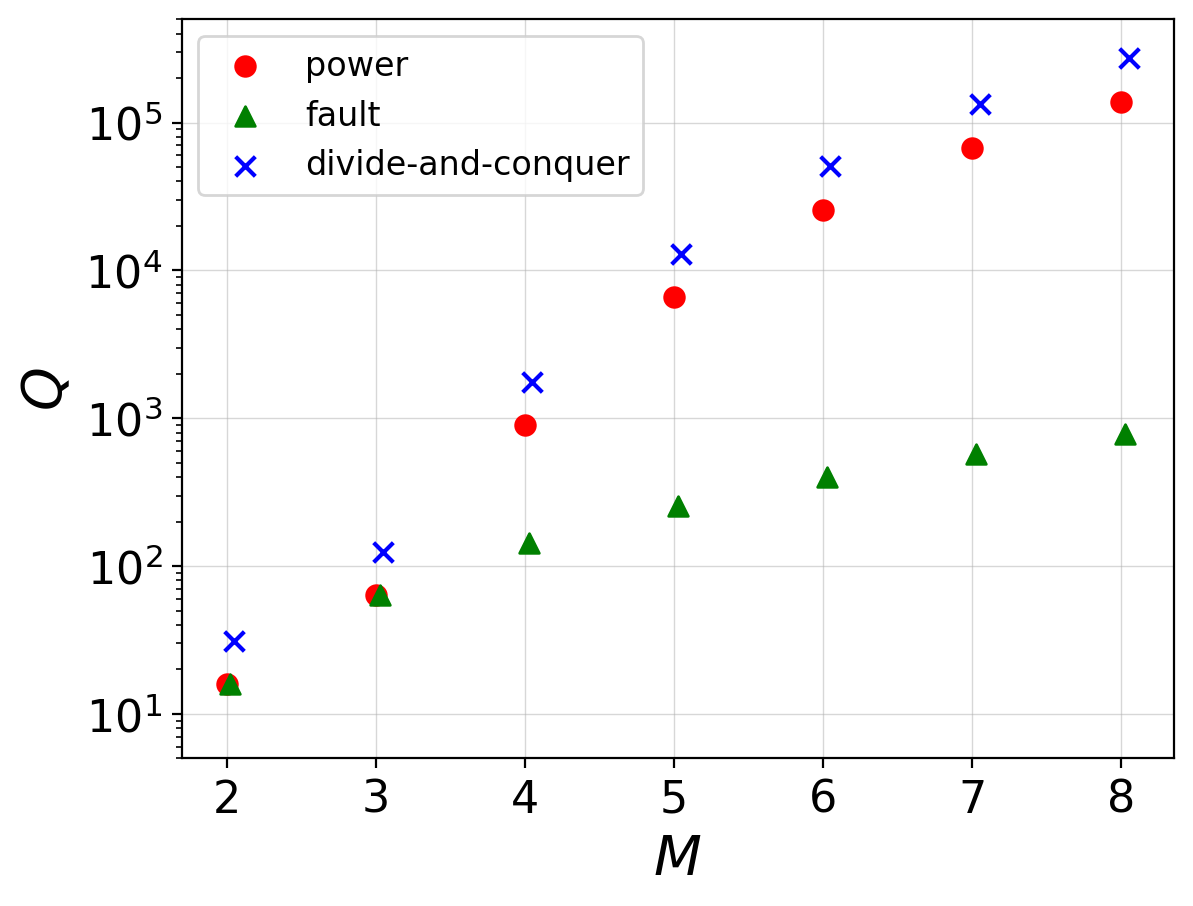}
    }
    \subfloat[reuse\label{fig:subspace-to-pauli_reuse}]{
        \includegraphics[width=0.48\textwidth]{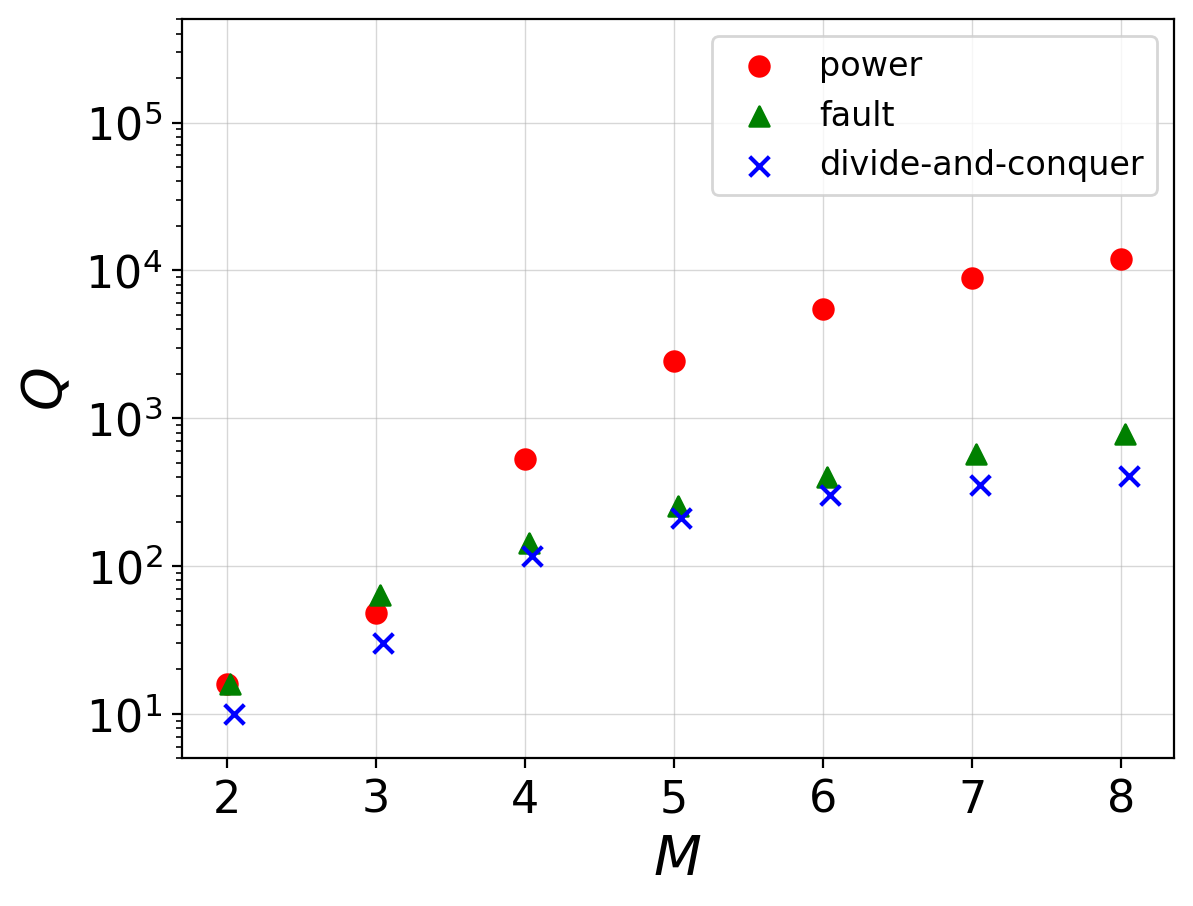}
    }
    \caption{
        The number of measurement queries in each subspace construction (a) without and (b) with reusing the expectation values of the same observable.
        The x-axis is the number of subspaces $M$.
        The y-axis is the number of measurement queries $Q$.
    }
    \label{fig:subspace-to-pauli}
\end{figure*}

On the maximum total shot counts examined in our numerical simulation, we remark that $\sim 10^{11}$ in total seems enough to suppress the shot noise effect, which is shown in Fig.~\ref{fig:local-stochastic-pauli_shots-to-diff} and Fig.~\ref{fig:local-stochastic-pauli_shots-to-stddev}.
This matches the available shot counts of the current and near-future quantum computers on which we can run $10^{5}\sim 10^{7}$ shots per quantum circuit with a realistic execution time~\cite{Yoshioka2022-gq, Yang2021, Yang2022, mooney2024characterization}, i.e. when we use $10^{3}$ quantum circuits to estimate the ground state energy, the total shot counts becomes $\sim 10^{11}$.

Assuming the estimated ground state energy deviates from the mean value $E_{0}$ to $E = E_{0} + \delta E$ under the finite shot counts $N_{\mathrm{s}}$, the sample variance $s^2$ of our numerical simulation becomes
\begin{equation}
    s^2 = \frac{1}{N_{\mathrm{samples}}}\sum_{k} \delta E(k)^2 \leq \max_{k}|\delta E(k)|^2.
\end{equation}
where $\delta E(k)$ denotes the deviation of the $k$-th sample among $N_{\mathrm{samples}}$ samples.
According to Ref.~\cite{Yoshioka2022-gq}, $|\delta E|$ is upper-bounded as
\begin{equation}
    \left|\delta E\right| \leq 4 \gamma Q\left\|\mathcal{S}_{0}^{-1}\right\|_{\mathrm{op}}N_{\mathrm{s}}^{-1/2}
\end{equation}
with $\gamma$ satisfying $E_{0} \leq\|H\|_{\mathrm{op}} \leq \gamma$, the number of measurement queries $Q$, and the operator norm of the inverse of $\mathcal{S}$.
Therefore, the standard deviation $s$ can also be upper-bounded as
\begin{equation}
\begin{split}
    s \leq \left|\delta E\right| \leq 4 \gamma Q\left\|\mathcal{S}_{0}^{-1}\right\|_{\mathrm{op}}N_{\mathrm{s}}^{-1/2} \\
    y\leq -\frac{1}{2}x + \log_{10}{4 \gamma Q\left\|\mathcal{S}_{0}^{-1}\right\|_{\mathrm{op}}}
\end{split}
\end{equation}
with $x=\log_{10}{N_{\mathrm{s}}}$ and $y=\log_{10}{s}$.
This well explains the scaling of the standard deviation in accordance with the number of samples in Fig.~\ref{fig:local-stochastic-pauli_shots-to-stddev}.

\subsection{Leveraging mitigation cost\label{sec:Leveraging_mitigation_overheads}}

Each type of subspace used in our numerical simulation has different advantages and overheads.
Below, we analyze their overhead in terms of the number of measurement queries, achievable estimation bias, and the scaling of the total sampling cost.

\subsubsection{The number of measurement queries\label{sec:The_number_of_measurement_queries}}

\begin{figure*}[htbp]
    \centering
    \subfloat[\label{fig:compare_ps_cfe_noise-free}]{
        \includegraphics[width=0.48\textwidth]{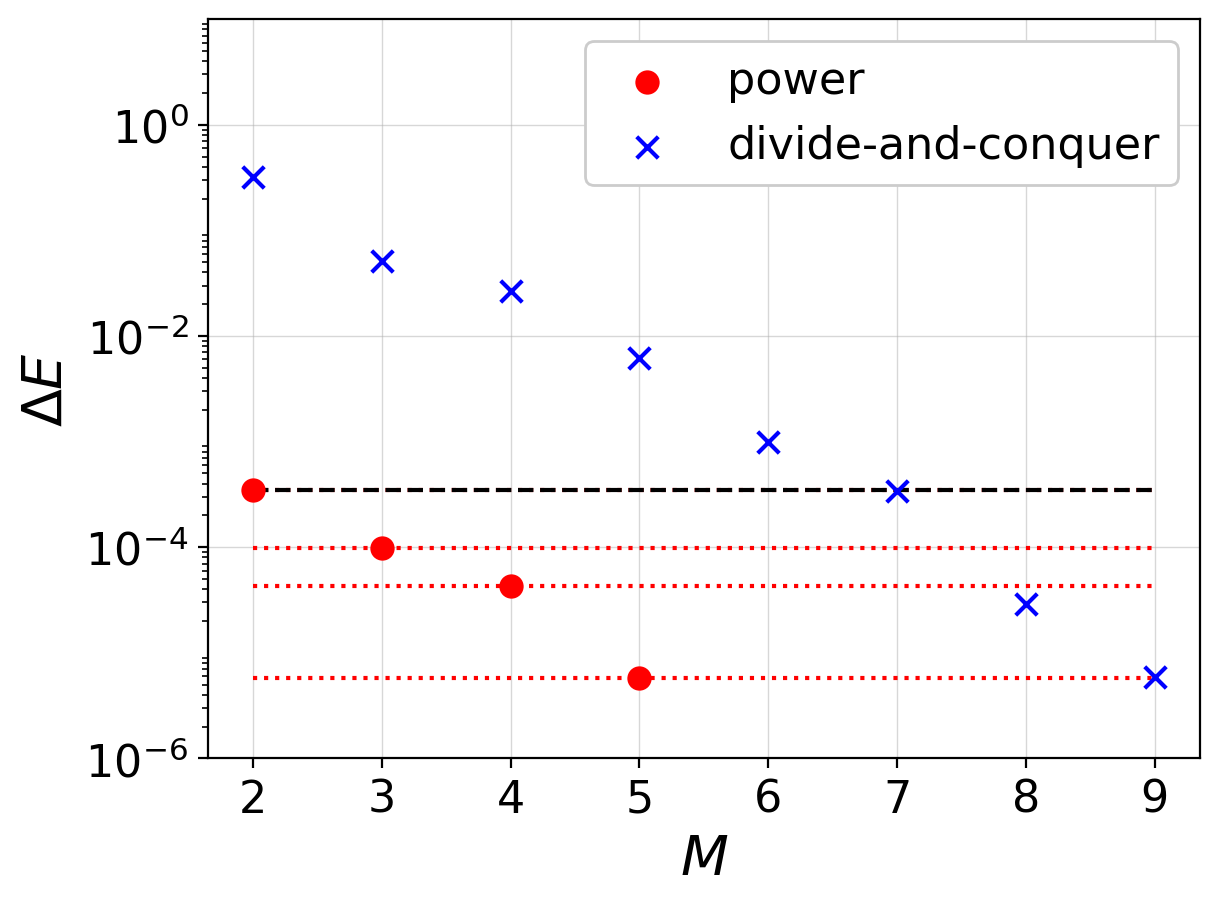}
    }
    \hfill
    \subfloat[\label{fig:cost-to-diff_regularise_original_M_noise-free}]{
        \includegraphics[width=0.48\textwidth]{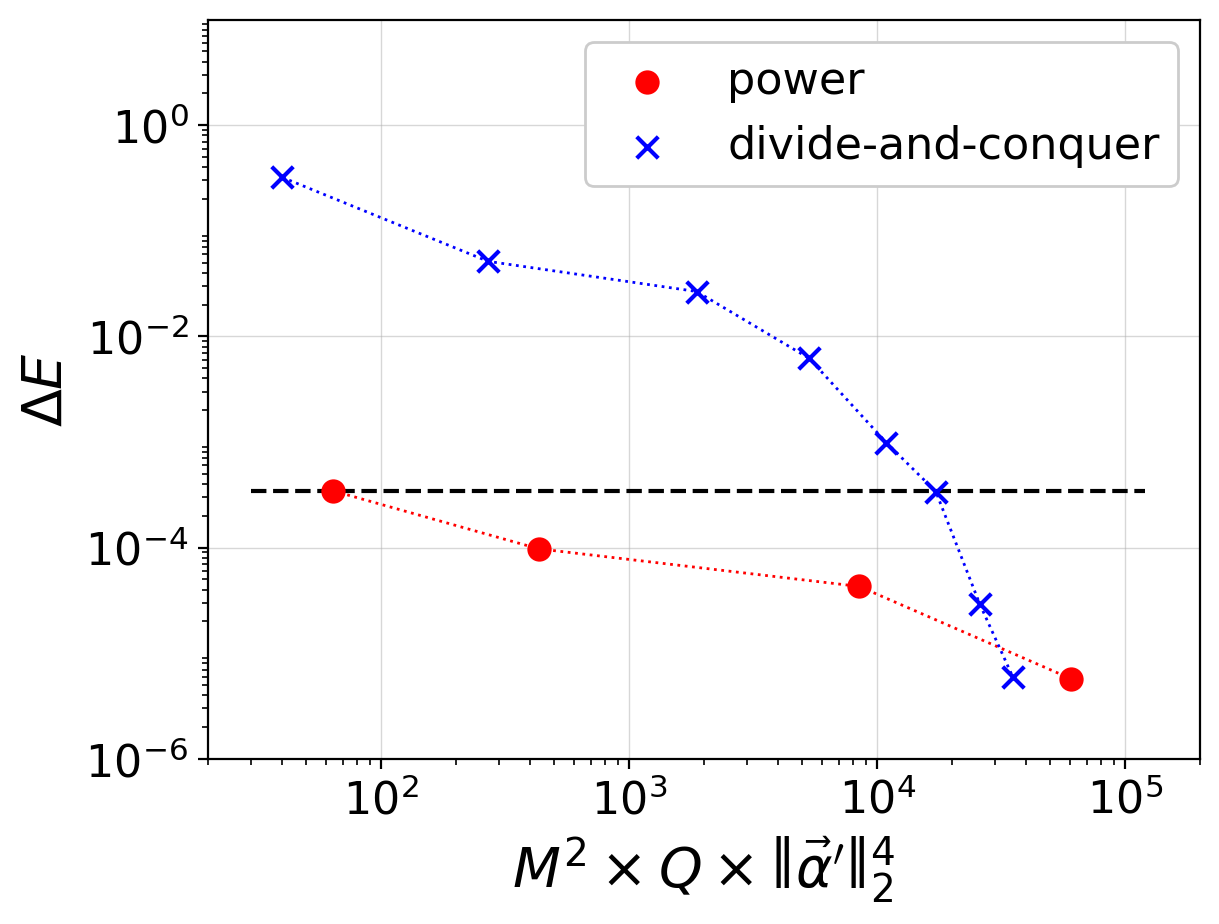}
    }
    \caption{
        (a) The estimation bias $\Delta E$ to the number of subspaces $M$ between the power subspace and the divide-and-conquer subspace.
            The circle red markers and the dotted red polyline represent the estimation bias by the power subspace.
            The cross blue markers and the dotted blue polyline represent the estimation bias by the divide-and-conquer subspace.
            The horizontal dotted red lines correspond to each estimation bias by the power subspace.
            The horizontal dashed black line shows the noise-free 8-qubit full-size VQE simulation.
        (b) The estimation bias $\Delta E$ to the measurement overhead in the metric of $M^{2} Q \|\vec{\alpha}^{\prime}\|_{2}^{4}$ when increasing the number of subspaces $M$ for the power subspace and the divide-and-conquer subspace.
        Each red and blue plot in (b) corresponds to each in (a) from the left to the right, since $M^{2} Q \|\vec{\alpha}^{\prime}\|_{2}^{4}$ increases monotonically to $M$ for both subspace constructions.
        The notation for other colored plots and lines is the same as (a).
    }
    \label{fig:cost-to-diff_1-bridge}
\end{figure*}

Figure~\ref{fig:subspace-to-pauli} compares the number of measurement queries $Q$ with and without the reusing strategy when simulating the eight-site Ising model in our setting.
We observe that the fault subspace and divide-and-conquer subspace are efficient in terms of $Q$, i.e. the number of running and measuring quantum states on actual quantum devices.
The key is to reuse measurement results with the same combination of Pauli observable and quantum state.

First, we focus on the advantage of the fault subspace compared with other subspace constructions without reusing measurement results.
As shown in Fig.~\ref{fig:subspace-to-pauli}(a), without reusing the measurement results with the same Pauli observable, $Q$ seems to scale in magnitude to the number of subspaces $M$, especially for the power subspace and the divide-and-conquer subspace, where the powers of the Hamiltonian are included in the subspaces.
The divide-and-conquer subspace with two subsystems in our setting consumes twice as much as the queries in the power subspace, which would also result in higher estimation variance.
In contrast, the increase of measurement queries for the fault subspace is much more moderate than the two other subspace constructions.
The number of measurement queries for the fault subspace $Q$ follows $O(M^{2}|H|)$, scaling polynomially to the number of subspaces $M$ and the number of Pauli observables $|H|$ in the Hamiltonian.

When it comes to reusing the measurement results with the same state and observable combination, $Q$ can be suppressed eminently smaller for the power subspace and the divide-and-conquer subspace, as shown in Fig.~\ref{fig:subspace-to-pauli}(b).
This is particularly effective for the divide-and-conquer subspace since the possible combinations of Pauli observable for a given Hamiltonian over $N$-site physical systems are drastically reduced from $4^N$ to $4^{N/2}$.
For example, $Q$ exceeds $10^{4}$ queries with five subspaces when not reusing the observables, whereas it only takes around $10^{2}$ queries when reusing the observables.
When we assign $10^{4}$ shots for each measurement query, the total shot count $N_{\mathrm{s}}$ is still less than $10^{7}$, which is already tractable for the current or near-future quantum devices.
Therefore, the divide-and-conquer subspace construction for Dual-GSE is also practical in terms of the measurement query overhead.
Note that in the fault subspace, the same combination of observables and quantum states is not likely to be used many times.

\subsubsection{The sampling cost to estimation bias\label{sec:The_sampling_cost_to_estimation_bias}}

Changing the metric, we next evaluate the sampling cost to achieve the same level of estimation bias between the power subspace and the divide-and-conquer subspace and verify the divide-and-conquer subspace demonstrates its performance as compatible with the power subspace.
Before we discuss the sampling cost, we first compare the estimation bias $\Delta E$ with different numbers of subspaces $M$ between the power subspace and the divide-and-conquer subspace without gate and shot noise.
In Fig.~\ref{fig:cost-to-diff_1-bridge}(a), we have plotted the estimation bias by the divide-and-conquer subspace up to $M=9$.
According to Fig.~\ref{fig:cost-to-diff_1-bridge}(a), the plots of both the power subspace and the divide-and-conquer subspace demonstrate the exponential decrease of estimation bias to the true ground state energy with the number of subspaces.
We also see that the estimation bias for $M=9$ by the divide-and-conquer subspace corresponds very closely to the estimation bias for $M=5$ by the power subspace.

One may conclude that the divide-and-conquer subspace sacrifices a greater number of subspaces to achieve the same level of estimation bias.
However, this does not necessarily imply that the divide-and-conquer subspace requires a larger sampling overhead.
According to Eq.~\eqref{eq:Ns_lower_bound_final} in Appendix~\ref{sec:appendix_Leveraging_Sampling_Cost_Regarding_Divide-and-conquer_Overhead}, the overall sampling cost can be represented as
\begin{equation}\label{eq:Ns_lower_bound_final_main_text}
\begin{split}
    N_{\mathrm{s}}
    \geq  \frac{ 4\gamma^{2} M^{2}Q \left\|\vec{\alpha}^{\prime}\right\|_{2}^{4} }
               { \epsilon^{2} },
\end{split}
\end{equation}
which is proportional to the factor $\displaystyle M^{2}Q\|\vec{\alpha}^{\prime}\|_{2}^{4}$ when fixing estimation bias denoted by $\epsilon$.
Note that $\gamma$ in Eq.~\eqref{eq:Ns_lower_bound_final_main_text} defined as $\displaystyle \gamma := \sum_{i}\left|h_{i}\right|$ for the given Hamiltonian $\displaystyle H = \sum_{i}h_{i}P_{i}$ stays the same between the power subspace and the divide-and-conquer subspace, and $\vec{\alpha}^{\prime}$ is the optimized vector associated with the diagonally normalized matrix $\mathcal{S}$ (see Appendix~\ref{sec:appendix_Leveraging_Sampling_Cost_Regarding_Divide-and-conquer_Overhead} for details).
Thus, to compare the mitigation overhead, one can further check the relation between this overhead factor $\displaystyle M^{2}Q\|\vec{\alpha}^{\prime}\|_{2}^{4}$ and the estimation bias $\Delta E$.

The correspondence between the factor $\displaystyle M^{2}Q\|\vec{\alpha}^{\prime}\|_{2}^{4}$ and the estimation bias $\Delta E$ is shown in Fig.~\ref{fig:cost-to-diff_1-bridge}(b).
Note that the measurement results are reused here, i.e. $Q$ follows the plots in Fig.~\ref{fig:subspace-to-pauli}(b).
From Fig.~\ref{fig:cost-to-diff_1-bridge}(b), we observe that the total sampling cost scales almost linearly to $\Delta E$ for the power subspace.
On the other hand, the increase in the sampling cost of the divide-and-conquer subspace becomes clearly moderate when $M\geq 5$, and the sampling cost of the divide-and-conquer subspace surpasses that of the power subspace when $M=9$.

This efficiency in the sampling cost by the divide-and-conquer subspace generally owes to its drastically fewer measurement queries, which is compatible with Fig.~\ref{fig:subspace-to-pauli}(b).
Besides, the faster decrease in $\Delta E$ in the range of $M\geq 5$ can also be explained by the suitable choice of subspace, implying the linear combination of subspaces with the optimized classical parameters $\vec{\alpha}^{\prime}$ is effectively reaching the domain close to the true ground state.
This suggests the significance of the choice of subspaces since a good subspace construction, taking the symmetry or the strength of entanglement into account, will effectively reach the ground state energy with moderate search space and sampling overhead as demonstrated in our case.

\section{Summary and Outlook \label{sec:conclusion}}

We have designed a resource-efficient generalized subspace expansion method, which we call Dual-GSE, by incorporating dual-state purification (DSP) and the divide-and-conquer strategy with the generalized subspace expansion (GSE).
The resource efficiency of Dual-GSE covers the reduction of the size requirement of quantum devices and the accurate and precise estimation of the ground state energy with practical sampling cost, thanks to the techniques of DSP, the divide-and-conquer strategy, and the reusing strategy of the same measurement results.
Further developed from GSE, Dual-GSE integrates these techniques in harmony, allowing for simulating even larger physical systems beyond the size of actual quantum devices.
As an extension of DSP, we also formalized and organized how to implement the quantum circuit for computing the expectation value of density matrices of higher power degrees in Dual-GSE in Appendix~\ref{sec:appendix_quantum_circuits_for_the_generalized_ansatz}.

The numerical simulations also support the practicality of our proposed method.
The accuracy of the estimated ground state energy of the Ising model is shown to be even better than that of noise-free simulation.
The standard deviation of the estimated ground state energy can also be suppressed with a reasonable amount of finite shot counts under realistic noise levels.
Furthermore, while the divide-and-conquer subspace method incurs the sampling cost for classically reintroducing the entanglement between the subsystems, we have also numerically demonstrated that the reusing strategy of Pauli expectation values significantly reduces the total sampling overhead with an appropriate choice of subspaces regarding the target problem.
In the Appendix sections, we have further comprehensively analyzed in detail the performance and overhead of Dual-GSE under different setups to support its practicality and revealed its detailed characteristics.

There are several future directions.
First, valid eigenvalues must be obtained from the generalized linear problem $\mathcal{H}\vec{\alpha}=E\mathcal{S}\vec{\alpha}$ in Dual-GSE, which may output physically invalid values due to the gate noise, finite shot noise, and other factors rendering the linear equation unstable.
While we regularize $\mathcal{H}$ and $\mathcal{S}$ to make the estimation stable in our numerical experiments, other preprocess on these matrices may also be incorporated depending on the properties we want to guarantee, e.g., the physicality of error-mitigated states or the estimation variance.

Besides, more efficient subspace constructions may possibly be explored.
The power of the Hamiltonian in the power subspace and the divide-and-conquer subspace will span many combinations of Pauli observables as the number of subspaces $M$ increases.
To address this issue, we may add to the subspaces such operators that essentially characterize the correlation among each site or subsystem instead of the whole system Hamiltonian itself.
This would be effective when merging subsystems with a small correlation length using the divide-and-conquer strategy on large physical systems of interest.
In addition, other recently proposed subspace pruning heuristics~\cite{o2024partitioned,boyd2024high} can also be taken into account to obtain an effective subspace construction.

The GSE-based methods, including our Dual-GSE, can be further expanded and integrated with other resource-efficient methods.
To reduce the space overhead of the purification subroutine used in GSE, an alternative straightforward candidate is to use classical shadow~\cite{huang2020predicting, Seif2023, hu2022logical}.
In particular, Seif et al.~\cite{Seif2023} provides a way to use classical shadow to compute $\operatorname{Tr}\left[\rho^{M}O\right]$ for an observable $O$ without state copies.
It is also worth exploring the connection between the divide-and-conquer subspace in Dual-GSE and other divide-and-conquer strategies, which also classically simulate the effect of entanglement when computing the expectation values~\cite{Mitarai2021, Mizuta2021-pw, Fujii2022, eddins2022doubling, Chen2022, Takeuchi2022divideconquer, Yuan2021, jing2024circuit, harrow2024optimal, harada2024doubly}. 
In particular, because the framework of hybrid tensor networks proposed by Ref.~\cite{Yuan2021} offers a systematic description of a broad range of classical-quantum hybrid ansatz quantum states, the unification of Dual-GSE and the hybrid tensor network may give rise to quite general and practical methodology to simulate significantly larger quantum systems as well as mitigating a significant amount of errors.

\begin{acknowledgments}
B.Y. acknowledges insightful discussions with Mirko Amico from IBM Quantum.

This project is supported by Moonshot R\&D, JST, Grant No.\,JPMJMS2061; MEXT Q-LEAP Grant No.\,JPMXS0120319794, and No.\, JPMXS0118068682 and PRESTO, JST, Grant No.\, JPMJPR2114, No.\, JPMJPR2119, and No.\,JPMJPF2221; JST ERATO Grant Number JPMJER2302, JST CREST Grant Number JPMJCR23I4, Japan.
\end{acknowledgments}

\onecolumngrid
\appendix

\section{Comparing Noisy Density Matrix \texorpdfstring{$\rho$}{rho} and Its Dual State \texorpdfstring{$\bar{\rho}$}{barrho} \label{sec:appendix_diff_rho_rho_dual}}

In this section, we see how the state $\rho$ and its dual state $\bar{\rho}$ get closer to each other as the depth of the quantum circuit increases under the stochastic Pauli error, and thereby, how well the DSP ansatz approximates the power of the density matrix.
We examine the difference between the density matrices in terms of their trace distance $\operatorname{D}(\rho, \sigma)=\|\rho-\sigma\|_1/2=\operatorname{Tr}\left[\sqrt{(\rho-\sigma)^{\dagger}(\rho-\sigma)}\right]/2$.

First, we changed the number of total layers $L$, which is the number of blocks in Fig.~\ref{fig:qc_vqe}, to see how the noisy $\rho$ and $\bar{\rho}$ get close to each other.
Under the local stochastic Pauli noise, we run a four-qubit noisy parameterized quantum circuit and its dual circuit with random parameters.
For each circuit depth $L$, the number of errors in the whole circuit is fixed to $0.5$, $1.0$, and $1.5$ by adjusting the noise level of each gate operation, and the trace distance between $\rho$ and $\bar{\rho}$ is averaged over 20 samples with different random parameters.

\begin{figure}[htbp]
    \subfloat[\label{fig:local-stochastic-pauli_depth-to-tracedist}]{
        \includegraphics[width=0.48\textwidth]{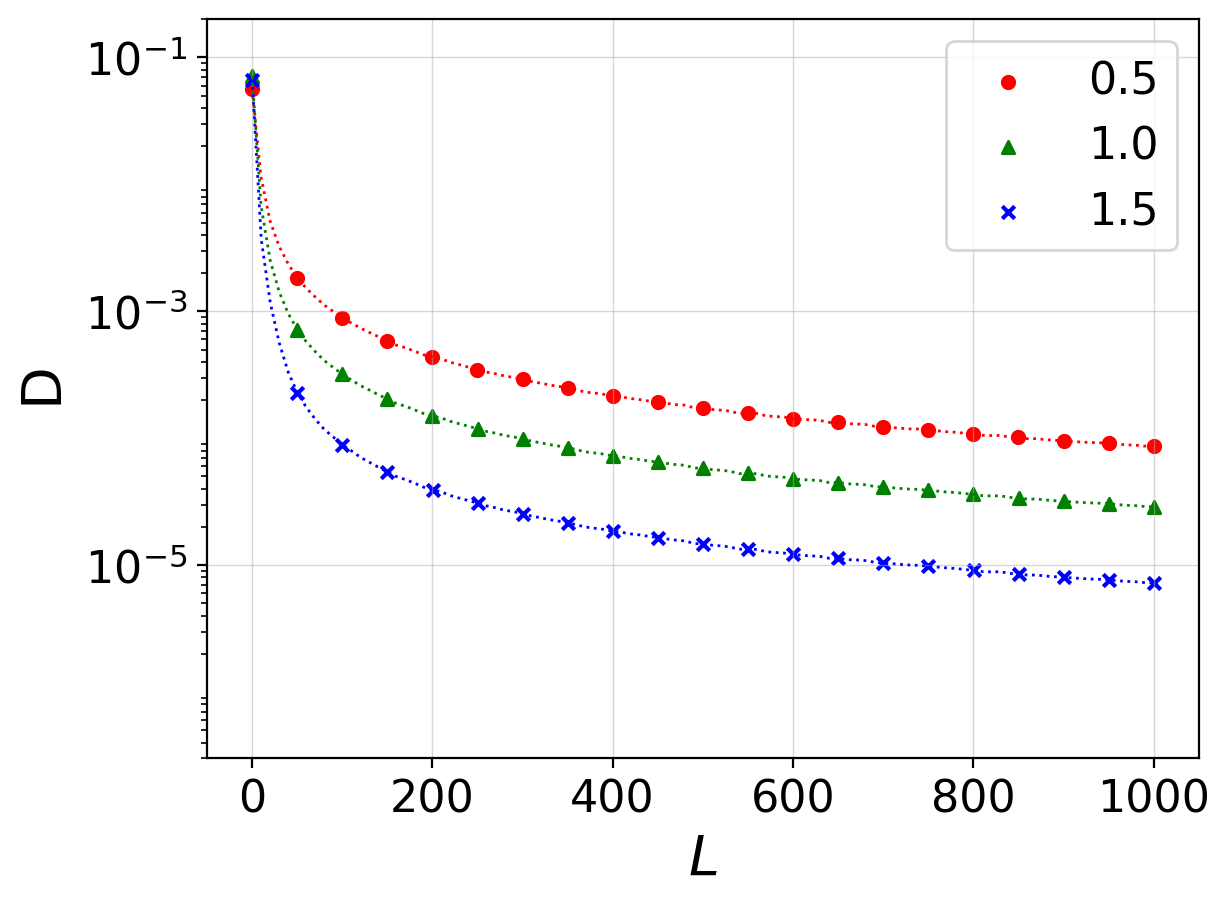}
    }
    \subfloat[\label{fig:local-stochastic-pauli_depth-to-tracedist-dm}]{
        \includegraphics[width=0.48\textwidth]{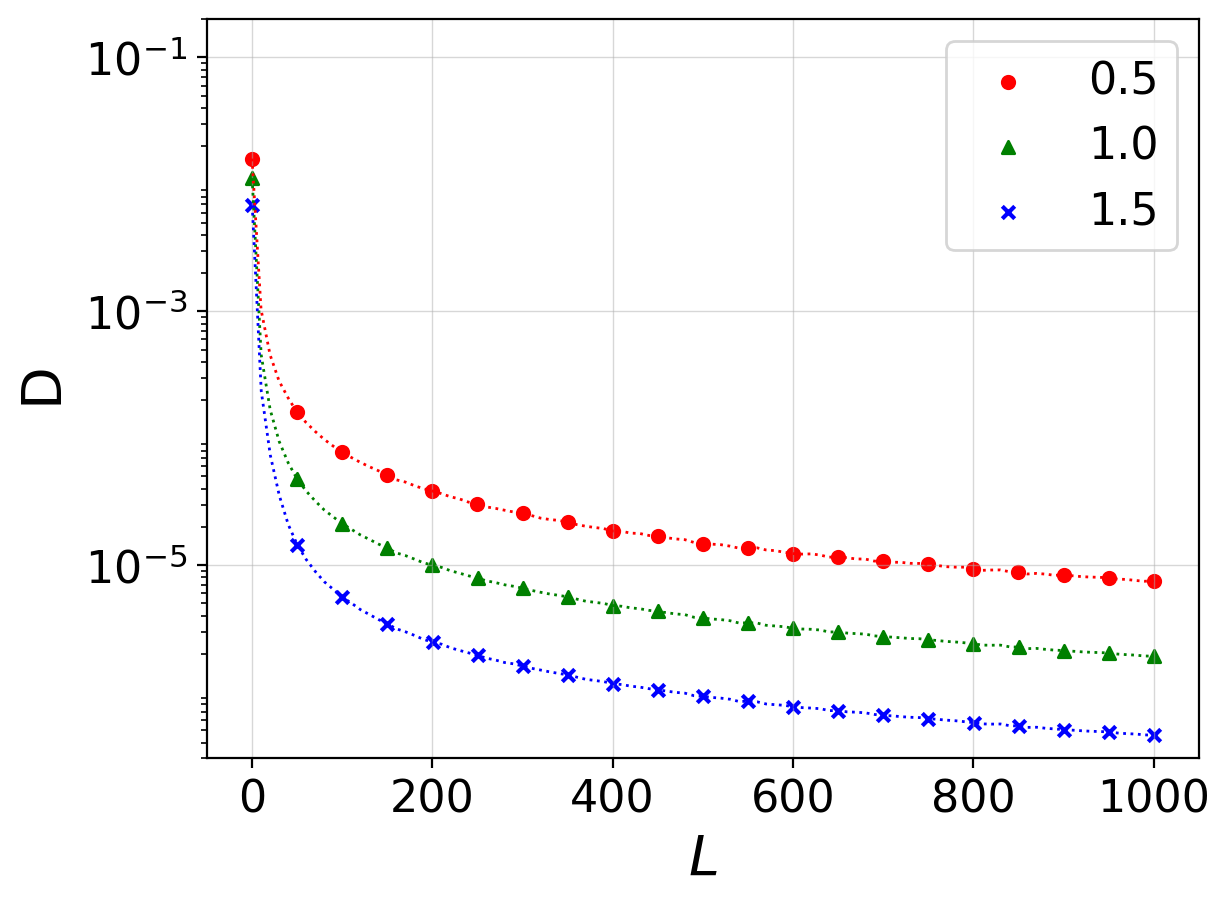}
    }
    \caption{
        The trace distance $\operatorname{D}$ to the different numbers of layers $L$ between 
        (a) $\rho$ and its dual state $\bar{\rho}$, and 
        (b) $\rho^{2}$ and its approximation $(\bar{\rho}\rho + \rho\bar{\rho}) / 2$ in a symmetrized form using $\bar{\rho}$.
        We average over 20 samples for each depth from $0$ to $1000$ layers with an interval of $10$ layers, which are shown by the dotted curves.
        The discrete plots visualized in the figure are the representative depths with an interval of $50$ layers.
        The red, green, and blue markers and dotted curves denote the setting of different numbers of errors in the whole circuit, $0.5$, $1.0$, and $1.5$.
    }
    \label{fig:local-stochastic-pauli_comparison}
\end{figure}

The results are shown in Fig.~\ref{fig:local-stochastic-pauli_comparison}.
When the circuit depth is large enough, the trace distance $\operatorname{D}$ between $\rho$ and $\bar{\rho}$ gradually diminishes.
This implies the difference in the state $\rho$ and its dual $\bar{\rho}$ becomes ignorable under the error schemes where the error mitigation techniques are expected to work in practice.
The reason the circuit containing more errors gives a shorter trace distance is likely to be that $\rho$ and $\bar{\rho}$ are both shrinking to the completely mixed state.

Consequently, we can see the DSP ansatz $(\bar{\rho}\rho+\rho\bar{\rho})/2$ gets closer to the square of density matrix $\rho^{2}$ from Fig.~\ref{fig:local-stochastic-pauli_depth-to-tracedist-dm} in terms of trace distance.
By taking the power of noisy states, the dominant eigenvector is amplified, and the trace distance between $(\bar{\rho}\rho+\rho\bar{\rho})/2$ and $\rho^{2}$ becomes around ten times shorter than $\rho$ and $\bar{\rho}$.
Therefore, we can numerically verify that $(\bar{\rho}\rho+\rho\bar{\rho})/2$ from DSP well approximates the power of noisy density matrix $\rho^{2}$.

We also remark that we have $\bar{\rho} = \rho$ under some depolarizing noises as we consider only the single-qubit and two-qubit gate operations in ansatz circuits.
In particular, we point out the global depolarizing channel $\mathcal{E}_{\mathrm{dep}}^{(\mathrm{global})}$ and the set of local depolarizing channels $\left\{\mathcal{E}_{\mathrm{dep}}^{(1)}, \mathcal{E}_{\mathrm{dep}}^{(2)}\right\}$ described below yield $\bar{\rho} = \rho$.

First, the global depolarizing channel $\mathcal{E}_{\mathrm{dep}}^{(\mathrm{global})}$ with depolarizing probability $p$ taking a density matrix $\rho$ with dimension $d$ is defined
\begin{equation}
\begin{split}
    \mathcal{E}_{\mathrm{dep}}^{(\mathrm{global})} \left(\rho\right) = \left(1-p\right)\rho + p\operatorname{Tr}\left[\rho\right]\frac{I}{d},
\end{split}
\end{equation}
where $I$ is the maximally mixed state with dimension $d$.
When the noise is assumed during the circuit, we assume the global depolarizing noise $\mathcal{E}_{\mathrm{dep}}^{(\mathrm{global})}$ is applied right after each gate operation.
It is clear that $\mathcal{E}_{\mathrm{dep}}^{(\mathrm{global})}$ commutes with any unitary operations, and its dual $\bar{\mathcal{E}}_{\mathrm{dep}}^{(\mathrm{global})}$ is equivalent to $\mathcal{E}_{\mathrm{dep}}^{(\mathrm{global})}$ as well.

When it comes to the set of local depolarizing channels $\left\{\mathcal{E}_{\mathrm{dep}}^{(1)}, \mathcal{E}_{\mathrm{dep}}^{(2)}\right\}$ that fit the locality of each gate operation, it also fulfills $\bar{\rho} = \rho$.
We define $\mathcal{E}_{\mathrm{dep}}^{(1)}$ and $\mathcal{E}_{\mathrm{dep}}^{(2)}$ respectively as follows, assuming they occur right after each gate operation.
For a single-qubit unitary operation, we assume the channel
\begin{equation}
\begin{split}
    \mathcal{E}_{\mathrm{dep}}^{(1)}\left(\rho\right) &= \left(1-p\right)\rho + p\operatorname{Tr}\left[\rho\right]\frac{I}{2}, \\
    \mathcal{E}_{\mathrm{dep}}^{(2)}\left(\rho\right) &= \left(1-p\right)\rho + p\operatorname{Tr}\left[\rho\right]\frac{I}{4}.
\end{split}
\end{equation}
The single-qubit depolarizing channel $\mathcal{E}_{\mathrm{dep}}^{(1)}$ commutes with single-qubit unitary operations and the two-qubit depolarizing channel $\mathcal{E}_{\mathrm{dep}}^{(2)}$ commutes with both single-qubit and two-qubit unitary operations.
For the dual channels of $\mathcal{E}_{\mathrm{dep}}^{(1)}$ and $\mathcal{E}_{\mathrm{dep}}^{(2)}$, it is also clear that $\bar{\mathcal{E}}_{\mathrm{dep}}^{(1)} = \mathcal{E}_{\mathrm{dep}}^{(1)}$ and $\bar{\mathcal{E}}_{\mathrm{dep}}^{(2)} = \mathcal{E}_{\mathrm{dep}}^{(2)}$.

Using the definitions and properties above, we can see $\bar{\rho} = \rho$ under both channels, $\mathcal{E}_{\mathrm{dep}}^{(\mathrm{global})}$ and $\left\{\mathcal{E}_{\mathrm{dep}}^{(1)}, \mathcal{E}_{\mathrm{dep}}^{(2)}\right\}$.
Recalling in Eq.~\eqref{eq:mathcal_U} and Eq.~\eqref{eq:mathcal_U_dual}, the noisy unitary operation $\mathcal{U}$ and its dual $\mathcal{U}_{\mathrm{dual}}$ are described by the composition of noisy channels $\{\mathcal{F}_{i}\}_{i=1}^{L}$ with $L$ layers of unitary operations $\{\mathcal{G}_{i}\}_{i=1}^{L}$ as
\begin{equation}
\begin{split}
    \mathcal{U}
    &= \mathcal{F}_{L} \circ \mathcal{G}_{L} \circ \mathcal{F}_{L-1} \circ \mathcal{G}_{L-1} \circ \cdots \circ \mathcal{F}_{1} \circ \mathcal{G}_{1}, \\
    \mathcal{U}_{\mathrm{dual}}
    &= \mathcal{G}_{L} \circ \bar{\mathcal{F}}_{L} \circ \mathcal{G}_{L-1} \circ \bar{\mathcal{F}}_{L-1} \circ \cdots \circ \mathcal{G}_{1} \circ \bar{\mathcal{F}}_{1},
\end{split}
\end{equation}
assuming $\{\mathcal{G}_{i}\}_{i=1}^{L}$ is the set of single-qubit or two-qubit unitary operations.
Replacing $\mathcal{F}_{l}$ with either $\mathcal{E}_{\mathrm{dep}}^{(\mathrm{global})}$ or $\mathcal{E}_{\mathrm{dep}}^{(k)}$where $k = 1,2$ depending on whether $\mathcal{G}_{l}$ is a single-qubit unitary operation or a two-qubit operation will straightforwardly result in $\mathcal{U}_{\mathrm{dual}} = \mathcal{U}$ as follows
\begin{equation}
\begin{split}
    \mathcal{U}_{\mathrm{dual}} 
    &= \mathcal{G}_{L} \circ \bar{\mathcal{F}}_{L} \circ \mathcal{G}_{L-1} \circ \bar{\mathcal{F}}_{L-1} \circ \cdots \circ \mathcal{G}_{1} \circ \bar{\mathcal{F}}_{1} \\
    &= \mathcal{G}_{L} \circ \mathcal{F}_{L} \circ \mathcal{G}_{L-1} \circ \mathcal{F}_{L-1} \circ \cdots \circ \mathcal{G}_{1} \circ \mathcal{F}_{1} \\
    &= \mathcal{F}_{L} \circ \mathcal{G}_{L} \circ \mathcal{F}_{L-1} \circ \mathcal{G}_{L-1} \circ \cdots \circ \mathcal{F}_{1} \circ \mathcal{G}_{1} \\
    &= \mathcal{U},
\end{split}
\end{equation}
which means $\bar{\rho} = \rho$.
Note that $\bar\rho = \rho$ would no longer hold in general for the local depolarizing error taking the form of the tensor product of two single-qubit depolarizing channels for two-qubit operations.

\section{Quantum Circuits for the General Dual-GSE Ansatz\label{sec:appendix_quantum_circuits_for_the_generalized_ansatz}}

In this section, we design quantum circuits to obtain expectation values of the general Dual-GSE ansatz $\rho_{\mathrm{DGSE}}$ defined in Sec.~\ref{sec:dgse_The_General_Ansatz_for_Dual-GSE}.
Depending on the parity of $L_{ik}$ and $L_{i'k'}$, $\rho_{\mathrm{DGSE}}$ is represented as
\begin{equation}\label{eq:rho_dgse_general}
    \rho_{\mathrm{DGSE}}=
    \begin{cases}
        \displaystyle \sum_{i,i'}\alpha_{i}^{*}\alpha_{i'} \sum_{k,k'} h_{k}^{(i)*}h_{k'}^{(i')} 
        \left(\tau_{kL_{ik}}^{(i)\dagger} \cdots \tau_{k3}^{(i)\dagger}\bar{\tau}_{k2}^{(i)\dagger}\tau_{k1}^{(i)\dagger}\right) 
        \left(\bar{\tau}_{k'1}^{(i')}\tau_{k'2}^{(i')}\bar{\tau}_{k'3}^{(i')} \cdots \bar{\tau}_{k'L_{i'k'}}^{(i')}\right) & \text{ for odd $L_{ik}$ and odd $L_{i'k'}$, } \\
        \displaystyle \sum_{i,i'}\alpha_{i}^{*}\alpha_{i'} \sum_{k,k'} h_{k}^{(i)*}h_{k'}^{(i')} 
        \left(\tau_{kL_{ik}}^{(i)\dagger} \cdots \tau_{k3}^{(i)\dagger}\bar{\tau}_{k2}^{(i)\dagger}\tau_{k1}^{(i)\dagger}\right) 
        \left(\bar{\tau}_{k'1}^{(i')}\tau_{k'2}^{(i')}\bar{\tau}_{k'3}^{(i')} \cdots \tau_{k'L_{i'k'}}^{(i')}\right) & \text{ for odd $L_{ik}$ and even $L_{i'k'}$, } \\
        \displaystyle \sum_{i,i'}\alpha_{i}^{*}\alpha_{i'} \sum_{k,k'} h_{k}^{(i)*}h_{k'}^{(i')} 
        \left(\tau_{kL_{ik}}^{(i)\dagger} \cdots \bar{\tau}_{k3}^{(i)\dagger}\tau_{k2}^{(i)\dagger}\bar{\tau}_{k1}^{(i)\dagger}\right) 
        \left(\tau_{k'1}^{(i')}\bar{\tau}_{k'2}^{(i')}\tau_{k'3}^{(i')} \cdots \tau_{k'L_{i'k'}}^{(i')}\right) & \text{ for even $L_{ik}$ and odd $L_{i'k'}$, } \\
        \displaystyle \sum_{i,i'}\alpha_{i}^{*}\alpha_{i'} \sum_{k,k'} h_{k}^{(i)*}h_{k'}^{(i')} 
        \left(\tau_{kL_{ik}}^{(i)\dagger} \cdots \bar{\tau}_{k3}^{(i)\dagger}\tau_{k2}^{(i)\dagger}\bar{\tau}_{k1}^{(i)\dagger}\right) 
        \left(\tau_{k'1}^{(i')}\bar{\tau}_{k'2}^{(i')}\tau_{k'3}^{(i')} \cdots \bar{\tau}_{k'L_{i'k'}}^{(i')}\right) & \text{ for even $L_{ik}$ and even $L_{i'k'}$. }
    \end{cases}
\end{equation}
To obtain the expectation value $\operatorname{Tr}\left[\rho_{\mathrm{DGSE}}O \right]$ of $\rho_{\mathrm{DGSE}}$ in terms of an observable $O$, we have to evaluate the following quantities:
\begin{equation}\label{eq:trace_dgse_general}
\begin{split}
    \operatorname{Tr}\left[ \left(\tau_{kL_{ik}}^{(i)\dagger} \cdots \tau_{k3}^{(i)\dagger}\bar{\tau}_{k2}^{(i)\dagger}\tau_{k1}^{(i)\dagger}\right) 
                            \left(\bar{\tau}_{k'1}^{(i')}\tau_{k'2}^{(i')}\bar{\tau}_{k'3}^{(i')} \cdots \bar{\tau}_{k'L_{i'k'}}^{(i')}\right) O \right] & \text{ for odd $L_{ik}$ and odd $L_{i'k'}$, } \\
    \operatorname{Tr}\left[ \left(\tau_{kL_{ik}}^{(i)\dagger} \cdots \tau_{k3}^{(i)\dagger}\bar{\tau}_{k2}^{(i)\dagger}\tau_{k1}^{(i)\dagger}\right) 
                            \left(\bar{\tau}_{k'1}^{(i')}\tau_{k'2}^{(i')}\bar{\tau}_{k'3}^{(i')} \cdots \tau_{k'L_{i'k'}}^{(i')}\right) O \right] & \text{ for odd $L_{ik}$ and even $L_{i'k'}$, } \\
    \operatorname{Tr}\left[ \left(\tau_{kL_{ik}}^{(i)\dagger} \cdots \bar{\tau}_{k3}^{(i)\dagger}\tau_{k2}^{(i)\dagger}\bar{\tau}_{k1}^{(i)\dagger}\right) 
                            \left(\tau_{k'1}^{(i')}\bar{\tau}_{k'2}^{(i')}\tau_{k'3}^{(i')} \cdots \tau_{k'L_{i'k'}}^{(i')}\right) O \right] & \text{ for even $L_{ik}$ and odd $L_{i'k'}$, } \\
    \operatorname{Tr}\left[ \left(\tau_{kL_{ik}}^{(i)\dagger} \cdots \bar{\tau}_{k3}^{(i)\dagger}\tau_{k2}^{(i)\dagger}\bar{\tau}_{k1}^{(i)\dagger}\right) 
                            \left(\tau_{k'1}^{(i')}\bar{\tau}_{k'2}^{(i')}\tau_{k'3}^{(i')} \cdots \bar{\tau}_{k'L_{i'k'}}^{(i')}\right) O \right] & \text{ for even $L_{ik}$ and even $L_{i'k'}$, }
\end{split}
\end{equation}
where the implementation of $\tau_{kl}^{(i)}$, $\bar{\tau}_{kl}^{(i)}$, and their conjugate transpose $\tau_{kl}^{(i)\dagger}$, $\bar{\tau}_{kl}^{(i)\dagger}$ is required.
All of these expectation values are available by running independent quantum circuits with $\displaystyle \left\lceil\frac{L_{ik} + L_{i'k'}}{2}\right\rceil$ state copies.
The following part introduces how to design those quantum circuits systematically.

Before providing specific circuit constructions for Eq.~\eqref{eq:trace_dgse_general}, we first review the measurement outcomes of the quantum circuit in Fig.~\ref{fig:qc_dgse_general_n_copies}, which is a generalization of Fig.~\ref{fig:qc_dgse_general_one_copy} from one state copy to $n$ state copies.
When $n=1$, the measurement outcomes of this circuit become
\begin{equation}\label{eq:trace_dgse_X_Y_one_copy}
\begin{split}
    \langle X\otimes \operatorname{P}_{\vec{0}}\rangle_{O} 
    &= \frac{1}{2} \operatorname{Tr}\left[\mathcal{U}_{1}^{\mathrm{(out)}}(V_{1}^{\dagger}V_{1}\mathcal{U}_{1}^{\mathrm{(in)}}(\rho_{\vec{0}})W_{1}^{\dagger}OW_{1})\operatorname{P}_{\vec{0}}\right] 
     + \frac{1}{2} \operatorname{Tr}\left[\mathcal{U}_{1}^{\mathrm{(out)}}(W_{1}^{\dagger}OW_{1}\mathcal{U}_{1}^{\mathrm{(in)}}(\rho_{\vec{0}})V_{1}^{\dagger}V_{1})\operatorname{P}_{\vec{0}}\right] \\
    &= \frac{1}{2} \operatorname{Tr}\left[(V_{1}\rho W_{1}^{\dagger})O(W_{1}\bar{\rho}V_{1}^{\dagger})\right] 
     + \frac{1}{2} \operatorname{Tr}\left[O(W_{1}\rho V_{1}^{\dagger})(V_{1}\bar{\rho}W_{1}^{\dagger})\right] \\
    &= \frac{1}{2} \operatorname{Tr}\left[\tau_{1} O\bar{\tau}_{1}^{\dagger}\right] + \frac{1}{2} \operatorname{Tr}\left[O\tau_{1}^{\dagger}\bar{\tau}_{1}\right]
     = \operatorname{Tr}\left[\frac{\bar{\tau}_{1}^{\dagger}\tau_{1} + \tau_{1}^{\dagger}\bar{\tau}_{1}}{2}O \right], \\
    \langle Y\otimes \operatorname{P}_{\vec{0}}\rangle_{O} 
    &= \frac{i}{2} \operatorname{Tr}\left[\tau_{1} O\bar{\tau}_{1}^{\dagger}\right] - \frac{i}{2} \operatorname{Tr}\left[O\tau_{1}^{\dagger}\bar{\tau}_{1}\right]
     = \operatorname{Tr}\left[\frac{\bar{\tau}_{1}^{\dagger}\tau_{1} - \tau_{1}^{\dagger}\bar{\tau}_{1}}{-2i}O \right],
\end{split}
\end{equation}
where $\rho_{1} = \mathcal{U}_{1}^{(\mathrm{in})}(|\vec{0}\rangle\langle\vec{0}|)$, $\bar{\rho}_{1} = \mathcal{U}_{1,\mathrm{dual}}^{(\mathrm{out})}(|\vec{0}\rangle\langle\vec{0}|)$, and $\mathcal{U}_{1,\mathrm{dual}}^{(\mathrm{out})}$ is the dual process of $\mathcal{U}_{1,\mathrm{rev}}^{(\mathrm{out})}$.
This allows for computing the following quantities
\begin{equation}
\begin{split}
    \langle X \otimes \operatorname{P}_{\vec{0}}\rangle_{O} - i \langle Y \otimes \operatorname{P}_{\vec{0}}\rangle_{O} &= \operatorname{Tr}\left[\bar{\tau}_{1}^{\dagger}\tau_{1} O \right], \\
    \langle X \otimes \operatorname{P}_{\vec{0}}\rangle_{O} + i \langle Y \otimes \operatorname{P}_{\vec{0}}\rangle_{O} &= \operatorname{Tr}\left[\tau_{1}^{\dagger}\bar{\tau}_{1} O \right],
\end{split}
\end{equation}
which are essentially extracting the first and the second term on the right-hand side of Eq.~\eqref{eq:trace_dgse_X_Y_one_copy}.
When it comes to the circuit with derangement operator $D_n$ for $n$ copies, these quantities become
\begin{equation}\label{eq:trace_dgse_X_Y_n_copies}
\begin{split}
    \langle X \otimes \operatorname{P}_{\vec{0}}^{\otimes n}\rangle_{O} - i \langle Y \otimes \operatorname{P}_{\vec{0}}^{\otimes n}\rangle_{O} &= \operatorname{Tr}\left[ \left(\bar{\tau}_{2}^{\dagger}\tau_{2}\right) \left(\bar{\tau}_{3}^{\dagger}\tau_{3}\right) \cdots \left(\bar{\tau}_{n-1}^{\dagger}\tau_{n-1}\right) \left(\bar{\tau}_{n}^{\dagger}\tau_{n}\right) \left(\bar{\tau}_{1}^{\dagger}\tau_{1}\right) O \right], \\
    \langle X \otimes \operatorname{P}_{\vec{0}}^{\otimes n}\rangle_{O} + i \langle Y \otimes \operatorname{P}_{\vec{0}}^{\otimes n}\rangle_{O} &= \operatorname{Tr}\left[ \left(\tau_{1}^{\dagger}\bar{\tau}_{1}\right) \left(\bar{\tau}_{2}^{\dagger}\tau_{2}\right) \cdots \left(\bar{\tau}_{n-1}^{\dagger}\tau_{n-1}\right) \left(\tau_{n}^{\dagger}\bar{\tau}_{n}\right) O \right].
\end{split}
\end{equation}
Since the measurement result of $\langle X + iY \rangle_{O} := \langle X \otimes \operatorname{P}_{\vec{0}}^{\otimes n}\rangle_{O} + i \langle Y \otimes \operatorname{P}_{\vec{0}}^{\otimes n}\rangle_{O}$ gives a simpler form of $2n$-th power degree, we use $\langle X + iY \rangle_{O}$ to the design quantum circuits for Eq.~\eqref{eq:trace_dgse_general}.

\begin{figure}
    \centering
             \begin{adjustbox}{width=0.99\textwidth}\begin{quantikz}
\lstick{$|+\rangle$}      & \qw                                    & \ctrl{1}   & \octrl{1}  & \qw & \cdots\qquad & \ctrl{3}   & \octrl{3}  & \ctrl{1} & \ctrl{1}                & \ctrl{1}           & \octrl{1}          & \qw & \cdots\qquad & \ctrl{3}           & \octrl{3}          & \qw                                 & \meter{X,Y}    \\
\lstick{$|\vec{0}\rangle$} & \gate{\mathcal{U}_{1}^{\mathrm{(in)}}} & \gate{W_{1}} & \gate{V_{1}} & \qw & \cdots\qquad & \qw        & \qw        & \gate{O} & \gate[3, nwires=2]{D_n} & \gate{W_{1}^{\dagger}} & \gate{V_{1}^{\dagger}} & \qw & \cdots\qquad & \qw                & \qw                & \gate{\mathcal{U}_{1}^{\mathrm{(in)}}} & \push{\ \langle\vec{0}|} \\
\vdots                    & \vdots                                 &            &            &     & \ddots\qquad &            &            &          &                         &                    &                    &     & \ddots\qquad &                    &                    & \vdots                              & \vdots         \\
\lstick{$|\vec{0}\rangle$} & \gate{\mathcal{U}_{n}^{\mathrm{(in)}}} & \qw        & \qw        & \qw & \cdots\qquad & \gate{W_{n}} & \gate{V_{n}} & \qw      &                         & \qw                & \qw                & \qw & \cdots\qquad & \gate{W_{n}^{\dagger}} & \gate{V_{n}^{\dagger}} & \gate{\mathcal{U}_{n}^{\mathrm{(out)}}} & \push{\ \langle\vec{0}|}
        \end{quantikz}\end{adjustbox}
    \caption{
        The generalized resource-efficient purification circuit with $n$ state copies.
        This is the generalization of Fig.~\ref{fig:qc_dgse_general_one_copy} to obtain the expectation value that includes the multiplication of 
        $\tau_{kl}^{(i)}=V_{kl}^{(i)}\rho_{kl}^{(i)}W_{kl}^{(i)\dagger}$, 
        $\bar{\tau}_{kl}^{(i)}=V_{kl}^{(i)}\bar{\rho}_{kl}^{(i)}W_{kl}^{(i)\dagger}$, 
        $\tau_{kl}^{(i)\dagger}=W_{kl}^{(i)}\rho_{kl}^{(i)}V_{kl}^{(i)\dagger}$, and 
        $\bar{\tau}_{kl}^{(i)\dagger}=W_{kl}^{(i)}\bar{\rho}_{kl}^{(i)}V_{kl}^{(i)\dagger}$ to the power of $2n$ in total.
        We also omit the unfocused superscripts and subscripts by simply writing $\tau_{l}$, $\bar{\tau}_{l}$, $\tau_{l}^{\dagger}$, and $\bar{\tau}_{l}^{\dagger}$ to keep equations and circuits simple.
    }
    \label{fig:qc_dgse_general_n_copies}
\end{figure}

\begin{figure}[htbp]
    \subfloat[$\tau_{l}^{\dagger} = W_{l}\rho_{l}V_{l}^{\dagger}$ \label{fig:qc_dgse_tau_dag}]{
        \begin{adjustbox}{width=0.32\textwidth, valign=b}
        \begin{quantikz}
            \lstick{$|+\rangle$}       & \qw                                    & \ctrl{1}     & \octrl{1} \slice{before controlled\\$O$ and $D_n$} & \qw \\
            \lstick{$|\vec{0}\rangle$} & \gate{\mathcal{U}_{l}^{\mathrm{(in)}}} & \gate{W_{l}} & \gate{V_{l}}                                       & \qw
        \end{quantikz}
        \end{adjustbox}
        \begin{adjustbox}{width=0.16\textwidth, valign=b}
            \includegraphics[]{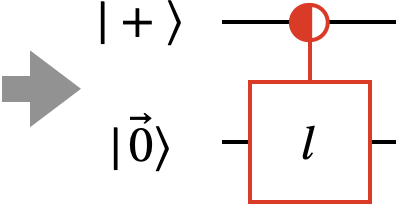}
        \end{adjustbox}
    }
    \hfill
    \subfloat[$\bar{\tau}_{l} = V_{l}\bar{\rho}_{l}W_{l}^{\dagger}$ \label{fig:qc_dgse_tau_dual}]{
        \begin{adjustbox}{width=0.32\textwidth, valign=b}
        \begin{quantikz}
            \slice{after controlled\\$O$ and $D_n$} & \ctrl{1}         & \octrl{1}        & \qw                                 & \meter{X,Y}    \\
                                                    & \gate{W_{l}^{\dagger}} & \gate{V_{l}^{\dagger}} & \gate{\mathcal{U}_{l}^{\mathrm{(out)}}} & \push{\ \langle\vec{0}|}
        \end{quantikz}
        \end{adjustbox}
        \begin{adjustbox}{width=0.16\textwidth, valign=b}
            \includegraphics[]{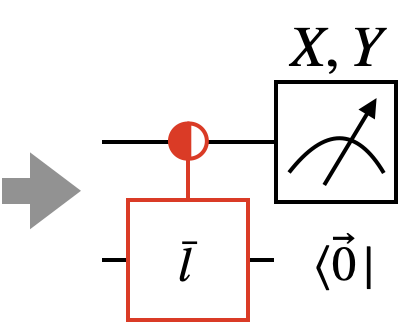}
        \end{adjustbox}
    }
    \hfill
    \subfloat[$\tau_{l} = V_{l}\rho_{l}W_{l}^{\dagger}$ \label{fig:qc_dgse_tau}]{
        \begin{adjustbox}{width=0.32\textwidth, valign=b}
        \begin{quantikz}
            \lstick{$|+\rangle$}       & \qw                                & \octrl{1} & \ctrl{1} \slice{before controlled\\$O$ and $D_n$} & \qw \\
            \lstick{$|\vec{0}\rangle$} & \gate{\mathcal{U}_{l}^{\mathrm{(in)}}} & \gate{W_{l}} & \gate{V_{l}}                                           & \qw 
        \end{quantikz}
        \end{adjustbox}
        \begin{adjustbox}{width=0.16\textwidth, valign=b}
            \includegraphics[]{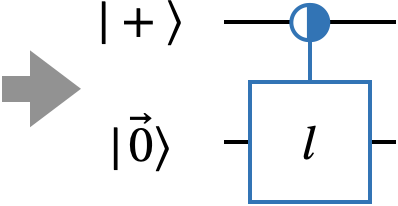}
        \end{adjustbox}
    }
    \hfill
    \subfloat[$\bar{\tau}_{l}^{\dagger} = W_{l}\bar{\rho}_{l}V_{l}^{\dagger}$ \label{fig:qc_dgse_tau_dual_dag}]{
        \begin{adjustbox}{width=0.32\textwidth, valign=b}
        \begin{quantikz}
            \slice{after controlled\\$O$ and $D_n$} & \octrl{1}         & \ctrl{1}        & \qw                                 & \meter{X,Y}    \\
                                                    & \gate{W_{l}^{\dagger}} & \gate{V_{l}^{\dagger}} & \gate{\mathcal{U}_{l}^{\mathrm{(out)}}} & \push{\ \langle\vec{0}|}
        \end{quantikz}
        \end{adjustbox}
        \begin{adjustbox}{width=0.16\textwidth, valign=b}
            \includegraphics[]{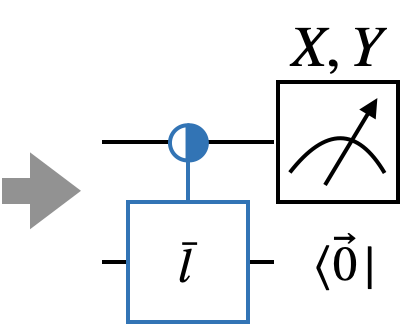}
        \end{adjustbox}
    }
    \caption{
        The fractions of quantum circuits to realize the quantum states
        (a) $\tau_{l}^{\dagger} = W_{l}\rho_{l}V_{l}^{\dagger}$, 
        (b) $\bar{\tau}_{l} = V_{l}\bar{\rho}_{l}W_{l}^{\dagger}$, 
        (c) $\tau_{l} = V_{l}\rho_{l}W_{l}^{\dagger}$, and 
        (d) $\bar{\tau}_{l}^{\dagger} = W_{l}\bar{\rho}_{l}V_{l}^{\dagger}$, 
        in $l$-th quantum register.
        The expectation value containing each state is obtained as $\langle X + iY\rangle_{O} := \langle X\otimes \operatorname{P}_{\vec{0}}\rangle_{O} + i\langle Y\otimes \operatorname{P}_{\vec{0}}\rangle_{O}$.
        To simplify the gate notation in Fig.~\ref{fig:qcs_dgse_general}, we color half-filled controlled gates red or blue.
        The red gates with left-filled and right-empty control first apply $|1\rangle$-controlled $W_{l}$ (or $W_{l}^{\dagger}$) then apply $|0\rangle$-controlled $V_{l}$ (or $V_{l}^{\dagger}$).
        The blue gates with left-empty and right-filled control first apply $|0\rangle$-controlled $W_{l}$ (or $W_{l}^{\dagger}$) then apply $|1\rangle$-controlled $V_{l}$ (or $V_{l}^{\dagger}$).
        This means 
        the red gate with label $l$ creates (a) $\tau_{l}^{\dagger}$, 
        the red gate with label $\bar{l}$ creates (b) $\bar{\tau}_{l}$, 
        the blue gate with label $l$ creates (c) $\tau_{l}$, and 
        the blue gate with label $\bar{l}$ creates (d) $\bar{\tau}_{l}^{\dagger}$.
    }
    \label{fig:qcs_dgse_tau}
\end{figure}

On the basis of the observation above, we now analyze which set of gate operations is in charge of creating $\tau_{kl}^{(i)}$, $\tau_{kl}^{(i)\dagger}$, $\bar{\tau}_{kl}^{(i)}$, and $\bar{\tau}_{kl}^{(i)\dagger}$ in Eq.~\eqref{eq:trace_dgse_general} when measuring $\langle X + iY \rangle_{O} := \langle X\otimes \operatorname{P}_{\vec{0}}\rangle_{O} + i\langle Y\otimes \operatorname{P}_{\vec{0}}\rangle_{O}$ on each state copy.
This can be realized by the fractions of quantum circuits in Fig.~\ref{fig:qcs_dgse_tau}, which focuses on the $l$-th quantum register for the state copies of Fig.~\ref{fig:qc_dgse_general_n_copies}.
To keep equations simple, we still omit unfocused superscripts and subscripts from $\tau_{kl}^{(i)}$, letting it simply be denoted by $\tau_{l}$.

First, $\tau_{l}$ is implemented as Fig.~\ref{fig:qcs_dgse_tau}(c), where measuring $\langle X + iY \rangle_{O}$ leads to the following state:
\begin{equation}
\begin{split}
    \operatorname{Tr}\left[V_{l}\mathcal{U}_{l}^{\mathrm{(in)}}(|\vec{0}\rangle\langle\vec{0}|)W_{l}^{\dagger}|\vec{0}\rangle\langle\vec{0}|\right] 
    = \operatorname{Tr}\left[V_{l}\rho_{l}W_{l}^{\dagger}|\vec{0}\rangle\langle\vec{0}|\right] 
    = \operatorname{Tr}\left[\tau_{l}|\vec{0}\rangle\langle\vec{0}|\right],
\end{split}
\end{equation}
where $\rho_{l} = \mathcal{U}_{l}^{(\mathrm{in})}(|\vec{0}\rangle\langle\vec{0}|)$.
Next, in Fig.~\ref{fig:qcs_dgse_tau}(a), we can include $\tau_{l}^{\dagger}$ in $\langle X + iY \rangle_{O}$ as
\begin{equation}
\begin{split}
    \operatorname{Tr}\left[W_{l}\mathcal{U}_{l}^{\mathrm{(in)}}(|\vec{0}\rangle\langle\vec{0}|)V_{l}^{\dagger}|\vec{0}\rangle\langle\vec{0}|\right] 
    = \operatorname{Tr}\left[W_{l}\rho_{l}V_{l}^{\dagger}|\vec{0}\rangle\langle\vec{0}|\right] 
    = \operatorname{Tr}\left[\tau_{l}^{\dagger}|\vec{0}\rangle\langle\vec{0}|\right],
\end{split}
\end{equation}
by flipping control conditions of $W_{l}$ and $V_{l}$ in the ancillary qubit from Fig.~\ref{fig:qcs_dgse_tau}(c).
To include $\bar{\tau}_{l}$ in $\langle X + iY \rangle_{O}$, we use Fig.~\ref{fig:qcs_dgse_tau}(b).
Adding this circuit to an arbitrary quantum state $|+\rangle\langle+|\otimes\rho$, $\langle X + iY \rangle_{O}$ becomes
\begin{equation}
\begin{split}
    \operatorname{Tr}\left[\mathcal{U}_{l}^{\mathrm{(out)}}(W_{l}^{\dagger}\rho V_{l})|\vec{0}\rangle\langle\vec{0}|\right] 
    = \operatorname{Tr}\left[\rho V_{l}\mathcal{U}_{l,\mathrm{dual}}^{\mathrm{(out)}}(|\vec{0}\rangle\langle\vec{0}|)W_{l}^{\dagger}\right] 
    = \operatorname{Tr}\left[\rho V_{l}\bar{\rho}_{l}W_{l}^{\dagger}\right]
    = \operatorname{Tr}\left[\rho \bar{\tau}_{l}\right],
\end{split}
\end{equation}
where $\bar{\rho}_{l} = \mathcal{U}_{l,\mathrm{dual}}^{(\mathrm{out})}(|\vec{0}\rangle\langle\vec{0}|)$ and $\mathcal{U}_{l,\mathrm{dual}}^{(\mathrm{out})}$ is the dual process of $\mathcal{U}_{l,\mathrm{rev}}^{(\mathrm{out})}$.
Finally, as for $\bar{\tau}_{l}^{\dagger}$, we can also flip the order of control conditions from Fig.~\ref{fig:qcs_dgse_tau}(b) to Fig.~\ref{fig:qcs_dgse_tau}(d), which computes
\begin{equation}
\begin{split}
    \operatorname{Tr}\left[\mathcal{U}_{l}^{\mathrm{(out)}}(V_{l}^{\dagger}\rho W_{l})|\vec{0}\rangle\langle\vec{0}|\right] 
    = \operatorname{Tr}\left[\rho W_{l}\mathcal{U}_{l,\mathrm{dual}}^{\mathrm{(out)}}(|\vec{0}\rangle\langle\vec{0}|)V_{l}^{\dagger}\right] 
    = \operatorname{Tr}\left[\rho W_{l}\bar{\rho}_{l}V_{l}^{\dagger}\right]
    = \operatorname{Tr}\left[\rho \bar{\tau}_{l}^{\dagger}\right].
\end{split}
\end{equation}
Now we find the rules that flipping the control condition of $W_{l}$ ($W_{l}^{\dagger}$) and $V_{l}$ ($V_{l}^{\dagger}$) will take conjugate transpose, and the circuits with $\mathcal{U}_{l}^{(\mathrm{in})}$ applied to the initial state create the computation state while the circuits with $\mathcal{U}_{l}^{(\mathrm{out})}$ followed by the measurement and post-selection create its dual state.

\begin{figure}[htbp]
    \centering
    \subfloat[odd $n$ and odd $n'$ \label{fig:qc_dgse_general_odd_odd}]{
        \begin{adjustbox}{width=0.48\textwidth}
            \includegraphics[]{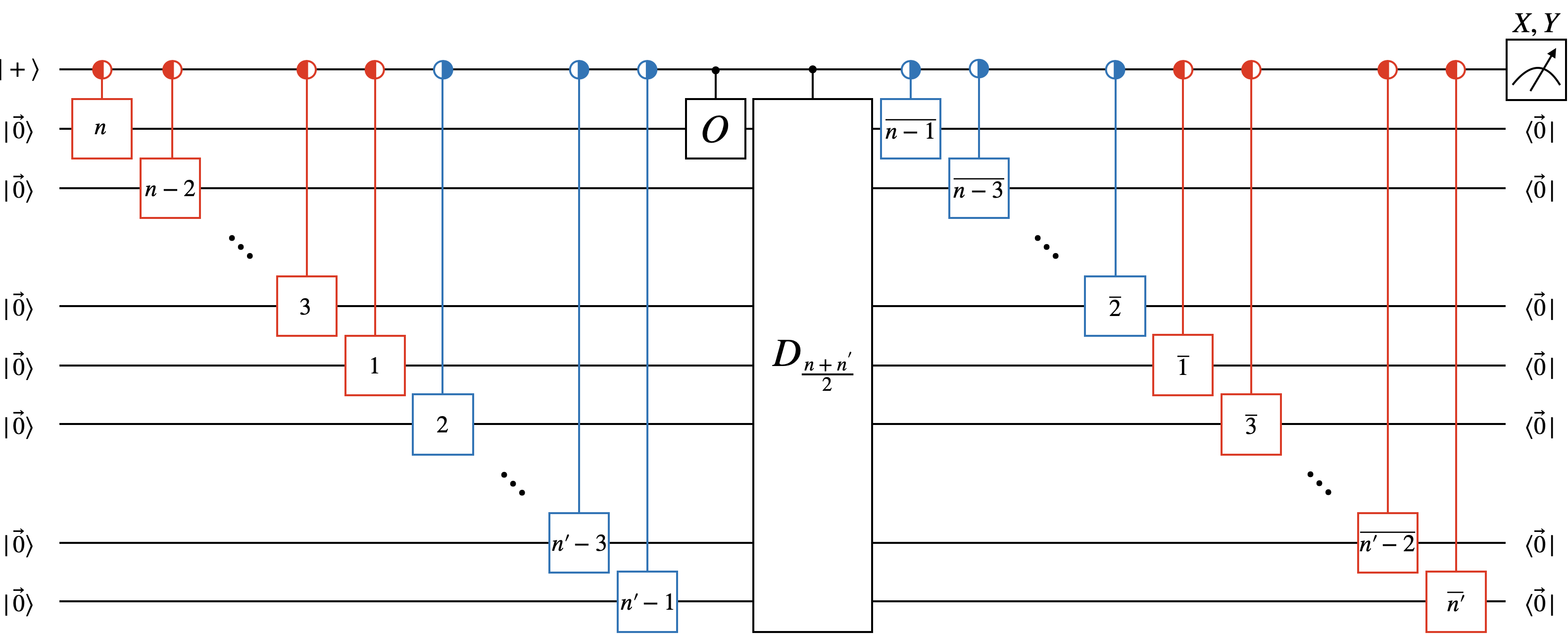}
        \end{adjustbox}
    }
    \hfill
    \subfloat[odd $n$ and even $n'$ \label{fig:qc_dgse_general_odd_even}]{
        \begin{adjustbox}{width=0.48\textwidth}
            \includegraphics[]{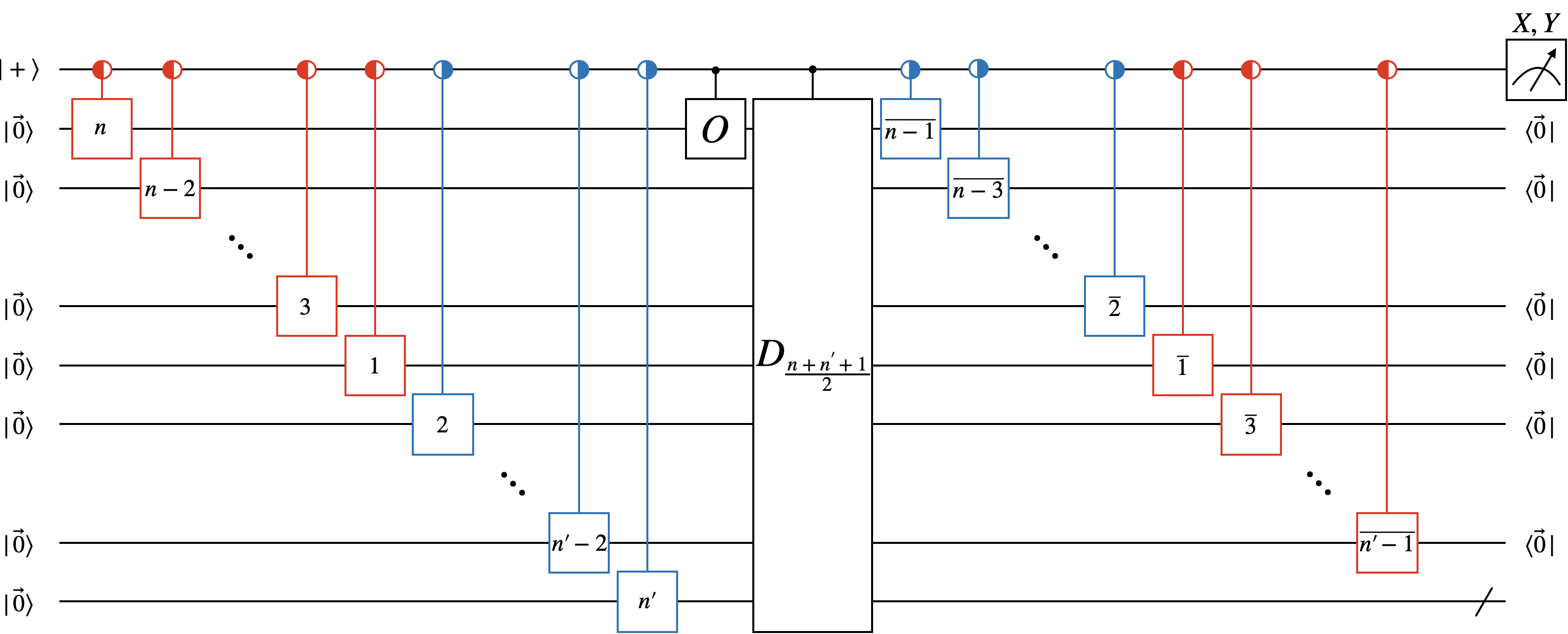}
        \end{adjustbox}
    }
    \hfill
    \subfloat[even $n$ and odd $n'$ \label{fig:qc_dgse_general_even_odd}]{
        \begin{adjustbox}{width=0.48\textwidth}
            \includegraphics[]{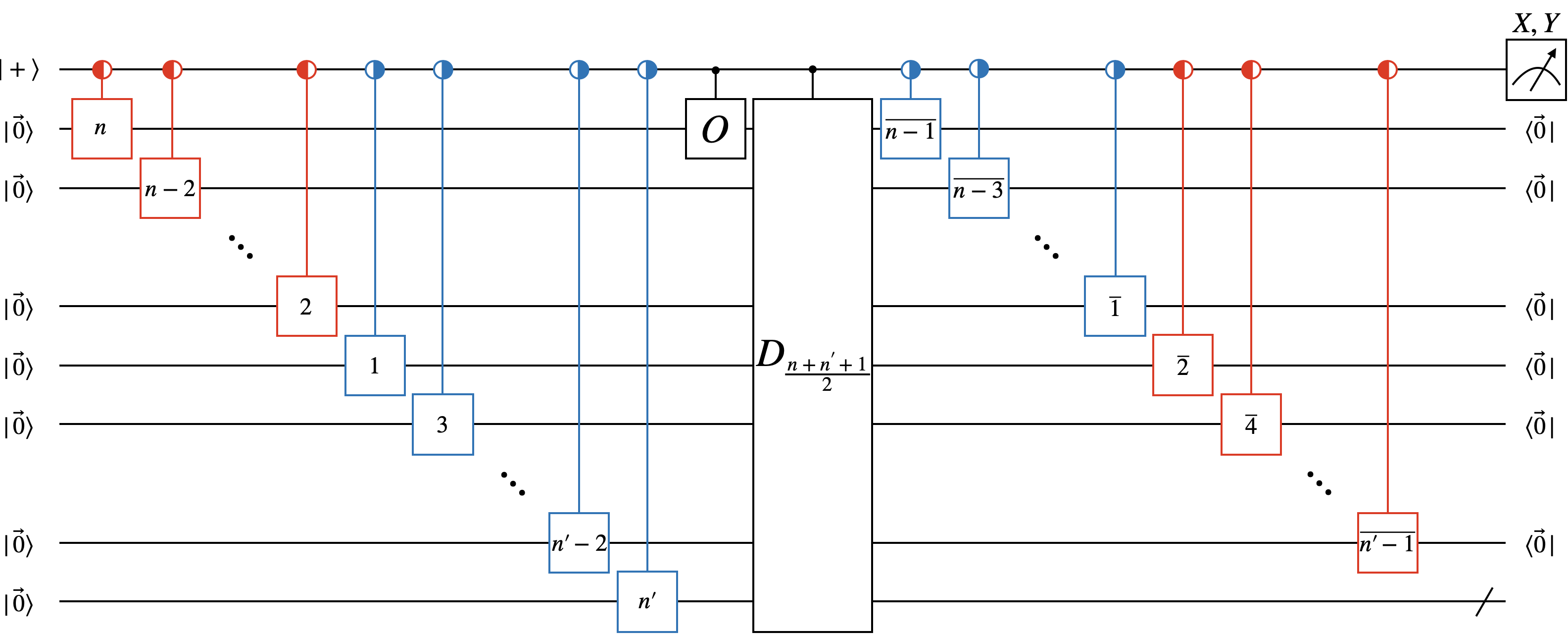}
        \end{adjustbox}
    }
    \hfill
    \subfloat[even $n$ and even $n'$ \label{fig:qc_dgse_general_even_even}]{
        \begin{adjustbox}{width=0.48\textwidth}
            \includegraphics[]{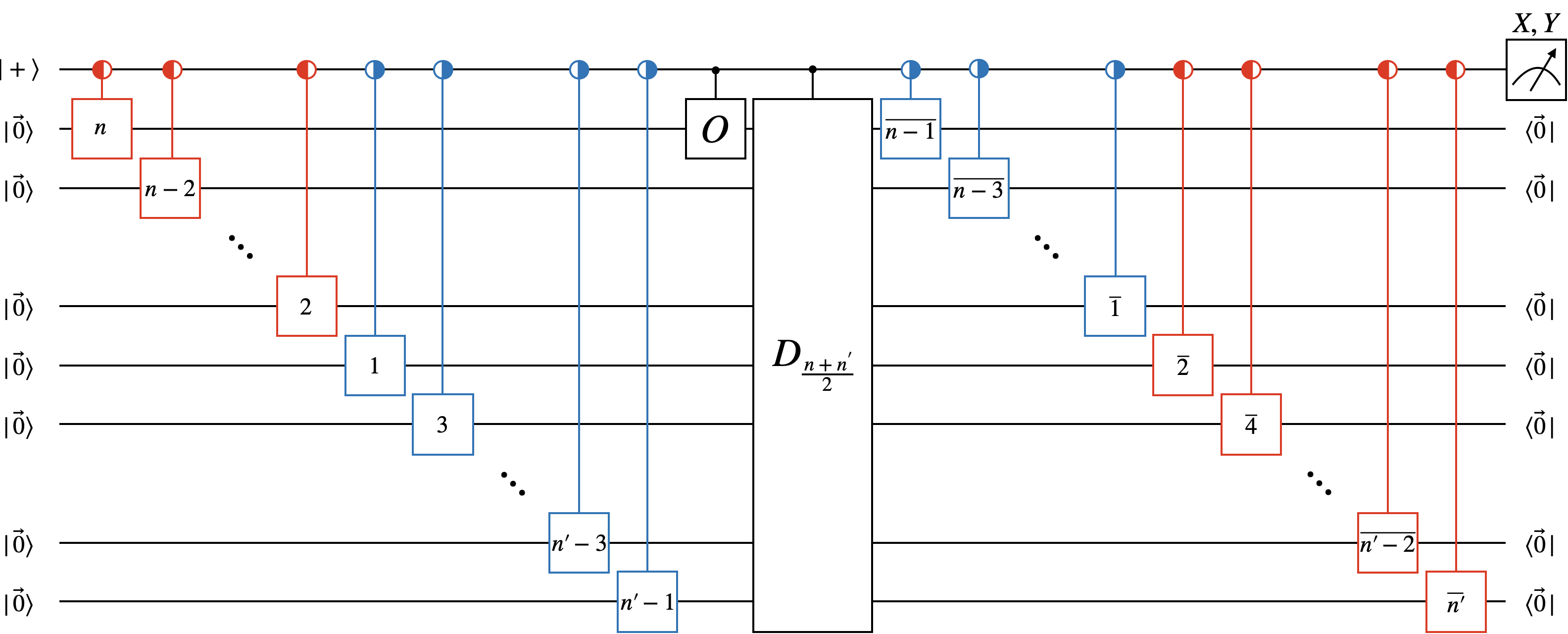}
        \end{adjustbox}
    }
    \caption{
        Quantum Circuits with $\displaystyle \left\lceil \frac{n+n'}{2}\right\rceil$ copies for computing 
        (a) $\operatorname{Tr}\left[ \left(\tau_{n}^{\dagger} \cdots \tau_{3}^{\dagger}\bar{\tau}_{2}^{\dagger}\tau_{1}^{\dagger}\right) 
                                     \left(\bar{\tau}_{1}\tau_{2}\bar{\tau}_{3} \cdots \bar{\tau}_{n'}\right) O \right]$, 
        (b) $\operatorname{Tr}\left[ \left(\tau_{n}^{\dagger} \cdots \tau_{3}^{\dagger}\bar{\tau}_{2}^{\dagger}\tau_{1}^{\dagger}\right) 
                                     \left(\tau_{1}\bar{\tau}_{2}\tau_{3} \cdots \bar{\tau}_{n'}\right) O \right]$,
        (c) $\operatorname{Tr}\left[ \left(\tau_{n}^{\dagger} \cdots \bar{\tau}_{3}^{\dagger}\tau_{2}^{\dagger}\bar{\tau}_{1}^{\dagger}\right) 
                                     \left(\bar{\tau}_{1}\tau_{2}\bar{\tau}_{3} \cdots \bar{\tau}_{n'}\right) O \right]$, and
        (d) $\operatorname{Tr}\left[ \left(\tau_{n}^{\dagger} \cdots \bar{\tau}_{3}^{\dagger}\tau_{2}^{\dagger}\bar{\tau}_{1}^{\dagger}\right) 
                                     \left(\tau_{1}\bar{\tau}_{2}\tau_{3} \cdots \bar{\tau}_{n'}\right) O \right]$.
        The red and blue controlled gates with half-filled controlled operations are defined in Fig.~\ref{fig:qcs_dgse_tau}.
        The symbol $|\vec{0}\rangle$ represents the initial quantum state with the all-zero bitstring.
        The symbol $\langle\vec{0}|$ represents the post-selection over the all-zero bitstring after measuring the quantum register on the computational basis.
    }
    \label{fig:qcs_dgse_general}
\end{figure}

Using these quantum circuit fractions to compute $\tau_{kl}^{(i)}$, $\tau_{kl}^{(i)\dagger}$, $\bar{\tau}_{kl}^{(i)}$, and $\bar{\tau}_{kl}^{(i)\dagger}$, we can design quantum circuits Fig.~\ref{fig:qcs_dgse_general} for the quantities in Eq.~\eqref{eq:trace_dgse_general}.
Let us simplify the notations $L_{ik}^{(i)}$ and $L_{i'k'}^{(i)}$ into $n$ and $n'$.
The circuits Fig.~\ref{fig:qcs_dgse_general}(a), Fig.~\ref{fig:qcs_dgse_general}(b), Fig.~\ref{fig:qcs_dgse_general}(c), and Fig.~\ref{fig:qcs_dgse_general}(d) respectively compute different cases in Eq.~\eqref{eq:expval_X_plus_iY} depending on the parity of $n$ and $n'$.
\begin{equation}\label{eq:expval_X_plus_iY}
    \langle X + iY \rangle_{O} = 
    \begin{cases}
        \displaystyle \operatorname{Tr}\left[ \left(\tau_{n}^{\dagger} \cdots \tau_{3}^{\dagger}\bar{\tau}_{2}^{\dagger}\tau_{1}^{\dagger}\right) 
                                              \left(\bar{\tau}_{1}\tau_{2}\bar{\tau}_{3} \cdots \bar{\tau}_{n'}\right) O \right] & \text{ for odd $n$ and odd $n'$, } \\[5pt]
        \displaystyle \operatorname{Tr}\left[ \left(\tau_{n}^{\dagger} \cdots \tau_{3}^{\dagger}\bar{\tau}_{2}^{\dagger}\tau_{1}^{\dagger}\right) 
                                              \left(\tau_{1}\bar{\tau}_{2}\tau_{3} \cdots \bar{\tau}_{n'}\right) O \right] & \text{ for odd $n$ and even $n'$, } \\[5pt]
        \displaystyle \operatorname{Tr}\left[ \left(\tau_{n}^{\dagger} \cdots \bar{\tau}_{3}^{\dagger}\tau_{2}^{\dagger}\bar{\tau}_{1}^{\dagger}\right) 
                                              \left(\bar{\tau}_{1}\tau_{2}\bar{\tau}_{3} \cdots \bar{\tau}_{n'}\right) O \right] & \text{ for even $n$ and odd $n'$, } \\[5pt]
        \displaystyle \operatorname{Tr}\left[ \left(\tau_{n}^{\dagger} \cdots \bar{\tau}_{3}^{\dagger}\tau_{2}^{\dagger}\bar{\tau}_{1}^{\dagger}\right) 
                                              \left(\tau_{1}\bar{\tau}_{2}\tau_{3} \cdots \bar{\tau}_{n'}\right) O \right] & \text{ for even $n$ and even $n'$, }
    \end{cases}
\end{equation}
where the expectation value $\langle X + iY \rangle_{O}$ is defined as
\begin{equation}
\begin{split}
    \langle X + iY \rangle_{O} := 
    \begin{cases}
        \displaystyle \langle X \otimes \operatorname{P}_{\vec{0}}^{\otimes \frac{n+n'}{2}} \rangle_{O} + i \langle Y \otimes \operatorname{P}_{\vec{0}}^{\otimes \frac{n+n'}{2}}\rangle_{O}, 
        & \text{ for odd $n$ and odd $n'$, } \\[5pt]
        \displaystyle \langle X \otimes \operatorname{P}_{\vec{0}}^{\otimes \frac{n+n'-1}{2}} \otimes I \rangle_{O} + i \langle Y \otimes \operatorname{P}_{\vec{0}}^{\otimes \frac{n+n'-1}{2}} \otimes I \rangle_{O},
        & \text{ for odd $n$ and even $n'$, } \\[5pt]
        \displaystyle \langle X \otimes \operatorname{P}_{\vec{0}}^{\otimes \frac{n+n'-1}{2}} \otimes I \rangle_{O} + i \langle Y \otimes \operatorname{P}_{\vec{0}}^{\otimes \frac{n+n'-1}{2}} \otimes I \rangle_{O},
        & \text{ for even $n$ and odd $n'$, } \\[5pt]
        \displaystyle \langle X \otimes \operatorname{P}_{\vec{0}}^{\otimes \frac{n+n'}{2}} \rangle_{O} + i \langle Y \otimes \operatorname{P}_{\vec{0}}^{\otimes \frac{n+n'}{2}} \rangle_{O}
        & \text{ for even $n$ and even $n'$. } 
    \end{cases}
\end{split}
\end{equation}
The number of copies required for each quantum circuit in Fig.~\ref{fig:qcs_dgse_general} to compute Eq.~\eqref{eq:expval_X_plus_iY} is $\displaystyle \left\lceil \frac{n + n'}{2} \right\rceil$.
This provides the design of quantum circuits and measurement processes for approximating the following general Dual-GSE ansatz
\begin{equation}\label{eq:rho_dgse_general_formal}
    \rho_{\mathrm{DGSE}} = \frac{{\mathcal{P}}^{\dagger} \mathcal{P}}{\operatorname{Tr}\left[{\mathcal{P}}^{\dagger} \mathcal{P}\right]} \text{ with }{\mathcal{P}}=\sum_{i}\alpha_{i} {\sigma}_{i},
\end{equation}
where $\displaystyle {\sigma}_{i} = \sum_{k} h_{k}^{(i)} \left(\prod_{l=1}^{L_{ik}} {\tau}_{kl}^{(i)}\right)$ for $h_{k}^{(i)} \in \mathbb{C}$ and $\tau_{kl}^{(i)}=V_{kl}^{(i)} \rho_{kl}^{(i)} W_{kl}^{(i)\dagger}$.
Note that $\tau_{kl}^{(i)}$ and $\bar{\tau}_{kl}^{(i)}$ do not have to be a physically valid density matrix in general, while $\rho_{\mathrm{DGSE}}$ almost becomes a physically valid quantum ansatz when the gap is small between the state $\rho$ and its dual state $\bar{\rho}$ used in the circuit implementations we have discussed.

\begin{figure}[htbp]
    \centering
    \subfloat[odd $n$ and odd $n'$ \label{fig:qc_dgse_general_odd_A_odd}]{
        \begin{adjustbox}{width=0.48\textwidth}
            \includegraphics[]{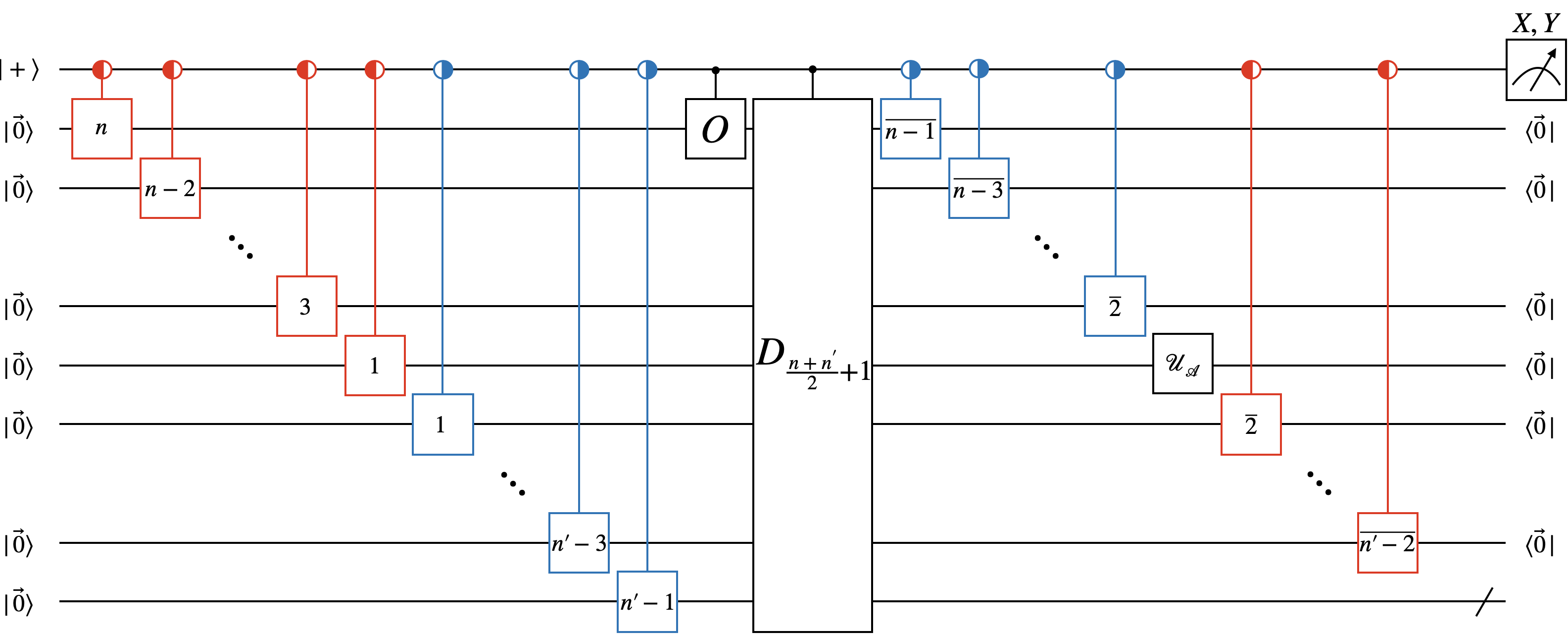}
        \end{adjustbox}
    }
    \hfill
    \subfloat[odd $n$ and even $n'$ \label{fig:qc_dgse_general_odd_A_even}]{
        \begin{adjustbox}{width=0.48\textwidth}
            \includegraphics[]{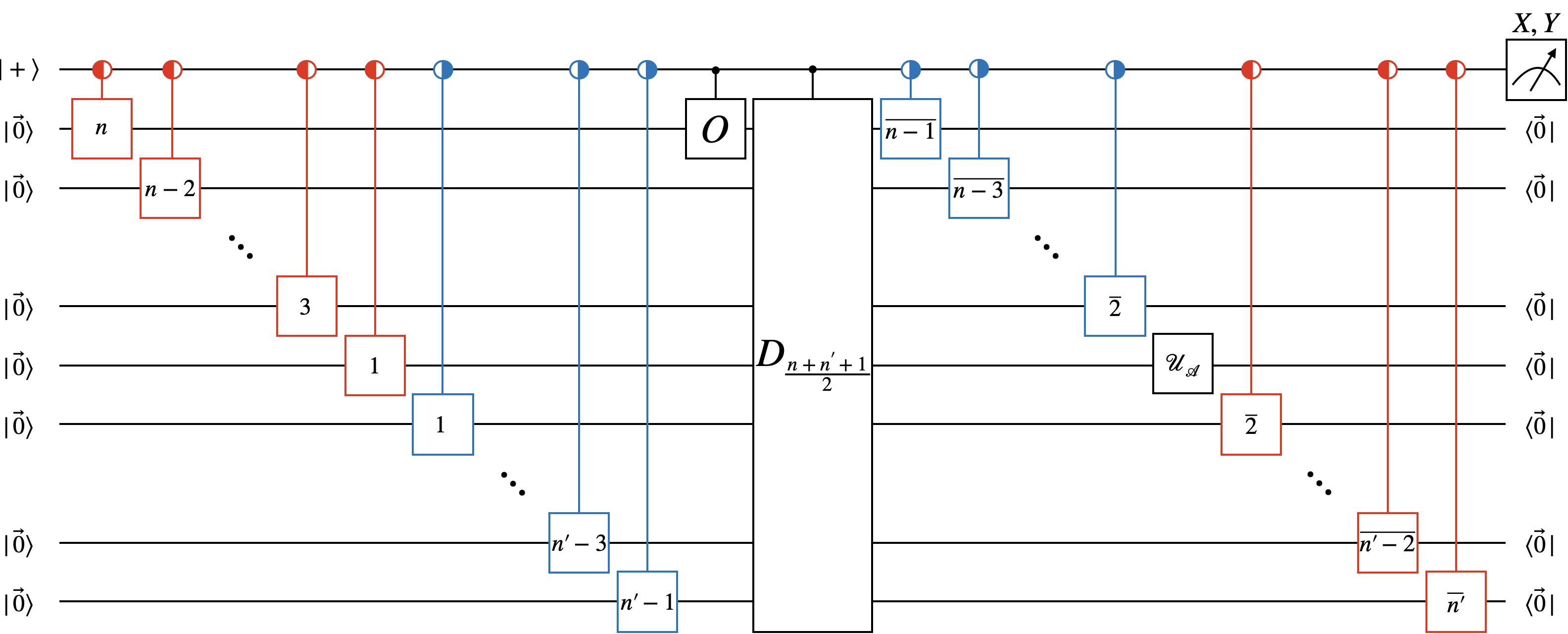}
        \end{adjustbox}
    }
    \hfill
    \subfloat[even $n$ and odd $n'$ \label{fig:qc_dgse_general_even_A_odd}]{
        \begin{adjustbox}{width=0.48\textwidth}
            \includegraphics[]{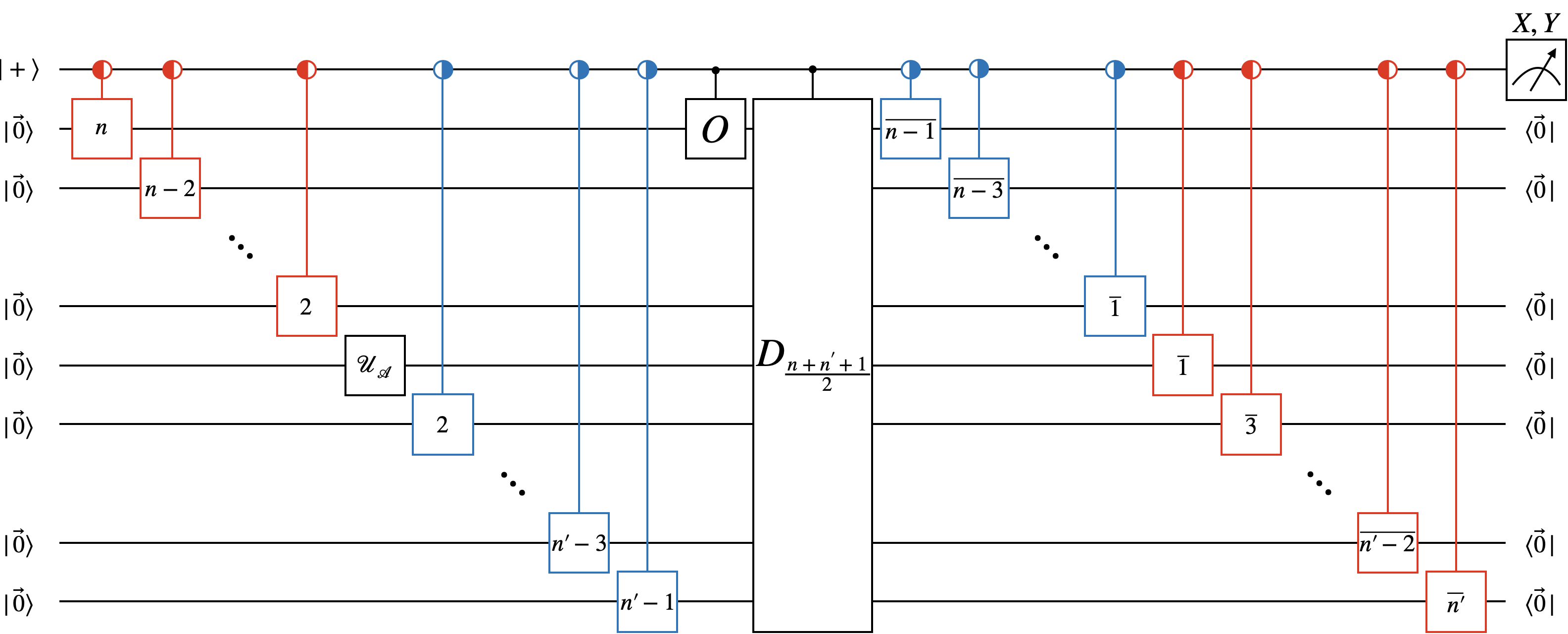}
        \end{adjustbox}
    }
    \hfill
    \subfloat[even $n$ and even $n'$ \label{fig:qc_dgse_general_even_A_even}]{
        \begin{adjustbox}{width=0.48\textwidth}
            \includegraphics[]{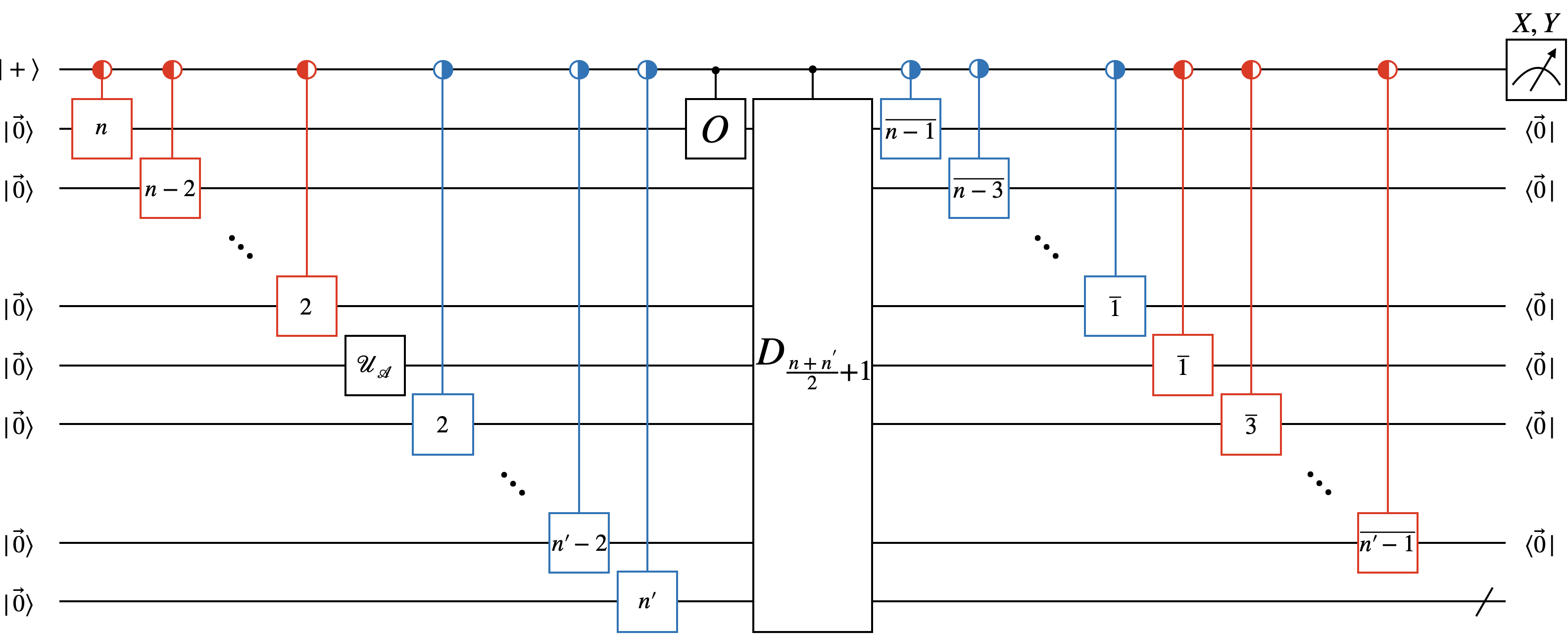}
        \end{adjustbox}
    }
    \caption{
        Quantum Circuits with $\displaystyle \left\lceil \frac{n+n'+1}{2}\right\rceil$ copies for computing 
        (a) $\operatorname{Tr}\left[ \left(\tau_{n}^{\dagger} \cdots \tau_{3}^{\dagger}\bar{\tau}_{2}^{\dagger}\tau_{1}^{\dagger}\right) \mathcal{A} 
                                     \left(\tau_{1}\bar{\tau}_{2}\tau_{3} \cdots \tau_{n'}\right) O \right]$, 
        (b) $\operatorname{Tr}\left[ \left(\tau_{n}^{\dagger} \cdots \tau_{3}^{\dagger}\bar{\tau}_{2}^{\dagger}\tau_{1}^{\dagger}\right) \mathcal{A} 
                                     \left(\tau_{1}\bar{\tau}_{2}\tau_{3} \cdots \bar{\tau}_{n'}\right) O \right]$,
        (c) $\operatorname{Tr}\left[ \left(\tau_{n}^{\dagger} \cdots \bar{\tau}_{3}^{\dagger}\tau_{2}^{\dagger}\bar{\tau}_{1}^{\dagger}\right) \mathcal{A} 
                                     \left(\bar{\tau}_{1}\tau_{2}\bar{\tau}_{3} \cdots \bar{\tau}_{n'}\right) O \right]$, and
        (d) $\operatorname{Tr}\left[ \left(\tau_{n}^{\dagger} \cdots \bar{\tau}_{3}^{\dagger}\tau_{2}^{\dagger}\bar{\tau}_{1}^{\dagger}\right) \mathcal{A} 
                                     \left(\bar{\tau}_{1}\tau_{2}\bar{\tau}_{3} \cdots \tau_{n'}\right) O \right]$,
        where $\mathcal{A}$ is prepared by a (noisy) gate operation $\mathcal{U}_{\mathcal{A}}$.
        The red and blue controlled gates with half-filled controlled operations are defined in Fig.~\ref{fig:qcs_dgse_tau}.
        The symbol $|\vec{0}\rangle$ represents the initial quantum state with the all-zero bitstring.
        The symbol $\langle\vec{0}|$ represents the post-selection over all-zero bitstring after measuring the quantum register on the computational basis.
    }
    \label{fig:qcs_dgse_general_with_A}
\end{figure}

We can further expand a positive semidefinite Hermitian $\mathcal{A}$ through Dual-GSE to approximate the following ansatz
\begin{equation} \label{eq:rho_dgse_general_formal_with_A}
    \rho_{\mathrm{DGSE}} = \frac{{\mathcal{P}}^{\dagger} \mathcal{A} \mathcal{P}}{\operatorname{Tr}\left[{\mathcal{P}}^{\dagger} \mathcal{A} \mathcal{P}\right]} \text{ with }{\mathcal{P}}=\sum_{i}\alpha_{i} {\sigma}_{i},
\end{equation}
with the same notations as Eq.~\eqref{eq:rho_dgse_general_formal}.
Likewise, the expectation values required in Eq.~\eqref{eq:rho_dgse_general_formal_with_A} are also classified as Eq.~\eqref{eq:expval_X_plus_iY_with_A} depending on the parity of $n$ and $n'$:
\begin{equation} \label{eq:expval_X_plus_iY_with_A}
    \langle X + iY \rangle_{O} = 
    \begin{cases}
        \displaystyle \operatorname{Tr}\left[ \left(\tau_{n}^{\dagger} \cdots \tau_{3}^{\dagger}\bar{\tau}_{2}^{\dagger}\tau_{1}^{\dagger}\right) \mathcal{A} 
                                       \left(\tau_{1}\bar{\tau}_{2}\tau_{3} \cdots \tau_{n'}\right) O \right] & \text{ for odd $n$ and odd $n'$, } \\[10pt]
        \displaystyle \operatorname{Tr}\left[ \left(\tau_{n}^{\dagger} \cdots \tau_{3}^{\dagger}\bar{\tau}_{2}^{\dagger}\tau_{1}^{\dagger}\right) \mathcal{A} 
                                       \left(\tau_{1}\bar{\tau}_{2}\tau_{3} \cdots \bar{\tau}_{n'}\right) O \right] & \text{ for odd $n$ and even $n'$, } \\[10pt]
        \displaystyle \operatorname{Tr}\left[ \left(\tau_{n}^{\dagger} \cdots \bar{\tau}_{3}^{\dagger}\tau_{2}^{\dagger}\bar{\tau}_{1}^{\dagger}\right) \mathcal{A} 
                                       \left(\bar{\tau}_{1}\tau_{2}\bar{\tau}_{3} \cdots \bar{\tau}_{n'}\right) O \right] & \text{ for even $n$ and odd $n'$, } \\[10pt]
        \displaystyle \operatorname{Tr}\left[ \left(\tau_{n}^{\dagger} \cdots \bar{\tau}_{3}^{\dagger}\tau_{2}^{\dagger}\bar{\tau}_{1}^{\dagger}\right) \mathcal{A} 
                                       \left(\bar{\tau}_{1}\tau_{2}\bar{\tau}_{3} \cdots \tau_{n'}\right) O \right] & \text{ for even $n$ and even $n'$, } 
    \end{cases}
\end{equation}
where the expectation value $\langle X + iY \rangle_{O}$ is defined as
\begin{equation}
\begin{split}
    \langle X + iY \rangle_{O} := 
    \begin{cases}
        \displaystyle \langle X \otimes \operatorname{P}_{\vec{0}}^{\otimes \frac{n+n'}{2}} \otimes I \rangle_{O} + i \langle Y \otimes \operatorname{P}_{\vec{0}}^{\otimes \frac{n+n'}{2}} \otimes I \rangle_{O}, 
        & \text{ for odd $n$ and odd $n'$, } \\[5pt]
        \displaystyle \langle X \otimes \operatorname{P}_{\vec{0}}^{\otimes \frac{n+n'+1}{2}} \rangle_{O} + i \langle Y \otimes \operatorname{P}_{\vec{0}}^{\otimes \frac{n+n'+1}{2}} \rangle_{O},
        & \text{ for odd $n$ and even $n'$, } \\[5pt]
        \displaystyle \langle X \otimes \operatorname{P}_{\vec{0}}^{\otimes \frac{n+n'+1}{2}} \rangle_{O} + i \langle Y \otimes \operatorname{P}_{\vec{0}}^{\otimes \frac{n+n'+1}{2}} \rangle_{O},
        & \text{ for even $n$ and odd $n'$, } \\[5pt]
        \displaystyle \langle X \otimes \operatorname{P}_{\vec{0}}^{\otimes \frac{n+n'}{2}} \otimes I \rangle_{O} + i \langle Y \otimes \operatorname{P}_{\vec{0}}^{\otimes \frac{n+n'}{2}}  \otimes I \rangle_{O}
        & \text{ for even $n$ and even $n'$. } 
    \end{cases}
\end{split}
\end{equation}
These quantities can be obtained by the quantum circuits in Fig.~\ref{fig:qcs_dgse_general_with_A}, which consume $\displaystyle \left\lceil \frac{n + n' + 1}{2} \right\rceil$ state copies.

\section{\texorpdfstring{Leveraging sampling cost regarding the post-selection probability of DSP circuits}{} \label{sec:appendix_Leveraging_sampling_cost_regarding_the_post-selection_probability_of_DSP_circuits}}

The post-selection overhead of the dual-state purification (DSP) circuits is reflected in the estimation variance using the finite shot.
In DSP, we do not actually ``post-select'' the right quantum state but rather compute and use the post-selection probability as classical data represented by $\mathrm{p}_{\vec{0}}$.
As a result, the overhead of not obtaining the right quantum state will make the matrix elements of $\mathcal{S}$ (and $\mathcal{H}$) in the generalized eigenvalue problem $\mathcal{H}\vec{\alpha} = E\mathcal{S}\vec{\alpha}$ smaller.
Hereafter, this overhead shows up as a large estimation variance, which can be described by the post-selection probability $\mathrm{p}_{\vec{0}}$.

To see this, we analytically derive the relation $N_{\mathrm{s}}\geq O\left(\mathrm{p}_{\vec{0}}^{-2}\right)$ between the post-selection probability $\mathrm{p}_{\vec{0}}$ and the sampling cost $N_{\mathrm{s}}$ required for keeping the estimation variance lower than a certain value when solving the generalized eigenvalue problem $H\vec{\alpha} = ES\vec{\alpha}$ in Dual-GSE.
While the post-selection probability $\mathrm{p}_{\vec{0}}$ will get worse as the circuit depth or noise level increases, $N_{\mathrm{s}}\geq O\left(\mathrm{p}_{\vec{0}}^{-2}\right)$ increases only in a quadratic speed of $\mathrm{p}_{\vec{0}}^{-1}$, which may not scale badly.

We use the fact that the sampling cost shows up as the operator norm of the inverse matrix $\mathcal{S}^{-1}$.
To see this, we refer to the discussion in Appendix S7 of Yoshioka et al.~\cite{Yoshioka2022-gq}, where they derived Eq.~\eqref{eq:delta_E_in_yoshioka_etal} described below, between the sampling cost and the matrix norm associated with the matrix $\mathcal{S}$.
Assuming the estimated ground state energy deviates from the mean value $E_{0}$ to $E = E_{0} + \delta E$ under the finite shot counts $N_{\mathrm{s}}$, the estimation bias $\epsilon = |\delta E|$ becomes
\begin{equation} \label{eq:delta_E_in_yoshioka_etal}
\begin{split}
    |\delta E| \leq \frac{4 \gamma M^{2}\left\|\mathcal{S}_0^{-1}\right\|_{\mathrm{op}}}{\sqrt{N_{\mathrm{s}}}},
\end{split}
\end{equation}
where $\gamma$ is defined as $\displaystyle \gamma = \sum_{i}\left|h_{i}\right|$ for the given Hamiltonian $\displaystyle H=\sum_{i}h_{i}P_{i}$, and thus satisfies $E_{0} \leq\|H\|_{\mathrm{op}} \leq \gamma$ for the theoretical noise-free ground state energy $E_{0}$ and $M$ subspaces.
Therefore, the number of samples required for estimating an observable below a certain estimation bias $\epsilon$ becomes
\begin{equation} \label{eq:N_s_in_yoshioka_etal}
\begin{split}
    N_{\mathrm{s}} \geq \frac{16\gamma^{2}M^{4}\|\mathcal{S}^{-1}\|_{\mathrm{op}}^{2}}{\epsilon^{2}}.
\end{split}
\end{equation}

We exploit the property that $\mathcal{S}$ is positive semidefinite, which comes from the definition of constraint on $\rho_{\mathrm{DGSE}}$ that $\operatorname{Tr}\left[\rho_{\mathrm{DGSE}}\right]=\vec{\alpha}^{\dagger}\mathcal{S}\vec{\alpha}=1$.
Using Hadamard's inequality~\cite{garling2007inequalities}, the determinant of $\mathcal{S}$ can be upper bounded by the product of diagonal elements of $\mathcal{S}$ as
\begin{equation}
\begin{split}
    \det\left(\mathcal{S}\right) \leq \prod_{i=1}^{M}\mathcal{S}_{ii}.
\end{split}
\end{equation}
Let the eigenvalues of $\mathcal{S}$ be $0 \leq \lambda_{\mathrm{min}}, \ldots, \lambda_{\mathrm{max}}$ in ascending order.
As $\det\left(\mathcal{S}\right) = \lambda_{\mathrm{min}} \cdots \lambda_{\mathrm{max}} \geq \lambda_{\mathrm{min}}^{M}$,
\begin{equation} \label{eq:lambda_min_upper_bound}
\begin{split}
    \lambda_{\mathrm{min}} \leq \left(\prod_{i=1}^{M}\mathcal{S}_{ii}\right)^{\frac{1}{M}}.
\end{split}
\end{equation}

Next, we define the post-selection probability $\mathrm{p}_{\vec{0}}:=\operatorname{Tr}\left[\bar{\rho}\rho\right]$.
For the power subspace, we have defined $\mathcal{S}$ in Eq.~\eqref{eq:S_ij} and Eq.~\eqref{eq:S_{i}=0_or_{j}=0}, i.e. 
\begin{equation}
\mathcal{S}_{ij} = 
\begin{cases}
    \displaystyle \operatorname{Tr}\left[\frac {\bar{\rho}\rho + \rho\bar{\rho}} {2} H^{i+j-4} \right] & \text{ for } i, j \geq 2, \\[10pt]
    \displaystyle \operatorname{Tr}\left[\frac {\rho + \bar{\rho}} {2} H^{i+j-3} \right] & \text{ for } i = 1 \text{ or } j = 1.
\end{cases}
\end{equation}
Using $\displaystyle \gamma = \sum_{i}\left|h_{i}\right|$ to upper bound each element $\mathcal{S}_{ij}$, $\lambda_{\mathrm{min}}$ in Eq.~\eqref{eq:lambda_min_upper_bound} can be upper bounded by $\mathrm{p}_{\vec{0}}$ as
\begin{equation} \label{eq:lambda_min_final}
\begin{split}
    \lambda_{\mathrm{min}} 
    &\leq \left( \operatorname{Tr}\left[I\right] \operatorname{Tr}\left[\bar{\rho}\rho\right] \operatorname{Tr}\left[ \frac {\bar{\rho}\rho + \rho\bar{\rho}} {2} H^{2} \right] \cdots \operatorname{Tr}\left[ \frac {\bar{\rho}\rho + \rho\bar{\rho}} {2} H^{2(M-2)} \right] \right)^{\frac{1}{M}} \\
    &\leq \left( \gamma^{\sum_{k=1}^{M-2}2k} \operatorname{Tr}\left[I\right] \operatorname{Tr}\left[\bar{\rho}\rho\right]^{M-1} \right)^{\frac{1}{M}} \\
    &\leq \left( \gamma^{\left(M-1\right)\left(M-2\right)} \operatorname{Tr}\left[I\right] \mathrm{p}_{\vec{0}}^{M-1} \right)^{\frac{1}{M}} \\
    &= O\left( \gamma^{M+\frac{1}{M}} \mathrm{p}_{\vec{0}}^{1-\frac{1}{M}} \right) = O\left( \mathrm{p}_{\vec{0}}^{1-\frac{1}{M}} \right).
\end{split}
\end{equation}
Finally, according to Eq.~\eqref{eq:N_s_in_yoshioka_etal}, the sampling cost $N_{\mathrm{s}}$ to achieve the estimation bias smaller than $\epsilon$ can be bounded as
\begin{equation}\label{eq:N_s_to_P_0}
\begin{split}
    N_{\mathrm{s}} 
    \geq \frac{16\gamma^{2}M^{2}\|\mathcal{S}^{-1}\|_{\mathrm{op}}^{2}}{\epsilon^{2}} = \frac{16\gamma^{2}M^{2}}{\epsilon^{2}}\frac{1}{\lambda_{\mathrm{min}}^{2}} 
    \gtrsim O\left(\mathrm{p}_{\vec{0}}^{-2} \right).
\end{split}
\end{equation}

By Eq.~\eqref{eq:N_s_to_P_0}, we have connected the sampling cost $N_{\mathrm{s}}$ and the post-selection probability $\mathrm{p}_{\vec{0}}$ by DSP.
Note that the post-selection process also occurs in ESD~\cite{Koczor2021} and VD~\cite{Huggins2021} circuits.
In the ESD circuit, the quantity $\operatorname{Tr}\left[\rho^{2}\right]$ is obtained by
\begin{equation}
\begin{split}
    \operatorname{Tr}\left[\rho^{2}\right] = 2 \mathrm{p}_{0}^{\prime} - 1,
\end{split}
\end{equation}
with the post-selection probability of $\mathrm{p}_{0}^{\prime}$ in the ancillary qubit.
This suggests that the post-selection overhead of Dual-GSE and conventional GSE scales in the same order, while Dual-GSE is advantageous up to a constant factor.
This is because $\mathrm{p}_{\vec{0}}$ for DSP scales almost twice larger than $\mathrm{p}_{0}^{\prime}$ for ESD under small noise rates, and thus the sampling cost $N_{\mathrm{s}}$ with ESD subroutine would be around four times larger than that with DSP subroutine, assuming $\operatorname{Tr}\left[\bar{\rho}\rho\right] \sim\operatorname{Tr}\left[\rho^{2}\right]$.

In conclusion, since we have revealed the overhead of Dual-GSE and conventional GSE scales in the same order, Dual-GSE can be said to be more ``resource-efficient'' in qubit (and gate) overhead than GSE.
In fact, the reduction of qubit overhead makes up for the additional circuit depth overhead in Dual-GSE.
In Appendix~\ref{sec:appendix_Comparison_of_GSE_with_Different_Noisy_Subroutines}, we demonstrate the advantage of adopting DSP circuits over ESD circuits under the noisy execution of the GSE process.

\section{Comparing Different Noisy Subroutines in GSE \label{sec:appendix_Comparison_of_GSE_with_Different_Noisy_Subroutines}}

Since dual-state purification (DSP)~\cite{Huo2022} computes $\operatorname{Tr}\left[\rho^{2}O\right]$ by doubling the depth of the circuit, one may wonder whether it would not surpass the performance of exponential suppression by derangement (ESD)~\cite{Koczor2021}.
In this section, we discuss the advantages of using DSP over ESD as a subroutine in Dual-GSE by numerically comparing the performance of these two subroutines.

As a qualitative observation, although ESD circuits can parallelize the VQE ansatz, the total number of gates used in VQE ansatzes to create $\operatorname{Tr}\left[\rho^{2}\right]$ stays the same as when using DSP.
In fact, the circuit depth required to implement the DSP process can usually be lighter than ESD.
The circuit implementation of DSP requires only the additional gate operations to insert the circuit for observable between the circuits of $\rho$ and $\bar{\rho}$.
In contrast, the ESD circuit requires doubling the size of the given ansatz and multiple swapping gates controlled by an ancillary qubit.
This is especially evident when computing the expectation value of low-degree observables because ESD requires the entire state swapping operation in its circuit construction.

Besides, since a single controlled-swap gate consists of five controlled-NOT gates~\cite{smolin1996five}, executing more controlled-NOT gates under the noise will drastically deteriorate the post-selection probability and estimation accuracy.
This overhead would be even worse when running on the sparsely connected qubits, which is the case for most of the current superconducting devices.
This may cause many more additional controlled-NOT operations to apply the controlled-SWAP operation among qubits far away.
To compute $\operatorname{Tr}\left[\rho^{2}O\right]$ on the linear nearest neighbor (LNN) qubit structure, ESD requires $O\left(n^{2}\right)$ controlled-NOT operations to the number of qubits $n$ ~\cite{ohkura2023leveraging}.
In contrast, it is shown that using DSP can ease this overhead to $O\left(n\right)$ for $\operatorname{Tr}\left[\rho^{2}O\right]$ on LNN structure~\cite{Huo2022}.
Moreover, even when using quantum devices with full connectivity and long coherent operation time, such as photonic and trapped ion implementation, it would be more demanding to reduce the qubit overhead than to parallelize the ansatzes to fit the coherent time.

Below, we justify this resource efficiency by demonstrating that using DSP with a deeper circuit depth has higher estimation accuracy and post-selection probability than using ESD under noisy execution.
First, we numerically compare the expectation value of the ground-state energy of a 4-qubit Ising Hamiltonian in a path structure.
We adopt the task of estimating the ground state energy of an Ising Hamiltonian $\displaystyle H=\sum_{k}h_{k}P_{k}$ by setting the ansatz for DSP and ESD as
\begin{itemize}
    \item $\displaystyle \langle H\rangle_{\mathrm{ESD}} = \frac{\sum_{k}h_{k}\operatorname{Tr}\left[\rho^{2}P_{k}\right]}{\operatorname{Tr}\left[\rho^{2}\right]}$ using ESD circuits, and 
    \item $\displaystyle \langle H\rangle_{\mathrm{DSP}} = \frac{\sum_{k}h_{k}\operatorname{Tr}\left[\left(\bar{\rho}\rho + \rho\bar{\rho}\right)P_{k}\right]}{\operatorname{Tr}\left[\bar{\rho}\rho + \rho\bar{\rho}\right]}$ using DSP circuits.
\end{itemize}
The ansatz $\rho$ is still defined as the circuit in Fig.~\ref{fig:qc_vqe} for four qubits and eight layers.
Note that we train the parameters of the circuit without noise in advance and run noisy ESD and DSP circuits for the GSE process.

During the execution of the DSP and ESD circuit, we have introduced two noise settings: 
\begin{itemize}
    \item[(1)] the local stochastic Pauli noise only, and
    \item[(2)] the T1/T2 thermal relaxation to controlled-NOT along with the local stochastic Pauli noise.
\end{itemize}
The error channels are applied after each gate operation in quantum circuits.
The noise level of local stochastic Pauli noise is set to $p_{1} = 1.0\times 10^{-4}, 3.0\times 10^{-4}, 1.0\times 10^{-3}, 3.0\times 10^{-3}, 1.0\times 10^{-2}$ for single-qubit operations and $p_{2} = 10p_{1}$ for two-qubit operations.
The T1 relaxation time is sampled randomly from a normal distribution by setting $50$ microseconds as its mean.
The T2 relaxation time is sampled randomly from a normal distribution by setting $70$ microseconds as its mean and forced to satisfy $T_{2}\leq 2T_{1}$.
The gate time of controlled-NOT operation is set to $200$ nanoseconds.

\begin{figure}[htbp]
    \centering
    \subfloat[\label{fig:noiselevel-to-diff_local-stochastic-pauli}]{
        \includegraphics[width=0.48\textwidth]{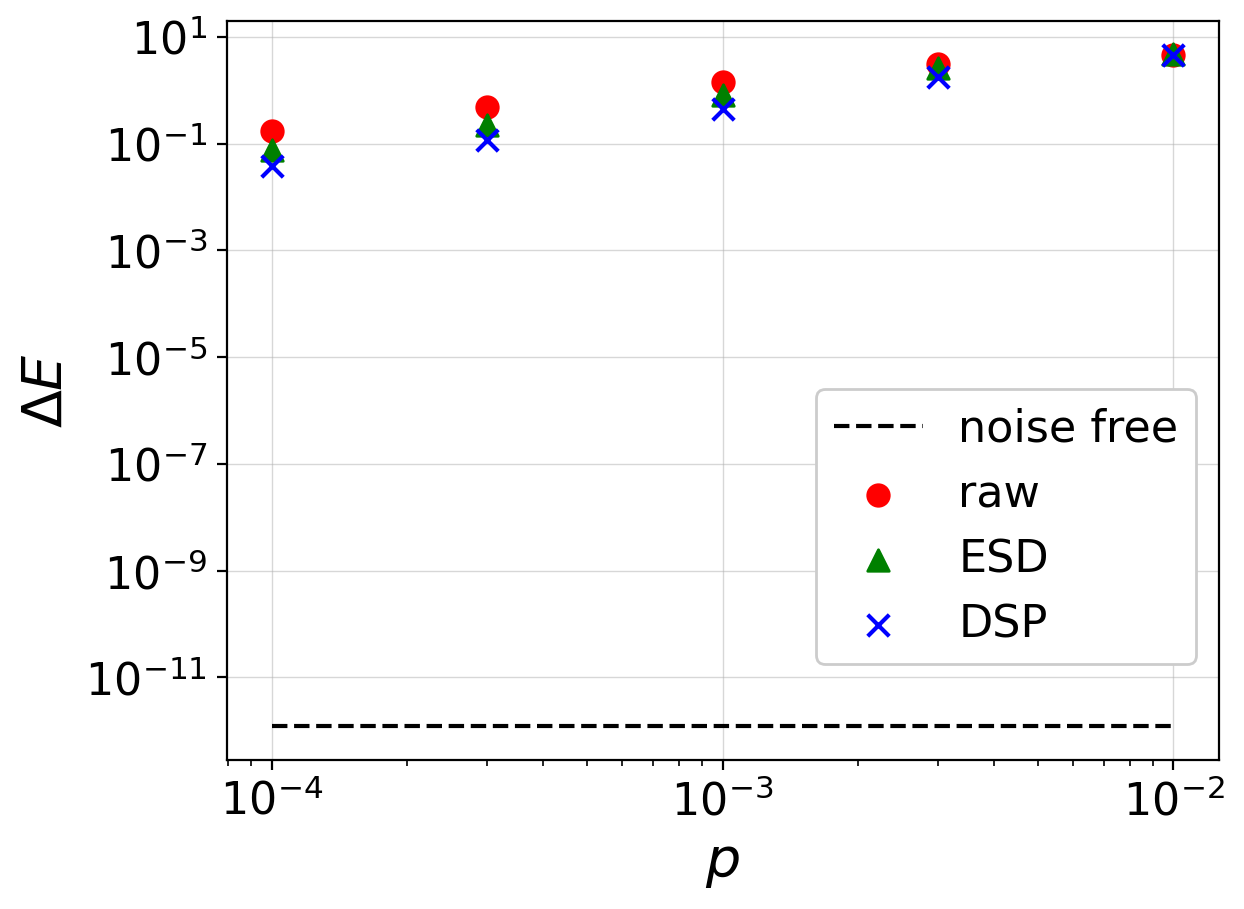}
    }
    \hfill
    \subfloat[\label{fig:noiselevel-to-diff_thermal-relaxation}]{
        \includegraphics[width=0.48\columnwidth]{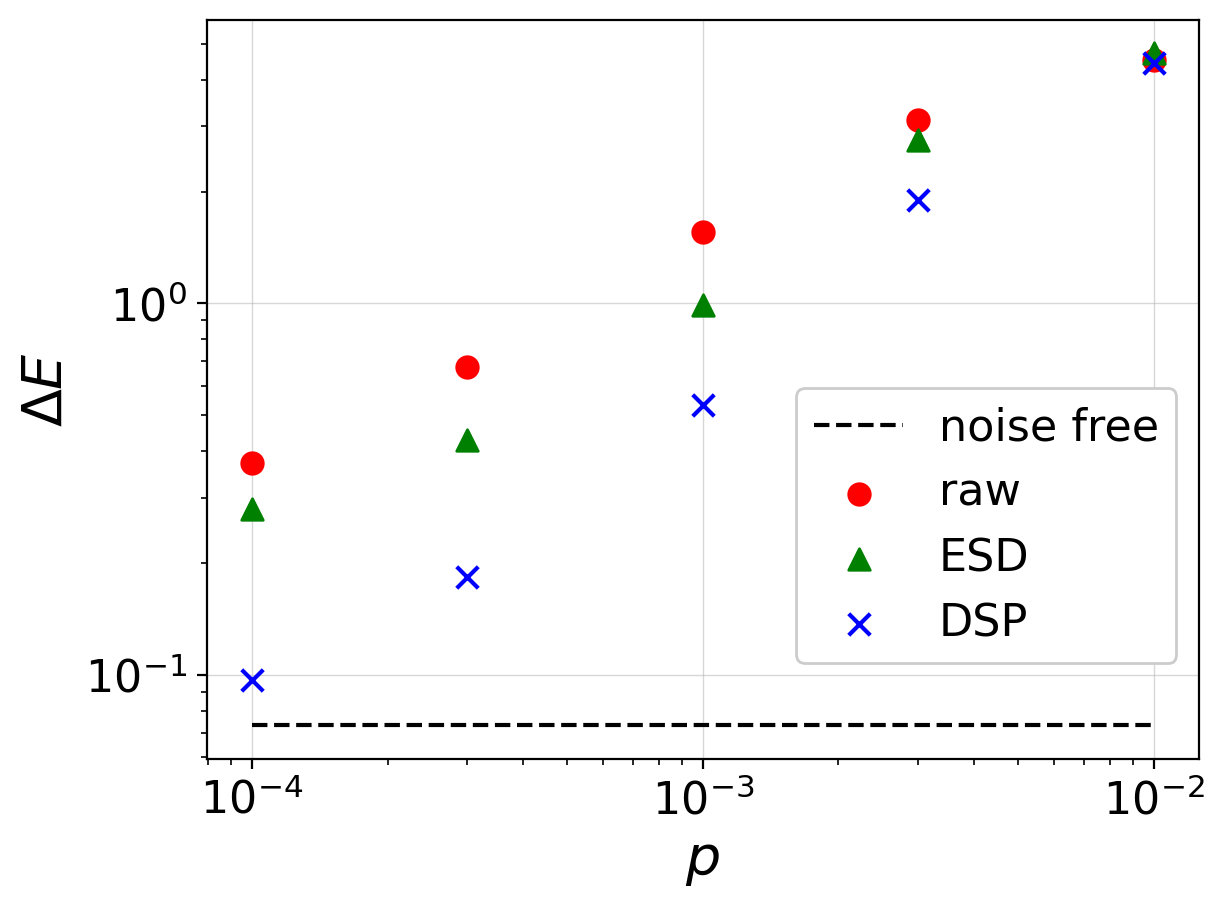}
    }
    \caption{
        The estimation error $\Delta E$ to different noise levels and different noise models (1) and (2).
        (a) The estimation error $\Delta E$ by ESD and DSP under the local stochastic errors only.
        (b) The estimation error $\Delta E$ by ESD and DSP under the local stochastic errors and the thermal relaxation.
    }
    \label{fig:noiselevel-to-diff}
\end{figure}

The simulation result of estimation error is shown in Fig.~\ref{fig:noiselevel-to-diff}.
Figure~\ref{fig:noiselevel-to-diff}(a) shows the result by ESD and DSP under noise setting (1).
The estimation errors by DSP seem to be smaller than ESD in all noise levels.
Figure~\ref{fig:noiselevel-to-diff}(b) shows the results under noise setting (2).
We observe that the estimation errors by DSP are clearly smaller than those by ESD for noise settings (2), even reaching the estimation error by the noise-free variational optimization in the small noise levels.

\begin{figure}[htbp]
    \centering
    \subfloat[\label{fig:noiselevel-to-IIII_local-stochastic-pauli}]{
        \includegraphics[width=0.48\columnwidth]{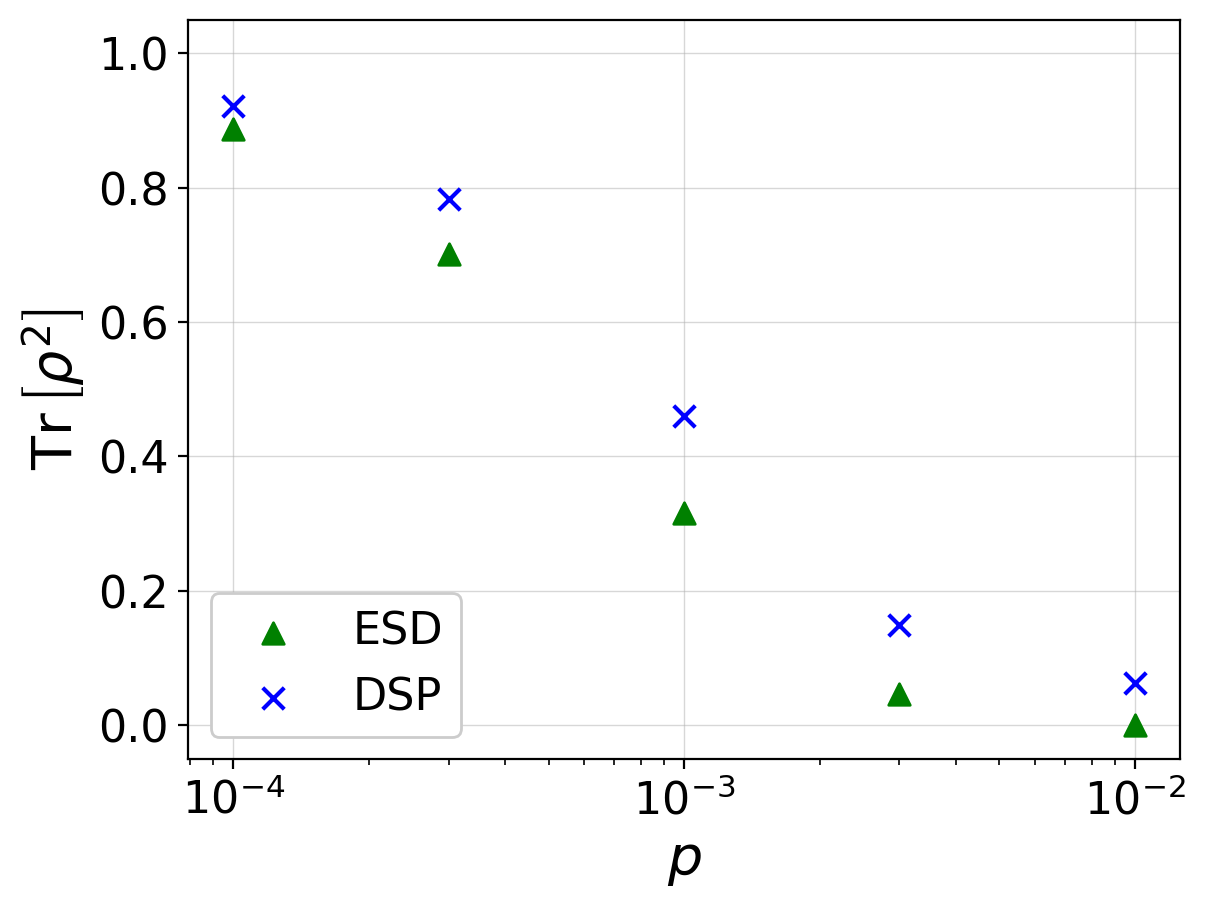}
    }
    \hfill
    \subfloat[\label{fig:noiselevel-to-IIII_thermal-relaxation}]{
        \includegraphics[width=0.48\columnwidth]{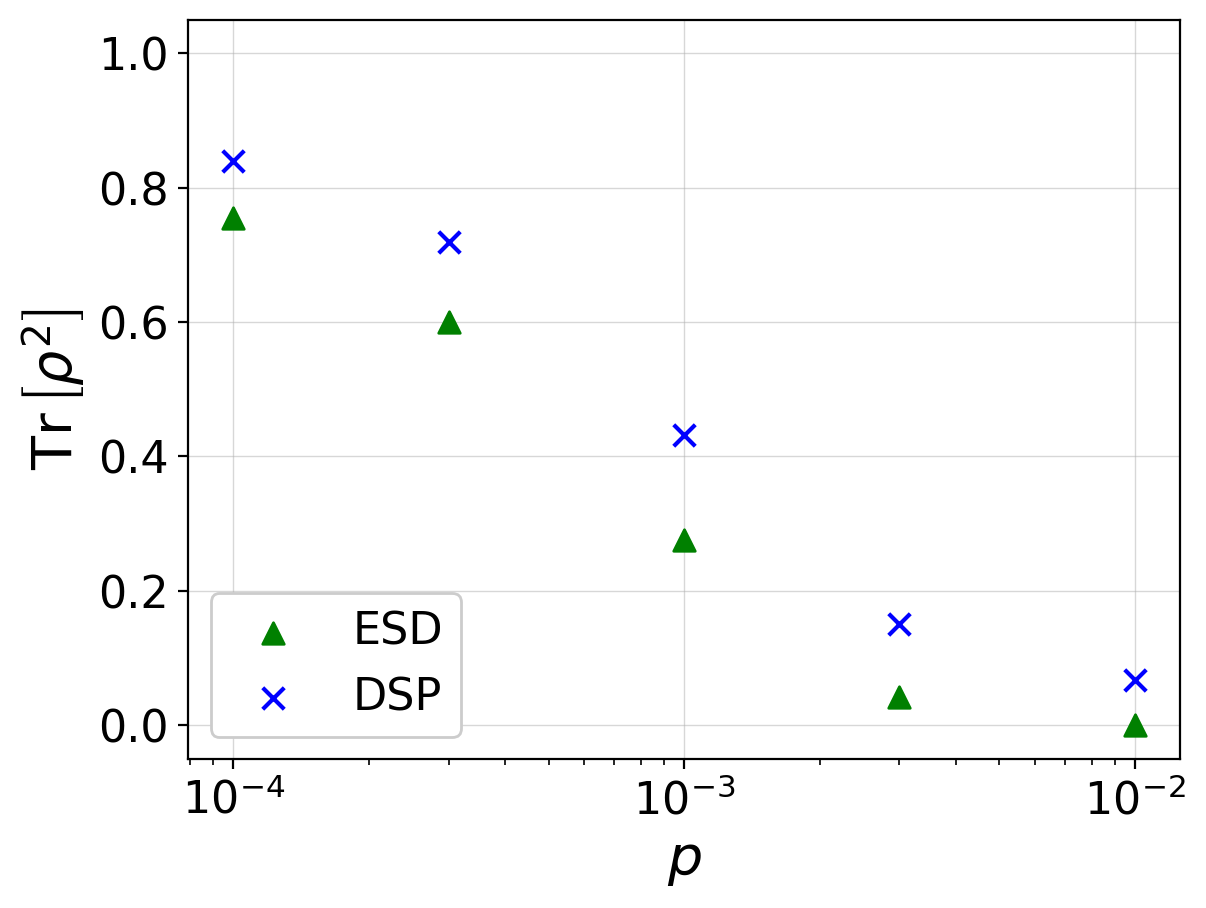}
    }
    \caption{
        The estimation of $\operatorname{Tr}\left[\rho^{2}\right]$ by ESD and DSP under the noise setting of (1) and (2), each corresponds to subfigure (a) and (b).
        The triangle green plots represent the purity estimation of $\rho$ by ESD and the cross blue plots represent the purity estimation of $\rho$ by DSP in the form of $\operatorname{Tr}\left[\left(\bar{\rho}\rho + \rho\bar{\rho}\right)/2\right]$.
    }
    \label{fig:noiselevel-to-expval-and-IIII}
\end{figure}

We also demonstrate that DSP has a higher post-selection probability over ESD in the setting above.
From Fig.~\ref{fig:noiselevel-to-expval-and-IIII}, we verify that the purity term $\operatorname{Tr}\left[\rho^{2}\right]$ of ESD gets worse than that of DSP for both case of noise settings (1) and (2).
Since the low post-selection probability will increase the required number of circuit executions to reduce the estimation variance, this implies that DSP is more resource-efficient than ESD in terms of post-selection probability under the noise.

\section{Leveraging Sampling Cost Regarding Divide-and-conquer Overhead \label{sec:appendix_Leveraging_Sampling_Cost_Regarding_Divide-and-conquer_Overhead}}

The divide-and-conquer subspace classically restores the effect of entanglement in the expectation value or ground state energy.
As pointed out in recent literature~\cite{eddins2022doubling, jing2024circuit, harrow2024optimal}, this classical postprocessing may require an exponential sampling cost to the quantity of entanglement virtually reintroduced.
Nevertheless, we remark that our divide-and-conquer strategy will reintroduce a reasonable amount of divided entanglement effect to final expectation values with a practically feasible overhead for models of practical interests.

Through the following analysis, we have clarified that the overhead of entanglement recovery by the divide-and-conquer subspace can be evaluated by the quartic power of two-norm of the optimized coefficients $\vec{\alpha}$ in $\mathcal{H}\vec{\alpha} = E\mathcal{S}\vec{\alpha}$.
We discuss this divide-and-conquer overhead separately from the noise overhead by first analyzing the noise-free sampling cost and then considering the noise.
Notably, we have derived the sampling cost in the form of $N_{\mathrm{s}}\geq O\left(\|\vec{\alpha}\|_{2}^{4}\right)$ in terms of the divide-and-conquer overhead.
Besides, the two norm $\|\vec{\alpha}\|_{2}$ increases as the number of subspaces $M$ increases, leading to the increase of sampling cost $N_{\mathrm{s}}$.
On the other hand, the divide-and-conquer method reduces the necessary observables to be measured, which contributes to the suppression of sampling cost $N_{\mathrm{s}}$.
We have numerically verified the relation between the divide-and-conquer overhead $\|\vec{\alpha}\|_{2}^{4}$ to the number of subspaces $M$ as well.
Finally, on the noise overhead, we consider the effect of depolarizing noise to see the sampling cost increases by a factor of $\left(1-p\right)^{-8}$ to the depolarizing rate $p$ for the divide-and-conquer approach with two symmetric subsystems.

\subsection{The divide-and-conquer overhead without gate noise and shot noise}

To focus on the divide-and-conquer overhead in $\vec{\alpha}$, we adopt the following Dual-GSE ansatz with a square of the ansatz $\rho_{\mathrm{local}}^{(\mathrm{AB})} = \rho_{\mathrm{local}}^{(\mathrm{A})}\otimes\rho_{\mathrm{local}}^{(\mathrm{B})}$ and expand it with $M$ subspaces as
\begin{equation} \label{eq:rho_dc2}
\begin{split}
    \rho_{\mathrm{DC2}}
    = \sum_{i,j = 1}^{M}\alpha_{i}^{*}\alpha_{j} 
        H^{i-1} \left(\rho_{\mathrm{local}}^{(\mathrm{AB})}\right)^{2} H^{j-1}
    &= \sum_{i,j = 1}^{M}
        \left( \operatorname{Tr}\left[ \left(\rho_{\mathrm{local}}^{(\mathrm{AB})}\right)^{2} H^{2(i-1)} \right]^{1/2} \alpha_{i}^{*} \right)
        \left( \operatorname{Tr}\left[ \left(\rho_{\mathrm{local}}^{(\mathrm{AB})}\right)^{2} H^{2(j-1)} \right]^{1/2} \alpha_{j} \right) \\
    &\qquad\qquad\qquad\qquad\qquad \times
        \frac { H^{i-1} \left(\rho_{\mathrm{local}}^{(\mathrm{AB})}\right)^{2} H^{j-1} } 
              { \operatorname{Tr}\left[ \left(\rho_{\mathrm{local}}^{(\mathrm{AB})}\right)^{2} H^{2(i-1)} \right]^{1/2}
                \operatorname{Tr}\left[ \left(\rho_{\mathrm{local}}^{(\mathrm{AB})}\right)^{2} H^{2(j-1)} \right]^{1/2} }.
\end{split}
\end{equation}
For a fair comparison and demonstration of the effect of the divide-and-conquer overhead in the optimized vectors, we normalize the diagonal elements of matrix $\mathcal{S}$.
Note that this subspace construction also corresponds to the Krylov subspace $\left\{\left(\rho_{\mathrm{local}}^{(\mathrm{AB})}\right)^{2} H^{i=1}\right\}_{i=1}^{M}$.
Based on Eq.~\eqref{eq:rho_dc2}, we also define $\vec{\alpha}_{i}^{\prime}$ and $\tilde{\mathcal{S}}$ as
\begin{equation} \label{eq:alpha_prime}
\begin{split}
    \vec{\alpha}_{i}^{\prime} = \operatorname{Tr}\left[ \left(\rho_{\mathrm{local}}^{(\mathrm{AB})}\right)^{2} H^{2(i-1)} \right]^{1/2} \vec{\alpha}_{i},
    \quad \text{and} \quad 
    \tilde{\mathcal{S}}_{ij}
    = \frac { \operatorname{Tr}\left[ \left(\rho_{\mathrm{local}}^{(\mathrm{AB})}\right)^{2} H^{i+j-2} \right] } 
            { \operatorname{Tr}\left[ \left(\rho_{\mathrm{local}}^{(\mathrm{AB})}\right)^{2} H^{2(i-1)} \right]^{1/2}
              \operatorname{Tr}\left[ \left(\rho_{\mathrm{local}}^{(\mathrm{AB})}\right)^{2} H^{2(j-1)} \right]^{1/2} } \leq 1,
\end{split}
\end{equation}
where the $\tilde{\mathcal{S}}_{ij} \leq 1$ comes from the Cauchy-Schwarz inequality for the weighted trace.
Below, we bound the sampling cost $N_{\mathrm{s}}$ by the optimized coefficients $\vec{\alpha}$ (in the form of $\vec{\alpha}^{\prime}$) from the generalized eigenvalue problem $\mathcal{H}\vec{\alpha} = E\mathcal{S}\vec{\alpha}$.

From Appendix S7 in Yoshioka et al.~\cite{Yoshioka2022-gq}, we have the following bound for the estimation bias $|\delta E|= |E - E_{0}|$, assuming $E$ as the estimated ground state energy and $E_{0}$ as the true one.
\begin{equation} \label{eq:ineq_yoshioka2022generalized_s7}
\begin{split}
    |\delta E|
    &= \left| \vec{\alpha}^{\dagger} \left( \delta\mathcal{H} - E_{0} \delta\mathcal{S} \right) \vec{\alpha} \right| \\
    &= \left| \vec{\alpha}^{\prime\dagger} \left( \delta\tilde{\mathcal{H}} - E_{0} \delta\tilde{\mathcal{S}} \right) \vec{\alpha}^{\prime} \right| \\
    &\leq \left\|\vec{\alpha}^{\prime}\right\|_{2}^{2} \left\| \left(\delta \tilde{\mathcal{H}} - E_{0} \delta \tilde{\mathcal{S}} \right) \right\|_{\mathrm{op}} \\
    &\leq \left\|\vec{\alpha}^{\prime}\right\|_{2}^{2} \left( \|\delta \tilde{\mathcal{H}}\|_{\mathrm{op}} + \left|E_{0}\right| \| \delta \tilde{\mathcal{S}} \|_{\mathrm{op}} \right),
\end{split}
\end{equation}
where we have used the Cauchy-Schwartz inequality between the second line and the third line.
Focusing on the fact that the observables and states that should be measured in $\mathcal{S}$ are also included in $\mathcal{H}$ by ansatz construction,
the shot counts are distributed only to the measurement queries in $\mathcal{H}$, and thus
\begin{equation} \label{eq:upper_bound_of_delta_H_prime}
\begin{split}
    \|\delta\tilde{\mathcal{H}}\|_{\mathrm{op}} 
    \lesssim \|\tilde{\mathcal{H}}\|_{\mathrm{op}} \left(\frac{N_{\mathrm{s}}}{Q}\right)^{-1/2}
    \leq \gamma \|\tilde{\mathcal{S}}\|_{\mathrm{op}} \left(\frac{N_{\mathrm{s}}}{Q}\right)^{-1/2}.
\end{split}
\end{equation}
The second inequality in Eq.~\eqref{eq:upper_bound_of_delta_H_prime} is obtained by using $\|\tilde{\mathcal{H}}\|_{\mathrm{op}}\leq \gamma \|\tilde{\mathcal{S}}\|_{\mathrm{op}}$ since $\tilde{\mathcal{H}}_{ij} \leq \gamma \tilde{\mathcal{S}}_{ij}$, where again $\displaystyle \gamma = \sum_{i}\left|h_{i}\right|$ for the given Hamiltonian $\displaystyle H = \sum_{i}h_{i}P_{i}$.
Using this property, we can further transform Eq.~\eqref{eq:ineq_yoshioka2022generalized_s7} to
\begin{equation}
\begin{split}
    |\delta E|
    \leq  2\gamma \left(\frac{N_{\mathrm{s}}}{Q}\right)^{-1/2} \|\tilde{\mathcal{S}}\|_{\mathrm{op}} \left\|\vec{\alpha}^{\prime}\right\|_{2}^{2}.
\end{split}
\end{equation}
Note that we have used $|E_{0}|\leq \gamma$ here.
Moreover, $\|\tilde{\mathcal{S}}\|_{\mathrm{op}}\leq \|\tilde{\mathcal{S}}\|_{\mathrm{F}}\leq M$, since $|\tilde{\mathcal{S}}_{ij}|\leq 1$.
This gives the upper bound of $|\delta E|$ as
\begin{equation}
\begin{split}
    |\delta E|
    \leq  2\gamma M\left(\frac{N_{\mathrm{s}}}{Q}\right)^{-1/2} \left\|\vec{\alpha}^{\prime}\right\|_{2}^{2}.
\end{split}
\end{equation}
Therefore, to suppress the estimation bias to $\epsilon$, it is enough to pay the sampling cost of
\begin{equation} \label{eq:Ns_lower_bound_final}
\begin{split}
    N_{\mathrm{s}}
    \geq  \frac{ 4\gamma^{2} M^{2}Q \left\|\vec{\alpha}^{\prime}\right\|_{2}^{4} }
               { \epsilon^{2} }.
\end{split}
\end{equation}
This implies the divide-and-conquer overhead to reintroduce the entanglement among physically separable subsystems scales in $O\left(\left\|\vec{\alpha}^{\prime}\right\|_{2}^{4}\right)$.

We numerically demonstrate how $\left\|\vec{\alpha}^{\prime}\right\|_{2}^{4}$ scales to the number of subspaces $M$ and see how the change of number of measurement queries in the divide-and-conquer subspace reduces this overhead.
Our numerical simulation here still adopts the 8-qubit 1D Ising Hamiltonian.
Taking the 8-qubit parameterized circuits of Fig.~\ref{fig:qc_vqe} as the undivided ansatz $\rho_{\mathrm{whole}}^{(\mathrm{AB})}$, we define the error-mitigated ansatz $\rho_{\mathrm{WHOLE}}$ without the divide-and-conquer strategy as 
\begin{equation} \label{eq:rho_whole}
\begin{split}
    \rho_{\mathrm{WHOLE}}
    = \sum_{i,j = 1}^{M}\alpha_{i}^{*}\alpha_{j} 
        H^{i-1} \left(\rho_{\mathrm{whole}}^{(\mathrm{AB})}\right)^{2} H^{j-1}.
\end{split}
\end{equation}
Accordingly, the numbers of measurement queries required for $\rho_{\mathrm{WHOLE}}$ and $\rho_{\mathrm{DC2}}$ are defined as $Q_{\mathrm{WHOLE}}$ and $Q_{\mathrm{DC2}}$ respectively.
Using these notations, we define the quantity $R$ as below to see how the reduction of measurement queries by the divide-and-conquer strategy will contribute to reducing the overall overhead of Dual-GSE.
\begin{equation} \label{eq:ratio_R}
\begin{split}
    R = \|\vec{\alpha}^{\prime}\|_{2}^{4} \times \frac{Q_{\mathrm{DC2}}}{Q_{\mathrm{WHOLE}}},
\end{split}
\end{equation}
where $\vec{\alpha}^{\prime}$ is the coefficients of the ``DC2'' subspace defined in Eq.~\eqref{eq:rho_dc2}.

\begin{figure}[htbp]
    \centering
    \includegraphics[width=0.5\columnwidth]{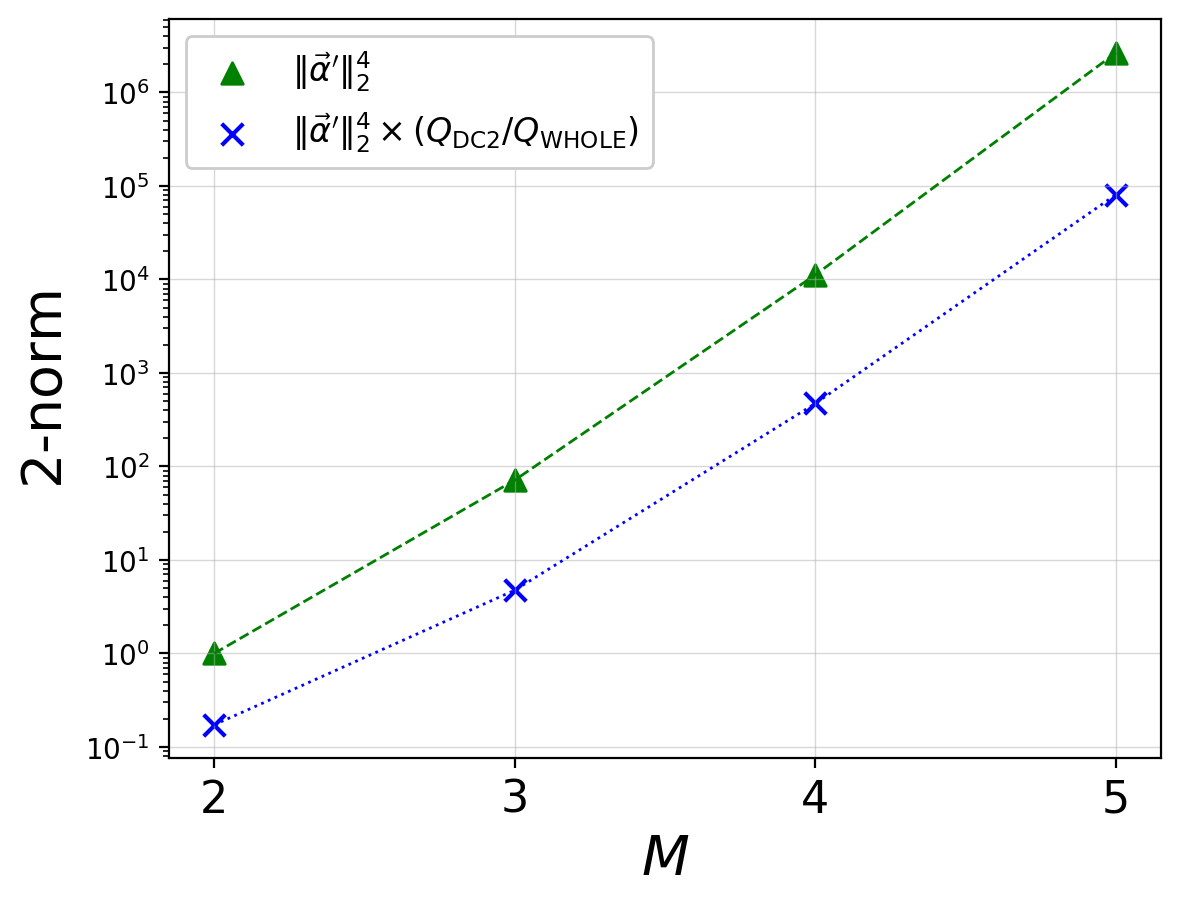}
    \caption{
        The quartic of two-norm of $\vec{\alpha}^{\prime}$ to the number of subspaces $M$ is shown in the green plots with the dashed fitting curve.
        The quantity $R$ defined in Eq.~\eqref{eq:ratio_R} to the number of subspaces $M$ is shown in the blue plots with the dotted fitting curve.
    }
    \label{fig:local-stochastic-pauli_cfe-subspace_subspace-to-ratio-2norm_noise-free}
\end{figure}

Under the above setting, we run the noise-free simulation to visualize the divide-and-conquer overhead with the two-norm of $\vec{\alpha}^{\prime}$.
Both the quantities $\left\|\vec{\alpha}^{\prime}\right\|_{2}^{4}$ and $R$ are plotted to different number of subspaces $M$ in Fig.~\ref{fig:local-stochastic-pauli_cfe-subspace_subspace-to-ratio-2norm_noise-free}.
As the sampling cost scales the quartic of $\left\|\vec{\alpha}^{\prime}\right\|_{2}$, we see the green plots rise harshly in Fig.~\ref{fig:local-stochastic-pauli_cfe-subspace_subspace-to-ratio-2norm_noise-free}.
In fact, this can be eased by the reduction of Pauli observables to be measured in the divide-and-conquer strategy, as discussed both in the last part of Section~\ref{sec:dgse_divide-and-conquer_subspace} and Section~\ref{sec:Leveraging_mitigation_overheads}.
This reduction effect is taken into account by the quantity $R$, which seems to be ten times smaller than $\left\|\vec{\alpha}^{\prime}\right\|_{2}^{4}$.
Overall, the numerical simulation also supports the theoretical analysis of the divide-and-conquer overhead discussed above and demonstrates the practical advantage of using the divide-and-conquer subspace at the same time by benefiting its measurement query reduction.

\subsection{\texorpdfstring{The noise overhead in $\left\|\vec{\alpha}^{\prime}\right\|_{2}$}{}}

When the gate noise comes in, we can accordingly evaluate its influence separately from the divide-and-conquer overhead. 
We here assume the noise shapes in the depolarizing channel, which is numerically justified by Tsubouchi et al.~\cite{Tsubouchi2022} that various types of noise converge to the global depolarizing channel.
Below, we see that the sampling cost increases $\left(1-p\right)^{-8}$ times larger to the depolarizing rate $p$ and that the noise overhead can be considered separately from the divide-and-conquer overhead.

We keep using the subspace construction of Eq.~\eqref{eq:rho_dc2}.
When the $d$-dimensional noisy ansatz $\tilde{\rho}_{\mathrm{local}}^{(\mathrm{AB})} = \tilde{\rho}_{\mathrm{local}}^{(\mathrm{A})} \otimes \tilde{\rho}_{\mathrm{local}}^{(\mathrm{B})}$ undergoes the $d$-dimensional depolarizing channel with probability $p$, each pure local state $\rho_{\mathrm{local}}^{(\mathrm{A})} = |\psi^{(\mathrm{A})}\rangle\langle\psi^{(\mathrm{A})}|$ and $\rho_{\mathrm{local}}^{(\mathrm{B})} = |\psi^{(\mathrm{B})}\rangle\langle\psi^{(\mathrm{B})}|$ becomes
\begin{equation}
\begin{split}
    \tilde{\rho}_{\mathrm{local}}^{(\mathrm{A})} &= \left(1 - p\right) |\psi^{(\mathrm{A})}\rangle\langle\psi^{(\mathrm{A})}| + p \frac{I^{(\mathrm{A})}}{d}, \quad \text{and} \\
    \tilde{\rho}_{\mathrm{local}}^{(\mathrm{B})} &= \left(1 - p\right) |\psi^{(\mathrm{B})}\rangle\langle\psi^{(\mathrm{B})}| + p \frac{I^{(\mathrm{B})}}{d}.
\end{split}
\end{equation}
Focusing on $\vec{\alpha}_{i}^{\prime}$, the coefficient of $\vec{\alpha}_{i}$ in Eq.~\eqref{eq:alpha_prime} decays into
\begin{equation}
\begin{split}
    \operatorname{Tr}\left[ \left(\tilde{\rho}_{\mathrm{local}}^{(\mathrm{AB})}\right)^{2} H^{2(i-1)} \right]
    &=  \sum_{k} h_{ik}
        \operatorname{Tr}\left[\left(\tilde{\rho}_{\mathrm{local}}^{(\mathrm{A})}\right)^{2}P_{ik}^{(\mathrm{A})}\right]
        \operatorname{Tr}\left[\left(\tilde{\rho}_{\mathrm{local}}^{(\mathrm{B})}\right)^{2}P_{ik}^{(\mathrm{B})}\right] \\
    &=  \left(1-p\right)^{4} \sum_{k} h_{ik}
        \operatorname{Tr}\left[\left(\rho_{\mathrm{local}}^{(\mathrm{A})}\right)^{2}P_{ik}^{(\mathrm{A})}\right]
        \operatorname{Tr}\left[\left(\rho_{\mathrm{local}}^{(\mathrm{B})}\right)^{2}P_{ik}^{(\mathrm{B})}\right] \\
    &=  \left(1-p\right)^{4} \operatorname{Tr}\left[ \left(\rho_{\mathrm{local}}^{(\mathrm{AB})}\right)^{2} H^{2(i-1)} \right],
\end{split}
\end{equation}
where we have defined $\displaystyle H^{2(i-1)} = \sum_{i}h_{ik}\left(P_{ik}^{(\mathrm{A})}\otimes P_{ik}^{(\mathrm{B})}\right)$.
Introducing the coefficients affected by the depolarizing noise in the ansatz as $\vec{\alpha}_{\mathrm{noisy}}^{\prime}$, the condition $\vec{\alpha}_{\mathrm{noisy}}^{\prime\dagger}\tilde{\mathcal{S}}_{\mathrm{noisy}}\vec{\alpha}_{\mathrm{noisy}}^{\prime}=1$ requires $\vec{\alpha}_{\mathrm{noisy}}^{\prime}$ to satisfy
\begin{equation} \label{eq:alpha_noisy}
\begin{split}
    \left\|\vec{\alpha}_{\mathrm{noisy}}^{\prime}\right\|_{2} = \left(1-p\right)^{-2}\left\|\vec{\alpha}^{\prime}\right\|_{2}.
\end{split}
\end{equation}
Note that $\tilde{\mathcal{S}}_{\mathrm{noisy}} = \tilde{\mathcal{S}}$ for depolarizing noise since the diagonal elements of matrix $\mathcal{S}$ are normalized in the same way as Eq.~\eqref{eq:alpha_prime}.
The relation of Eq.~\eqref{eq:alpha_noisy} implies the sampling overhead $N_{\mathrm{s}}$ in Eq.~\eqref{eq:Ns_lower_bound_final} is amplified $\left(1-p\right)^{-8}$ times larger than noise-free scenario.

\section{Effect of Shot Noise on Observable Estimation \label{sec:appendix_effect_of_shot_noise_on_observable_estimation}}

When we compute the ground state energy through the generalized eigenvalue problem $\mathcal{H}\vec\alpha = E\mathcal{S}\vec\alpha$, each matrix element $\mathcal{S}_{ij}$ and $\mathcal{H}_{ij}$ is deviated by finite shot noise.
This deviation is modelled as $\mathcal{\bar{S}}_{ij} = \mathcal{S}_{ij} + \delta\mathcal{S}_{ij}$ and $\mathcal{\bar{H}}_{ij} = \mathcal{H}_{ij} + \delta\mathcal{H}_{ij}$, where $\mathcal{S}_{ij}, \mathcal{H}_{ij}$ are the shot-noise-free expectation values, and $\delta\mathcal{S}_{ij}, \delta\mathcal{S}_{ij}$ are the shot noise deviations.
Below is the procedure for adding the shot noise effect in our numerical simulation.
\begin{enumerate}
    \item Collect all combinations of Pauli observables and quantum states required for the classical postprocessing. Compute their single-shot variance in advance.
    \item In accordance with the single-shot variance, generate the Gaussian noise for each Pauli observable and add it to the expectation value without the shot noise effect.
    \item Sum up the noisy expectation values for Pauli observables to make the matrix elements $\mathcal{\bar{S}}_{ij}$ and $\mathcal{\bar{H}}_{ij}$.
    \item Solve the generalized eigenvalue problem $\mathcal{\bar{H}}\vec\alpha = E\mathcal{\bar{S}}\vec\alpha$ to obtain the mitigated expectation value $E$ under both the gate and the finite shot noise.
\end{enumerate}
To generate the Gaussian noise, we discuss the single-shot variance of Pauli estimators and also theoretically analyze the single-shot variance of matrix elements $\mathcal{\bar{S}}_{ij}$ and $\mathcal{\bar{H}}_{ij}$ under different choices of the subspace.

\subsection{Single-shot variance of Pauli observables using DSP}

Let $\operatorname{Var}\left((X\otimes \operatorname{P}_{\vec{0}})_{P_{k}}\right)$ be the single-shot variance of Pauli observable $P_{k}$ through the DSP circuit for the symmetried ansatz.
This can be expanded into $\operatorname{Var}\left((X\otimes \operatorname{P}_{\vec{0}})_{P_{k}}\right) = \langle (X\otimes \operatorname{P}_{\vec{0}})_{P_{k}}^2\rangle - \langle X\otimes \operatorname{P}_{\vec{0}}\rangle_{P_{k}}^2$,
where $\langle X\otimes \operatorname{P}_{\vec{0}}\rangle_{P_{k}}$ is equivalent to the estimator of $P_{k}$ through the DSP circuit and thus $\displaystyle\operatorname{Tr}\left[\frac {\rho_{i}\bar{\rho}_{j} + \bar{\rho}_{j}\rho_{i}} {2} P_{k}\right]$, while the value $\langle (X\otimes \operatorname{P}_{\vec{0}})_{P_{k}}^2\rangle = \langle I\otimes \operatorname{P}_{\vec{0}}\rangle_{P_{k}}$ becomes $\displaystyle\operatorname{Tr}\left[\frac {\rho_{i}\bar{\rho}_{j} + P_{k}\rho_{i} P_{k}\bar{\rho}_{j}} {2}\right]$.
To prove this, we compute the expectation value of the observable $X\otimes \operatorname{P}_{\vec{0}}$ and $I\otimes \operatorname{P}_{\vec{0}}$ on the quantum state $\rho_{k}$ in Fig.~\ref{fig:qc_dsp_Pk}.
First, quantum state $\rho_{k}$ right before the measurement process in DSP circuit becomes
\begin{equation}
\begin{split}
    \rho_{k} 
    &= \frac{1}{2}
         \left(|0\rangle\langle0|\otimes \mathcal{U}_{j, \mathrm{rev}}\left(\mathcal{U}_{i}(|\vec{0}\rangle\langle\vec{0}|)\right)
    \quad+|0\rangle\langle1|\otimes \mathcal{U}_{j, \mathrm{rev}}\left(\mathcal{U}_{i}(|\vec{0}\rangle\langle\vec{0}|)P_{k}\right)\right.\\
    &~~~~+\left.|1\rangle\langle0|\otimes \mathcal{U}_{j, \mathrm{rev}}\left(P_{k}\mathcal{U}_{i}(|\vec{0}\rangle\langle\vec{0}|)\right)
    +|1\rangle\langle1|\otimes \mathcal{U}_{j, \mathrm{rev}}\left(P_{k}\mathcal{U}_{i}(|\vec{0}\rangle\langle\vec{0}|)P_{k}\right)\right).
\end{split}
\end{equation}
Performing measurement with the observables $X\otimes \operatorname{P}_{\vec{0}}$ and $I\otimes \operatorname{P}_{\vec{0}}$ on $\rho_{k}$, respectively, we obtain the expectation values
\begin{equation}
\begin{split}
    \langle I\otimes \operatorname{P}_{\vec{0}}\rangle_{P_{k}}
    &= \frac{1}{2}\operatorname{Tr}\left[\mathcal{U}_{j, \mathrm{rev}}\left(\mathcal{U}_{i}(|\vec{0}\rangle\langle\vec{0}|)\right)|\vec{0}\rangle\langle\vec{0}| 
                                       + \mathcal{U}_{j, \mathrm{rev}}\left(P_{k}\mathcal{U}_{i}(|\vec{0}\rangle\langle\vec{0}|)P_{k}\right)|\vec{0}\rangle\langle\vec{0}|\right]
    = \frac{1}{2}\operatorname{Tr}\left[\rho_{i}\bar{\rho}_{j} + \rho_{i}P_{k}\bar{\rho}_{j}P_{k}\right], \\
    \langle X\otimes \operatorname{P}_{\vec{0}}\rangle_{P_{k}}
    &= \frac{1}{2}\operatorname{Tr}\left[\mathcal{U}_{j, \mathrm{rev}}\left(\mathcal{U}_{i}(|\vec{0}\rangle\langle\vec{0}|)P_{k}\right)|\vec{0}\rangle\langle\vec{0}| 
                                       + \mathcal{U}_{j, \mathrm{rev}}\left(P_{k}\mathcal{U}_{i}(|\vec{0}\rangle\langle\vec{0}|)\right)|\vec{0}\rangle\langle\vec{0}|\right]
    = \frac{1}{2}\operatorname{Tr}\left[\left(\rho_{i}\bar{\rho}_{j} + \bar{\rho}_{j}\rho_{i}\right)P_{k}\right],
\end{split}
\end{equation}
where $\bar{\rho}=\mathcal{U}_{j,\mathrm{dual}}(|\vec{0}\rangle\langle\vec{0}|)$ with $\mathcal{U}_{j,\mathrm{dual}}$ as a dual process of $\mathcal{U}_{j}$.
Therefore, the single-shot variance of Pauli observable $P_{k}$ is described as
\begin{equation}
\begin{split}
    \operatorname{Var}\left((X\otimes \operatorname{P}_{\vec{0}})_{P_{k}}\right) 
    = \langle (X\otimes \operatorname{P}_{\vec{0}})_{P_{k}}^2\rangle - \langle X\otimes \operatorname{P}_{\vec{0}}\rangle_{P_{k}}^2
    = \operatorname{Tr}\left[\frac {\rho_{i}\bar{\rho}_{j} + \rho_{i}P_{k}\bar{\rho}_{j}P_{k}} {2}\right] - \operatorname{Tr}\left[\frac {\rho_{i}\bar{\rho}_{j} + \bar{\rho}_{j}\rho_{i}} {2} P_{k}\right]^2.
\end{split}
\end{equation}
We also use these facts to theoretically analyze the single-shot variance of $\mathcal{S}_{ij}$ and $\mathcal{H}_{ij}$ for each type of subspace below.

\begin{figure}
    \begin{adjustbox}{width=0.5\textwidth}\begin{quantikz}
        \lstick{$|+\rangle$} & \qw                               & \ctrl{1}                & \qw                                                               & \meter{I, X}        \\
        \lstick{$|0\rangle$} & \gate[3, nwires=2]{\mathcal{U}_{i}} & \gate[3, nwires=2]{P_{k}} & \gate[3, nwires=2]{\mathcal{U}_{j,\mathrm{rev}}} \slice{$\rho_{k}$} & \push{\ \langle 0|} \\
        \vdots               &                                   &                         &                                                                   & \vdots              \\
        \lstick{$|0\rangle$} &                                   &                         &                                                                   & \push{\ \langle 0|}
    \end{quantikz}\end{adjustbox}
    \caption{   
        The quantum circuit for computing $\langle X\otimes\operatorname{P}_{\vec{0}}\rangle_{P_{k}}$ and $\langle I\otimes\operatorname{P}_{\vec{0}}\rangle_{P_{k}}$ through the DSP process with indirect measurement.
        Here, $\rho_{k}$ denotes the quantum state before the measurement process.
    }
    \label{fig:qc_dsp_Pk}
\end{figure}

\subsection{Single-shot variance matrix elements in the power subspace}

We first discuss the variance of estimators under the following type of power subspace $\{\sigma_{i}\}_{i} = \{I\}\cup\{\rho H^{k-2}\}_{k=2,\ldots,M}$.
Note that for $i \geq 2$, $\sigma_{i} = \rho H^{i-2}$ and $\bar\sigma_{i} = \bar{\rho} H^{i-2}$.
The expectation value of using symmetrized ansatz in the element $\mathcal{S}_{ij}$ for $i,j \geq 2$ is computed as follows.
\begin{equation}
    \mathcal{S}_{i j} 
    =\operatorname{Tr}\left[H^{i-2}\frac{\bar{\rho}_{j}\rho_{i} + \rho_{i}\bar{\rho}_{j}}{2}H^{j-2}\right]
    =\operatorname{Tr}\left[\frac{\bar{\rho}_{j}\rho_{i}+ \rho_{i}\bar{\rho}_{j}}{2}H^{i+j-4}\right].
\end{equation}
Then, if we decompose the power of a Hamiltonian into $H^{i+j-4} = \sum_{k}{c}_{k}P_{k}$ with Pauli strings $P_{k}$ and their coefficients ${c}_{k}\in\mathbb C$, $\mathcal{S}_{i j}$ can be expanded by the summation of expectation values of those physical observables:
\begin{equation}
    \mathcal{S}_{i j} 
    =\sum_{k} {c}_{k} \operatorname{Tr}\left[\frac{\bar{\rho}_{j} \rho_{i}+\rho_{i} \bar{\rho}_{j}}{2}P_{k}\right] 
    =\sum_{k} {c}_{k}\langle X \otimes \operatorname{P}_{\vec{0}}\rangle_{P_{k}}.
\end{equation}
By using the same notation, the single-shot variance of $\mathcal{S}_{ij}$ is deduced as follows.
\begin{equation}
\label{eq:var_S_ij}
\begin{split}
    \operatorname{Var}(\mathcal{S}_{ij}) 
    &=\displaystyle \sum_{k} |{c}_{k}|^2\operatorname{Var}\left((X\otimes \operatorname{P}_{\vec{0}})_{P_{k}}\right) \\
    &=\displaystyle \sum_{k} |{c}_{k}|^2\left(\langle (X\otimes \operatorname{P}_{\vec{0}})^2\rangle_{P_{k}} - \langle X\otimes \operatorname{P}_{\vec{0}}\rangle_{P_{k}}^2\right) \\
    &=\displaystyle \sum_{k} |{c}_{k}|^2\left(\operatorname{Tr}\left[\frac {\rho_{i}\bar{\rho}_{j} + \rho_{i}P_{k}\bar{\rho}_{j}P_{k}} {2}\right] - \operatorname{Tr}\left[\frac {\bar{\rho}_{j}\rho_{i} + \rho_{i}\bar{\rho}_{j}} {2} P_{k}\right]^2\right).
\end{split}
\end{equation}
This also holds for $\mathcal{H}_{i j} $ and $\operatorname{Var}(\mathcal{H}_{ij})$ by just replacing $H^{i+j-4}$ with $H^{i+j-3}$.

\subsection{Single-shot variance matrix elements in the fault subspace}

Next, we analyze the variance of observables for the fault subspace $ \{\rho(\lambda_{k}\epsilon)\}_{k=1,2,\ldots,M}$.
In this case, since $\sigma_{i} = \rho_{i} = \rho(\lambda_{i}\epsilon)$, $\mathcal{S}_{i j} $ and $\operatorname{Var}(\mathcal{S}_{ij})$ become the following forms.
\begin{equation}
    \mathcal{S}_{i j} 
    =\operatorname{Tr}\left[\frac{\bar{\rho}_{j} \rho_{i}+\rho_{i} \bar{\rho}_{j}}{2}\right]
    =\operatorname{Tr}\left[\frac{\bar{\rho}(\lambda_{j}\epsilon) \rho(\lambda_{i}\epsilon)+\rho(\lambda_{i}\epsilon) \bar{\rho}(\lambda_{j}\epsilon)}{2}\right]
    =\operatorname{Tr}\left[\bar{\rho}(\lambda_{j}\epsilon) \rho(\lambda_{i} \epsilon) \right],
\end{equation}
\begin{equation}
\begin{split}
    \text{Var}(\mathcal{S}_{ij}) &= \text{Tr}\left[\frac{\bar{\rho}(\lambda_{j}\epsilon) \rho(\lambda_{i}\epsilon) + \rho(\lambda_{i}\epsilon)\bar{\rho}(\lambda_{j}\epsilon)}{2}\right]- \text{Tr}\left[\frac{\bar{\rho}(\lambda_{j}\epsilon) \rho(\lambda_{i}\epsilon) + \rho(\lambda_{i}\epsilon) \bar{\rho}(\lambda_{j}\epsilon)}{2}\right]^2 \\
    &= \text{Tr}\left[\bar{\rho}(\lambda_{j}\epsilon) \rho(\lambda_{i}\epsilon)\right]- \text{Tr}\left[\bar{\rho}(\lambda_{j}\epsilon) \rho(\lambda_{i}\epsilon)\right]^2.
\end{split}
\end{equation}
Therefore, assuming the Hamiltonian is given by $H = \sum_{h}c_{h}P_{h}$, $\mathcal{H}_{ij} $ and $\operatorname{Var}(\mathcal{H}_{ij})$ is computed as follows, respectively.
\begin{equation}
    \mathcal{H}_{ij} 
    =\operatorname{Tr}\left[\frac{\bar{\rho}_{j} \rho_{i}+\rho_{i} \bar{\rho}_{j}}{2}H\right]
    =\sum_{h} c_{h}\operatorname{Tr}\left[\frac{\bar{\rho}(\lambda_{j}\epsilon) \rho(\lambda_{i}\epsilon)+\rho(\lambda_{i}\epsilon) \bar{\rho}(\lambda_{j}\epsilon)}{2}P_{h}\right],
\end{equation}
\begin{equation}
\begin{split}
    \operatorname{Var}(\mathcal{H}_{ij}) 
    &=\displaystyle \sum_{h} |c_h|^2\left(\operatorname{Tr}\left[\frac {\rho(\lambda_{i}\epsilon)\bar{\rho}(\lambda_{j}\epsilon) + \rho(\lambda_{i}\epsilon) P_{h}\bar{\rho}(\lambda_{j}\epsilon)P_{h}} {2}\right] 
    - \operatorname{Tr}\left[\frac {\bar{\rho}(\lambda_{j}\epsilon)\rho(\lambda_{i}\epsilon) + \rho(\lambda_{i}\epsilon)\bar{\rho}(\lambda_{j}\epsilon)} {2} P_{h}\right]^2\right).
\end{split}
\end{equation}

\subsection{Single-shot variance matrix elements in the divide-and-conquer subspace}

Finally, we discuss the variance of estimators under the divide-and-conquer subspace $\{I^{(\mathrm{AB})}\}\cup\{(\rho^{(\mathrm{A})}\otimes\rho^{(\mathrm{B})})H^{k-2}\}_{k=2,\ldots,M}$.
Recalling the definition of the divide-and-conquer subspace in Eq.~\eqref{eq:dgse_dc_H_ij}, the QEM ansatz is described as
\begin{equation}
    \rho^{(\mathrm{AB})}=\sum_{i,j} \alpha_{i}^{*} \alpha_{j} C_{i}^{\dagger} \left(\frac{\bar{\rho}_{j}^{(\mathrm{A})} \rho_{i}^{(\mathrm{A})}+\rho_{i}^{(\mathrm{A})} \bar{\rho}_{j}^{(\mathrm{A})}}{2} \otimes \frac{\bar{\rho}_{j}^{(\mathrm{B})} \rho_{i}^{(\mathrm{B})}+\rho_{i}^{(\mathrm{B})} \bar{\rho}_{j}^{(\mathrm{B})}}{2}\right) C_{j}.
\end{equation}
We set $C_{1} = I$ and $C_{i} = H^{i-2}$ for $i\geq2$ in our numerical experiments.
In this case, the expectation value of each element in matrix $\mathcal{S}$ becomes
\begin{equation}
\begin{split}
    \mathcal{S}_{i j}
    &=\operatorname{Tr}\left[\left(\frac{\bar{\rho}_{j}^{(\mathrm{A})} \rho_{i}^{(\mathrm{A})}+\rho_{i}^{(\mathrm{A})} \bar{\rho}_{j}^{(\mathrm{A})}}{2} \otimes \frac{\bar{\rho}_{j}^{(\mathrm{B})} \rho_{i}^{(\mathrm{B})}+\rho_{i}^{(\mathrm{B})} \bar{\rho}_{j}^{(\mathrm{B})}}{2}\right)H^{i+j-4}\right] \\
    &=\sum_{k} {c}_{k} \operatorname{Tr}\left[\frac{\bar{\rho}_{j}^{(\mathrm{A})} \rho_{i}^{(\mathrm{A})}+\rho_{i}^{(\mathrm{A})} \bar{\rho}_{j}^{(\mathrm{A})}}{2}P_{k}^{(\mathrm{A})}\right] \operatorname{Tr}\left[\frac{\bar{\rho}_{j}^{(\mathrm{B})} \rho_{i}^{(\mathrm{B})}+\rho_{i}^{(\mathrm{B})} \bar{\rho}_{j}^{(\mathrm{B})}}{2}P_{k}^{(\mathrm{B})}\right] \\
    &=\sum_{k} {c}_{k}\langle X\otimes\operatorname{P}_{\vec{0}}^{(\mathrm{A})}\rangle_{P_{k}^{(\mathrm{A})}}\langle X\otimes\operatorname{P}_{\vec{0}}^{(\mathrm{B})}\rangle_{P_{k}^{(\mathrm{B})}},
\end{split}
\end{equation}
where we assume the power of Hamiltonian $H^{i+j-4}$ is expanded as $H^{i+j-4} = \sum_{k}{c}_{k}\left(P_{k}^{(\mathrm{A})}\otimes P_{k}^{(\mathrm{B})}\right)$.
As for the variance of this expectation value, the property of variance of the product of two independent random variables $\operatorname{Var}(X Y)=\operatorname{Var}(X) \operatorname{Var}(Y)+\operatorname{Var}(X)(\operatorname{E}(Y))^2+\operatorname{Var}(Y)(\operatorname{E}(X))^2$ is applied to $\displaystyle\operatorname{Var}\left(\mathcal{S}_{i j}\right)$ as follows:
\begin{equation}
\begin{array}{lll}
    \displaystyle\operatorname{Var}\left(\mathcal{S}_{i j}\right) 
    &\displaystyle=\sum_{k}\left|{c}_{k}\right|^2&\displaystyle \operatorname{Var}\left((X\otimes\operatorname{P}_{\vec{0}}^{(\mathrm{A})})_{P_{k}^{(\mathrm{A})}}(X\otimes\operatorname{P}_{\vec{0}}^{(\mathrm{B})})_{P_{k}^{(\mathrm{B})}}\right) \\
    &\displaystyle=\sum_{k}\left|{c}_{k}\right|^2&\displaystyle \left(\operatorname{Var}\left((X\otimes\operatorname{P}_{\vec{0}}^{(\mathrm{A})})_{P_{k}^{(\mathrm{A})}}\right) \operatorname{Var}\left((X\otimes\operatorname{P}_{\vec{0}}^{(\mathrm{B})})_{P_{k}^{(\mathrm{B})}}\right)\right.\\
    &\displaystyle&\displaystyle \left.+\langle X\otimes\operatorname{P}_{\vec{0}}^{(\mathrm{B})}\rangle_{P_{k}^{(\mathrm{B})}}^2\operatorname{Var}\left((X\otimes\operatorname{P}_{\vec{0}}^{(\mathrm{A})})_{P_{k}^{(\mathrm{A})}}\right)\right.\\
    &\displaystyle&\displaystyle \left.+\langle X\otimes\operatorname{P}_{\vec{0}}^{(\mathrm{A})}\rangle_{P_{k}^{(\mathrm{A})}}^2\operatorname{Var}\left((X\otimes\operatorname{P}_{\vec{0}}^{(\mathrm{B})})_{P_{k}^{(\mathrm{B})}}\right)\right).
\end{array}
\end{equation}
In the same way as the power subspace, we can also compute $\mathcal{H}_{ij}$ and $\operatorname{Var}(\mathcal{H}_{ij})$ by replacing $H^{i+j-4}$ with $H^{i+j-3}$.

\section{Numerical Simulation on Different Model and Noise Settings\label{sec:appendix_Numerical_Simulation_on_Different_Model_and_Noise_Settings}}

We have also verified that our method works well for different models and noise settings.
In addition to the setting in the main text, we have performed the following numerical experiments on estimating the ground state energy of the Ising Hamiltonian with different graph structures.
\begin{itemize}
    \item \textit{Different graph structure.} On the 8-qubit 2D cluster graph structure, we examine the estimation bias for each subspace, particularly for the divide-and-conquer subspace, with two ways of subsystem divisions: two 4-qubit cycle graphs and two 4-qubit path graphs.
    \item \textit{Different noise models.} We introduce the amplitude damping noise applied to the ansatz on the 8-qubit 2D cluster graph structure. 
    We also introduce the coherent drift noise applied to the ansatz on the 8-qubit 1D cluster graph structure.
    \item \textit{Different noise levels.} We explore the estimation bias when taking the ansatz of the original 8-qubit 1D path graph structure but with higher error rates of the local stochastic Pauli noise defined in the main text.
\end{itemize}

\subsection{On the different Hamiltonian: 8-qubit 2D cluster graph structure\label{sec:appendix_On_the_different_Hamiltonian:_8-qubit_2D_cluster_graph_structure}}

\begin{figure}[htbp]
    \centering
    \includegraphics[width=0.4\textwidth]{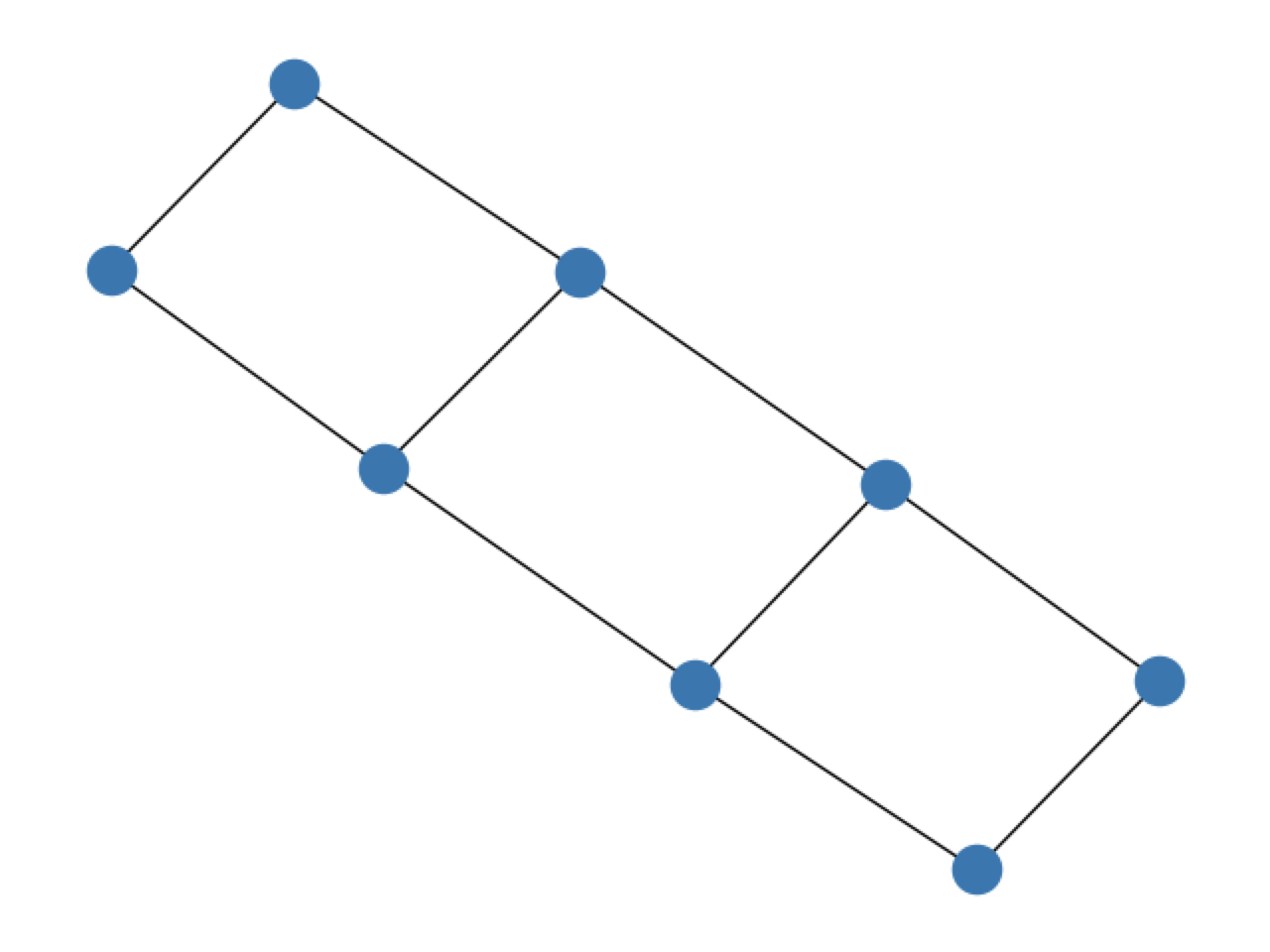}
    \caption{
        The cluster graph structure of 8-qubit 2D Ising Hamiltonian in our additional numerical demonstration.
    }
    \label{fig:graph_8-qubits_referee2}
\end{figure}

To demonstrate that our method also works for the 2D Ising model, we have considered the graph structure of Fig.~\ref{fig:graph_8-qubits_referee2} and verified that the energy estimation works well for the power subspace, the fault subspace, and the divide-and-conquer subspace.
In particular, for the divide-and-conquer subspace, we performed two ways of graph division shown in Fig.~\ref{fig:local-stochastic-pauli_dc_subspace-to-diff_referee2}(a) and Fig.~\ref{fig:local-stochastic-pauli_dc_subspace-to-diff_referee2}(b), each reintroducing the effect of entanglement among two and four edges between the divided two subsystems.
The quantum circuits used for constructing the noisy ansatz $\rho$ have the same layered structure as that in Fig.~\ref{fig:qc_vqe}. 
Meanwhile, the controlled-Z operations are applied according to the graph structure of the connectivity of $Z_{i}Z_{j}$ terms in the given Hamiltonian.
Below, we review the detailed results obtained from our additional numerical simulation.

\begin{figure}[htbp]
    \centering
    \subfloat[power subspace\label{fig:local-stochastic-pauli_power-subspace_subspace-to-diff_referee2}]{
        \includegraphics[width=0.48\textwidth]{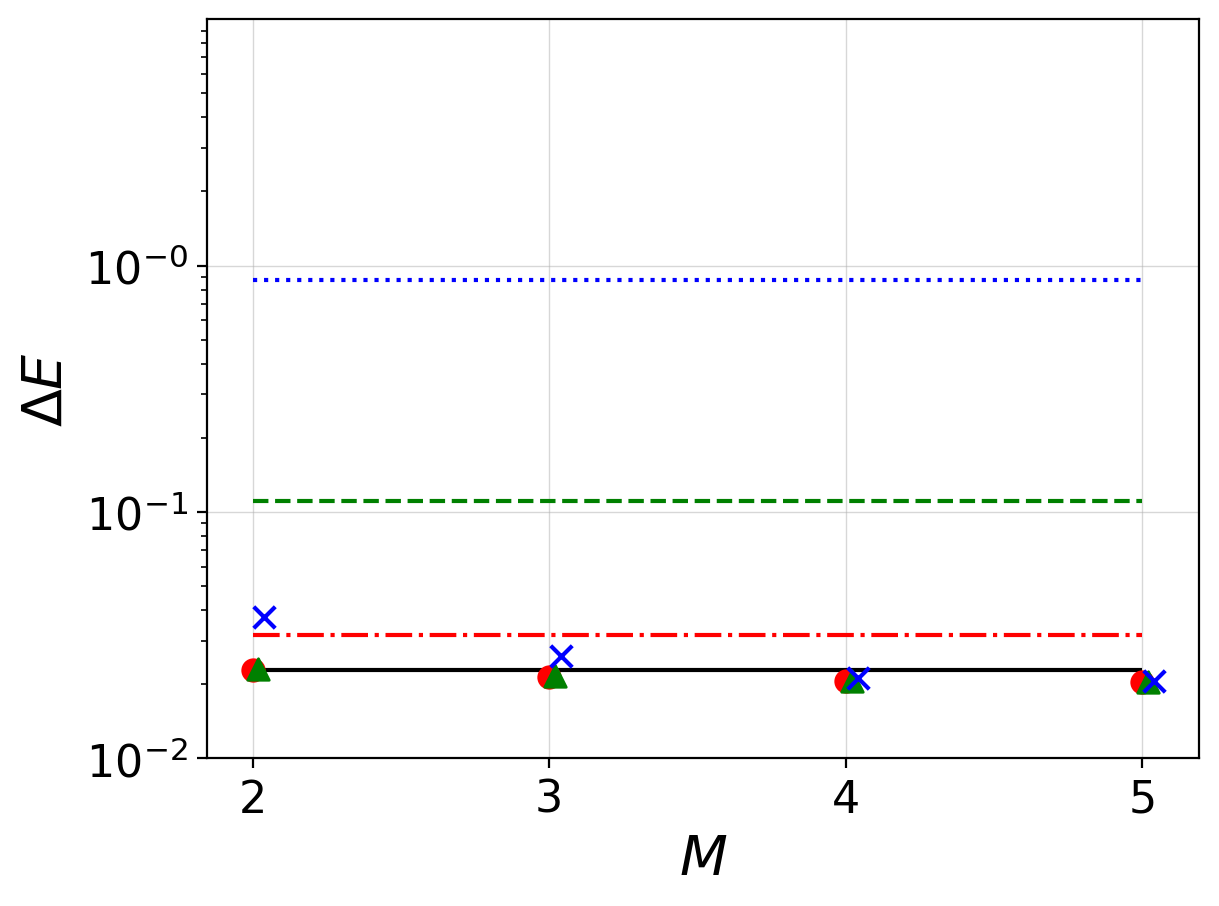}
    }
    \subfloat[fault subspace\label{fig:local-stochastic-pauli_fault-subspace_subspace-to-diff_referee2}]{
        \includegraphics[width=0.48\textwidth]{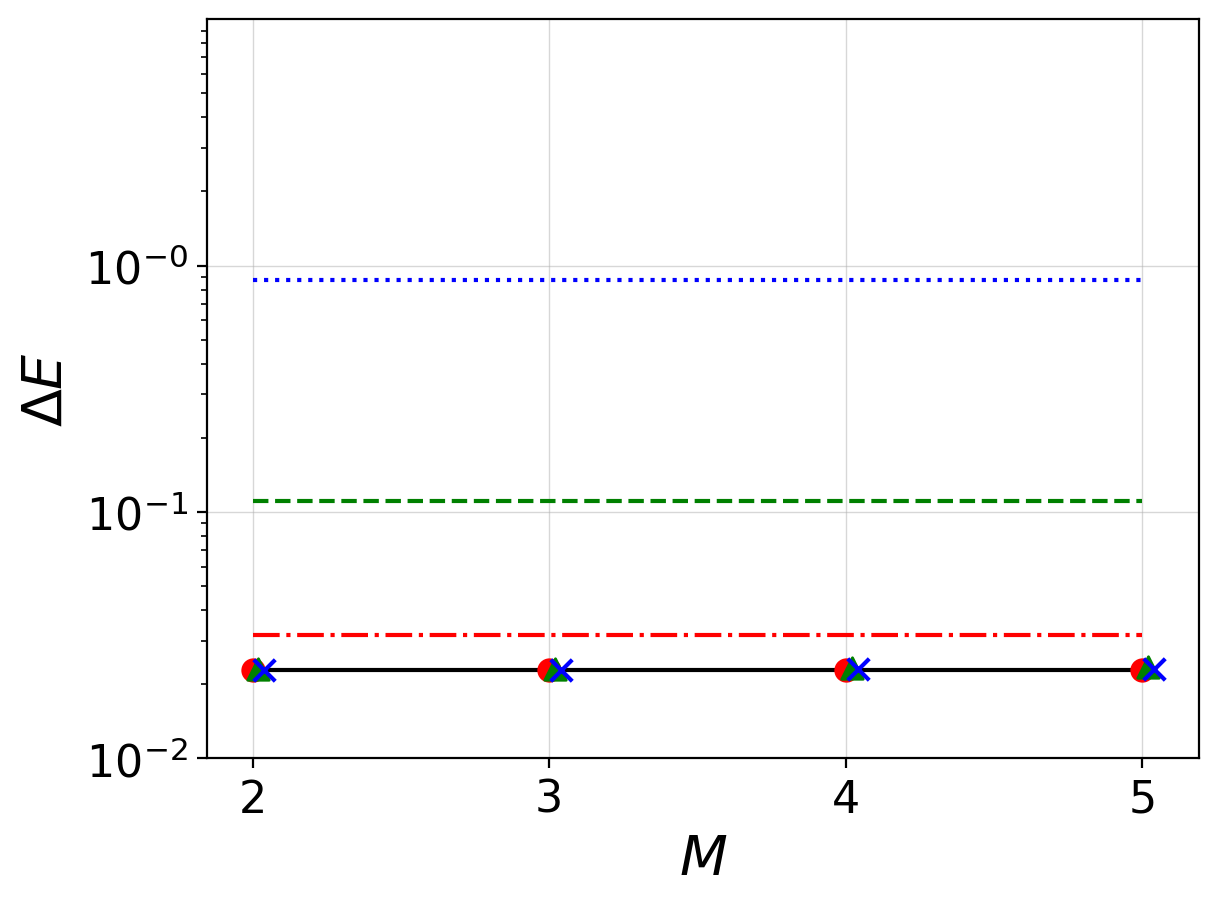}
    }
    \caption{
        The estimation bias $\Delta E$ from the theoretical ground state energy under the local stochastic Pauli error defined in Eq.~\eqref{eq:lsp_1} and Eq.~\eqref{eq:lsp_2}.
        The horizontal black line denotes the inherent estimation bias under the noise-free optimization of variational quantum circuits using the ansatz over the whole eight-qubit system.
        In every figure, the cross blue markers and the dotted blue line represent the estimation bias with and without Dual-GSE under the local stochastic noise with magnitude $2.0\times 10^{-4}$, the triangle green markers and the dashed green line with magnitude $2.0\times 10^{-5}$, and the circle red markers and dash-dot red line with magnitude $2.0\times 10^{-6}$.
    }
    \label{fig:local-stochastic-pauli_power-fault_subspace-to-diff_referee2}
\end{figure}

First, the results by the power subspace and fault subspace are shown in Fig.~\ref{fig:local-stochastic-pauli_power-fault_subspace-to-diff_referee2}(a) and Fig.~\ref{fig:local-stochastic-pauli_power-fault_subspace-to-diff_referee2}(b), respectively.
We can see from Fig.~\ref{fig:local-stochastic-pauli_power-fault_subspace-to-diff_referee2}(a) that the decrease of estimation error as the number of subspaces $M$ increases in the power subspace, which surpasses the estimation bias of noise-free simulation of the quantum circuit with optimized parameters (i.e. the horizontal black line).
This supports the ability of the power subspace to mitigate coherent errors in terms of algorithmic error.
We also see in Fig.~\ref{fig:local-stochastic-pauli_power-fault_subspace-to-diff_referee2}(b) that the estimation bias reaches the noise-free simulation even for small $M$, demonstrating the advantage of using the fault subspace when the available number of subspaces is limited.

\begin{figure}[htbp]
    \centering
    \subfloat[dividing two edges\label{fig:graph_8-qubits_2-bridges_referee2}]{
        \includegraphics[width=0.4\textwidth]{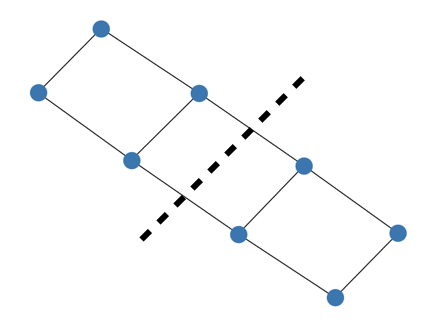}
    }
    \subfloat[dividing four edges\label{fig:graph_8-qubits_4-bridges_referee2}]{
        \includegraphics[width=0.4\textwidth]{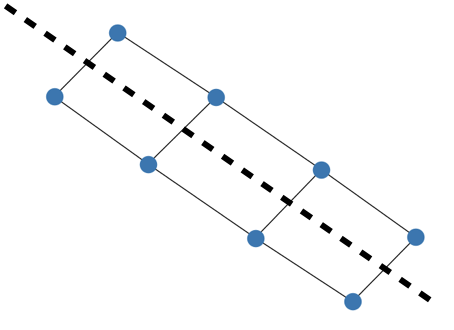}
    }
    \hfill
    \subfloat[divide-and-conquer subspace using (a)\label{fig:local-stochastic-pauli_cfe-subspace_subspace-to-diff_2-bridges_referee2}]{
        \includegraphics[width=0.48\textwidth]{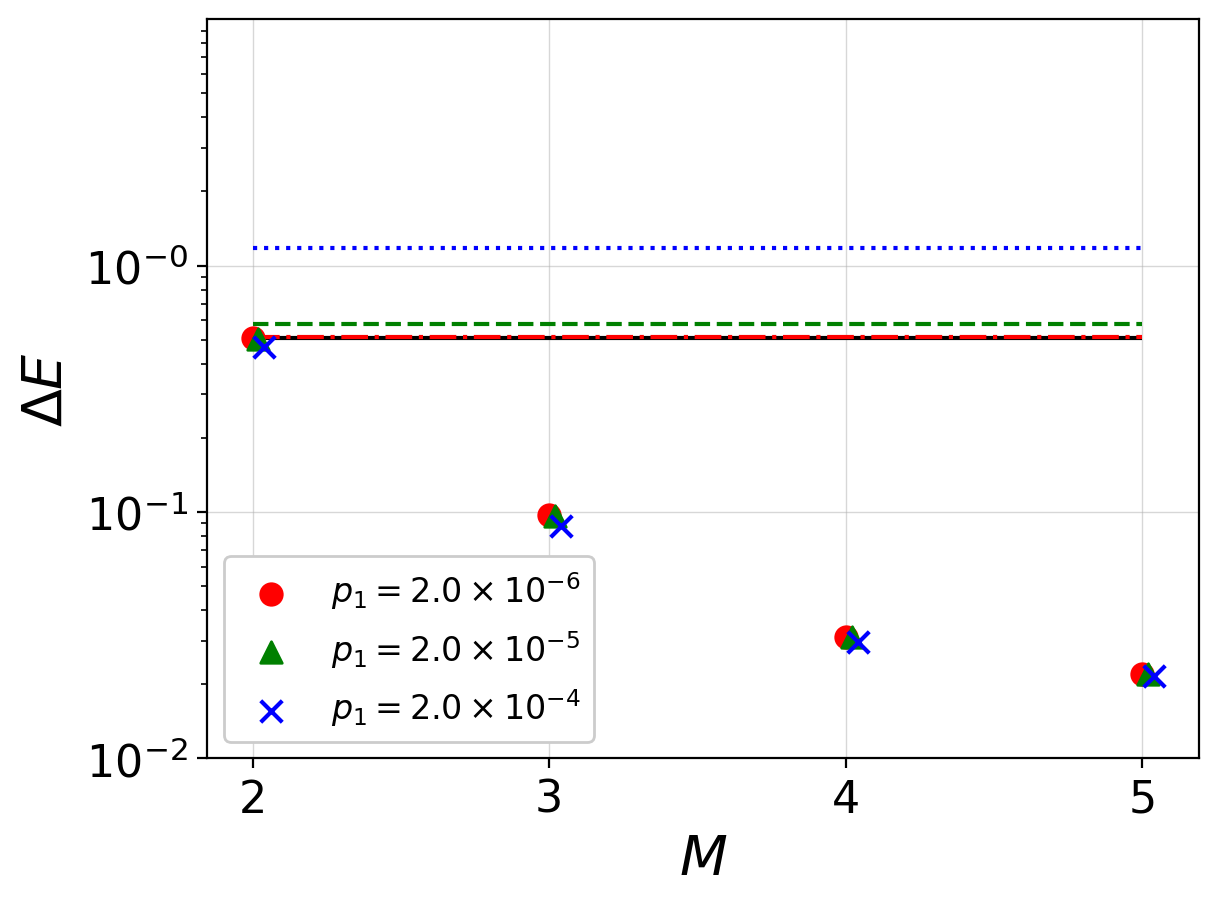}
    }
    \subfloat[divide-and-conquer subspace using (b)\label{fig:local-stochastic-pauli_cfe-subspace_subspace-to-diff_4-bridges_referee2}]{
        \includegraphics[width=0.48\textwidth]{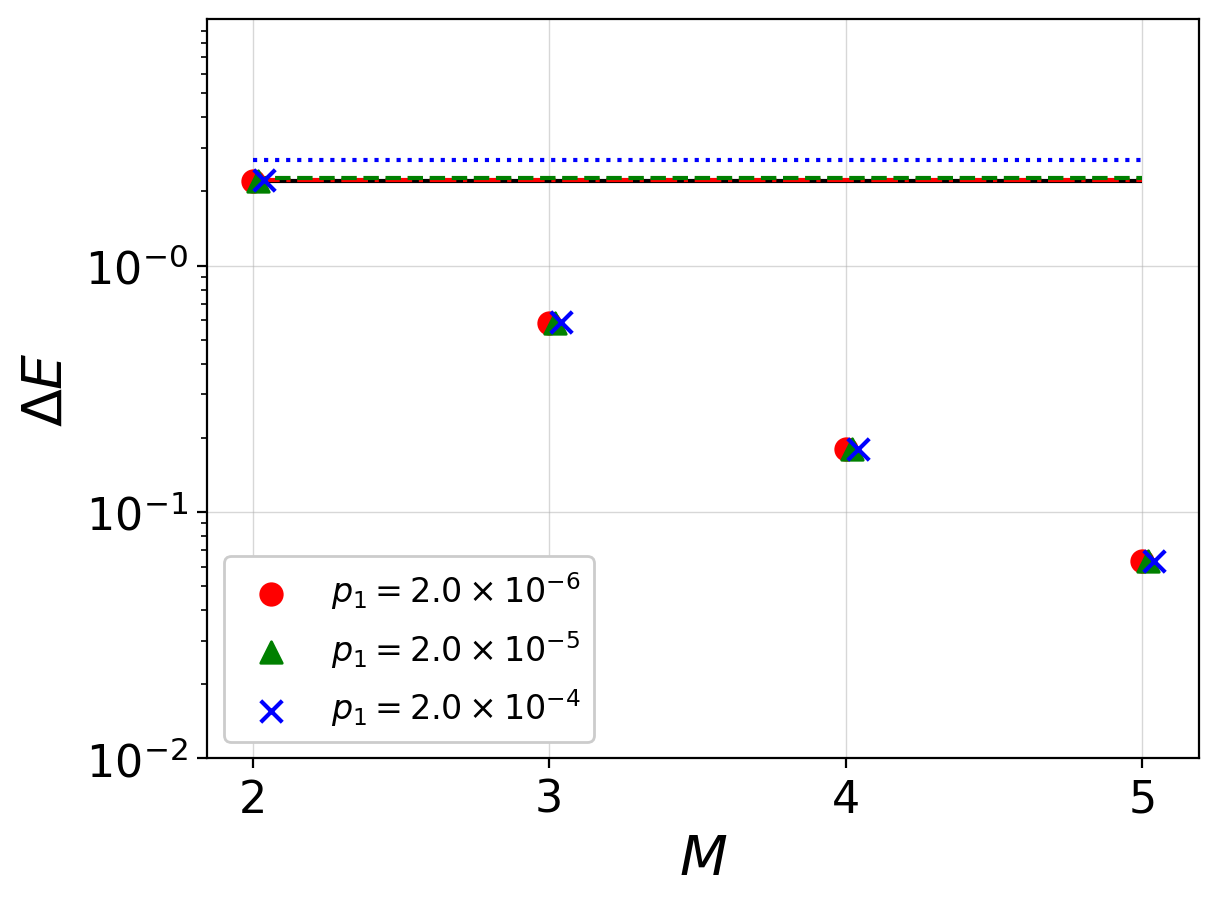}
    }
    \caption{
        Each figure (a) and (b) respectively represents the way to divide 8-qubit 2D Ising Hamiltonian into two pieces of (a) two 4-qubit cycle graphs and (b) two 4-qubit path graphs.
        Each figure (c) and (d) plots the estimation bias $\Delta E$ from the theoretical ground state energy under the local stochastic Pauli error using the divide-and-conquer subspace with two symmetric local systems divided as figure (a) and (b), respectively.
        The notations for the plots and horizontal lines are the same as Fig.~\ref{fig:local-stochastic-pauli_power-fault_subspace-to-diff_referee2}.
    }
    \label{fig:local-stochastic-pauli_dc_subspace-to-diff_referee2}
\end{figure}

Next, Fig.~\ref{fig:local-stochastic-pauli_dc_subspace-to-diff_referee2}(c) and Fig.~\ref{fig:local-stochastic-pauli_dc_subspace-to-diff_referee2}(d) show the estimation bias by the divide-and-conquer subspace with each division method of whole 2D Hamiltonian shown in Fig.~\ref{fig:local-stochastic-pauli_dc_subspace-to-diff_referee2}(a) and Fig.~\ref{fig:local-stochastic-pauli_dc_subspace-to-diff_referee2}(b) respectively.
We observe that the divide-and-conquer subspace also reduces the estimation bias according to the increase of $M$, which implies for both ways of graph division, the divide-and-conquer subspace reintroduces the effect of entanglement between the two divided separable parties to the expectation value.
We also observe that the unmitigated estimation biases and the mitigated plots in Fig.~\ref{fig:local-stochastic-pauli_dc_subspace-to-diff_referee2}(c) are more significant than those in Fig.~\ref{fig:local-stochastic-pauli_dc_subspace-to-diff_referee2}(d).
This is because the former division divides more edges and thus should take more entanglement into account classically.

\begin{figure}[htbp]
    \centering
    \subfloat[\label{fig:cost-to-diff_regularise_original_M_noise-free_2-bridge}]{
        \includegraphics[width=0.48\textwidth]{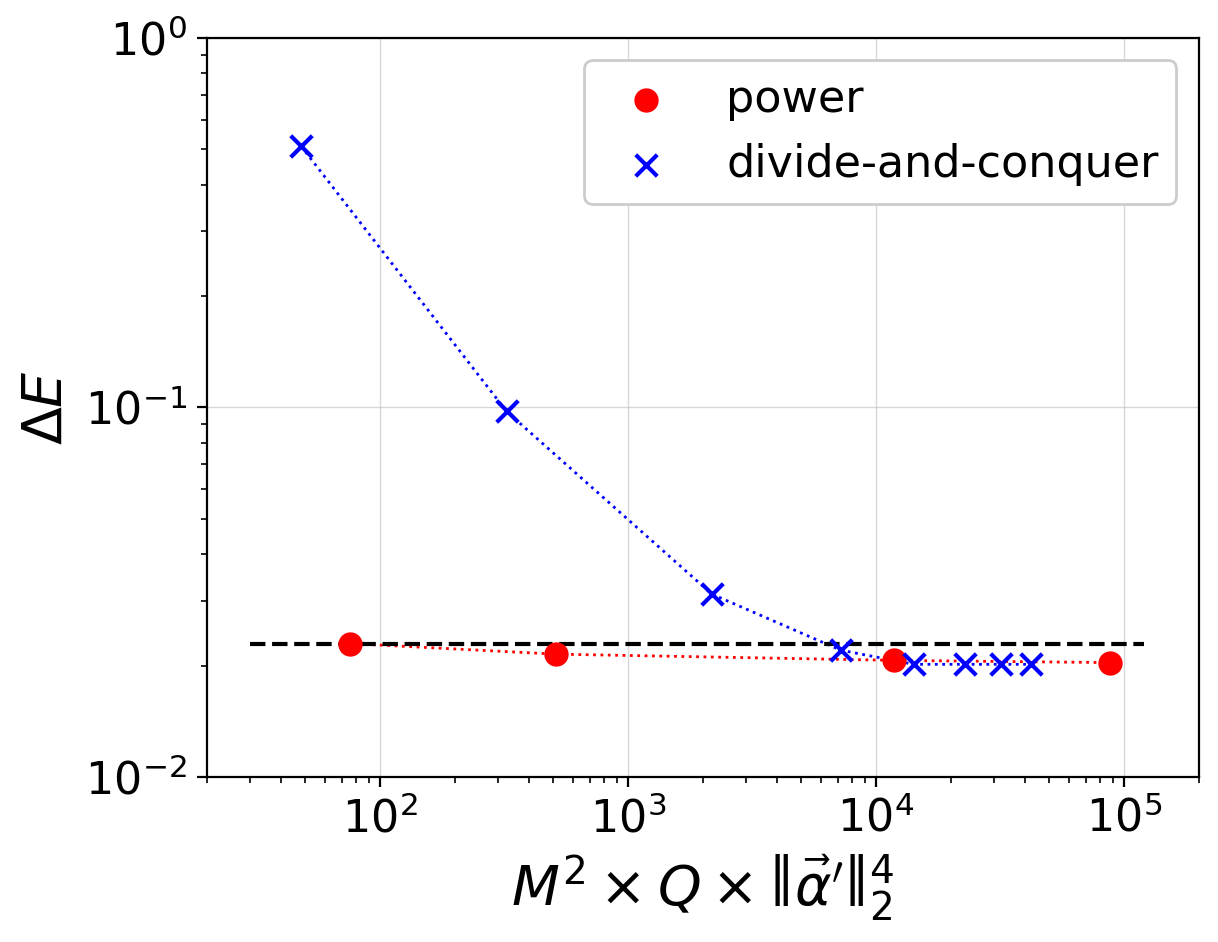}
    }
    \subfloat[\label{fig:cost-to-diff_regularise_original_M_noise-free_4-bridge}]{
        \includegraphics[width=0.48\textwidth]{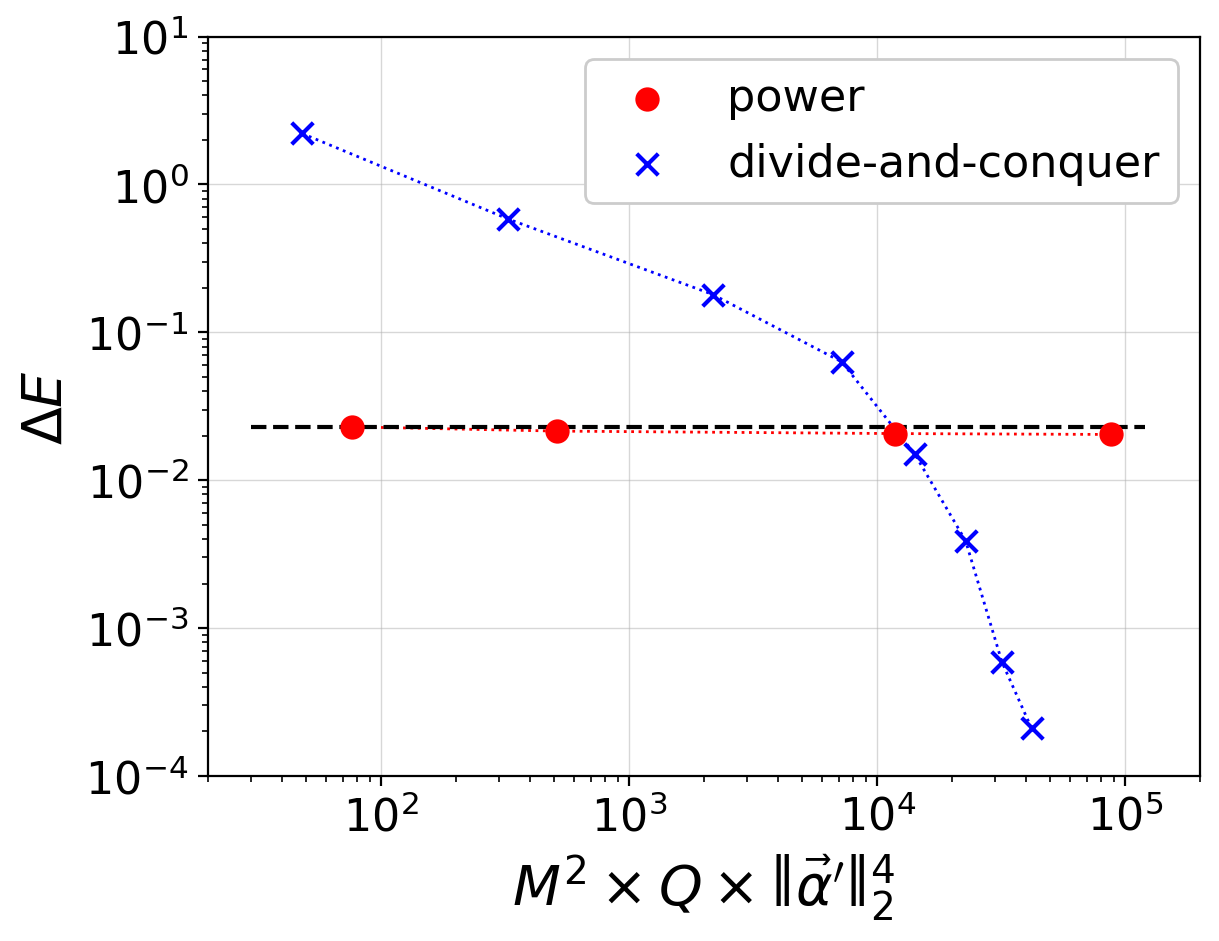}
    }
    \caption{
        The noise-free estimation bias $\Delta E$ from the theoretical ground state energy to the sampling overhead in the metric of $M^{2} Q \|\vec{\alpha}^{\prime}\|_{2}^{4}$ when increasing the number of subspaces $M$, $M=2,3,4,5$ for the power subspace and $M=2,3,\ldots,9$ for the divide-and-conquer subspace.
        Each figure (a) and (b) corresponds to each graph division in Fig.~\ref{fig:local-stochastic-pauli_dc_subspace-to-diff_referee2}(a) and Fig.~\ref{fig:local-stochastic-pauli_dc_subspace-to-diff_referee2}(b), respectively.
        In each figure, the horizontal dashed black line shows the noise-free 8-qubit full-size VQE simulation.
        The circle red markers and the dotted red polyline represent the estimation bias by the power subspace.
        The cross blue markers and the dotted blue polyline represent the estimation bias by the divide-and-conquer subspace.
    }
    \label{fig:cost-to-diff_2-bridge_4-bridge}
\end{figure}

We have also numerically investigated the sampling overhead of the divide-and-conquer subspace for these two ways of graph division.
Figure~\ref{fig:cost-to-diff_2-bridge_4-bridge} shows the estimation bias $\Delta E$ to the sampling overhead $M^{2} Q \|\vec{\alpha}^{\prime}\|_{2}^{4}$ for these two graph divisions in the same way as we have discussed for path graph structure of Ising Hamiltonian in Section~\ref{sec:Leveraging_mitigation_overheads}.
All results for both the power and divide-and-conquer subspace in Fig.~\ref{fig:cost-to-diff_2-bridge_4-bridge} surpass the estimation bias by the noise-free whole-system VQE ansatz with a sufficient number of subspaces.
From Fig.~\ref{fig:cost-to-diff_2-bridge_4-bridge}(a), we see $\Delta E$ by the graph division in Fig.~\ref{fig:local-stochastic-pauli_dc_subspace-to-diff_referee2}(a) decreases to the same level as that by the power subspace when $M\geq 5$.
From Fig.~\ref{fig:cost-to-diff_2-bridge_4-bridge}(b), we see $\Delta E$ by the graph division in Fig.~\ref{fig:local-stochastic-pauli_dc_subspace-to-diff_referee2}(b) further surpasses the estimation bias by the power subspace when $M\geq 6$ with much less sampling overhead.
Compared with Fig.~\ref{fig:cost-to-diff_2-bridge_4-bridge}(a) where the divide-and-conquer estimation is not going far beyond the VQE estimation, the increment of subspaces in Fig.~\ref{fig:cost-to-diff_2-bridge_4-bridge}(b) is effectively expanding the search space to efficiently improve the accuracy.

The result in Fig.~\ref{fig:cost-to-diff_2-bridge_4-bridge} supports the significance of the appropriate choice of subspace construction.
Particularly, the drastic decrease of estimation bias by the divide-and-conquer subspace for $M\geq 6$ in Fig.~\ref{fig:cost-to-diff_2-bridge_4-bridge}(b) is suggestive in the following two points.
First, it implies that this divide-and-conquer subspace construction is likely to cover the true ground state in its search space.
Second, it also implies that this subspace construction can enhance the stability of the generalized eigenvalue problem $\mathcal{H} \vec{\alpha} = E \mathcal{S} \vec{\alpha}$ to reach the solution closer to the true ground state energy.
The result above also gives a positive prospect to practically use the divide-and-conquer strategy to reduce the device-size overhead with modest sampling overhead despite the recent analysis on its limitation~\cite{harada2024doubly,harrow2024optimal,jing2024circuit}.
Meanwhile, the difference in the divide-and-conquer performance between Fig.~\ref{fig:cost-to-diff_2-bridge_4-bridge}(a) and Fig.~\ref{fig:cost-to-diff_2-bridge_4-bridge}(b) might also be caused by the instability and matrix regularization process of the generalized eigenvalue problem.

\subsection{On the different noise model applied to the ansatz: amplitude damping \label{sec:appendix_On_the_different_noise_model_applied_to_the_ansatz:_amplitude_damping}}

We also checked that our method works well when the given ansatz undergoes the amplitude damping noise, one of the typical non-unital noises.
The amplitude damping model for single-qubit gates with error rate $p_{1}$ is defined as follows:
\begin{equation} \label{eq:amplitude-damping_1qubit}
\begin{split}
    \mathcal{E}_{\mathrm{amp},p_{1}}^{(1)}\left(\rho\right) 
    = E_{0}\rho E_{0}^{\dagger} + E_{1}\rho E_{1}^{\dagger}, \quad
    \text{where} \quad 
    E_{0} 
    = \left[\begin{array}{cc}
        1 & 0 \\
        0 & \sqrt{1-p_{1}}
    \end{array}\right], \quad 
    E_{1} 
    = \left[\begin{array}{cc}
        0 & \sqrt{p_{1}} \\
        0 & 0
    \end{array}\right].
\end{split}
\end{equation}
For the noise level, we choose $p_{1} = 2.0 \times 10^{-6}, 2.0\times 10^{-5}, 2.0\times 10^{-4}$.
In two-qubit gate operations, the error channel is set as the tensor product of two local single-qubit error channels for each qubit with the error rate $p_{2}$.
\begin{equation} \label{eq:amplitude-damping_2qubit}
\begin{split}
    \mathcal{E}_{\mathrm{amp},p_{2}}^{(2)} = \mathcal{E}_{\mathrm{amp},p_{2}}^{(1)}\otimes \mathcal{E}_{\mathrm{amp},p_{2}}^{(1)}.
\end{split}
\end{equation}
Again, we set $p_{2} = 10p_{1}$ in our model.

\begin{figure}[htbp]
    \centering
    \subfloat[power subspace\label{fig:amplitude-damping_power-subspace_subspace-to-diff_2-bridges}]{
        \includegraphics[width=0.32\textwidth]{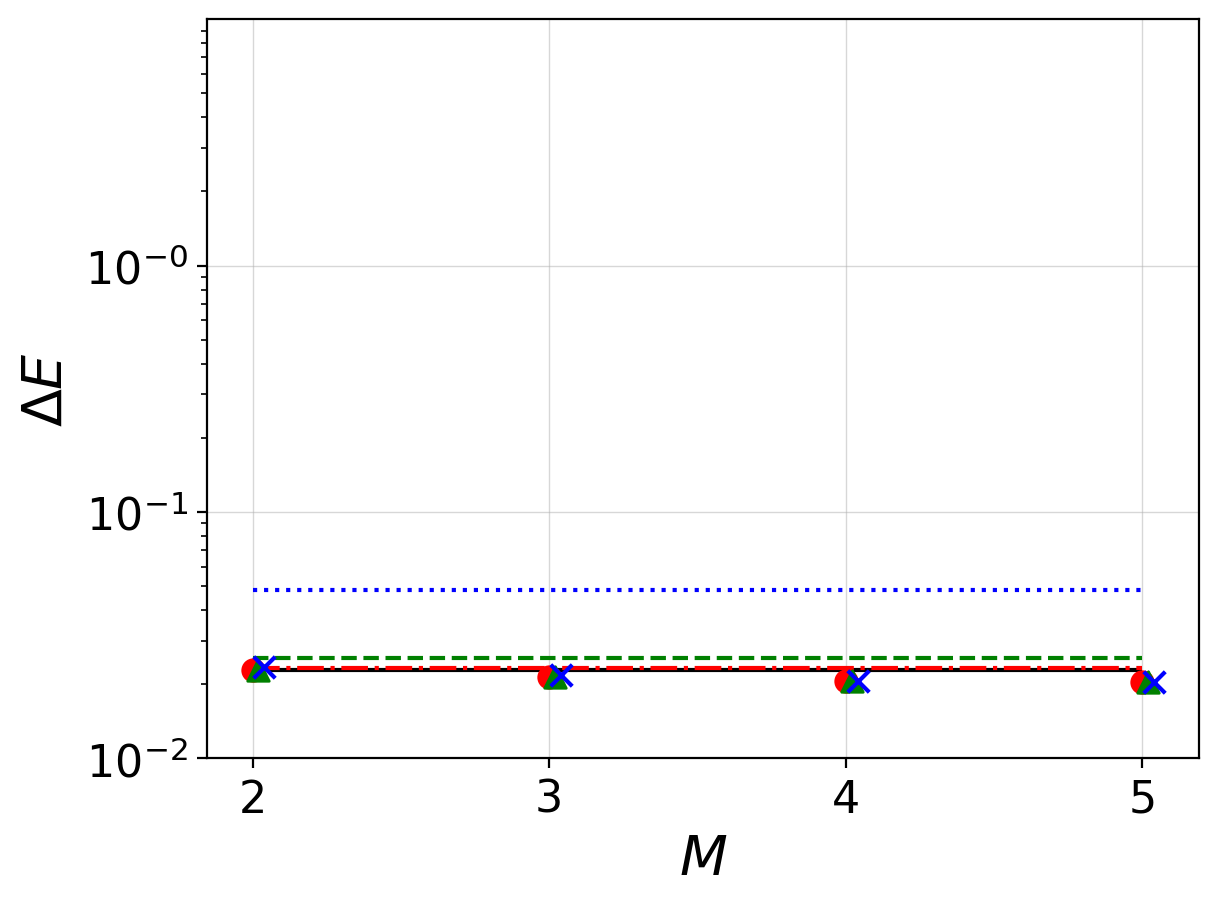}
    }
    \subfloat[fault subspace\label{fig:amplitude-damping_fault-subspace_subspace-to-diff_2-bridges}]{
        \includegraphics[width=0.32\textwidth]{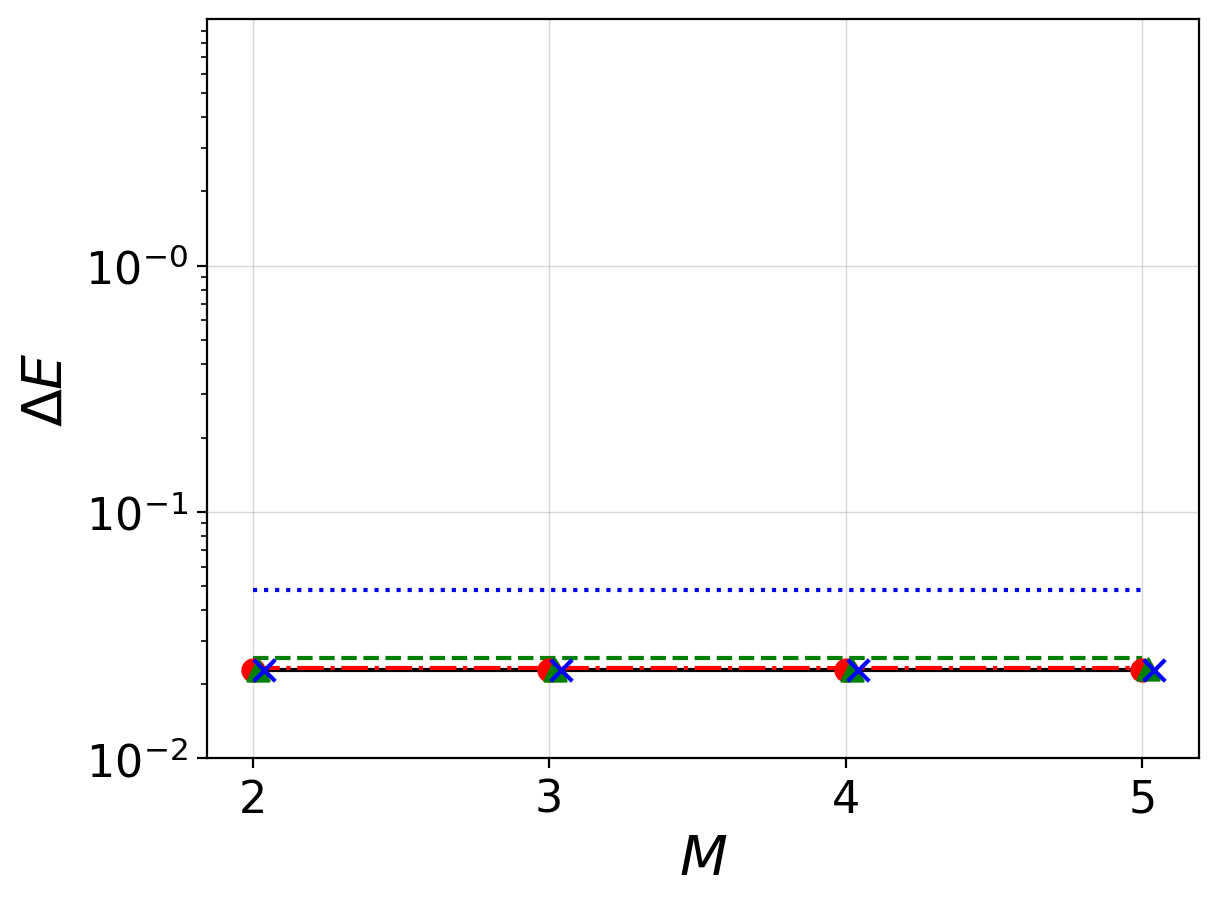}
    }
    \subfloat[divide-and-conquer subspace\label{fig:amplitude-damping_cfe-subspace_subspace-to-diff_2-bridges}]{
        \includegraphics[width=0.32\textwidth]{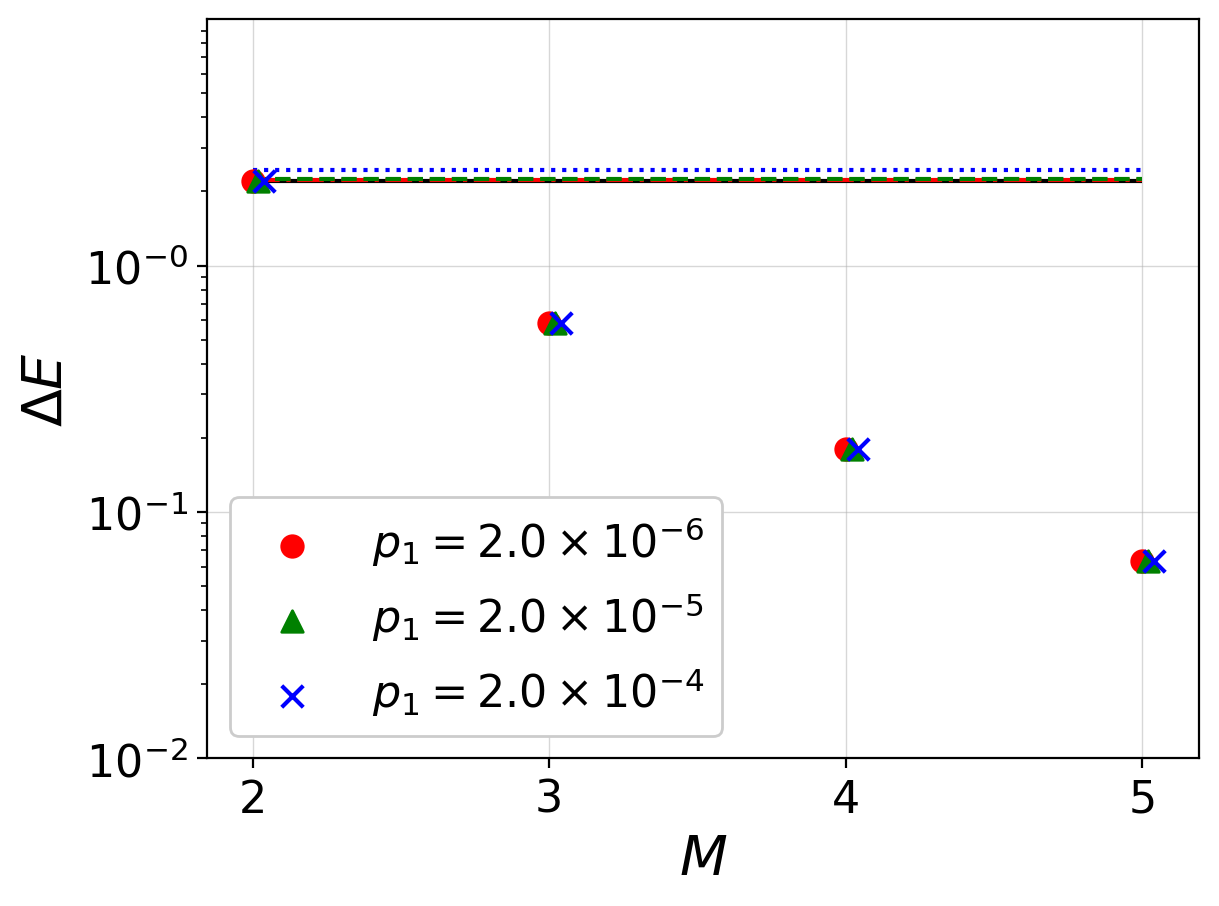}
    }
    \caption{
        The estimation bias $\Delta E$ from the theoretical ground state energy under the amplitude damping noise defined by Eq.~\eqref{eq:amplitude-damping_1qubit} and Eq.~\eqref{eq:amplitude-damping_2qubit}.
        The notations are the same as Fig.~\ref{fig:local-stochastic-pauli_power-fault_subspace-to-diff_referee2} while the result in (c) for the divided-and-conquer subspace is obtained with subsystems divided as Fig.~\ref{fig:local-stochastic-pauli_dc_subspace-to-diff_referee2}(a).
    }
    \label{fig:amplitude-damping_subspace-to-diff_2-bridges}
\end{figure}

The result is shown in Fig.~\ref{fig:amplitude-damping_subspace-to-diff_2-bridges}.
We observe that the estimation bias by the power subspace slightly surpasses the noise-free simulation and decreases according to the number of subspaces.
The estimation bias by the fault subspace also overlaps with the bias by the noise-free simulation, showing the same mitigation performance as the local stochastic noise.
Finally, the divide-and-conquer subspace seems to reduce the estimation bias exponentially to the number of subspaces.
These results imply that Dual-GSE also works well for different noise models.

\subsection{On the different noise model applied to the ansatz: coherent drift \label{sec:appendix_On_the_different_noise_model_applied_to_the_ansatz:_coherernt_drift}}

In addition, we also performed additional numerical simulations by taking the noisy ansatz under drifted rotation angles in $Rx$, $Rz$, and $CZ$ gates.
We randomly sampled the drift angles from $[0,p_{1}]$ for $Rx$ and $Rz$ gates, and from $[0,p_{2}]$ for each qubit of in $CZ$ gates with $p_{2}=10p_{1}$.
The setting of noise level $p_{1}$ is taken among $\{2.0\times 10^{-4}, 2.0\times 10^{-3}, 2.0\times 10^{-2}\}$.
The drifting gates are added right after the noise-free quantum gates are applied.

\begin{figure}[htbp]
    \centering
    \subfloat[power subspace\label{fig:coherent_power-subspace_subspace-to-diff}]{
        \includegraphics[width=0.32\textwidth]{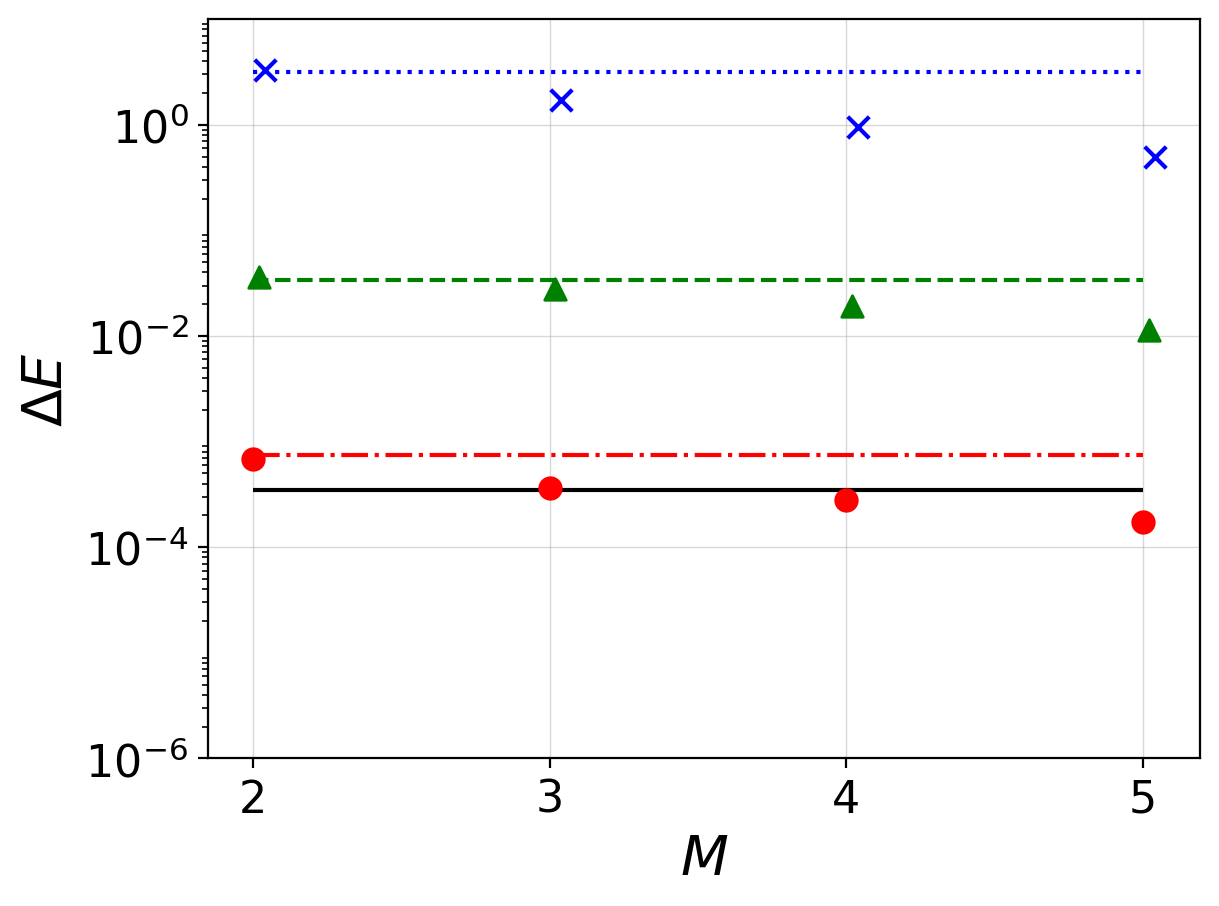}
    }
    \subfloat[fault subspace\label{fig:coherent_fault-subspace_subspace-to-diff}]{
        \includegraphics[width=0.32\textwidth]{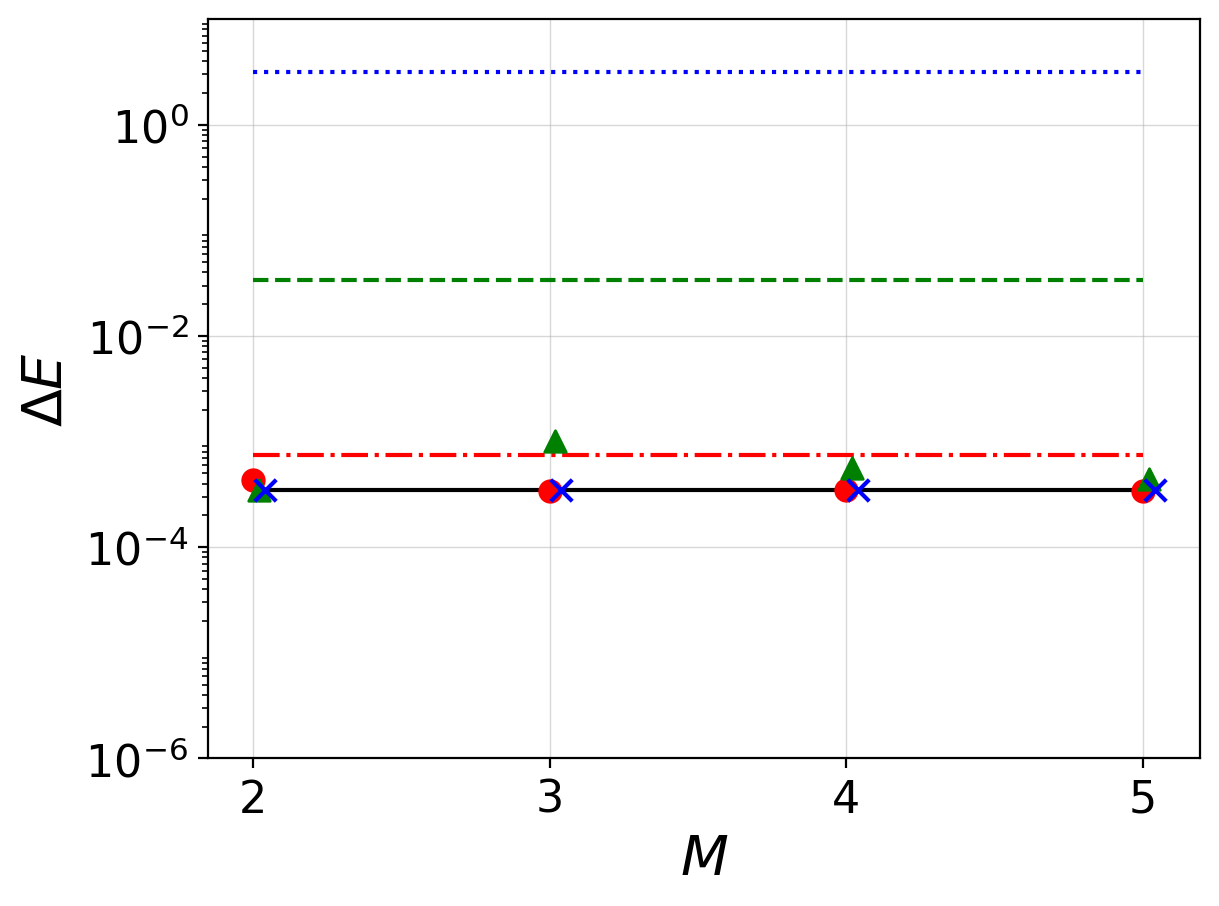}
    }
    \subfloat[divide-and-conquer subspace\label{fig:coherent_cfe-subspace_subspace-to-diff}]{
        \includegraphics[width=0.32\textwidth]{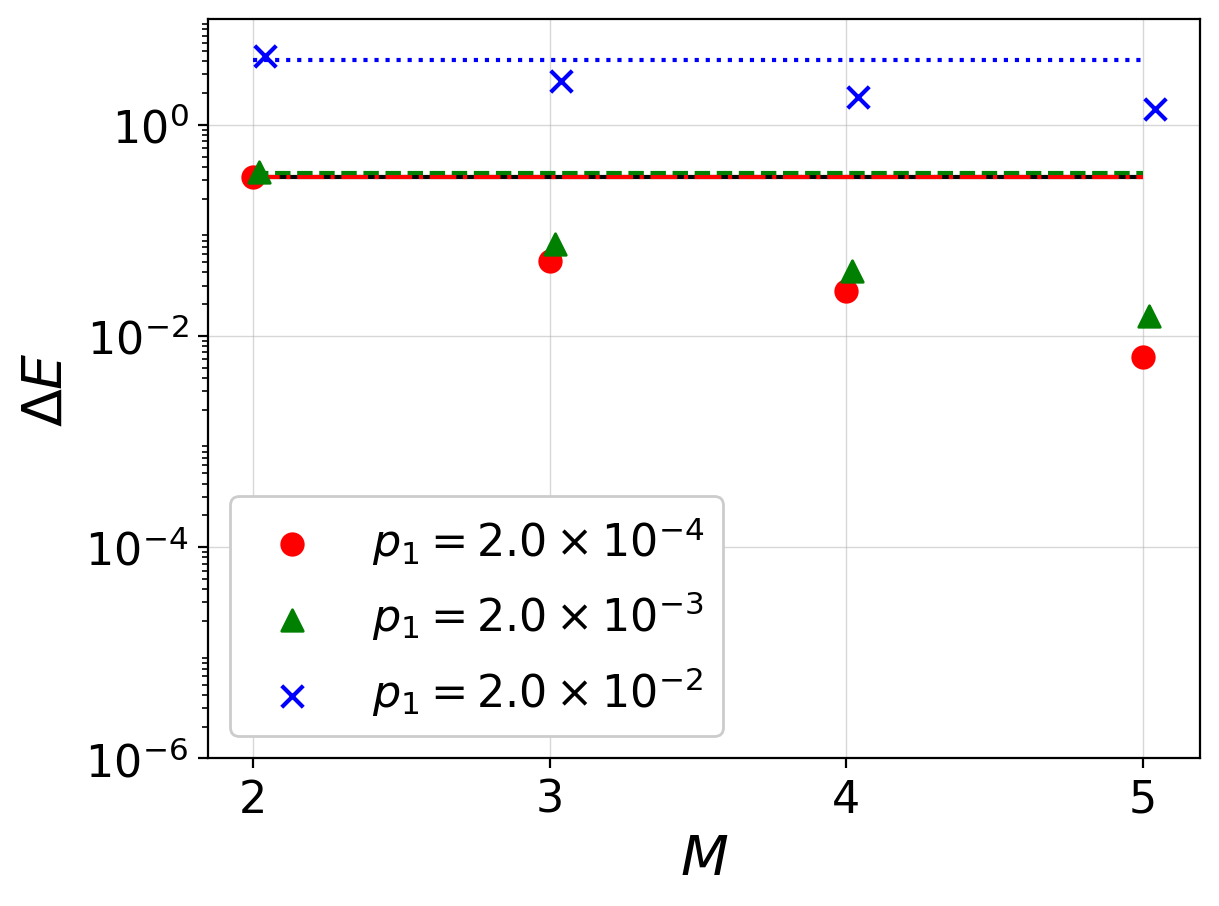}
    }
    \caption{
        The estimation bias $\Delta E$ from the theoretical ground state energy under the coherent error as the inaccurate gate angles.
        The subsystems used in the divided-and-conquer subspace are based on Fig.~\ref{fig:path_graphs}.
        The horizontal black line denotes the inherent estimation bias under the noise-free optimization of variational quantum circuits using the ansatz (a), (b): over the whole eight-qubit system and (c): over the tensor product of two four-qubit subsystems.
        In every figure, the cross blue markers and the dotted blue line represent the estimation bias with and without Dual-GSE under the coherent error with magnitude $2.0\times 10^{-2}$, the triangle green markers and the dashed green line with magnitude $2.0\times 10^{-3}$, and the circle red markers and dash-dot red line with magnitude $2.0\times 10^{-4}$.
    }
    \label{fig:coherent_subspace-to-diff}
\end{figure}

The estimation bias by each subspace is shown in Fig.~\ref{fig:coherent_subspace-to-diff}.
For both the power subspace and the divide-and-conquer subspace, we see a stable decrease in estimation bias according to the number of subspaces $M$.
We also see that the estimation bias by the fault subspace gets very close to the noise-free simulation.
These features are common with those in the mitigation results of ansatz under stochastic noise.
Thus, we can also state that our method can effectively mitigate coherent errors, particularly for both algorithmic errors and drifted rotation angles.

\subsection{On different noise levels applied to the ansatz \label{sec:appendix_On_different_noise_levels_applied_to_the_ansatz}}

We here highlight that the noise level $p_{1}$ we set is for the error rate in single-qubit gates.
For the two-qubit gates, as we have written in Section~\ref{sec:Simulation_results_without_shot_noise}, the error rate is set to $10p_{1}$, i.e. $10$ times larger than that of single-qubit gates.
In our numerical simulations, we set $p_2 = 10 p_{1}$ with $p_{1}=2.0 \times 10^{-6}, 2.0\times 10^{-5}, 2.0\times 10^{-4}$, which is equivalent to adding $0.0704$, $0.704$, and $7.04$ errors to one quantum circuit with eight qubits and eight layers respectively.
The error rate $10^{-2}\sim10^{-3}$ of two-qubit gates well reflects the error rates of the current quantum devices~\cite{Yang2021, Yang2022, mooney2024characterization}, and the error rates $10^{-4}\sim10^{-6}$ in our setting would be likely to be witnessed in the future.

\begin{figure}[htbp]
    \centering
    \subfloat[power subspace\label{fig:local-stochastic-pauli_power-subspace_subspace-to-diff_2x1e-2}]{
        \includegraphics[width=0.32\textwidth]{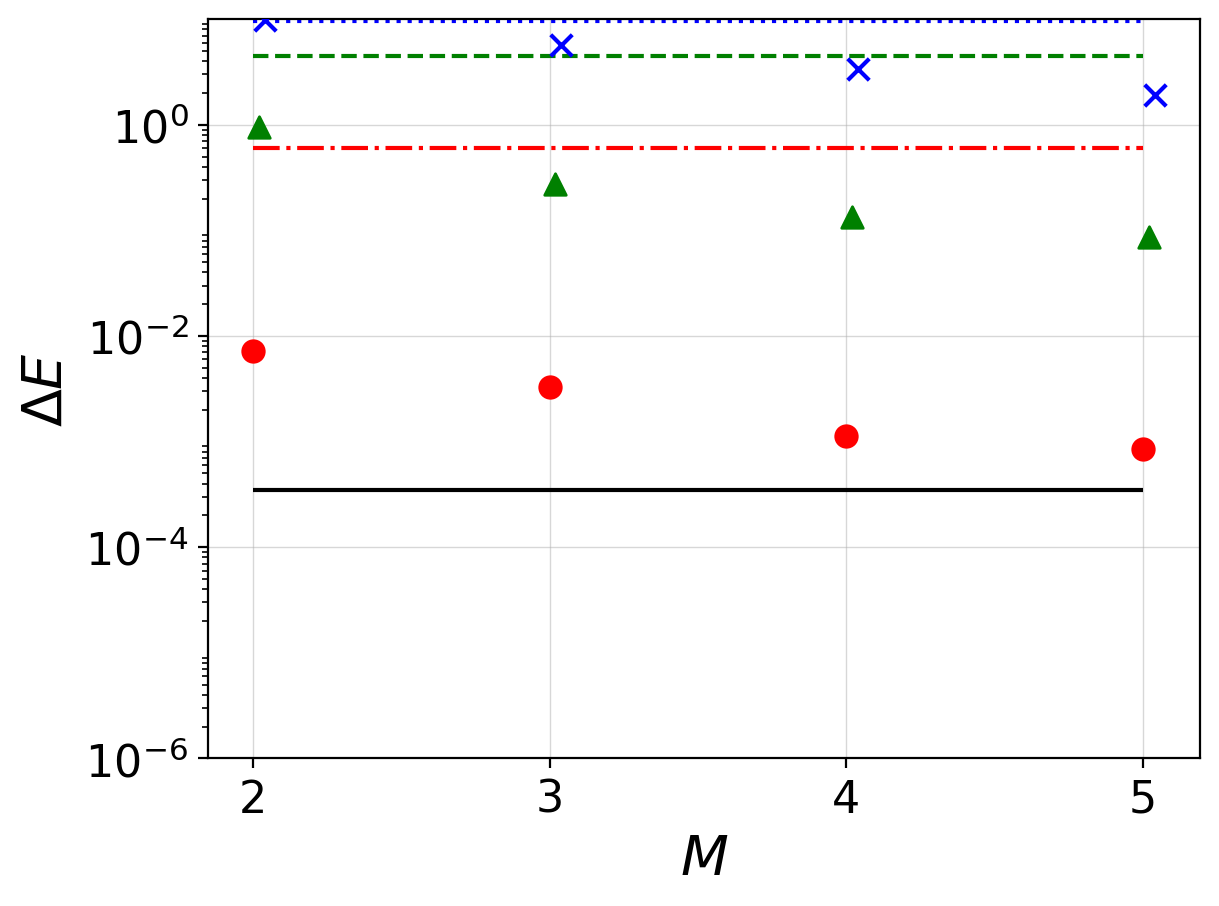}
    }
    \subfloat[fault subspace\label{fig:local-stochastic-pauli_fault-subspace_subspace-to-diff_2x1e-2}]{
        \includegraphics[width=0.32\textwidth]{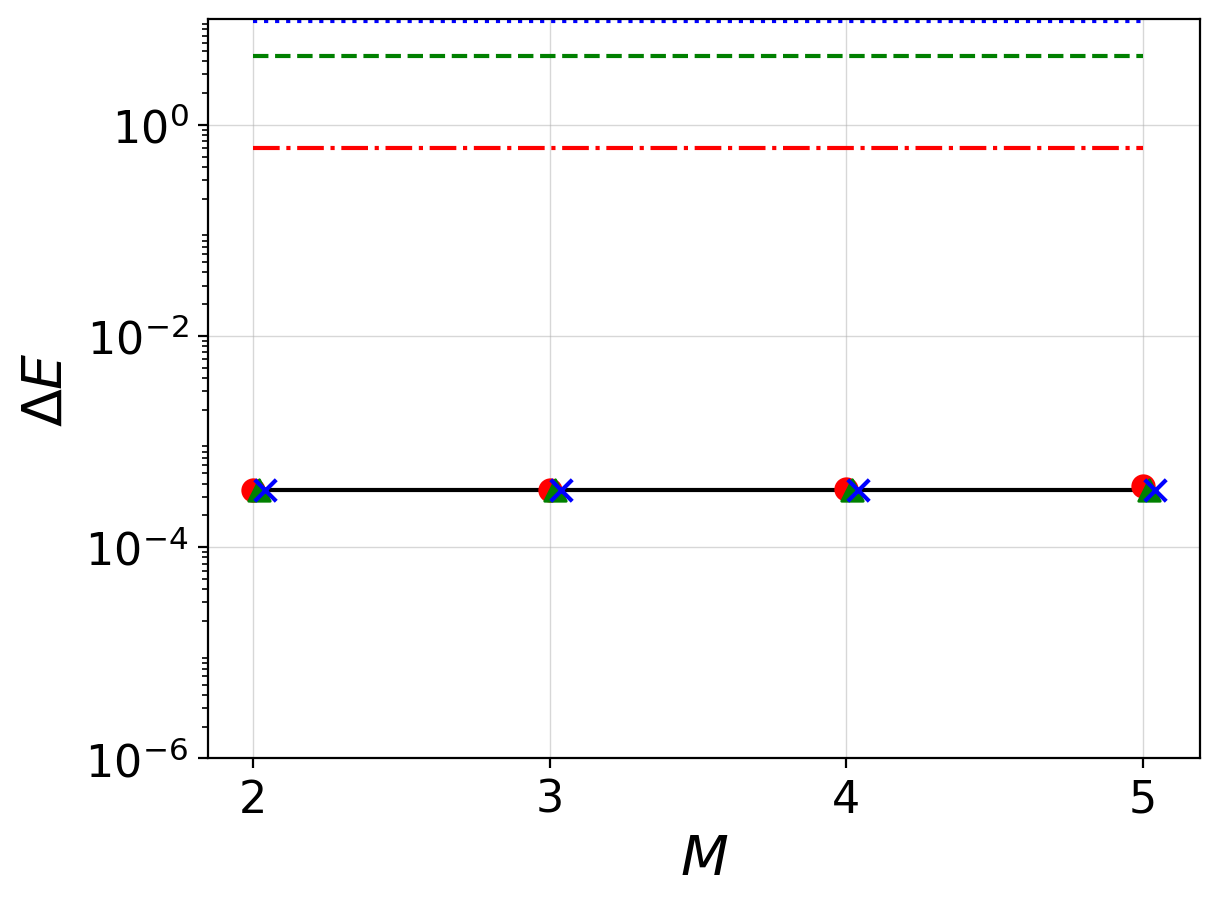}
    }
    \subfloat[divide-and-conquer subspace\label{fig:local-stochastic-pauli_cfe-subspace_subspace-to-diff_2x1e-2}]{
        \includegraphics[width=0.32\textwidth]{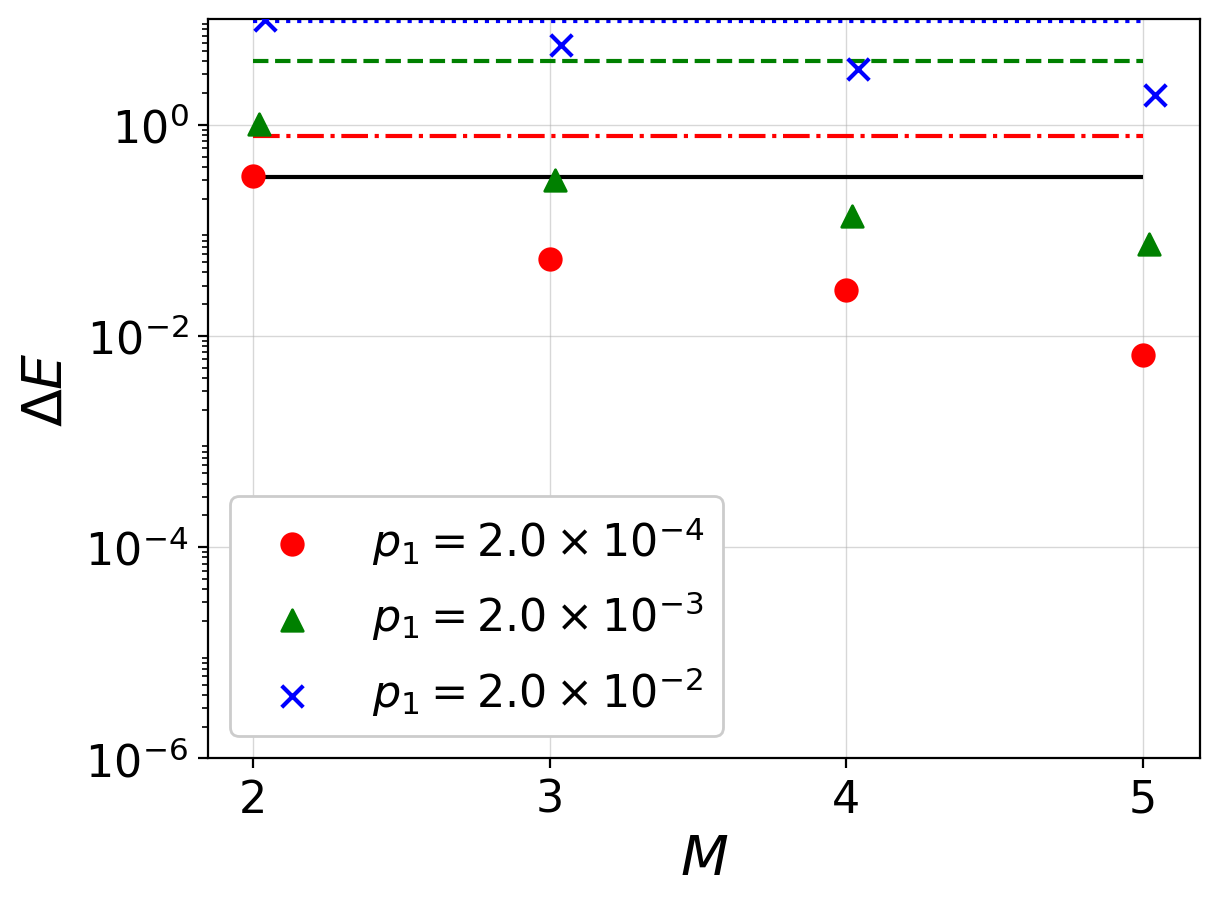}
    }
    \caption{
        The estimation bias $\Delta E$ from the theoretical ground state energy under the local stochastic Pauli error.
        The horizontal black line denotes the inherent estimation bias under the noise-free optimization of variational quantum circuits using the ansatz (a), (b): over the whole eight-qubit system and (c): over the tensor product of two four-qubit subsystems based on Fig.~\ref{fig:path_graphs}.
        In every figure, the cross blue markers and the dotted blue line represent the estimation bias with and without Dual-GSE under the local stochastic noise with magnitude $2.0\times 10^{-2}$, the triangle green markers and the dashed green line with magnitude $2.0\times 10^{-3}$, and the circle red markers and dash-dot red line with magnitude $2.0\times 10^{-4}$.
    }
    \label{fig:local-stochastic-pauli_subspace-to-diff_2x1e-2}
\end{figure}

Here, we see that our method still works with ansatz under higher noise rates.
We checked the estimation bias for the 8-qubit 1D Ising model, the same setting as used in Section~\ref{sec:Simulation_results_without_shot_noise}, when the noise level of single-qubit error is $p_{1}=2.0 \times 10^{-4}, 2.0\times 10^{-3}, 2.0\times 10^{-2}$.
The simulation results are shown in Fig.~\ref{fig:local-stochastic-pauli_subspace-to-diff_2x1e-2}.
Notably, the fault subspace still returns the estimation bias close to the noise-free simulation, regardless of the number of subspaces.
For both the power subspace and the divide-and-conquer subspace, the estimation bias is suppressed according to the number of subspaces.
Focusing on the divide-and-conquer subspace, the estimation bias for the noise level $p_{1} = 2.0\times 10^{-3}$ surpasses the noise-free bias with separable ansatz from $M=3$, meaning that the effect of entanglement between subsystems is still reintroduced with a small number of subspaces.

Again, since the current quantum devices have already reached the two-qubit gate error rate around or less than $10^{-2}$, the noise assumption $p_{1} = 2.0\times 10^{-3}$ (i.e. $p_{2} = 2.0\times 10^{-2}$) seems to be the noise level large enough for numerical simulation.
In the near future, it will be likely to obtain quantum devices with two-qubit error rates much less than $10^{-2}$, where our noise assumption of $p_{1}=2.0 \times 10^{-k}$ for $k>3$ would better fit the device noise.
Overall, both numerical simulation and the future prospects of real devices support the practicality of our method.

\twocolumngrid

\bibliography{main}

\end{document}